%% file: main.tex
\documentclass[%
superscriptaddress,
 amsmath,amssymb,
 aps,
]{revtex4-2}

\usepackage{graphicx}
\usepackage{xcolor}
\usepackage{ulem}
\usepackage{dcolumn}
\usepackage{bm}
\usepackage[mathlines]{lineno}
\usepackage{hyperref}

\begin{document}

\preprint{}

\title{EIC Physics from An All-Silicon Tracking Detector}

\author{John Arrington}
\affiliation{Lawrence Berkeley National Laboratory, Berkeley, CA 94720, USA}

\author{Reynier Cruz-Torres}
\affiliation{Lawrence Berkeley National Laboratory, Berkeley, CA 94720, USA}

\author{Winston DeGraw}
\affiliation{University of California, Berkeley, Berkeley CA 94720, USA}

\author{Xin Dong}
\affiliation{Lawrence Berkeley National Laboratory, Berkeley, CA 94720, USA}

\author{Leo Greiner}
\affiliation{Lawrence Berkeley National Laboratory, Berkeley, CA 94720, USA}

\author{Samuel Heppelmann}
\affiliation{University of California, Davis, Davis CA 95616, USA} 
\affiliation{Lawrence Berkeley National Laboratory, Berkeley, CA 94720, USA}

\author{Barbara Jacak}
\affiliation{Lawrence Berkeley National Laboratory, Berkeley, CA 94720, USA}
\affiliation{University of California, Berkeley, Berkeley CA 94720, USA}

\author{Yuanjing Ji}
\affiliation{Lawrence Berkeley National Laboratory, Berkeley, CA 94720, USA}

\author{Matthew Kelsey}
\affiliation{Wayne State University, Detroit, MI 48202, USA}
\affiliation{Lawrence Berkeley National Laboratory, Berkeley, CA 94720, USA}

\author{Spencer R. Klein}
\affiliation{Lawrence Berkeley National Laboratory, Berkeley, CA 94720, USA}

\author{Yue Shi Lai}
\affiliation{Lawrence Berkeley National Laboratory, Berkeley, CA 94720, USA}

\author{Grazyna Odyniec}
\affiliation{Lawrence Berkeley National Laboratory, Berkeley, CA 94720, USA}

\author{Sooraj Radhakrishnan}
\affiliation{Kent State University, Kent, OH 44242, USA}
\affiliation{Lawrence Berkeley National Laboratory, Berkeley, CA 94720, USA}

\author{Ernst Sichtermann}
\affiliation{Lawrence Berkeley National Laboratory, Berkeley, CA 94720, USA}

\author{Youqi Song}
\affiliation{University of California, Berkeley, Berkeley CA 94720, USA}

\author{Fernando Torales Acosta}
\affiliation{University of California, Berkeley, Berkeley CA 94720, USA}

\author{Lei Xia}
\affiliation{University of Science and Technology of China, Hefei, Anhui Province 230026, China}

\author{Nu Xu}
\affiliation{Lawrence Berkeley National Laboratory, Berkeley, CA 94720, USA}

\author{Feng Yuan}
\affiliation{Lawrence Berkeley National Laboratory, Berkeley, CA 94720, USA}

\author{Yuxiang Zhao}
\affiliation{Institute of Modern Physics, Chinese Academy of Sciences, Lanzhou, Gansu Province 730000, China}

\date{\today}


\begin{abstract}

The proposed electron-ion collider has a rich physics program to study the internal structure of protons and heavy nuclei. This program will impose strict requirements on detector design. This paper explores how these requirements can be satisfied using an all-silicon tracking detector, by consideration of three representative probes: heavy flavor hadrons, jets, and exclusive vector mesons.
\end{abstract}

\maketitle
\vskip -0.2 in
\tableofcontents

\input{intro}

\input{all_si_tracker}


\section{Physics Studies and Performance Requirements}

\subsection{Heavy Quarks}\label{sec:heavy-quarks}
\input{charm_intro_recon}
\input{charm_physics}

\input{jets}

\subsection{Exclusive Vector Mesons}

\subsubsection{Physics Introduction}

Exclusive production of vector mesons is an important channel for imaging light and heavy nuclei. The overall production cross section is directly sensitive to the gluon density in the target. The Good-Walker paradigm relates the cross sections for coherent and incoherent photoproduction to the spatial distribution of gluons in the nucleus and to event-by-event fluctuations in the nuclear configuration respectively.  The $Q^2$ evolution of these cross sections can provide a detailed picture of the nuclei over a range of length scales.   

Exclusive vector-meson production occurs when an incident photon fluctuates to a $q\overline q$ pair which then scatters elastically from the nuclear target, emerging as a real vector meson. Vector mesons are color singlets, so the scattering must involve at least two gluons and is usually described in terms of Pomeron exchange. The Pomeron has the same quantum numbers as the vacuum, so this scattering is elastic. At lower photon energies, Reggeon exchange may also contribute; Reggeons represent meson trajectories, so are mostly quarks, so can transfer a wider range of quantum numbers, including charge.

The exact meaning of exclusivity will depend on the analysis under consideration.   In incoherent photoproduction, the breakup of the nuclear target produces additional particles.  These are generally in the far-forward region, so should not cause confusion.  There can also be parton radiation from the photon before it interacts with the Pomeron; this is the resolved component of the photon.  Because of the constraints of spin and color neutrality, this involves the gluonic component of the photon; the radiation is smaller for vector meson final states than for other processes such as jets and open heavy flavor production \cite{Xu:1999qy}.  Good detector acceptance is required to be able to separate these non-exclusive resolved processes from direct production. 
  
For the EIC, the greatest interest is to use high-energy photoproduction via Pomeron exchange to probe the gluon distributions in nuclear targets. Because the Pomeron involves the exchange of at least two gluons, the relationship between cross section and the spatial dependence of the gluons in a nucleus is determined in a manner similar to that used to probe GPDs in a proton~\cite{Toll:2012mb,Adamczyk:2017vfu,Klein:2019qfb,Klein:2018grn}. The two-dimensional Fourier transform of $d\sigma/dt$, where $t$ is the squared four-momentum transfer from the target, gives the two-dimensional (transverse to the photon direction) density of interaction sites within the nucleus in the infinite-momentum frame:
\begin{equation}
    \frac{d\sigma}{dt} \propto \int_0^{t_{\rm max}} p_T dp_T J_0(bp_T) \sqrt{\frac{d\sigma_{\rm coh}}{dt}},
\end{equation}
where $b$ is the impact parameter of the struck parton within the nucleus, and $J_0$ is a Bessel function. 

The incoherent cross section on proton and nuclear targets is sensitive to event-by-event fluctuations in the nuclear target, due to variations in the nucleon positions (for $A>1$), and to the presence of gluonic hot spots~\cite{Mantysaari:2016ykx,Klein:2019qfb}.  For a given model of event-by-event fluctuations, one can predict $d\sigma/dt$ for incoherent production. Fluctuations at small distance scales are related to the cross section at large $|t|$.   The incoherent cross section should also evolve with collision energy~\cite{Cepila:2018zky}.  At low energies, the incoherent cross section should rise with increasing collision energies.  However, in the ultra-high energy limit, the nucleus will look like a black disk, with no event-by-event fluctuations.   So, as the collision energy rises, the incoherent cross section should reach a maximum and then decrease~\cite{Cepila:2018zky,Cepila:2019dzw}.   The energy at which the cross section is a maximum depends on the vector-meson mass. For the $\rho$, it might be within the range of the EIC. It would manifest itself as a rapidity-dependent variation of the ratio of the incoherent to the coherent cross section.   Although the effects of gluonic hot spots may be somewhat washed out in $e$+$A$ collisions, they may still be visible at larger $|t|$, or in lighter nuclei.

The main kinematics variables for exclusive vector-meson production are the $x$ and $Q^2$ of the struck gluons.  The kinematics is dominated by the case where the two gluons have very different $x$ values~\cite{Flett:2019pux,Klein:2020fmr}
and the large-$x$ gluon dominates the momentum transfer; the other gluon is often treated as a spectator.  Then, the Bjorken-$x$ can be determined from the rapidity $y$ of the vector meson.  For photoproduction~\cite{Klein:1999qj}, away from threshold,
\begin{equation}
    x=\frac{M_V}{2\gamma m_p}\exp{(-y)},
    \label{eq:rapiditytox}
\end{equation}
where $M_V$ is the vector-meson mass, $m_p$ is the proton mass, and $\gamma$ is the Lorentz boost of the ion.   Large $Q^2$ will shift these reactions slightly~\cite{Lomnitz:2018juf}.   The reaction $Q^2$ comes from the hard scales determined by the photon $Q^2$ and the vector meson mass:
\begin{equation}
Q^2 = (Q_\gamma^2 + M_V^2)/4.
\end{equation}
To study production at low and moderate $Q^2$, where phenomena like the colored-glass condensate are most visible, lighter mesons are required. 

Equation~\ref{eq:rapiditytox} highlights a key requirement for a tracking detector at an EIC: wide rapidity coverage, to cover a wide range in $x$, up to the kinematic limits.  Eq.~\ref{eq:rapiditytox} fails near threshold, i.e. as $x\rightarrow 1$, because it neglects the proton mass, but more detailed calculations show that, for 18 GeV electrons on 275 GeV protons, the high-$x$ limit is near $y=4$. For coherent photoproduction on ions, the maximum $x$ is about 0.03, since coherence must be maintained over a coherence distance $l_c=2k/(M_V^2+Q^2)$ which is larger than the nuclear diameter.  This limits coherent production to $y<2$, while incoherent production extends up to $y\approx 3$; the maximum is lower than for protons because of the lower per-nucleon ion energy. For ions, Fermi momentum leads to incoherent production with $x>1$. This is an active area of study at Jefferson lab~\cite{Ye:2018jth}.  The large $x$/ large $|y|$ region is also critical for studying meson production via Reggeon exchange \cite{Klein:2019avl}; Reggeon exchange allows for a much wider range of final states (including exotica) than photon-Pomeron fusion. 

The relationship between $x$ and $y$ depends on the ion-beam energy, so it is possible to shift the rapidity of a desired $x$ value by changing the ion energy.  This may be of value for studying near-threshold photoproduction, where the low photon energies correspond to forward production.  This is important for studying production via 3-gluon exchange, and for searches for pentaquarks or similar exotica~\cite{Klein:2019avl}. 

The minimum $x$ corresponds to the maximum photon energy, which we will take to be the electron beam energy, 18 GeV, even though the photon flux is drops rapidly near threshold.  For both $e$+$p$ and $e$+$A$ collisions this appears near $y=-4$.  Because of the different per-nucleon beam energies, this corresponds to $x$ values of about $6\times 10^{-5}$ and $10^{-4}$ respectively (at $Q^2\approx M_V^2/4$).  For electroproduction, with $Q^2>1$ GeV$^2$, the rapidity values are slightly shifted. 

Other detector requirements are more channel specific.  To address them, we use events simulated using the eSTARlight Monte Carlo generator~\cite{Lomnitz:2018juf}.  eSTARlight simulates the production of a variety of vector mesons.  It uses $e$+$p$ cross sections based on parametrized HERA and fixed-target data, including their $Q^2$ dependence.  $e$+$A$ cross sections are calculated using the $e$+$p$ cross sections and a quantum Glauber calculation.  The angular distributions of the decays are also based on HERA data, with $s$-channel helicity conservation holding for photoproduction, but with a rising fraction of longitudinally-polarized photons as the $Q^2$ rises.

The detector simulations were done for the detector described above, using a GEANT3 model, using the EICROOT simulation framework.  Simulations were done with uniform magnetic fields of 1.5 and 3.0~T.   

We consider four mesons: the $\rho$, $\phi$, the $J/\psi$ and the $\Upsilon$ family. Each illustrates different aspects of the detector requirements.  The $\psi'$ has also been studied, but present no specific problems, so we skip them here.  These all produce simple final states, with two leptons or charged mesons from the vector-meson decay, plus the scattered electron and the scattered and/or dissociated nuclear target.  An intact target is only marginally scattered, so, except for very light nuclei is not detectable.   If the target dissociates, it may leave remnants which will be visible in the far-forward detectors.  Observing these remnants is critical for determining if an event was coherent or incoherent photoproduction.  
These remnants typically have rapidity near the beam rapidity, so will be studied with a set of far-forward detectors that observe charged and neutral particles with rapidity above 5 \cite{Parker:2020xxt}. 

\subsubsection{$\rho$ production}

The $\rho$ is the lightest vector meson, so it allows studies down to the lowest possible $x$ values, albeit at low $Q^2$, and at a cost of a somewhat more complicated wave function than the $\phi$.  However, it is experimentally much simpler, and will be of value in regions where the $\phi$ may not be detectable. 

\begin{figure}[htb]
\centering
\includegraphics*[width=0.75\textwidth]{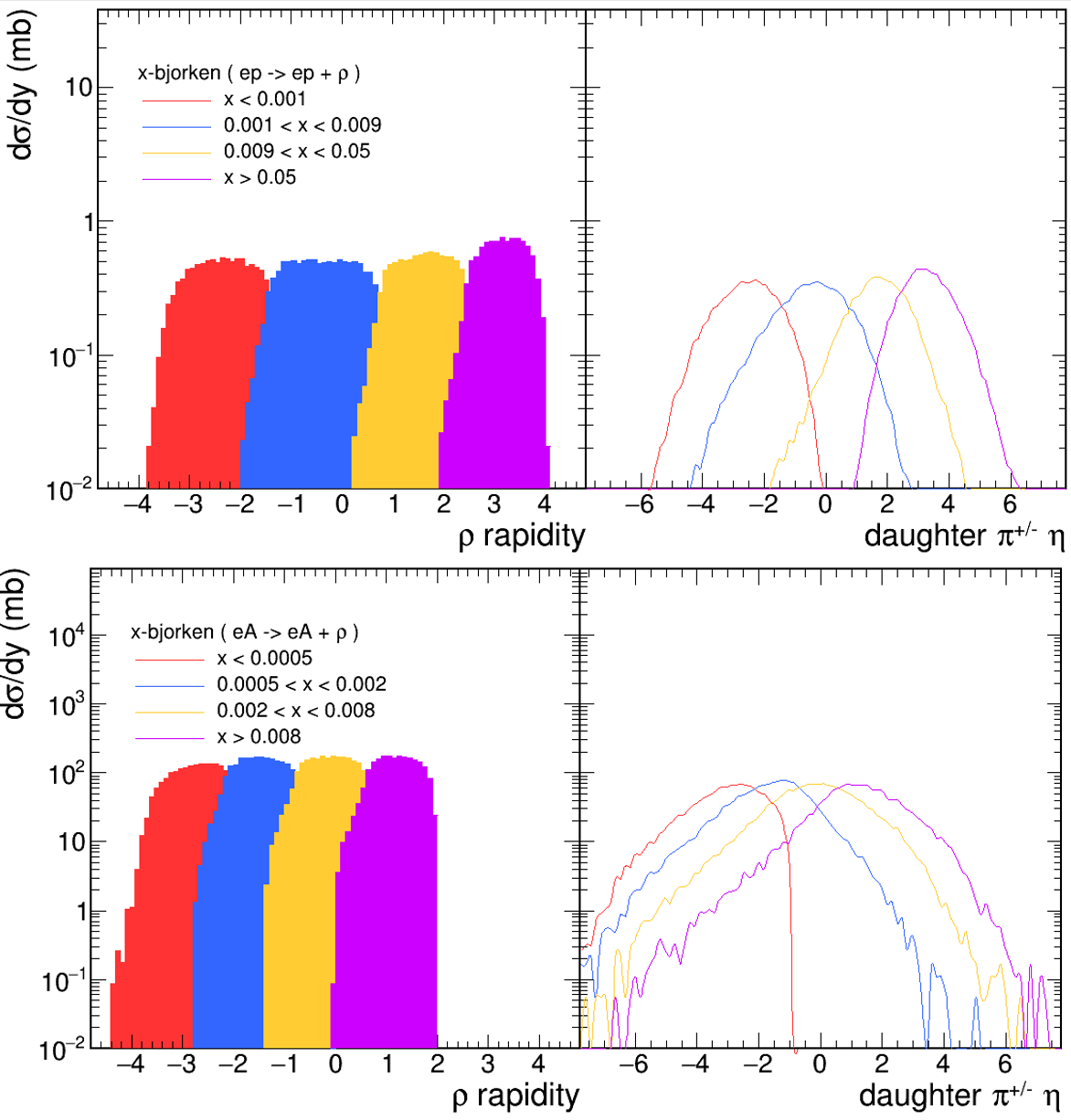}%
\caption{(left) Rapidity distributions for $\rho$ photoproduction at an EIC and (right) pseudorapidity distributions of the daughter $\pi^\pm$ from $\rho$ decay for four different $x_B$ ranges; in all cases, the highest range is at the left.  The top two panels are for collisions of 18 GeV electrons on 275 GeV protons, while the bottom are for 18 GeV electrons on 100 GeV/n Au nuclei.  In both plots, the distributions are broken up based on the $x_B$ value of the struck parton, with the largest $x_B$ on the right sides of the plots. 
}
\label{fig:rhodsdy}
\end{figure}

The left panels of Fig.~\ref{fig:rhodsdy} shows the rapidity distribution for $\rho$ photoproduction in $e$+$A$ and $e$+$p$ collisions.  The distribution is broken up to show subsamples with different $x$ values, clearly showing the relationship between rapidity and $x$.  $\rho$ photoproduction in $e$+$A$ occurs from rapidity -4 to 2, while the $e$+$p$ distribution extends from rapidity -4 to 4.

The right panels show the pseudorapidity distributions of the daughter pions for the same subsamples in $x$.  Most of the daughter pions are distributed within 1 unit in pseudorapidity of the rapidity of the parent $\rho$.  As a rough rule of thumb the acceptance in pseudorapidity should be one unit wider than the rapidity of the produced $\rho$.  This argues for a pseudorapidity coverage that extends to $\pm 5$.  This extended coverage is also important in measurements of the $J/\psi$ polarization, needed to separate the transverse and longitudinal components of the $\gamma p$ or $\gamma A$ cross section~\cite{Ivanov:2004ax}.

Unfortunately, this broad pseudorapidity coverage is precluded by the EIC beampipe design, because the non-zero crossing angle limits how close one can come to the forward direction. It may be possible to instrument some areas at larger rapidity, but with incomplete azimuthal coverage, by taking advantage of areas above and below the beampipe; because the crossing angle occurs in the horizontal plane, space is more constricted horizontally than vertically. 

One way to expand the coverage at large $x$ would be to run at a lower ion beam energy.  This will shift the rapidity distribution, with the $x\rightarrow 1$ moving toward mid-rapidity, where the detection efficiency is higher.

\subsubsection{$\phi$ production}

The $\phi$ is the lighter of the two mesons highlighted in the 2012 EIC White Paper~\cite{Accardi:2012qut}.  It decays to $K^+K^-$ 49.2\% of the time. The charged kaon decay seems attractive, but detection is greatly complicated because of the low-$Q$ value of the decay.  The charged kaons are produced with momenta of only 135 MeV/c in the $\phi$ rest frame.  For an at-rest $\phi$, the kaon velocities are only $v \approx 0.2 c$, and so they have rather large specific energy loss and are easily stopped. Unless the $\phi$ are Lorentz boosted, either longitudinally due to being produced away from $y=0$, or because they are produced at large $Q^2$, leading to a large $p_T$, the decays may not be visible.  

\begin{figure}[htb]
\centering
\includegraphics*[width=0.9\textwidth]{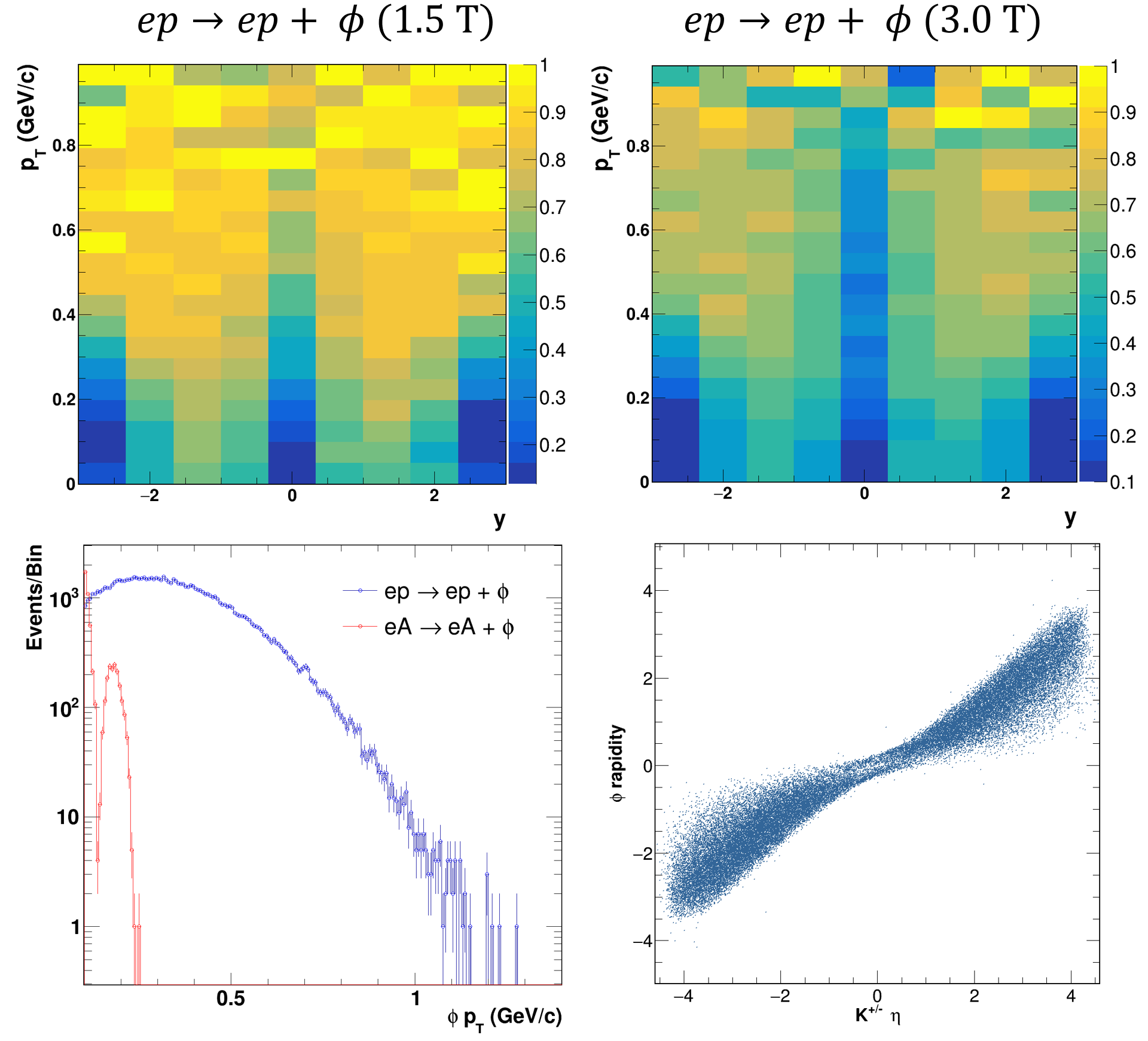}
\caption{(top) The detection efficiency for $\phi\rightarrow K^+K^-$ in $ep$ collisions in (left) a 1.5 T magnetic field and (right) a 3.0 T magnetic field, as a function of $\phi$ rapidity and $p_T$.  The $z$ axis is efficiency, from 0 to 1. The efficiencies at a given $p_T$ and rapidity would be very similar for $eA$ collisions.  (bottom left) The $p_T$ distribution for coherent $\phi$ production in $e$+$p$ (blue) and $eAu$ (red) collisions at an EIC, normalized to contain the same number of events.  The $eAu$ production is at much lower $p_T$ because of the larger size of the target.  This plot is for all $Q^2$, but is dominated by photoproduction, with $Q^2$ near 0. 
(bottom right) Scatter plot showing the relationship between $\phi$ rapidity and kaon daughter pseudorapidity.   A kaon pseudorapidity range $|\eta|<4$ covers the phi rapidity range $|y|<3$.  
}
\label{fig:phi}
\end{figure}

Figure~\ref{fig:phi} (bottom left) shows the $p_T$ spectra for coherently-produced $\phi$ in $e$+$p$ and $e$+$A$ collisions, again simulated in eSTARlight.  This is for head-on collisions, with no crossing angle. For $e$+$A$ collisions, the bulk of the production is at $p_T < 100$ MeV/c, while for $e$+$p$, most of the production is in the 100 to 750 MeV/c range.

The two top panels show the $\phi$ reconstruction efficiency for the all-silicon detector  in 1.5 and 3.0~T fields, as a function of $p_T$ and rapidity.  The efficiency is uniformly lower in the 3.0~T field, likely because the tracks curl up more tightly.  For both, the efficiency generally decreases as $p_T$ decreases, because the kaons are too soft to be reconstructed.
The low efficiency near $y=0$ is because of the very low kaon momentum; at larger $|y|$, the kaon momenta are boosted; the higher velocity reduces the kaon specific energy loss, $dE/dx$, so the kaons can more easily penetrate the beampipe and silicon layers, even with the larger column density due to the angle of incidence.
The $y\approx 0$ reduction covers almost all of the $e$+$A$ $p_T$ range for both magnetic fields, while at 3.0~T, it severely affects most production in $e$+$p$ collisions.   For electroproduction, the $p_T$ range is higher, but, especially at 3.0~T, the efficiency is likely to be reduced for electroproduction as well as photoproduction.  It would be very difficult to design a detector with better acceptance in this region, due to the required beampipe thickness and large kaon $dE/dx$.  The efficiency also drops at very large $|y|$ and low-to-modest $p_T$, because, as can be seen in Fig.~\ref{fig:phi} (left), a pseudorapidity acceptance out to $|\eta|<4$ only provides good coverage out to $\phi$ rapidity $|y|<3$ or so. It is possible that adding timing to the tracking detectors would lead to considerably improved momentum resolution for these low-momentum tracks~\cite{Klein:2020sts}.

The alternate decay modes $\phi\rightarrow K_L K_S$ (34.0\% branching ratio) and $l^+l^-$ (with branching ratio $3\times10^{-4}$ each for $ee$ and $\mu\mu$) seem unattractive, due respectively to the difficulty in reconstructing the $K_L$ and the low branching ratio.

\subsubsection{$J/\psi$ production}

\begin{figure}[htb]
\includegraphics*[width=0.9\textwidth]{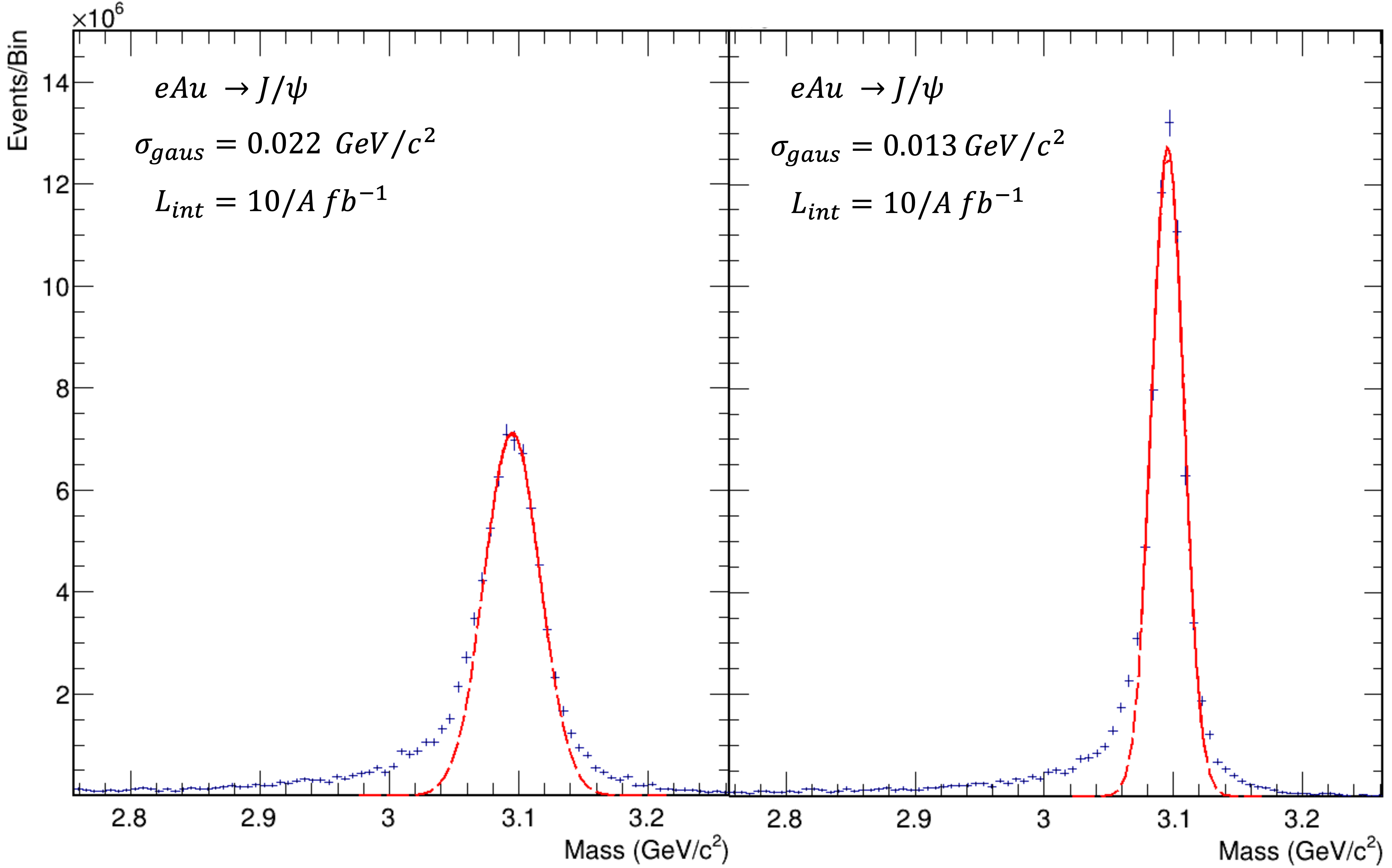}%
\caption{The reconstructed $J/\psi\rightarrow e^+e^-$ mass peak for the all-silicon detector, in (left) 1.5~T and (right) 3.0~T magnetic fields, for $e$+$A$ collisions with an integrated luminosity of 10 fb$^{-1}$/A.  Although most of the peak is well fit by a Gaussian, The higher magnetic field improves the resolution by almost a factor of two. There are also significant shoulders outside the Gaussian, from less well reconstructed events.  The shoulder is larger on the low-mass side, because of electron bremsstrahlung in the detector material.  With the higher field, the resolution (sigma) is more than a factor of 2 better, allowing better resolution of the peak.  The Gaussian fits were done only to the central part of the peak, as shown by the solid part of the red curves.
}
\label{fig:jpsi}
\end{figure}

The $J/\psi$ is the other meson highlighted in the 2012 EIC White Paper~\cite{Accardi:2012qut}.  It can be reconstructed from its decays into either the $\mu^+\mu^-$ or $e^+e^-$ final state.  Although the $J/\psi$ reconstruction is straightforward, bremsstrahlung from electrons can produce a low-mass shoulder in the $e^+e^-$ mass spectrum if the detector is too thick.  Figure~\ref{fig:jpsi} shows the reconstructed $J/\psi$ spectrum expected from the all-silicon detector for 1.5 and 3.0~T magnetic fields.  The peaks were fitted to a Gaussian function, with resolutions for the two fields of 22 MeV/$c^2$ and 13 MeV/$c^2$ respectively.  Although the Gaussian describes the peak well, there significant shoulders are visible.  The shoulders have two components: events that are less well reconstructed, and events where the electron underwent bremsstrahlung in the beampipe or detector material.  The latter only contributes to the low-mass shoulder.  The bremsstrahlung contribution should be the same for the two fields. 

Although a low-mass shoulder is visible in both spectra,  the peaks stand out clearly. The Gaussians are fitted to the data above the $J/\psi$ peak, and the portion of the lower mass data where the peak is above the shoulder, as indicated by the darker black line.  

The origin of the shoulders is demonstrated in Fig. \ref{fig:jpsi}, which shows the dilepton mass $M_{ee}$ vs. $p_T$. In addition to the pileup around $M_{J/\psi}$, there is a clear diagonal band with lower $M_{ee}$ but higher $p_T$. This band is expected due to bremsstrahlung.  If one of the electrons radiates a photon and loses energy while traversing the detector, the pair will be reconstructed with lower pair mass, but higher $p_T$~\cite{Schmidke2019}.  These events can be rejected by cuts on pair $p_T$ and $M_{ee}$, or with a photon veto in the calorimeter.  There is a further unsimulated source for these events, the decay $J/\psi\rightarrow e^+e^-\gamma$.  For photon energies above 100 MeV, the branching ratio for this channel is 0.88\% \cite{Zyla:2020zbs}, or 15\% of the rate to $e^+e^-$.

\begin{figure}[htb]
\includegraphics*[width=0.5\textwidth]{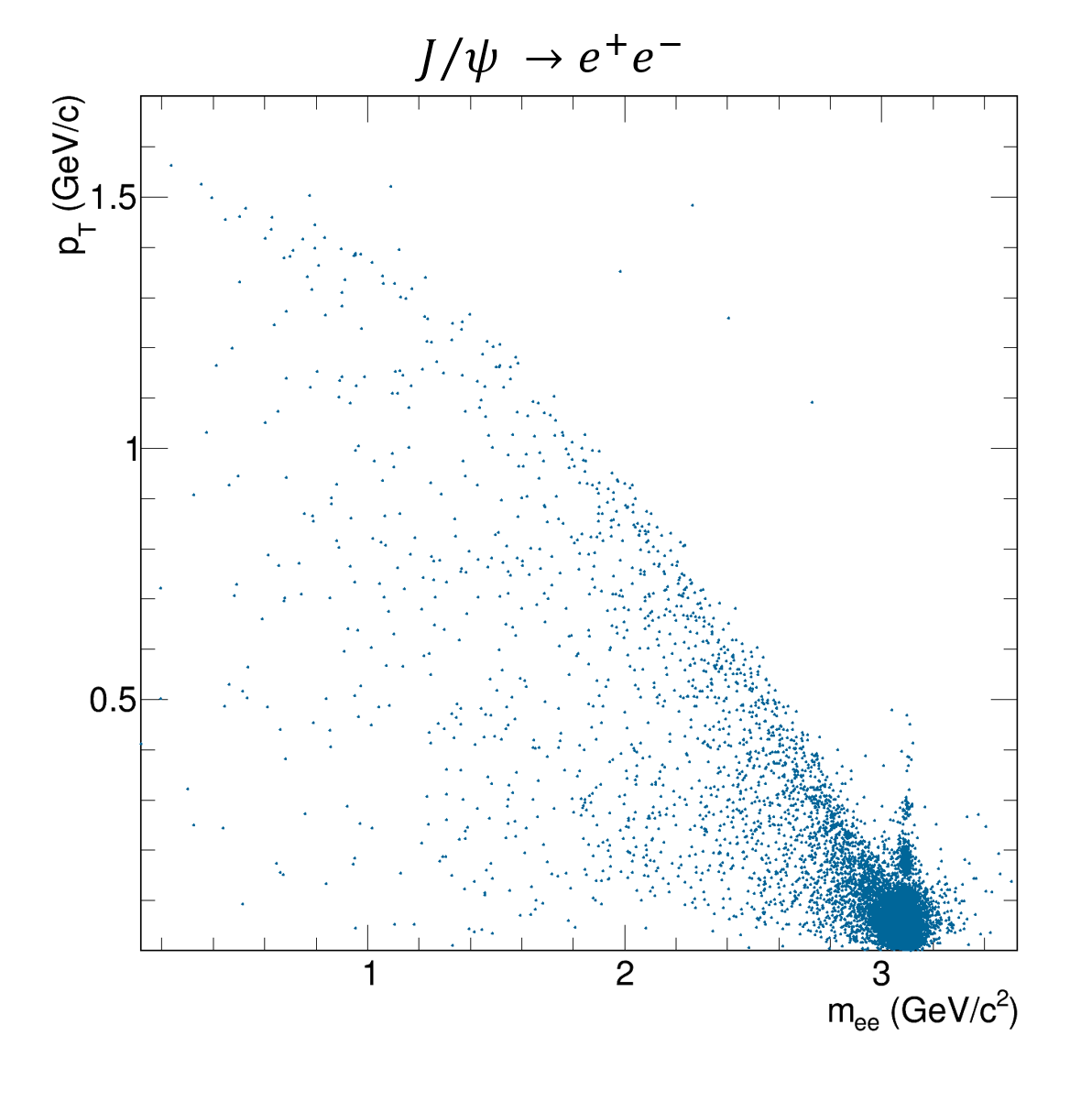}%
\label{fig:jpsi_m_ee}
\caption{A scatter plot showing the relationship between the reconstructed $J/\psi\rightarrow e^+e^-$ mass and transverse momentum in $e$+$p$ collisions in a 3.0~T field.  The vast majority of events are reconstructed with $M_{ee}\approx M_{J/\psi}$, with a $p_T$ distribution expected for coherent photoproduction.  There is also a clear diagonal band extending to lower $M_{ee}$ but higher $p_T$.  In these events, one of the leptons radiated a photon in the detector material, leading to the lower pair invariant mass and higher $p_T$. 
}
\end{figure}

\subsubsection{The $\Upsilon$ family}

\begin{figure}[htb]
\centering
\includegraphics*[width=0.85\textwidth]{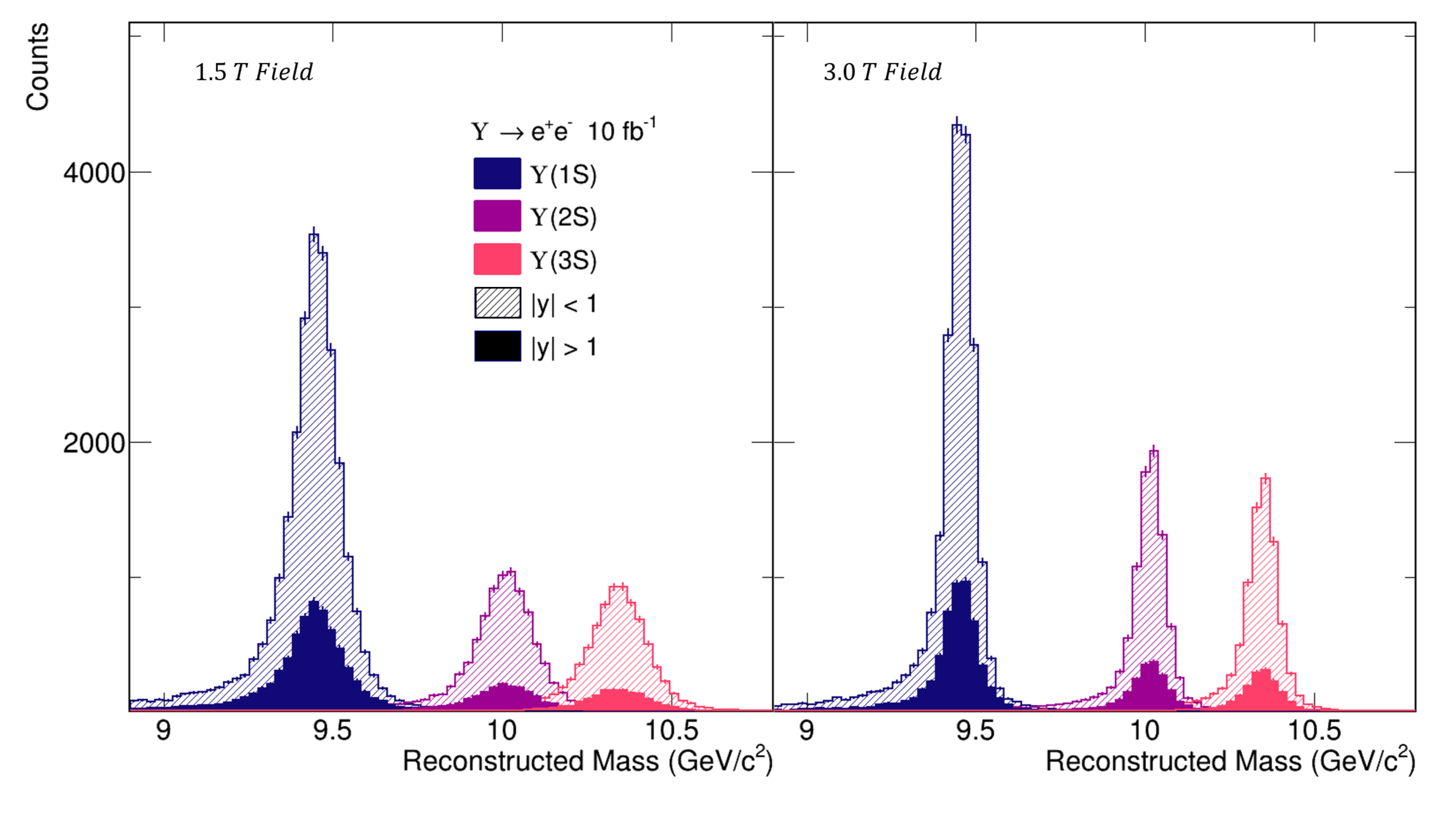}
\caption{The reconstructed $e^+e^-$ mass spectrum in the Upsilon region, with the $\Upsilon(1S)$, $\Upsilon(2S)$ $\Upsilon(3S)$ peaks, in the ratio predicted by eSTARlight, in 10 fb$^{-1}$ of data, in (left) 1.5~T and (right) 3.0~T fields. The spectra are divided into two rapidity ranges, $|y|<1$ and $|y|>1$.  Because the Upsilons are heavy, most of the production occurs in $|y|<2$.  
}
\label{fig:Upsilon}
\end{figure}

The three $\Upsilon$ states are relatively heavy, $\sim10$~GeV/$c^2$, but with rather small mass splittings - 563 MeV between the first and second, and 331 MeV between the second and third. Good momentum resolution is required to effectively separate the three states. Figure~\ref{fig:Upsilon} shows the $e^+e^-$ mass spectrum expected from the three Upsilon states, in two different rapidity ranges.  Although the Bjorken-$x$ ranges are different for positive and negative rapidity, the detector resolution should be similar at $+y$ and $-y$.  Table~\ref{tab:Upsilon} shows the resolutions extracted from a Gaussian fit to the $\Upsilon(1S)$ peaks. The resolution is about 40\% better in the 3.0~T field and in both cases worsens by about 20\% at larger $|y|$. Nonetheless, either magnetic field option provides adequate separation over the full range in $y$. 

\begin{table}[htb]
    \centering
    \caption{Mass resolution, for a Gaussian fit to the $\Upsilon (1S)$ peak shown in Fig.~\ref{fig:Upsilon}.}
    \begin{tabular}{c|cc}
~~~Rapidity~~~ &    ~~~1.5~T field~~~ &  ~~~3.0~T field~~~ \\
\hline \hline
$|y|<1$         & 66 MeV/$c^2$ & 40 MeV/$c^2$ \\
$|y|>1$         & 79 MeV/$c^2$ & 51 MeV/$c^2$ \\

    \end{tabular}
    \label{tab:Upsilon}
\end{table}

\subsubsection{Vector meson conclusions}

A full EIC program will include studies of the vector mesons considered here along with other vector mesons, likely including the $\rho'$ and other excited states.  In most cases, the proposed silicon detector exceeds the requirements for vector meson reconstruction. There are two problematic areas - inadequate coverage in pseudorapidity to cover light vector meson over the full range of Bjorken-$x$, and problematic reconstruction of very soft kaons from $\phi$ decays.  Rapidity is closely linked to Bjorken-$x$, so limitations in rapidity will reduce acceptance at large and small $x$. The large $x$ limitation is problematic both for parton measurements, and for searches for exotica, including the XYZ states~\cite{Klein:2019avl} and pentaquarks, and for backward production of mesons. However, the pseudorapidity coverage of tracking detectors is largely limited by the positioning of the beampipe.

\section{Summary and Conclusions}

The EIC will be a high-luminosity, variable-energy collider with a broad and compelling physics program. Completing this program will require detectors capable of making precision measurements of many physics channels over a wide kinematic range. We have presented a design for an all-silicon EIC tracking detector and shown that it provides tracking with high resolution and good vertex reconstruction over a large kinematic range, enabling the EIC's broad physics. We have also performed physics simulations for a variety of reactions for which tracking is critical, including measurements of heavy quarks, jets, and exclusive vector mesons. In most cases, the proposed all-silicon tracker provides the necessary performance, and we have identified cases where improved performance could enhance the kinematic coverage or enable new measurements.

Our simulations have demonstrated the feasibility of a broad program of heavy-quark studies that allow for measurements of gluonic PDFs, TMDs, and helicity distributions in nucleons and nuclei, as well as cold nuclear matter effects (see Sec.~\ref{sec:heavy-quarks}). The proposed tracking system based on the ultra-thin (0.3\%\,$X_0$ per layer) and fine-pitch (10$\times$10 \textmu m$^2$) MAPS sensor technology provides the momentum resolution as well as vertex reconstruction performance necessary to enable these studies, with potential for improved sensitivity in low momentum and/or charm baryon reconstruction with further reduced detector thickness.

The charged-jet energy and angular resolution performance of the all-silicon tracker was studied. Jets can be reconstructed with a resolution parameter $R=1.0$, as the multiplicity in the collisions is small. 
While there is a significant resolution loss when jet constituents are not all properly included in charged jets, the charged jet resolutions with the silicon tracker are nevertheless encouraging.
Finally, we studied azimuthal differences between jets and the scattered electron, which can provide access to parton transport in nuclear matter, the quark TMD PDF and the Sivers effect in transversely polarized e-p collisions, and the charged-jet fragmentation function, which should be sensitive to the hadronization process. 

The silicon detector 
can likewise reconstruct important vector meson decays.  Our simulations considered reconstructions of $\rho$, $\phi$, $J/\psi$ and $\Upsilon$ decays, but these are representative of other vector mesons decays, including the $\psi'$ and $\rho^{*0}$ decays.  The momentum resolution will allow us to cleanly separate the different $\Upsilon$ states in the dilepton spectrum.  The major limitation of the proposed tracker is the limited pseudorapidity coverage.  Broad coverage is important to reconstruct the photoproduction and electroproduction over the full range of Bjorken$-x$.  Unfortunately, this limitation appears to come primarily from the current beampipe design; the non-zero beam crossing angle leads to a ``X" shaped interaction region, limiting the length of the free (for detector) region in $|z|$, thereby limiting the acceptance of any tracking detector design. A secondary consideration is that the detector should have low mass, to minimize bremsstrahlung in dielectron events. 

For $D^0$, $\Lambda_c$ and most quarkonia reconstruction, a stronger magnetic field choice would be preferred as it enables better momentum and therefore mass resolution for these resonances. However, the impact 
on the low $p_T$ threshold due to the increased magnetic field strength is much 
less than the gain in signal significance 
due to the better mass resolution. For $D^0$ mesons, the $p_T$-integrated signal significance is improved by $\sim$50\% comparing 3.0~T vs. 1.5~T magnetic field setting (See Sec.~\ref{sec:sim:fast}). For $D^{*+}\rightarrow D^0\pi^{+}$ reconstruction, the 
$p_T$ threshold for the soft pion may cause  acceptance loss for low $p_T$ $D^{*+}$ mesons. Our initial study shows reconstruction using fewer tracking layers for the soft pions enables $D^{*+}$ to be still a viable channel 
in the 3.0~T magnetic field configuration (See Sec.~\ref{sec:sim:fast}).  The efficiency for $\phi\rightarrow K^+K^-$ is significantly higher for the 1.5~T field than at 3.0~T, because the higher field causes the tracks to curl up more tightly. 

One limitation, common to all of these analyses is that the Bjorken-$x$ of the struck parton is strongly correlated with the final state rapidity, with interactions at very large or very small $x$ corresponding to large $|y|$, where detector acceptance may be limited by the interaction region geometry. This, along with other issues identified in the simulations presented here, will require further examination of the tracking system or overall spectrometer design to see if modifications or optimizations can improve these measurements.

The all-silicon tracker geometry presented in this document will be revised as details of the EIC overall detector
are established and silicon-pixel R\&D efforts progress,
$e.g.$ optimizing an asymmetric tracker geometry in case the nominal interaction point is shifted away from ($0,0,0$).
The electron and hadron beams at the EIC are expected to cross each other at an angle of 25 mrad in the interaction region. Since the solenoid axis will be aligned with the electron beam direction, this crossing angle translates into a 25 mrad angular difference between the magnetic field and the hadron-beam direction, which causes the momentum resolution in the hadron direction to develop a dependence on the azimuthal angle. For tracks with $\phi\approx 0$ (in a coordinate system with the $z$ axis aligned along the hadron beampipe), the field integral is smaller, and the momentum resolution degrades. Conversely, for tracks with $\phi\approx \pi$ the field integral is larger, and the momentum resolution improves. This effect is more significant at higher pseudorapidities and hadron momenta. For example, for hadrons with $\eta\approx3.6$ and $p\approx50\,{\rm GeV}/c$, this effect can lead to an improvement (deterioration) of the momentum resolution with respect to the nominal momentum resolution by $\approx 40\%$ ($\approx 90\%$) on either side of the beam pipe in the horizontal plane. While we quantified the effect of the beam-crossing angle on the momentum resolution, we have not propagated the resulting azimuthal dependence in the physics studies presented here. This is in line with other studies for the Yellow Report, but does present an important area for continued study at forward pseudorapidities.

In summary, we have presented a concept all-silicon tracker that makes use of state-of-the-art technology, and have demonstrated the applicability of such tracker to fulfill several physics studies at the EIC.

\section{Acknowledgements}
This work is supported in part by the U.S. Department of Energy, Office of Science, Office of Nuclear Physics, under contract numbers DE-AC02-05CH11231 and DE-FG02-89ER40531, 
the DOE Office of Science Distinguished Scientists Fellow Award MPO/IEWO606342,
the DOE National Nuclear Security Administration under the Nuclear Science Security Consortium award DE-NA0003180,
by the EIC Generic Detector R\&D program under project eRD16, and by the Strategic Priority Research Program of Chinese Academy of Sciences, under grant number XDB34000000.

\bibliography{main}
\clearpage
\appendix

\end{document}

%% file: intro.tex
\section{Introduction, and Physics at the EIC}
The Electron Ion Collider (EIC) is a planned US-based facility that will make precision measurements of the collisions of electrons with polarized protons 
and ions over a large mass range
to study Quantum Chromodynamics (QCD) \cite{Accardi:2012qut,Aschenauer:2017jsk}.   The EIC will explore a very wide range of physics topics, including - among others - the spin structure of protons and light nuclei, the partonic structure of light and heavy ions, parton transport in nuclear matter, and the hadronization process. Electrons with energies up to 18 GeV will collide with protons up to 275 GeV and ions with energies up to 110 GeV per nucleon, at luminosities up to $10^{34}$ cm$^{-2}$ s$^{-1}$~\cite{eRHIC:preCDR}.   

The broad physics program and high energies and luminosities impose significant requirements on the EIC detectors.  This paper describes requirements for charged-particle tracking, and studies of how the requirements can be met with state-of-the-art silicon pixel detectors. 
The requirements 
discussed here are driven by
three promising physics probes at the EIC: 
heavy quarks, jets and exclusive vector mesons. High precision measurements addressing the
physics questions listed above will require wide pseudorapidity coverage, excellent momentum resolution, and low mass to limit bremsstrahlung from electrons and positrons. We 
have developed a conceptual design for an all-silicon tracking system and simulated its implementation, including a first approximation of detector support structures and services. The paper explores the three probes and quantifies how well the design meets the requirements.

The paper is organized as follows: Section II describes the proposed all-silicon tracker. Section III presents the physics goals, detector requirements and simulation studies demonstrating the performance for heavy quarks, jets and exclusive vector mesons.  Section IV presents the summary and conclusions of these studies.

%% file: all_si_tracker.tex
\section{All-Silicon Tracker Concept Design}
\label{sec:sim:full}
\subsection{Requirements on EIC Tracker}

EIC detectors have been conceptualized as  general-purpose instruments surrounding the interaction point (IP) and embedded in a solenoidal magnetic field with
maximum field strengths of either 1.4 or 3.0\,T.
Approximately 2.5\,m along the $z$ axis and a radial extent of 80\,cm are allocated for the innermost tracking system, which will be surrounded by other sub-detectors including particle identification (PID) detectors and electromagnetic and hadronic calorimeters.
Tracking and vertexing systems at the EIC must have wide kinematic coverage, good momentum resolution and
secondary vertex separation capabilities, in order to carry out the EIC physics programs. There are, in general, two types of tracking/vertexing detector designs being considered: 1) a hybrid system composed of silicon-pixel layers for vertexing plus outer gas detectors ($e.g.$ Time Projection Chamber (TPC) and Micro-Pattern Gaseous Detectors (MPGD)) and 2) an all-silicon tracker for both momentum and vertexing measurements.
Recent R\&D work showed that the all-silicon tracker can deliver a comparable 
or better momentum resolution than hybrid concepts, while keeping the radial dimension quite compact. The compact all-silicon tracker leaves much more open space for the outer PID detectors to enhance their PID performance. If even more space were needed, this could be achieved by making the all-silicon tracker even more compact, albeit at the expense of some resolution loss.

\subsection{Geometry}

\begin{figure}[htbp]
    \centering
    \includegraphics[width=0.495\textwidth]{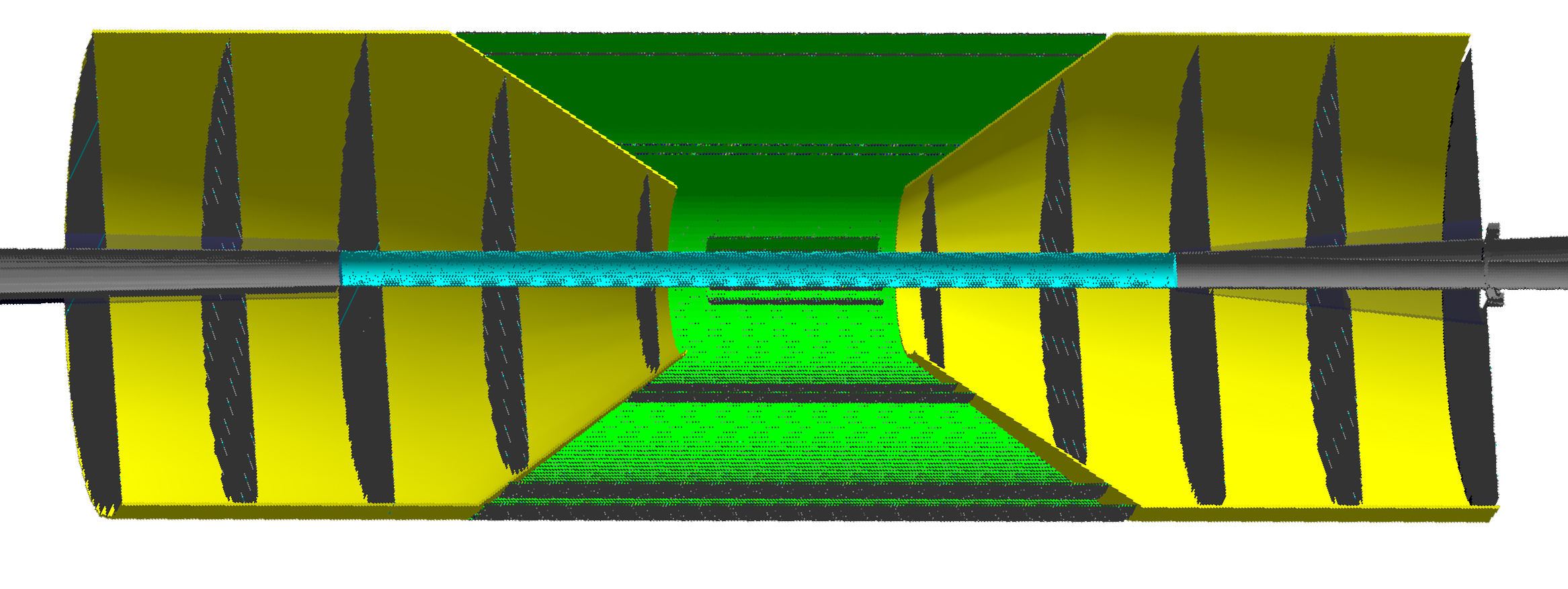}
    \includegraphics[width=0.495\textwidth]{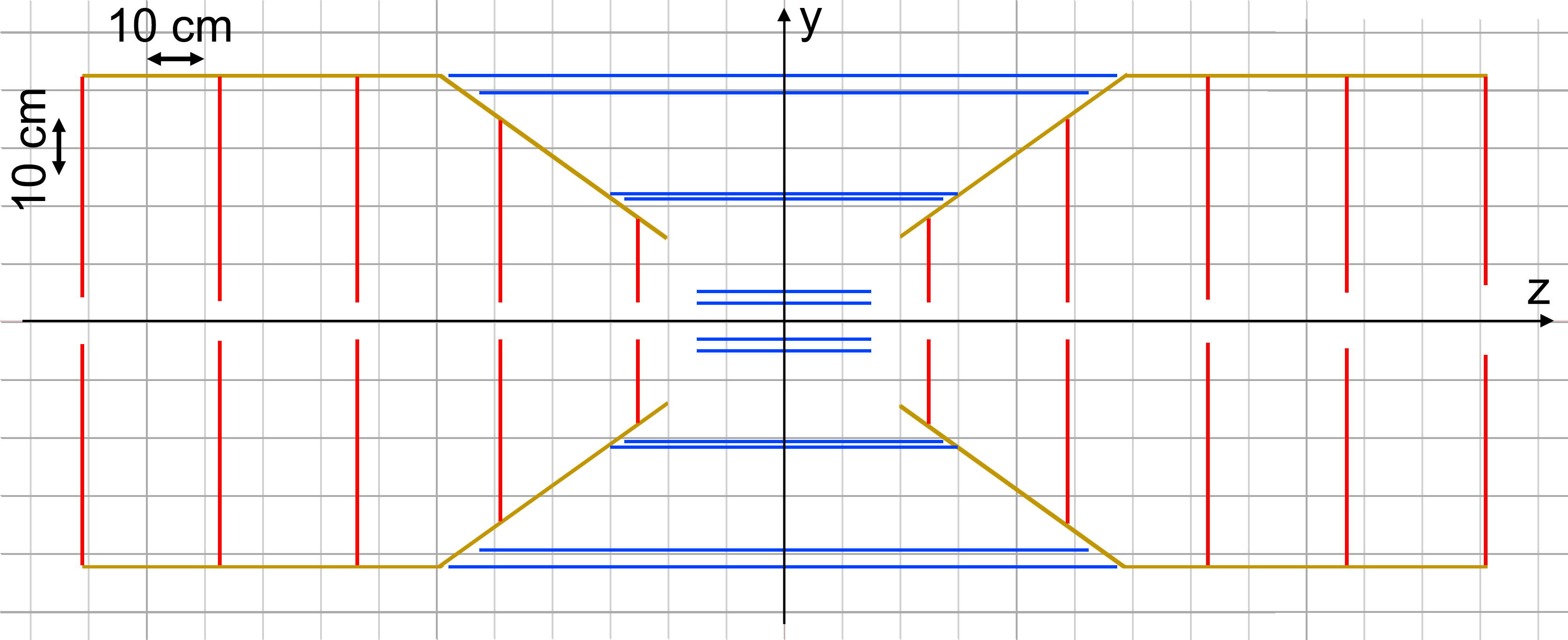}
    \caption{All-silicon tracker geometry.
    Left: GEANT4 schematic of the tracker cross section. The barrel, disks, and support structure correspond to the green, dark-gray, and yellow components, respectively. The beryllium section of the beam pipe is shown in cyan. The rest of the beam pipe, which takes into account the expected electron-hadron-beam crossing angle is shown in light-gray.
    Right: detector schematic (side view). The barrel layers, disks, and support structure are represented in blue, red, and yellow, respectively. See text for details.}
    \label{fig:all_si_schematic}
\end{figure}

A schematic of the all-silicon tracker concept considered in this work is shown in Fig.~\ref{fig:all_si_schematic}.
This detector, which has been designed within a generic EIC-detector R\&D effort~\cite{eRD16},
corresponds to a cylindrical tracker with radius of 43.2\,cm and length of 242\,cm along the $z$ direction, wrapped around the beam pipe and centered at the nominal IP (which corresponds to $(x,y,z)=(0,0,0)$). In the region $-79.8 < z < 66.8$\,cm, the current design of the beam pipe corresponds to a beryllium cylinder of radius of 3.17\,cm and thickness of 760~\textmu m. Outside of this region, the beam pipe fans out to take into account the beam-crossing angle of $\approx$25\,mrad. Inside the beam pipe, a vacuum is simulated. The rest of the geometry is embedded in an air volume.

The tracker coverage for low values of pseudorapidity, $\eta\equiv-{\rm ln}\big(\tan(\theta/2)\big)$ (where $\theta$ is the polar angle in a coordinate system with the $z$ axis aligned along the beam pipe) is provided by a barrel with 6 layers.
The radii at which these layers are located and their corresponding lengths along $z$ are summarized on Table~\ref{tab:all_si:barrel} and illustrated in Fig.~\ref{fig:all_si_schematic} (right). They are laid out in three double layers to provide redundancy, and the middle double layer is placed equidistantly between the inner and outer double layers to measure hits in the vicinity of the sagitta to optimize the momentum resolution. The pairing of the barrel layers also has the benefit of reducing the number of stave designs.
The vertexing capabilities of the detector are driven primarily by the
first two layers. The innermost layer was placed as close to the beam pipe as possible, and the position of the second layer was varied until the optimal vertexing performance was found.

Coverage at larger absolute values of pseudorapidity is provided by 5 disks in each direction.
These disks are assembled by adding rectangular staves in parallel and giving each a length that satisfies that fits within a circle or radius $R$, the disk outer radii presented in Table~\ref{tab:all_si:disks}, along with
their $z$ positions and inner radii;
see Fig.~\ref{fig:all_si_schematic} (right).
Given the rectangular geometry of the staves, the hole through which the beam pipe passes is shaped as a square of side equal to twice the inner radii presented in Table~\ref{tab:all_si:disks}.
While the disks on either side of the $x-y$ plane are positioned at the same distance from the center of the detector and their outer radii are the same, their inner radii are optimized to be as close to the beam pipe as possible. Thus, the acceptance limit at high $|\eta|$ is given by the beam-pipe geometry.
Given the asymmetric nature of the EIC collisions ($i.e.$ electrons colliding with protons or nuclei with different lab-frame energies), a potential future improvement is to optimize the disk layout separately for the forward and backward regions.
An odd number of disks is favored to measure hits in the vicinity of the sagitta, thus achieving a better resolution.
The transition between the barrel and the disks occurs at
$|\eta|\approx 1.1$.

\begin{figure}[htbp]
    \centering
    \includegraphics[width=0.40\textwidth]{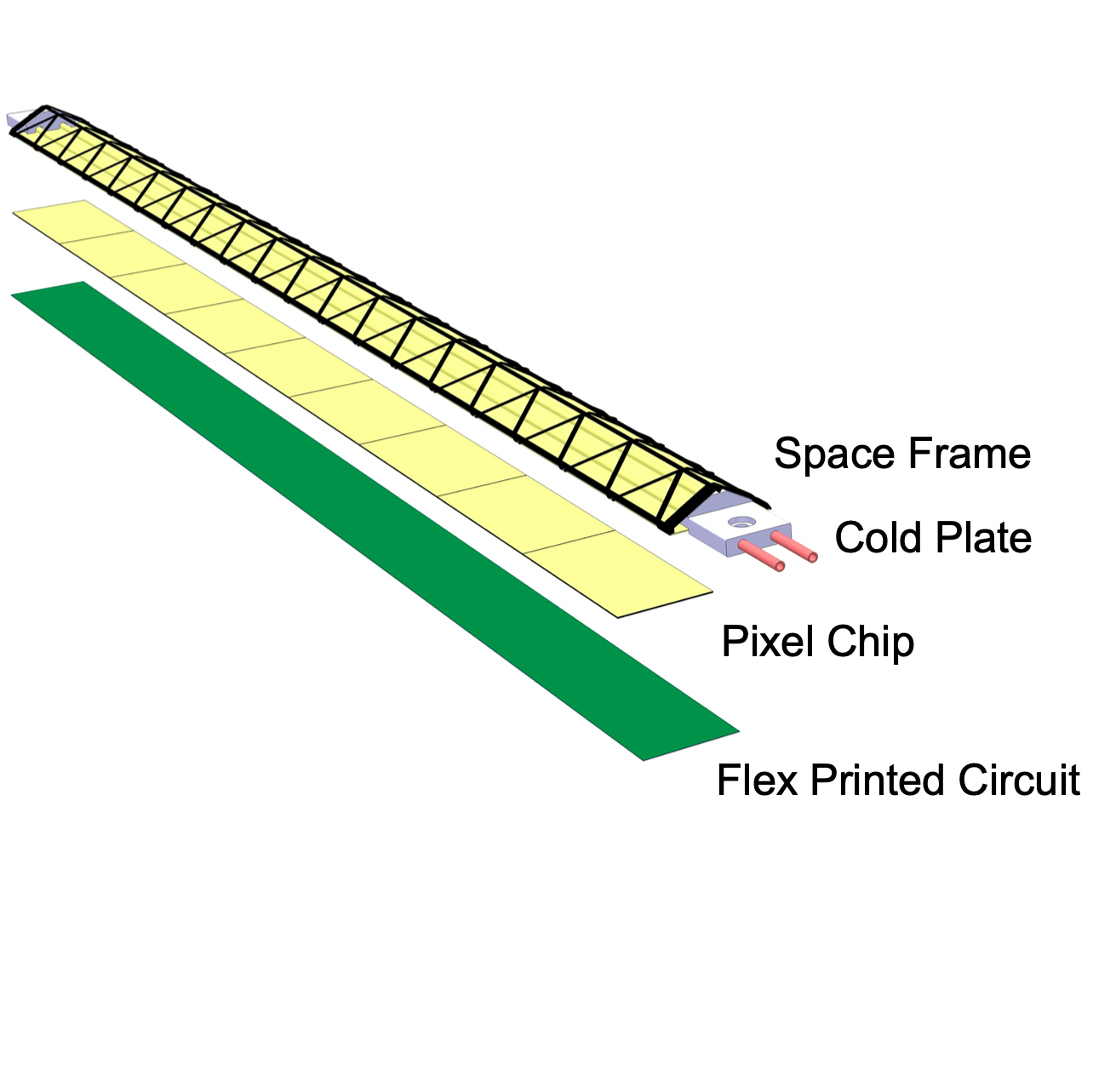}
    \includegraphics[width=0.59\textwidth]{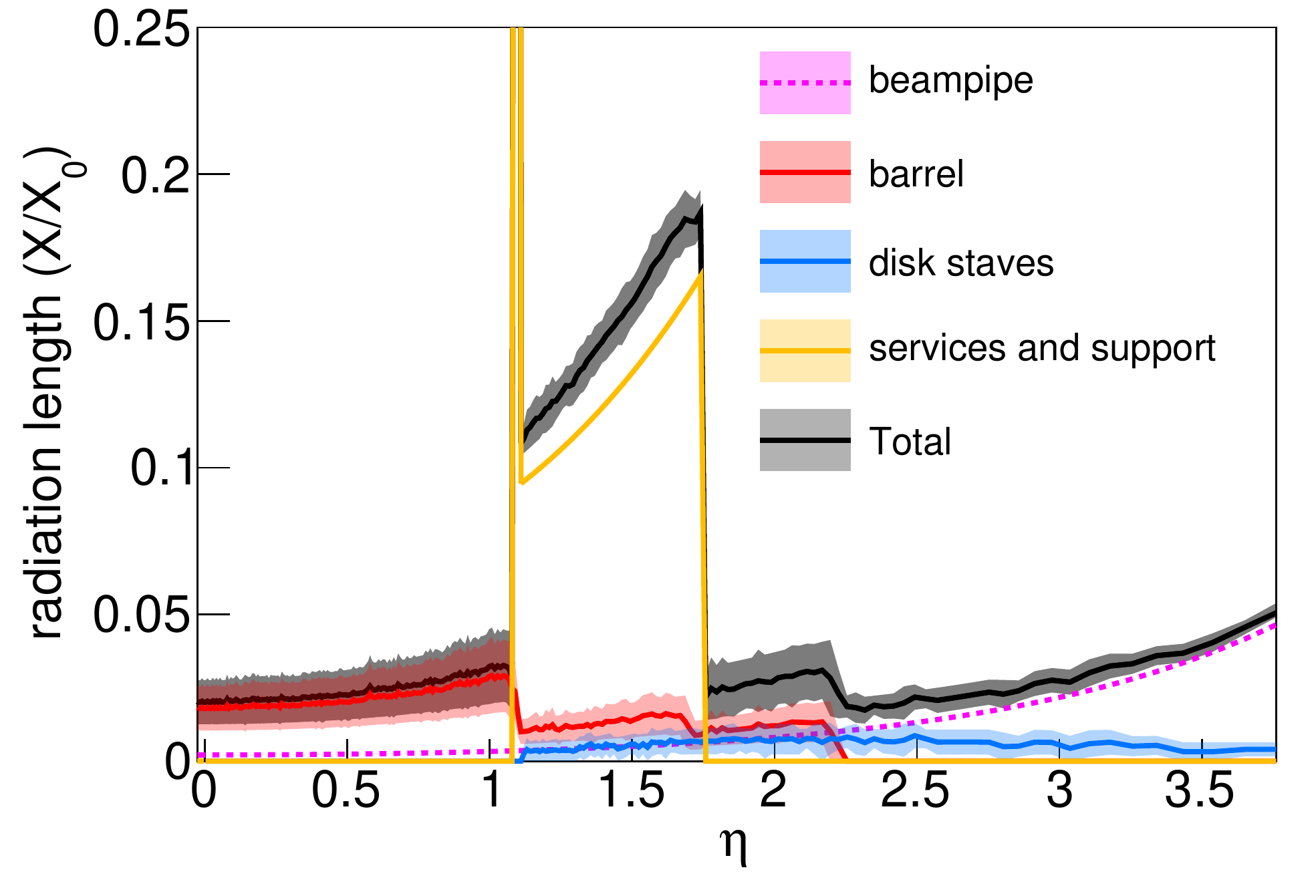}
    \caption{Detector material budget.
    Left: schematic of the ALICE ITS2 inner-barrel staves used in the all-silicon tracker design presented here. Schematic taken from Fig.\,1.3 in~\cite{Abelevetal:2014dna}.
    Right: all-silicon tracker material scan.
    The dashed magenta line corresponds to the material from the beam pipe.
    The barrel and disk contributions are shown in red and blue, respectively.
    The aluminum support structure is shown in yellow.
    The total contribution is shown in black.
    See text for details.
    }
    \label{fig:material}
\end{figure}

\begin{table}
    \parbox{.45\linewidth}{
        \centering
\caption{Barrel-layer radii and lengths.}
\label{tab:all_si:barrel}
\begin{tabular}{c|cc}
~~Barrel~~  & ~~radius~~    & ~~length along z~~        \\
layer   & {[}cm{]}  & {[}cm{]}              \\
  \hline \hline
1       & 3.30      & 30                    \\
2       & 5.70      & 30                    \\
\hline
3       & 21.00     & 54                    \\
4       & 22.68     & 60                    \\
\hline
5       & 39.30     & 105                   \\
6       & 43.23     & 114                    
\end{tabular}
}
\parbox{.45\linewidth}{
\centering
\caption{Disk $z$ position and inner and outer radii.}
\label{tab:all_si:disks}
\begin{tabular}{c|ccc}
Disk  		& ~~z position~~	& outer 			& inner 			\\
~~number		& {[}cm{]}      & ~~radius {[}cm{]}~~	& ~~radius {[}cm{]}~~	\\
\hline \hline
-5			& -121		    & 43.23			    & 4.41			    \\
-4			& -97		    & 43.23			    & 3.70			    \\
-3			& -73		    & 43.23			    & 3.18			    \\
-2			& -49		    & 36.26			    & 3.18			    \\
-1			& -25		    & 18.50			    & 3.18			    \\
\hline
1           & 25            & 18.50			    & 3.18 			    \\
2           & 49            & 36.26			    & 3.18 			    \\
3           & 73            & 43.23			    & 3.50 			    \\
4           & 97            & 43.23			    & 4.70 			    \\
5           & 121           & 43.23             & 5.91
\end{tabular}
}
\end{table}

Both the barrel layers and the disks are made up of realistic staves modeled after the ALICE-ITS2-upgrade inner-barrel staves~\cite{Abelevetal:2014dna,Keil:2015vta,Reidt:2016ysg}
and shown in Fig.~\ref{fig:material} (left).
Besides the active silicon volume, each stave includes components such as carbon-fiber support structures and water cooling pipes, which combined correspond to an average material budget of 0.3$\% \,X_0$ per stave.
The total amount of material that these staves contribute to the all-silicon tracker geometry is shown in Fig.~\ref{fig:material} (right)
Since the staves create a periodic but $\phi$-varying structure (where $\phi$ corresponds to the azimuth), the geometry is scanned around the azimuth for a fixed $\eta$, and the minimum and maximum amounts of material found define the boundaries of the uncertainty band.
With the current configuration, the material budget contributed by the barrel and disk staves is $<5\%\,X_0$.

The attributes of the sensor used in the simulations are taken from the eRD25 and EIC Silicon Consortium~\cite{eRD25} descriptions of the projected properties of an EIC specific Monolithic Active Pixel Sensor (MAPS) currently under development.
The sensor silicon pixels have a pitch of 10$\times$10\,\textmu m$^2$ (corresponding to a point resolution of 10/$\sqrt{12}$ \textmu m) and silicon thickness of 50 \textmu m. While this simulation effort uses 0.3$\% \,X_0$ for the inner two tracking layers, there are ongoing R\&D efforts to use stitched, thinned and bent air-cooled silicon to allow the vertexing layers to become as thin as 0.05$\% \,X_0$~\cite{its3det}. 

As part of the EIC Yellow Report effort, projections were generated for both the radiation length of staves and discs~\cite{leo} based on the eRD25 EIC specific sensor and for the services (location and composition) and mechanical supports~\cite{leo2} that would be required to complete a tracking detector. These projections, which were only available after most of the work presented here, are referenced for completeness and are reasonably consistent with what is used in the shown simulation for material in the tracking detectors acceptance.

The detector is complemented with a simplistic conical aluminum support structure with a thickness of 5\,mm which is tapered for $z>58$\,cm. As shown in Fig.~\ref{fig:material} (right), this support structure adds a significant amount of material 
to the detector. However, the projective design concentrates this material into a narrow pseudorapidity range at $|\eta|=1.1$.
More realistic support structures (likely made of carbon-fiber composite) and services are still to be implemented.  
An earlier notional all-silicon detector extended out to a radius of 75 cm \cite{Klein:2020sts}; a larger detector can offer improved momentum resolution, at a cost in larger silicon area, and hence cost.  This study also considered the possibility of using timing to improve resolution for low-momentum particles, but this is difficult with the smaller 43 cm lever arm.

\subsection{Performance}
\label{sec:performance}

This geometry was implemented in GEANT4 and
studied within the full Monte-Carlo framework for detector simulation, Fun4All~\cite{Fun4All,Pinkenburg:2011zza,Pinkenburg:2005zza}.
Performance studies were carried out by generating charged particles ($e.g.$ pions, electrons, protons, and muons) from the nominal IP in the momentum range $0 < p < 30$ GeV$/c$ and over the entire detector acceptance ($i.e.$ $|\eta|<4$ and $0<\phi<2\pi$).
The two magnetic fields considered for the simulations correspond to solenoidal field maps for the BeAST~\cite{Beast1,Beast2} and BaBar~\cite{Babar} magnets, with peak intensities of 3.0~T and 1.4~T, respectively.
The hits resulting from the interaction between the generated particles and the detector (which was setup with a hit efficiency at 100\%) were combined into tracks, and differences between the variables generated and reconstructed in the simulation (labeled `truth' and `reco', respectively) were used to characterize various detector resolutions. 
Pattern recognition for combining hits
into reconstructed tracks 
is seeded
using truth-track information. Thus, final efficiency studies are not feasible at the moment and will be carried out when more realistic seeding algorithms are implemented.

\begin{figure}[htbp]
    \centering
    \includegraphics[width=0.45\textwidth]{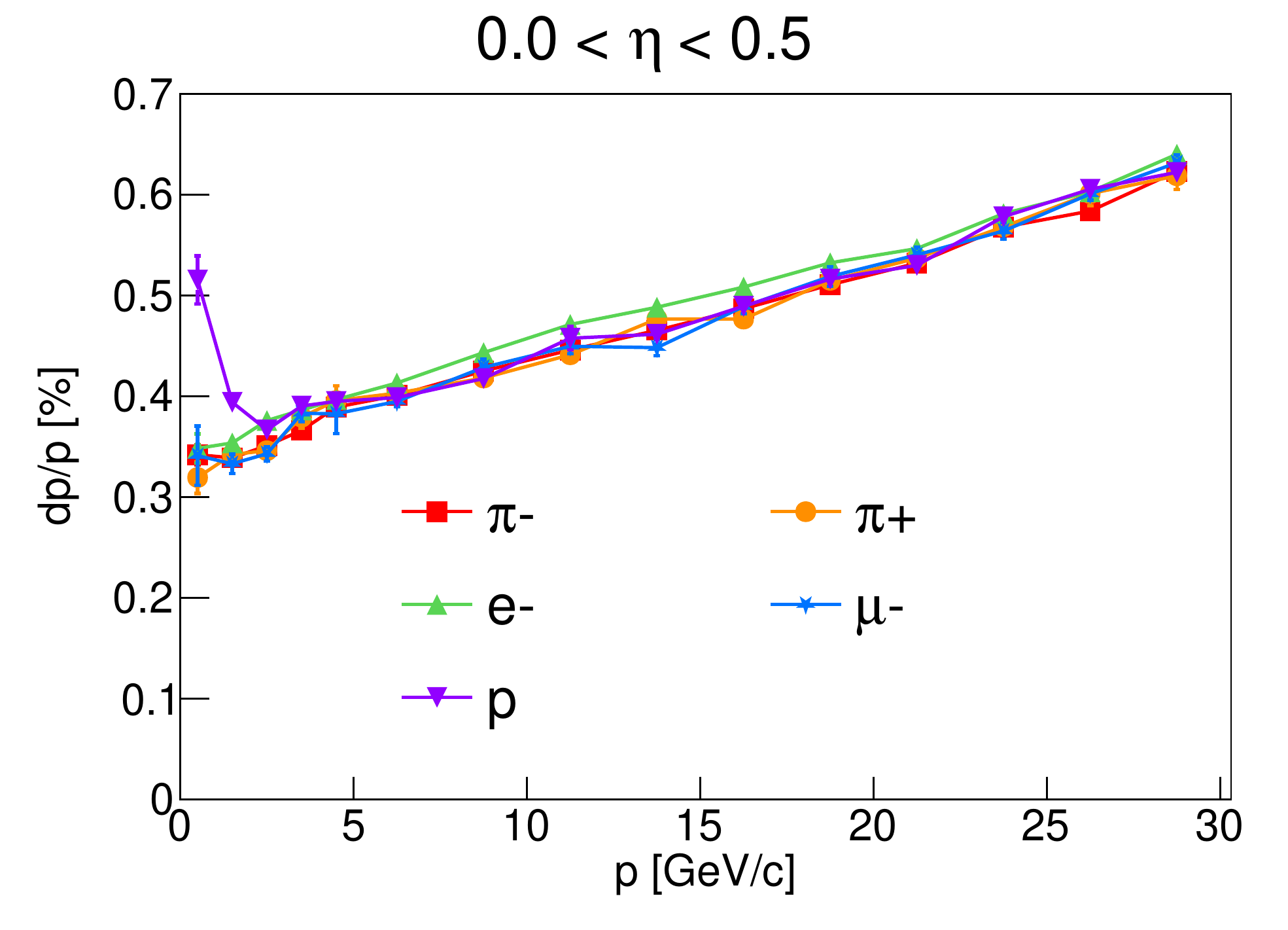}
    \includegraphics[width=0.45\textwidth]{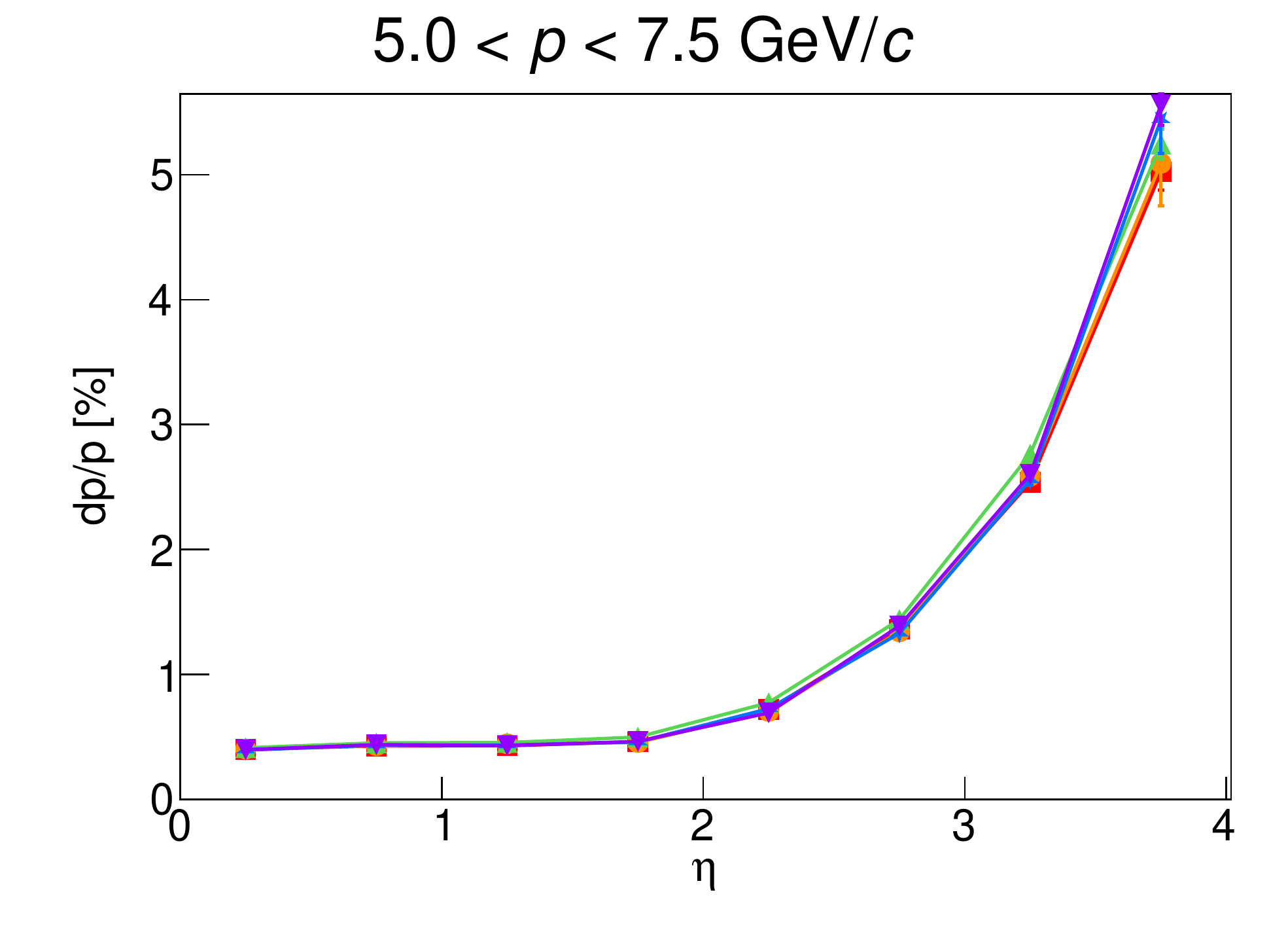}
    \caption{
    Momentum resolution for different particles in the 3.0~T magnetic field.
    Left: $dp/p$ as a function of momentum in the $0 < \eta < 0.5$ range.
    Right: $dp/p$ as a function of pseudorapidity in the $5.0 < p < 7.5$\,GeV/$c$ range.
    See text for details.}
    \label{fig:diff_particles}
\end{figure}

\begin{figure}[htbp]
    \centering
    \includegraphics[width=0.95\textwidth]{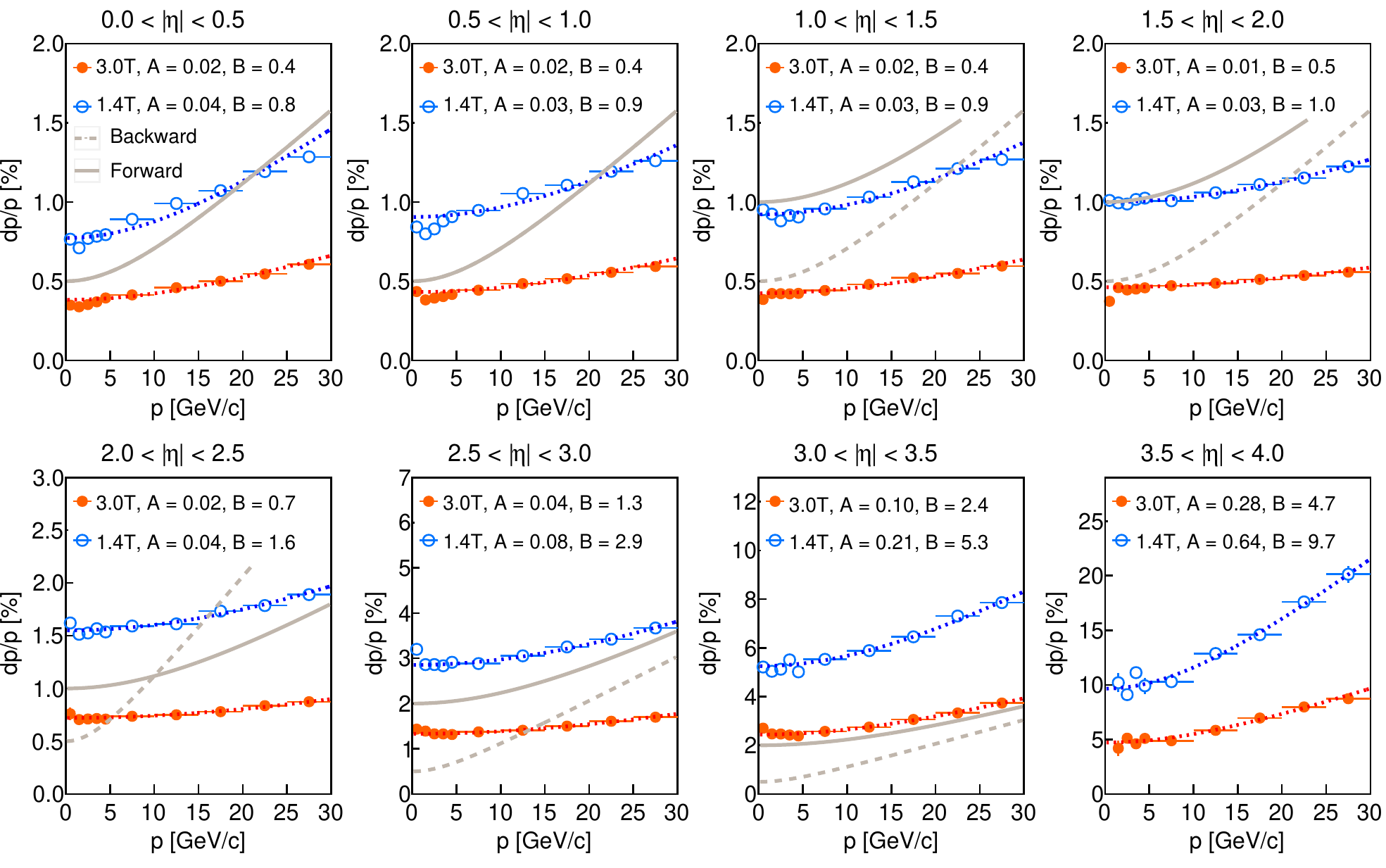}
    \caption{Momentum resolution as a function of momentum for several
    pseudorapidity bins. The markers correspond to the resolutions extracted
    from the simulations, and the lines correspond to fits to such resolution curves.
    The orange (filled) and blue (open) circles correspond to simulations carried out with the BeAST (3.0~T) and BaBar (1.4~T) field maps respectively.
    The functional form used in the fits is $dp/p=Ap\oplus B$, and the parameters $A \ [\%/({\rm GeV}/c)]$ and $B \ [\%]$ are given in the plots. The EIC physics requirements~\cite{DMtable:2020} are shown as gray lines for $|\eta|<3.5$. In the cases where the forward and backward requirements are different, the backward requirements are shown as dashed lines.}
    \label{fig:all_si:mom_res_param}
\end{figure}

\begin{figure}[htbp]
    \centering
    \includegraphics[width=0.95\textwidth, clip]{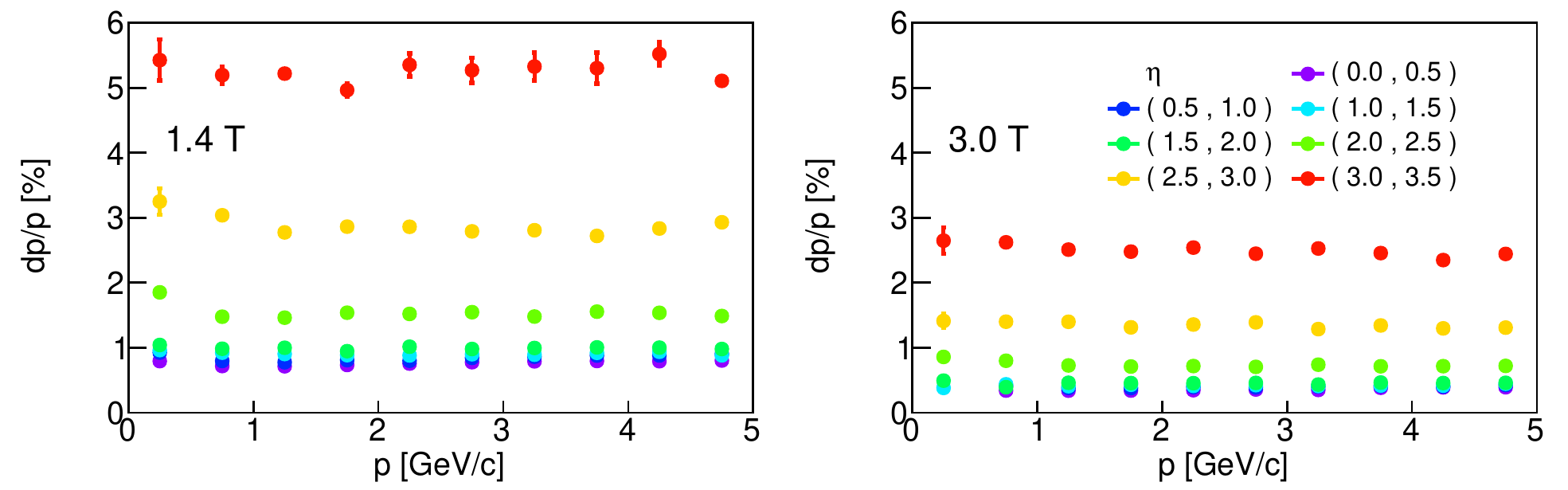}
    \caption{Momentum resolution as a function of momentum for several
    pseudorapidity bins in the low-momentum region. Left: 1.4~T magnetic field. Right: 3.0~T magnetic field.}
    \label{fig:all_si:mom_res_lowp}
\end{figure}

The momentum resolution, $dp/p$, is defined as the standard deviation of the $(|\vec{p}_{\rm truth}|-|\vec{p}_{\rm reco}|)/|\vec{p}_{\rm truth}|$ distribution, where $|\vec p|$ is the absolute value of the particle momentum.
Figure~\ref{fig:diff_particles} (left) shows the momentum resolution as a function of momentum for charged pions, electrons, muons, and protons in the pseudorapidity range $0.0 < \eta < 0.5$ in the 3.0~T magnetic field.
The multiple-scattering effect is more pronounced for protons below 3-4~GeV/$c$, significantly worsening the resolution at low momentum.  For other particles and for high-momentum protons, the rise with increasing momenta originates primarily from the decreasing sagitta for stiffer tracks, with the electron momentum resolution systematically above that for other particles in most of the studied range.
Figure~\ref{fig:diff_particles} (right) shows the momentum-resolution results as a function of pseudorapidity in the momentum range $5.0<p<7.5$\,GeV$/c$.
The momentum resolution is approximately constant up to $\eta\approx 2$, and then quickly rises.
It is worth emphasizing that this pseudorapidity value corresponds to $\theta \approx 15^{\circ}$, implying that the rising part of the resolution represents less than 17\% of the polar-angle acceptance.
Overall, the performance is very similar for the studied particles, and further results will be shown only for pions.

More detailed momentum-resolution studies for pions are shown for both magnetic-field settings in Fig.~\ref{fig:all_si:mom_res_param}. As expected from the leading-order $\sim1/B$ dependence of the momentum resolution, doubling the magnetic-field intensity
improves the momentum resolution by a factor of $\approx 2$.
Momentum resolutions are typically parametrized by the function
$dp/p = A \cdot p \oplus B$,
where $A$ and $B$ are fit parameters and $\oplus$ indicates sum in quadrature. The resulting fits and the fit parameters are shown in the figure.
Also shown as gray lines are the requirements determined by the physics working groups in the EIC Yellow Report effort~\cite{DMtable:2020}. 
In the case of the 3.0\,T field, the initial tracker design satisfies the physics requirements over the entire $0 < p < 30$\,GeV$/c$ range for $-2.0 < \eta < 3.0$ but falls short, especially for lower-momentum particles, outside of this range.
At 1.4\,T, there is no pseudorapidity range where the tracker can satisfy the requirements over the entire momentum range.
While the momentum resolutions are better with the 3.0\,T magnet, the 1.4\,T magnet has advantages: such a magnet already exists (reducing the cost of the detector) and the smaller field intensity allows lower-transverse-momentum particles to be measured.
Momentum resolutions for particles with $p<5\,{\rm GeV}/c$ are shown in Fig.~\ref{fig:all_si:mom_res_lowp}.

\subsection{Further momentum resolution optimization}
\label{sec:optimization}

\begin{figure}[htbp]
    \centering
    \includegraphics[page=1,width=0.85\textwidth]{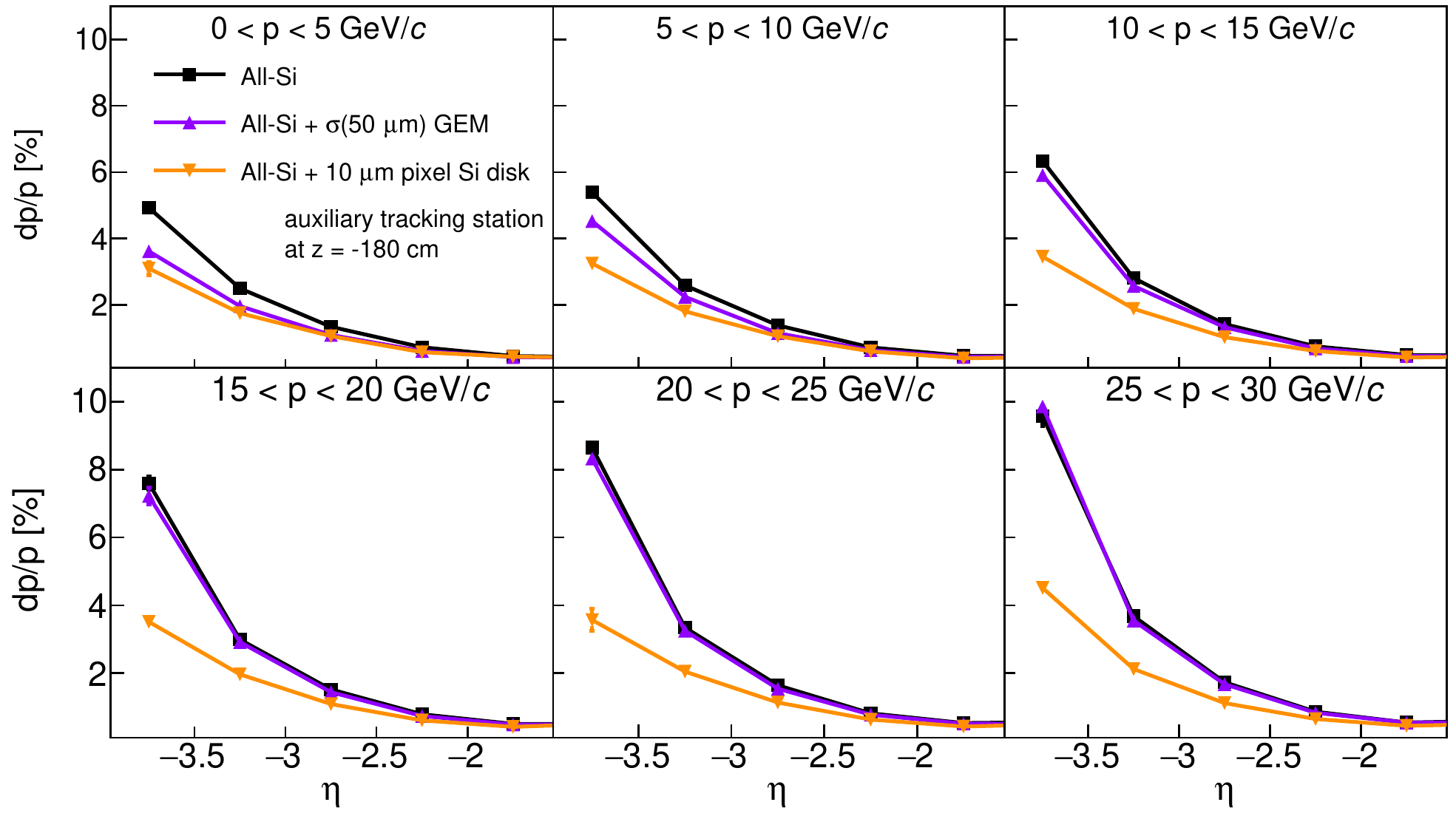}
    \caption{Momentum resolution as a function of pseudorapidity in the backward region. The all-silicon tracker standalone performance is shown in black squares. The purple triangles and orange inverted triangles describe the momentum resolution achieved by complementing the all-silicon tracker with a 50\,\textmu${\rm m}$-resolution GEM station or a 10\,\textmu${\rm m}$-pixel silicon disk at $z=-180$\,cm, respectively. In this region, the all-silicon tracker complemented with a silicon disk offers a significantly better performance. See text for details.}
    \label{fig:gem_back}
\end{figure}

To study whether the momentum resolution could be improved at forward and backward pseudorapidities, where there is tension between the tracker performance and physics requirements, the detector was complemented with additional tracking stations taking into consideration the space currently projected to be available according to the Central Detector/Integration \& Magnet Working Group~\cite{ayk}. Placing such complementary trackers away from the interaction point increases the field integral, $\int B\cdot\mathrm{d}l$, thus improving the momentum resolution, and no specific detectors are planned to be installed between the backward station and the all-silicon tracker. We examined the impact of adding methane-based gas electron multiplier (GEM) detectors or additional silicon disks in the backward and forward regions at $z = -180$ and 300\,cm, respectively.
The resulting momentum resolutions in the backward region are shown in Fig.~\ref{fig:gem_back}.
Complementing the all-silicon tracker with a 50\,\textmu${\rm m}$-resolution GEM station yields a small improvement, mainly for low-momentum particles, while a 10\,\textmu${\rm m}$-pixel silicon disk significantly improves the momentum resolution in the far-backward region (negative $\eta$), by a factor of two or more for high-momentum particles.

\begin{figure}[htbp]
    \centering
    \includegraphics[page=2,width=0.85\textwidth]{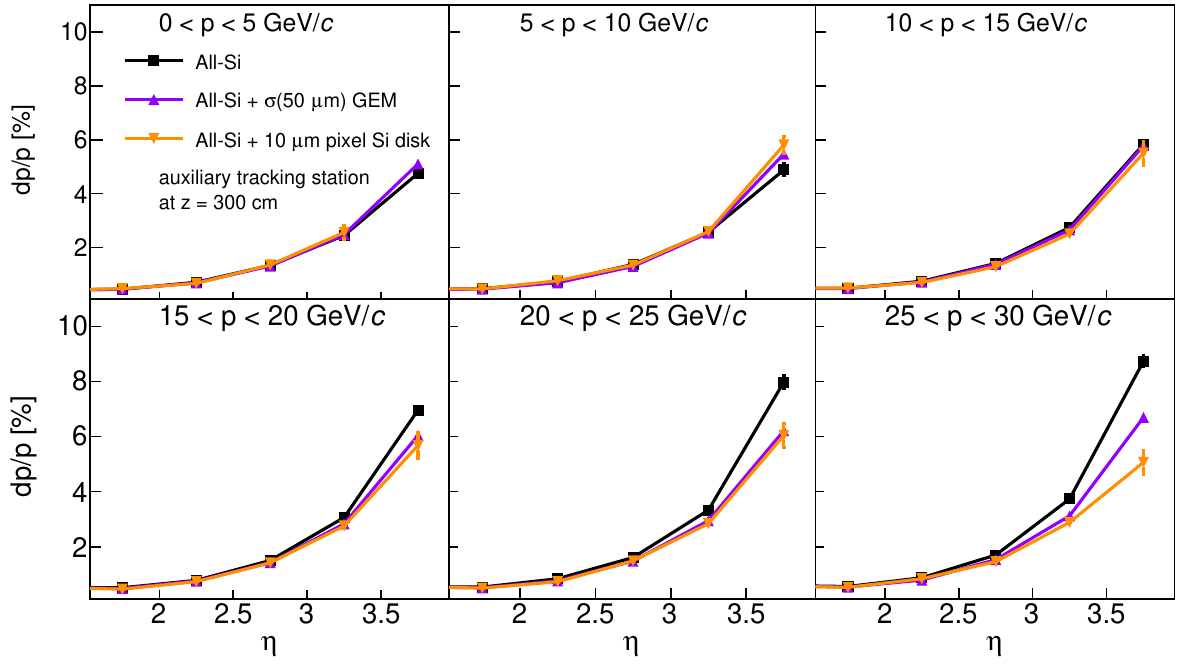}
    \caption{Momentum resolution as a function of pseudorapidity in the forward region. The all-silicon tracker standalone performance is shown in black squares. The purple triangles and orange inverted triangles describe the momentum resolution achieved by complementing the all-silicon tracker with a 50\,\textmu${\rm m}$-resolution GEM station or a 10\,\textmu${\rm m}$-pixel silicon disk at $z=300$\,cm, respectively.
    In this region, a RICH detector is placed in the space between the all-silicon tracker and the complementary tracking station.
    The all-silicon tracker complemented with a GEM offers a performance comparable to that of the tracker complemented with a silicon disk (except in the highest momentum bin shown). See text for details.}
    \label{fig:gem_forw}
\end{figure}

In the forward region, the complementary station is placed behind a Ring Imaging Cherenkov (RICH) detector. The RICH material budget was provided by the PID detector working group~\cite{cisbani} and it corresponds to a dual-radiator (aerogel and C$_2$F$_6$) device.
The resulting momentum resolutions are shown in Fig.~\ref{fig:gem_forw}.
In this region, a 50\,\textmu${\rm m}$-resolution GEM provides a momentum-resolution enhancement comparable to that of a 10\,\textmu${\rm m}$-pixel silicon disk except in the highest momentum bin. Both provide a small improvement in the resolution in the far-forward region for momenta above 5 GeV/$c$.
In practice, the magnetic-field lines are expected to be shaped in such a way that bending inside the $\sim150$-cm-long RICH will be minimal, and the momentum-resolution improvement will be smaller in the forward region. More detailed magnetic-field simulations are needed to study this effect.
The complementary tracking stations were simulated with acceptance in the region $|\eta|>1.2$. Nevertheless, it can be seen in Figs.~\ref{fig:gem_back}~and~\ref{fig:gem_forw} that their impact is negligible for $|\eta|<2.5$. Thus, smaller tracking stations can be constructed to complement the tracker in the EIC.

\subsection{Pointing resolution}
\label{sec:pointing}

\begin{figure}[htbp]
    \centering
    \includegraphics[width=0.95\textwidth,page=3]{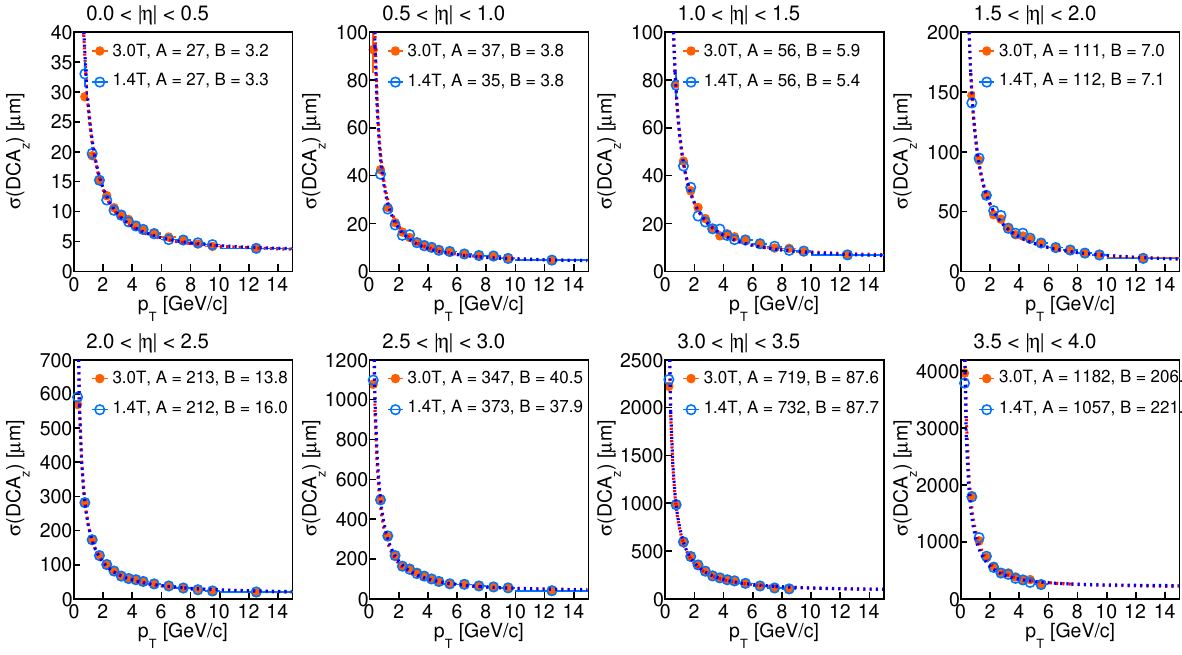}
    \caption{Longitudinal DCA resolution vs. transverse momentum for several pseudorapidity bins.
    The circles correspond to the resolutions extracted from the simulations, and the lines correspond to fits to the resolution of the form $\sigma({\rm DCA}_z)=A/p_T\oplus B$,
    with the parameters
    $A \ [$\textmu${\rm m} \cdot {\rm GeV}/c]$ and $B \ [$\textmu${\rm m}]$ given in the figure. 
    The orange, filled and blue, open circles correspond to simulations using the BeAST (3.0~T) and BaBar (1.4~T) field maps.}
    \label{fig:all_si:dca_z_res_param}
\end{figure}

\begin{figure}[htbp]
    \centering
    \includegraphics[width=0.95\textwidth,page=4]{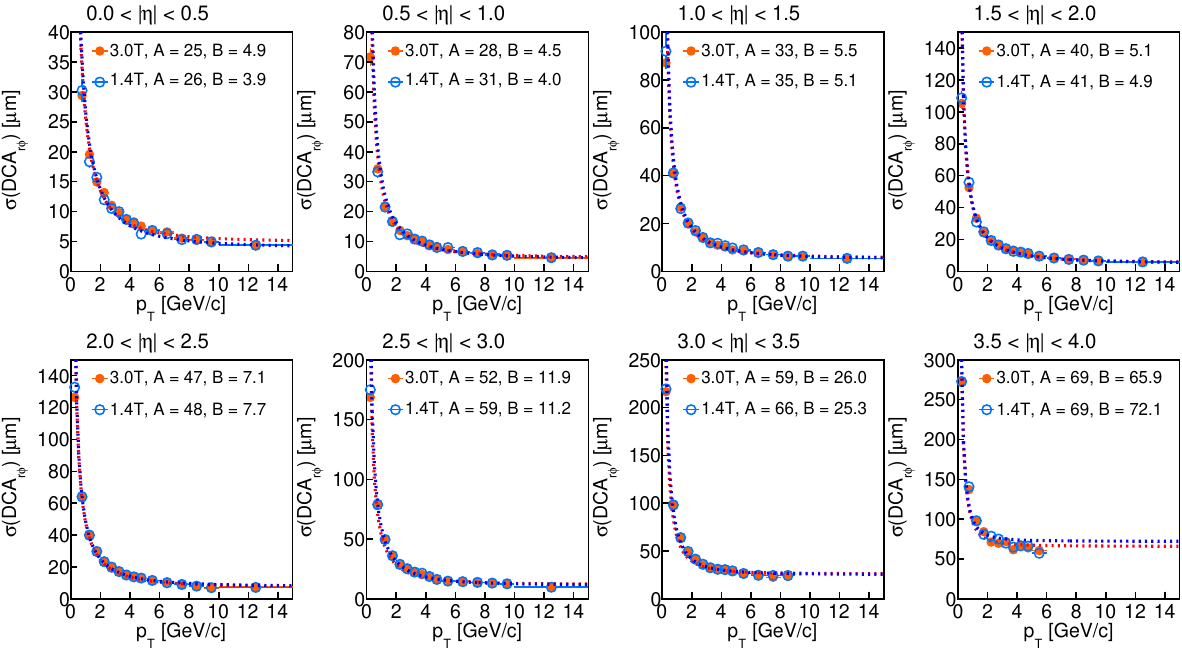}
    \caption{Transverse DCA resolution vs. transverse momentum for several pseudorapidity bins.
     The circles correspond to the resolutions extracted from the simulations, and the lines correspond to fits to the resolution of the form $\sigma({\rm DCA}_{r\phi})=A/p_T\oplus B$,
     with the parameters
    $A \ [$\textmu${\rm m} \cdot {\rm GeV}/c]$ and $B \ [$\textmu${\rm m}]$ given in the figure. 
    The orange, filled and blue, open circles correspond to simulations using the BeAST (3.0~T) and BaBar (1.4~T) field maps.}
    \label{fig:all_si:dca_t_res_param}
\end{figure}

In addition to measuring the momenta of particles, the silicon tracker must be able to reconstruct secondary vertices and project track trajectories to the outer detector systems.
The Distance of Closest Approach (DCA) is defined as the spatial separation between the primary vertex and the reconstructed track projected back to the $z$ axis (DCA$_{z}$) or to the $x-y$ plane (DCA$_{r\phi}$). The DCA resolutions were determined as the standard deviation of
normal functions fitted to the DCA$_{z}$ and DCA$_{r\phi}$ distributions.
DCA-resolution results as a function of transverse momentum ($p_T$) for pions are shown in Figs.~\ref{fig:all_si:dca_z_res_param}~and~\ref{fig:all_si:dca_t_res_param}.
The resulting distributions were characterized via fits with the functional form $\sigma({\rm DCA})=A/p_T\oplus B$.
The fits and fit parameters are presented in the figures. It is clear that the DCA resolutions are insensitive to the magnetic field.

\begin{figure}[htbp]
    \centering
    \includegraphics[width=0.45\textwidth,page=3, trim={0 3mm 0 12mm}, clip]{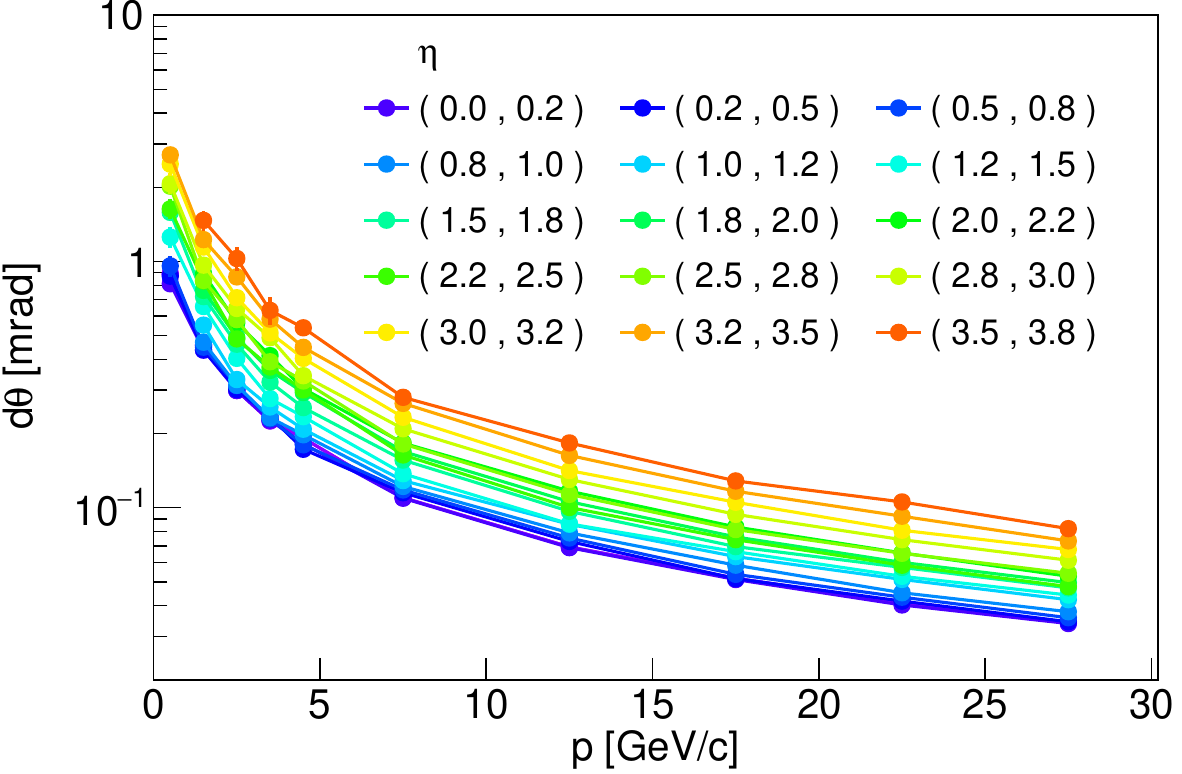}
    \includegraphics[width=0.45\textwidth,page=4, trim={0 3mm 0 12mm}, clip]{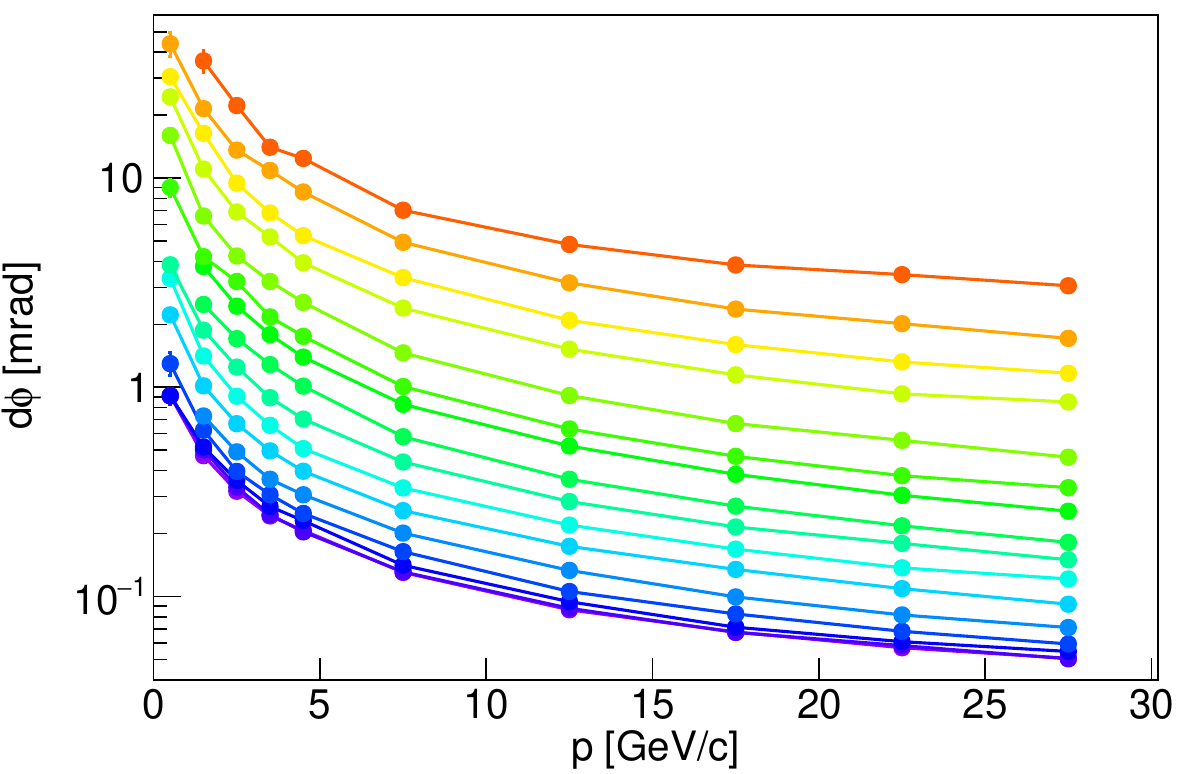}
    \caption{Polar (left) and azimuthal (right) angular resolutions at the vertex as a function of momentum for several pseudorapidity bins for pions in the BeAST (3.0~T) magnetic field. These distributions are largely insensitive to the magnetic-field.}
    \label{fig:ang_res}
\end{figure}

The polar and azimuthal angular resolutions are determined as the standard deviation of
normal functions fitted to the $\Delta \theta \equiv \theta_{truth}-\theta_{reco}$ and $\Delta \phi \equiv \phi_{truth}-\phi_{reco}$ distributions, respectively.
Figure~\ref{fig:ang_res} shows the polar and azimuthal angular resolutions at the vertex as a function of momentum for several pseudorapidity bins. While these graphs were extracted from simulations with the BeAST (3.0~T) magnetic field, the angular resolutions are largely insensitive to the magnetic field.

\begin{figure}[htbp]
    \centering
    \includegraphics[width=0.45\textwidth, page=1, trim={0 3mm 0 12mm}, clip]{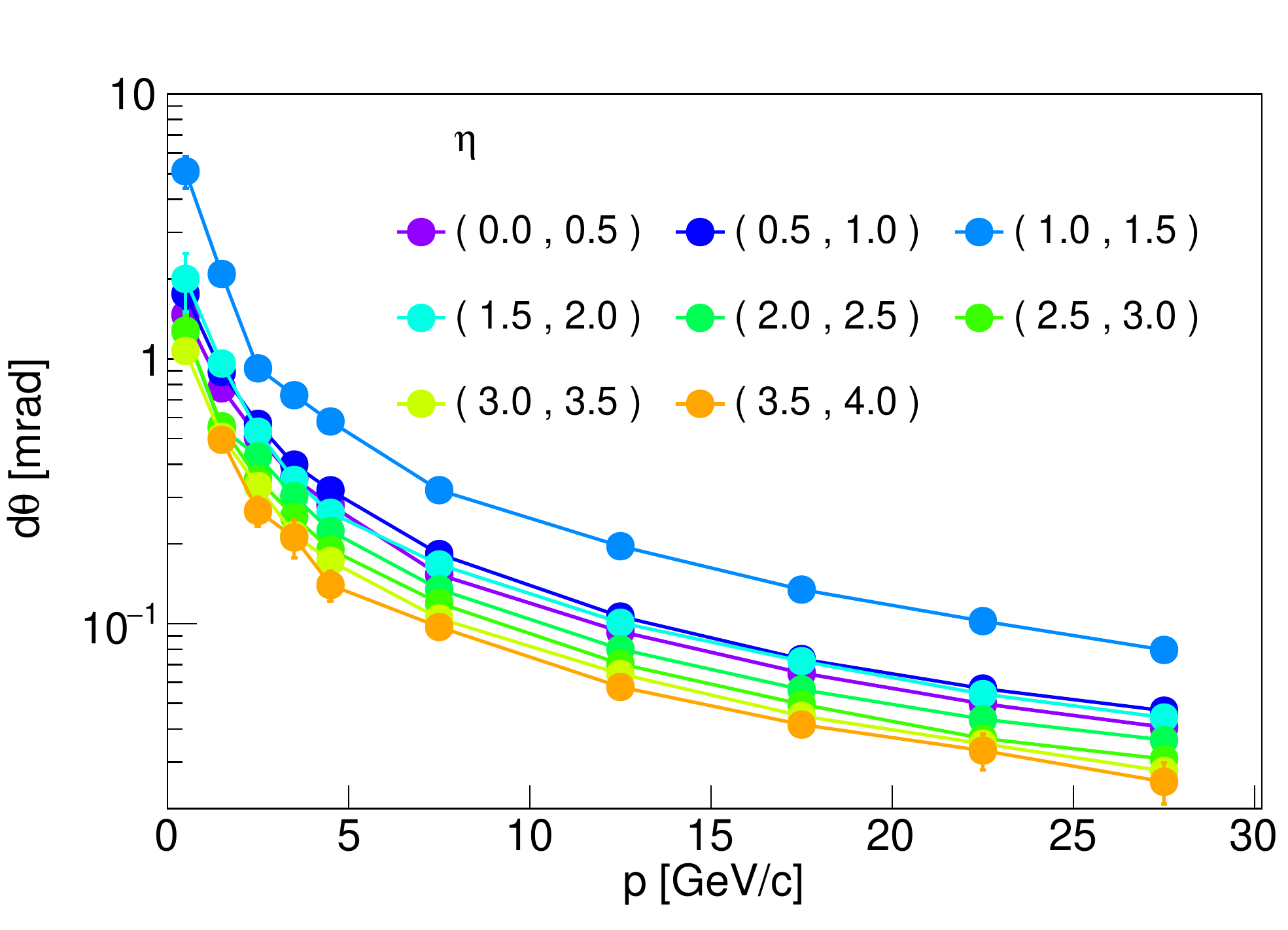}
    \includegraphics[width=0.45\textwidth, page=2, trim={0 3mm 0 12mm}, clip]{figs/results_res_at_PID_10um_separate.pdf}
    \caption{Polar (left) and azimuthal (right) angular resolutions at the location of PID detectors as a function of momentum for several pseudorapidity bins for pions in the BeAST (3.0~T) magnetic field. These distributions are overall insensitive to the magnetic-field intensity.}
    \label{fig:ang_res_at_pid}
\end{figure}

An important function of an EIC general-purpose tracker is aiding in particle identification (PID). Specifically, a good angular resolution is needed at the spatial coordinates corresponding to the entrance of Cherenkov detectors, since these detectors rarely measure the trajectory of tracks.
To study this resolution, the reconstructed momenta were projected onto a cylindrical surface of radius equal to 50\,cm and length along the $z$ axis of 260\,cm and were compared to the truth information at the same location.
Figure~\ref{fig:ang_res_at_pid} shows the resulting angular resolutions.

\begin{figure}[htbp]
    \centering
    \includegraphics[width=0.45\textwidth]{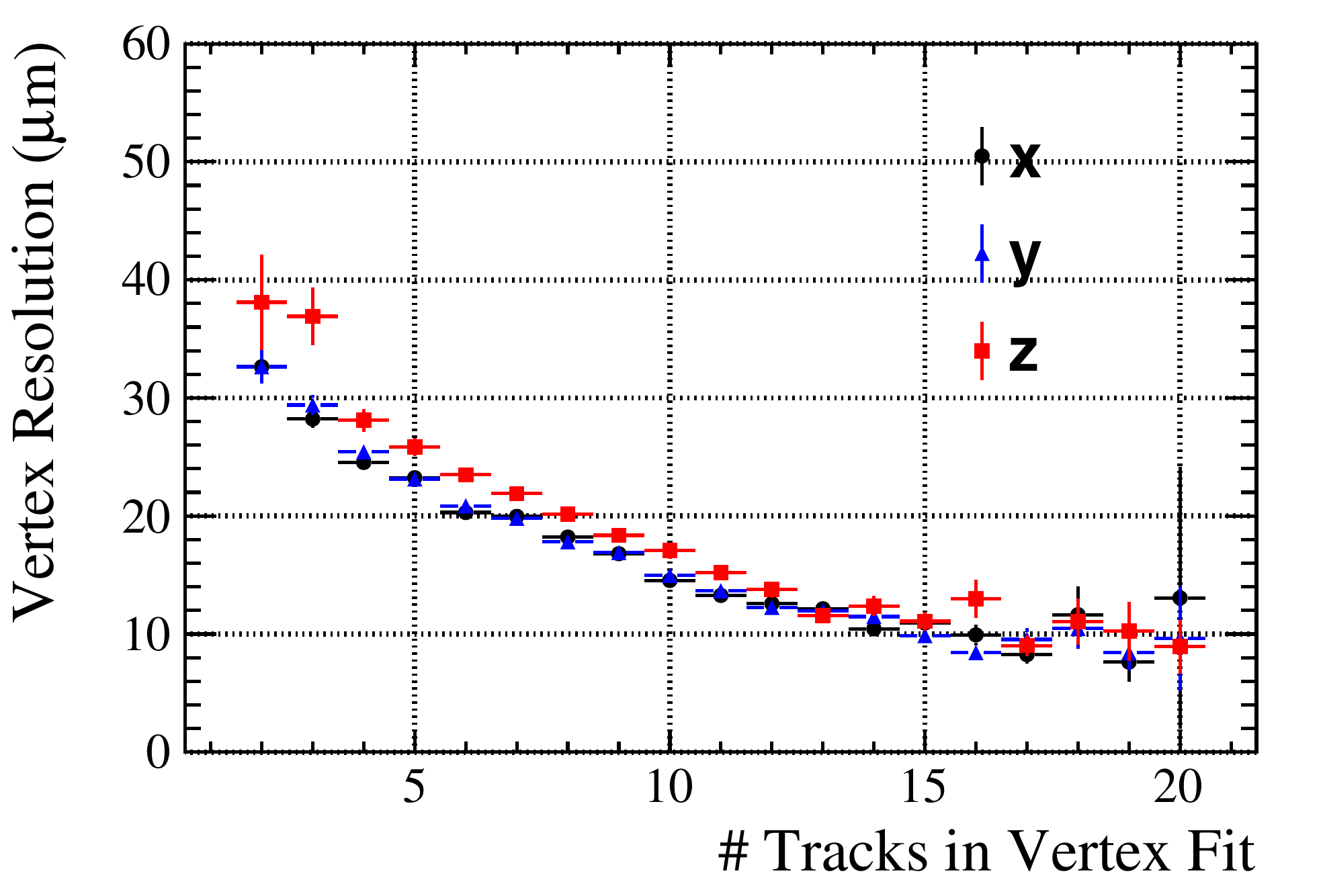}
    \hspace{0.2in}
    \includegraphics[width=0.45\textwidth]{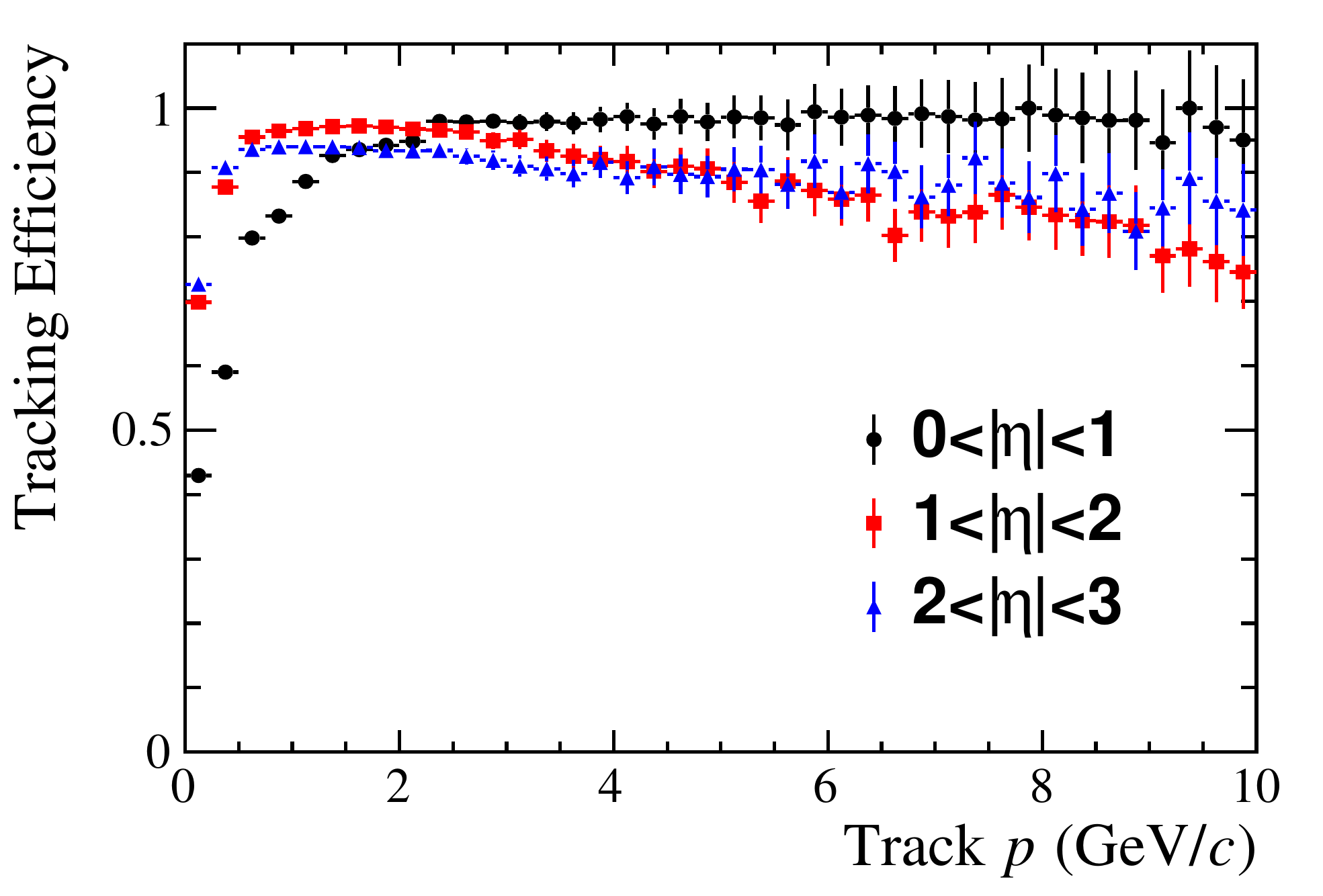}
    \caption{Left: primary vertex resolution determined in the full simulation setup with PYTHIA $e$+$p$ events at 18$\times$275 GeV collisions with an event level selection of $Q^{2}>$1 GeV$^{2}$. Right: Tracking efficiency determined in the full simulation for three different $\eta$ regions. Both figures incorporate events generated in a 3.0~T magnetic field.}
    \label{fig:fullsim:vtx}
\end{figure}

Primary vertex resolutions in three dimensions and 
a first look at 
tracking efficiency are determined by generating PYTHIA $e$+$p$ events at 18$\times$275 GeV collisions, reconstructing the final-state particles in the full-simulation detector, and fitting the vertex residual distribution with respect to the generated truth vertex with a Gaussian function. Figure~\ref{fig:fullsim:vtx}, left plot, shows the resulting primary vertex resolution as a function of track multiplicity in events with $Q^{2}>$1 GeV$^{2}$. One can see the event primary vertex resolution is typically $\sim$25\,$\mu$m at a multiplicity of $\sim$5, the average number of charged particles produced within the acceptance of tracking detectors in these collisions. Figure~\ref{fig:fullsim:vtx}
right plot shows the charged pion tracking efficiency in different $\eta$ regions, which shows reasonable tracking efficiencies over a broad kinematic region.
As stated before, pattern recognition is seeded
using truth-track information. As a result, the extracted quantity constitutes a best-case-scenario and is expected to describe the detector performance only in very-low-multiplicity events.
These tracking efficiencies obtained from the full simulation were applied in the following performance projection studies through fast simulation.

The $p_T$ threshold can be estimated from $p_T\,[{\rm GeV}/c] = 0.3 \cdot B\,[{\rm T}] \cdot r/2\,[{\rm m}]$, derived from Newtonian mechanics.
Here, $r$ corresponds to the radius of curvature of the track, following a circular trajectory in the magnetic field of intensity $B$. The threshold for a track to reach the outer layer of the detector ($r\geq 43.23$\,cm) in a 3.0\,T (1.4\,T) uniform solenoidal magnetic field corresponds to $p_T \geq$ 195 (90) MeV/$c$ in this estimate. We can consider a lower threshold corresponding to particles that reach the third barrel layer ($r\geq 21.00$\,cm), since three is the minimum number of hits needed for a momentum reconstruction. Clearly, not reaching the outer layers has a negative impact on the resolution of such particles. In this case, the threshold in a 3.0\,T (1.4\,T) uniform solenoidal magnetic field 
would correspond to $p_T \geq$ 95 (44) MeV/$c$.
However, energy-loss and multiple scattering, in particular for non-relativistic particles, lead to higher thresholds than the values estimated above.
We have incorporated the efficiencies from full Fun4All simulations in our heavy quark studies and conservative values for the thresholds in the studies of jets, while the vector meson studies are based directly on full simulations albeit in the older 
EICroot framework~\cite{EICroot}.

In this section, simulations were carried out with magnetic-field maps
incorporating a gradual decrease in the magnetic-field strength with increasing distance from the nominal interaction point in the $z$ direction.
Furthermore, the BaBar magnet, which is a candidate solenoid for the EIC, peaks at $B = 1.4$\,T. In some of the studies presented in this document, parametrizations from perfectly-solenoidal fields determined for $B = 1.5$\,T are used.
The change between $B = 1.4$\,T and $1.5$\,T leads to a 7$\%$ difference in the momentum resolution. Additionally, differences between realistic field maps and uniform solenoids lead to $\sim10\%$ differences at high-$|\eta|$. These differences should not affect the conclusions reached in each section.

%% file: charm_intro_recon.tex
\subsubsection{Physics Introduction}\label{sec:hq:intro}

\begin{figure}[htb]
    \centering
    \includegraphics[width=0.5\textwidth, height=6.5cm]{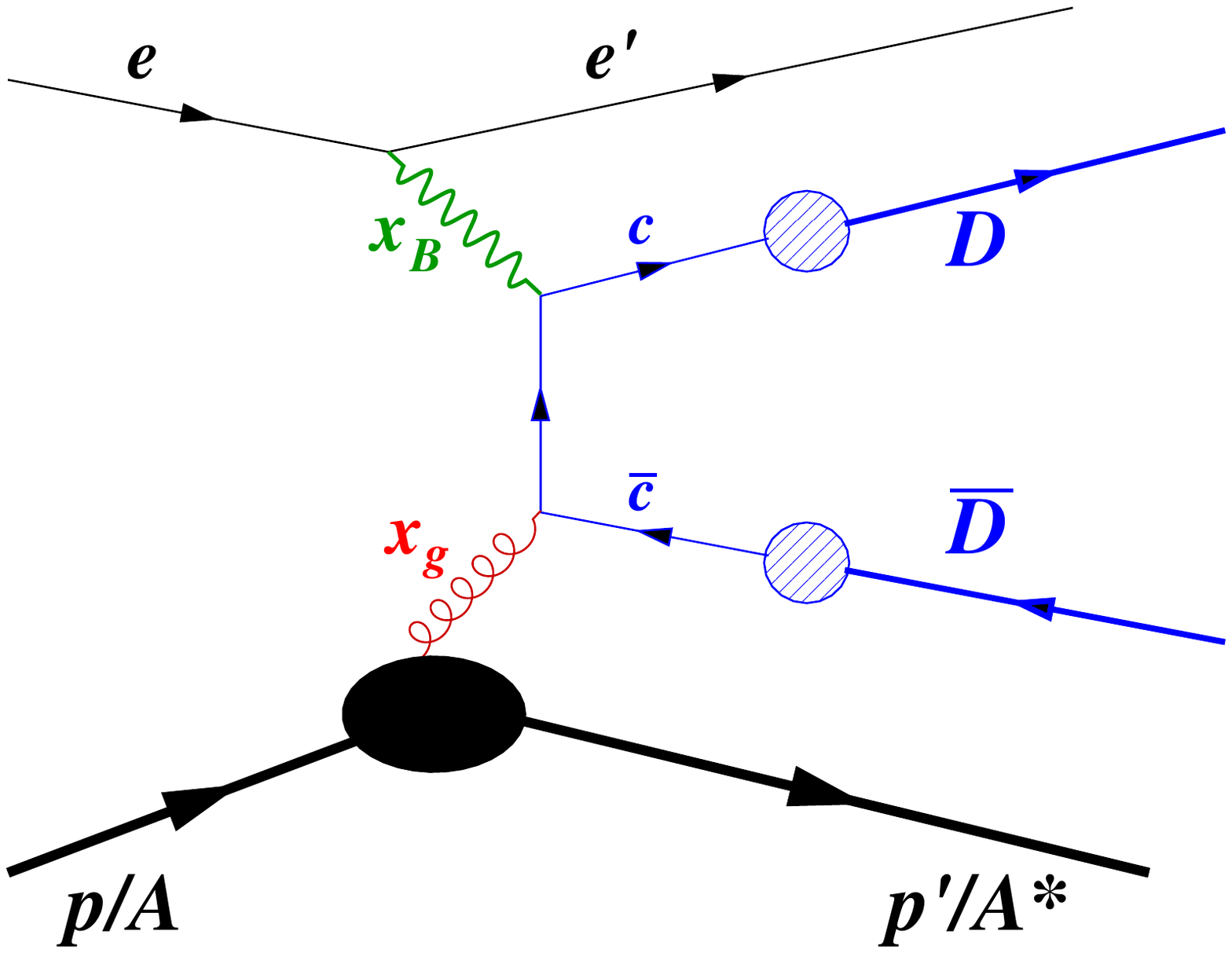}
    \caption{Leading-order diagram for charm-anti-charm pair production in $e$+$p$/A deep inelastic scatterings (DIS).}
    \label{fig:diagram}
\end{figure}

Heavy quarks are produced through photon-gluon fusion (PGF) at the leading order in high-energy $e$+$p$/A deep inelastic collisions: $\gamma^*+g\rightarrow Q+\overline{Q}$, see Fig.~\ref{fig:diagram}. Therefore, heavy-quark production via deep inelastic scattering (DIS) has unique sensitivity to the gluon distributions in the nucleon or nucleus, and can elucidate the QCD dynamics 
of heavy compared to light flavor quarks.

There have been prior studies of
heavy-flavor measurements to investigate gluon distributions at the EIC~\cite{Aschenauer:2017oxs,Chudakov:2016ytj,Wong:2020xtc}. In the study in Ref.~\cite{Aschenauer:2017oxs}, charm events were tagged by secondary charged kaon tracks while Ref.~\cite{Chudakov:2016ytj} reported secondary vertex reconstruction of the decays of charm mesons. 
In the past few years, there have been rapid developments in both the EIC machine design/performance projections as well as  experimental detector R\&D, including tracking and vertexing. 
In this section, we include these new developments in the charm-meson simulations to quantify the physics reach offered by the all-silicon tracker concept presented in Sec.~\ref{sec:sim:full}.
We report secondary vertex reconstruction performance for charm-hadron decays to hadronic channels. We also discuss QCD background contributions in EIC collisions. The study sets a list of detector performance requirements, especially on the vertexing/tracking detectors. The GEANT-based simulation of the all-silicon-tracker described in Sec.~\ref{sec:sim:full} is used to 
benchmark a fast simulation, which then provides performance projections for the following physics objectives:

\begin{itemize}
    \item Inclusive heavy-flavor hadron production in unpolarized $e$+$p$/A collisions to constrain gluon (nuclear) parton distribution functions (PDFs) in nucleons and nuclei, especially in the large Bjorken-$x$ ($x_B$) region ($x_B\gtrsim0.1$).
    \item Heavy-flavor hadron pair (e.g. $D$+$\overline{D}$) production to constrain gluon transverse momentum dependent (TMD) PDFs in both unpolarized and transversely-polarized experiments.
    \item Heavy-flavor hadron double spin asymmetry ($A_{\rm LL}$) measurement to constrain the gluon helicity distributions ($\Delta g/g$).
    \item Heavy-flavor hadrochemistry (abundance between different heavy-flavor hadron states) studies to better understand heavy-quark hadronization as well as the impact of cold nuclear matter effects in $e$+$A$ collisions.
\end{itemize}

Our analysis re-affirms the potential impact of heavy-flavor measurements to constrain the gluon distribution functions in nuclear targets reported in Refs.~\cite{Aschenauer:2017oxs,Chudakov:2016ytj}. However, we now include detailed detector response simulations when evaluating the physics capabilities of the measurements.  The charm structure function of protons measured at the HERA $e$+$p$ collider~\cite{Abramowicz:1900rp} demonstrated the powerful reach of 
heavy-flavor measurements to constrain the gluon distribution in the proton. 
Measurements at the EIC will also probe the gluon distributions in nuclei, as also shown in Refs.~\cite{Aschenauer:2017oxs,Chudakov:2016ytj}.

Heavy-quark (hadron) pair production in DIS has attracted great attention in the last few years~\cite{Boer:2010zf,Metz:2011wb,Dominguez:2012ad,Dominguez:2011br,Zhu:2013yxa,Boer:2016fqd,Zhang:2017uiz,delCastillo:2020omr,Kang:2020xgk}.
Reconstructing the total transverse momentum of the pair can probe the TMD gluon distribution in the nucleon/nucleus. The TMD parton distribution provides an important aspect of the nucleon/nucleus tomography~\cite{Collins:2011zzd}, revealing the momentum distribution of partons not only in the longitudinal direction but also in the transverse direction. Among the TMD gluon distributions, previous studies have focused on the so-called linearly-polarized gluon distribution in a unpolarized nucleon~\cite{Boer:2010zf,Boer:2016fqd} and the gluon Sivers function in a transversely polarized nucleon~\cite{Accardi:2012qut}. The key for these proposed measurements at the EIC relies on the precision of the total transverse momentum of the hadron pair reconstructed from the decay products, along with the sensitivity to the gluon TMD from the nucleon/nucleus targets. This will be complementary to the dijet production at the EIC, which has also been explored~\cite{Zheng:2018ssm, Dumitru:2018kuw}.

The gluon helicity distribution is one of the major topics for the EIC. It can be well constrained from inclusive polarized structure function measurements over a wide kinematic range. Its impact on the gluon-spin contribution to the proton spin has been well documented in the EIC White Paper~\cite{Accardi:2012qut}. 
Extensive studies in the literature indicate that inclusive single-jet and dijet production provide complementary constraints on the gluon helicity distribution~\cite{Hinderer:2017ntk, Boughezal:2018azh, Page:2019gbf, Borsa:2020ulb,Borsa:2020yxh}. 
Our analysis of heavy-flavor production in DIS at the EIC shows that this measurement will improve an earlier measurement from the COMPASS experiment~\cite{Adolph:2012ca} and provide complementary constraints on the gluon helicity distribution function, in particular, in the moderate $x_B$ region.

Heavy-quark hadronization is a critical component in understanding the experimental heavy-flavor hadron production data. Recent experimental data on charm baryon $\Lambda_c^+$ production in hadronic collisions at RHIC and LHC show the $\Lambda_c^+/D^0$ is considerably larger than the fragmentation baseline constrained by the $e^+$+$e^-$ and $e$+$p$ data in the low to intermediate $p_T$ region~\cite{Adam:2019hpq,Acharya:2017kfy,Aaij:2013mga}. Several Monte Carlo hadronization models, e.g. color reconnection in PYTHIA~\cite{Bierlich:2015rha, Christiansen:2015yqa}, rope in DIPSY~\cite{Flensburg:2011kk} etc. have been studied in order to understand the $\Lambda_c^+/D^0$ ratio data. 
$e$+$p$/A collisions at high luminosity EIC experiments will allow us to have a better control on the initial condition compared to hadronic collisions, and enable detailed investigation in $\Lambda_c^+$ baryon production and how hadronization plays a role from $e^+e^-$ to hadronic collisions.

\subsubsection{Charm-Hadron Reconstruction}\label{sec:sim}

\paragraph{PYTHIA Event Generator:}\label{para:sim:setup}

Electron-proton ($e$+$p$) collisions are generated using the PYTHIA v6.4 event generator~\cite{Sj_strand_2006} with the explicit parameters used in these studies documented in~\cite{PYTHIA6}. Events are generated with vector-meson diffractive and resolved processes, semi-hard QCD 2$\rightarrow$2 scattering, 
neutral boson scatterings off heavy quarks within the proton, and photon-gluon fusion. The latter two are predominately responsible for heavy-quark production in $e$+$p$ collisions. Radiative corrections are not included for these studies except where explicitly mentioned.

\begin{figure}[htb]
    \centering
    \includegraphics[width=0.4\textwidth]{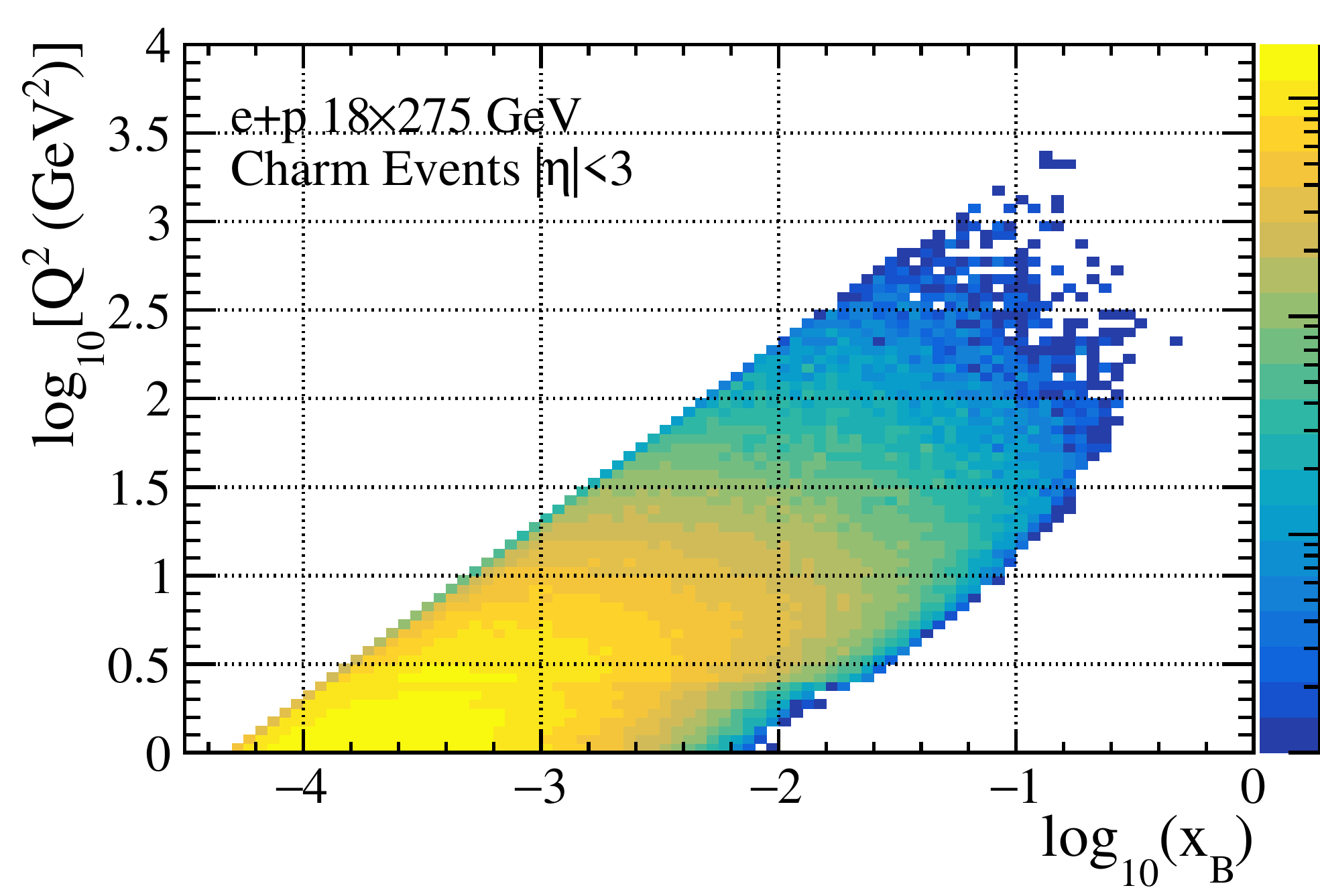}
    \includegraphics[width=0.4\textwidth]{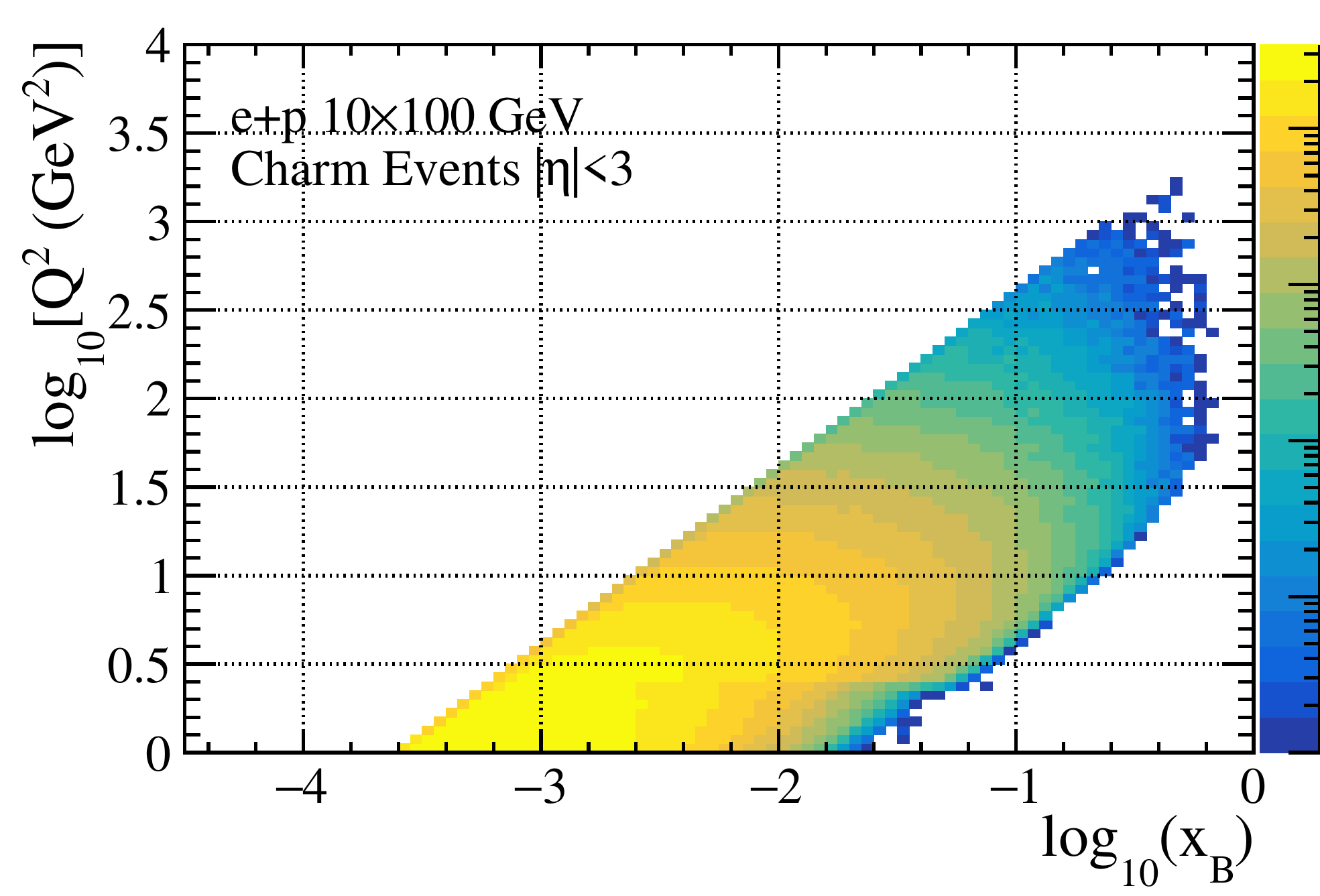}   
    \includegraphics[width=0.4\textwidth]{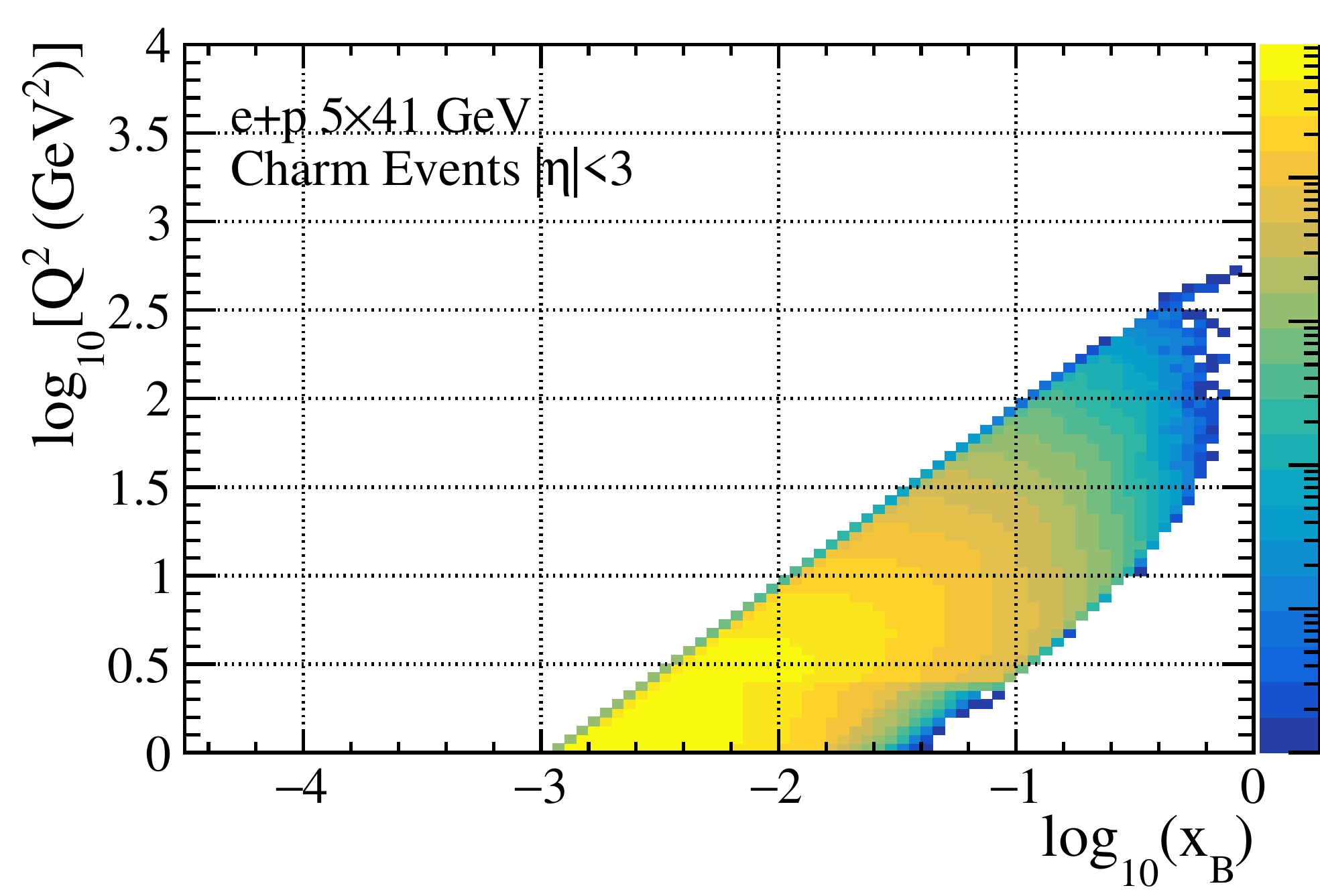}       
    \caption{Two-dimensional $\log_{10}(Q^{2})$ versus $\log_{10}(x_B)$ coverage for three beam-energy configurations in which each event contains at least one charm hadron with pseudorapidity $|\eta|<3$: 18$\times$275\,GeV, 10$\times$100\,GeV and 5$\times$41\,GeV, respectively.}
    \label{fig:sim:phasespace}
\end{figure}

In the one-photon-exchange approximation, an incoming electron of four momentum $e$ scatters into a final state $e'$ via the emission of a virtual photon of four momentum
$q=e-e'$, which subsequently interacts with the 
hadron beam with four momentum $p$.
We follow the ``HERA convention'', in which the hadron beam momentum is along the positive $z$ direction.
Several kinematical variables are typically used to characterize 
the scattering process.
The Bjorken scaling variable is defined as
$x_B \equiv Q^2/(2p \cdot q)$
and $Q^2\equiv -q^2$ is minus the square of the four momentum transfer.
The inelasticity is defined as $y\equiv p \cdot q /(p \cdot e)$.

The EIC program will run at multiple $e$+$p$(A) beam energy configurations and we have simulated several center-of-mass (CM) energies. Figure~\ref{fig:sim:phasespace} shows the $Q^{2}$ and $x_B$ reach for $e$+$p$ collisions in the 18$\times$275 GeV (electron and proton beam energies with head-on collisions), 10$\times$100 GeV, and 5$\times$41 GeV beam configurations in which at least one charm hadron is produced within pseudorapidity $|\eta|<3$ in the event. From these studies we can see that as the CM energy decreases at fixed-$Q^2$, the kinematic reach shifts to larger values of $x_B$.

\begin{figure}[htb]
    \centering
    \includegraphics[width=0.4\textwidth]{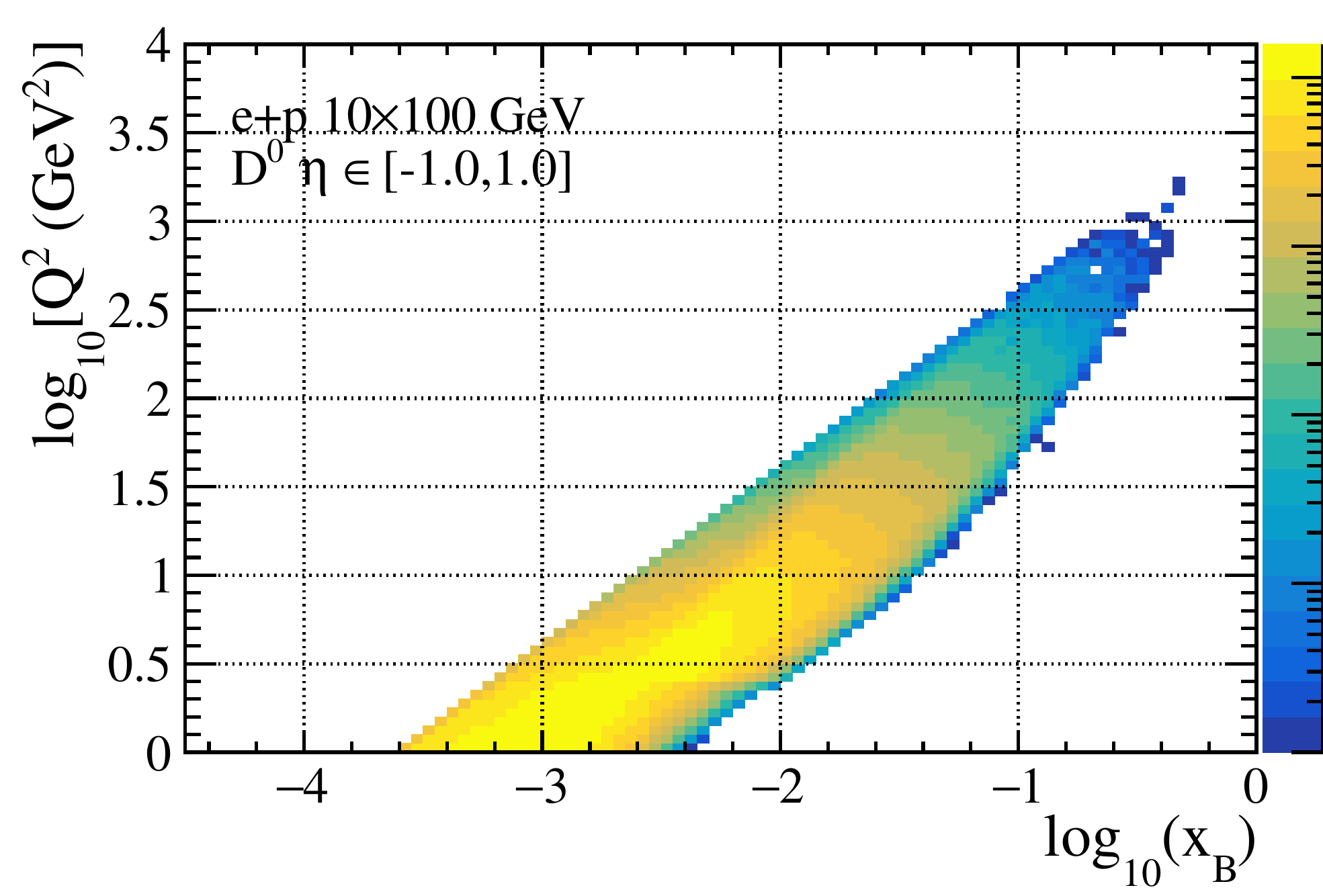}
    \includegraphics[width=0.4\textwidth]{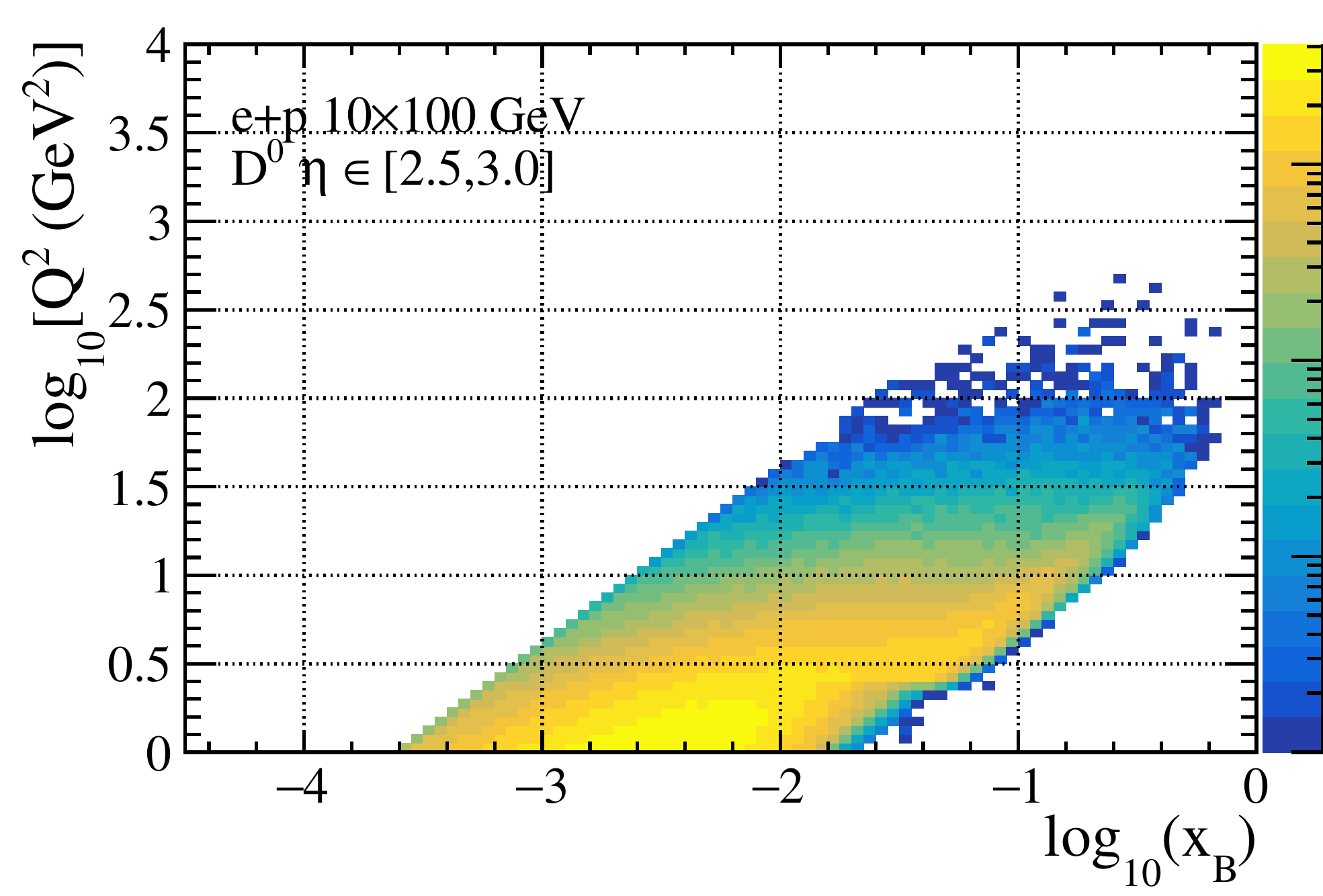}
    \caption{Two-dimensional $\log_{10}(Q^{2})$ versus $\log_{10}(x_B)$ coverage for $e$+$p$ collisions at 10$\times$100 GeV with events containing $D^{0}$ mesons in $-1<\eta<1$ (left) and $2.5<\eta<3.0$ (right), respectively.}
    \label{fig:sim:phasespace1}
\end{figure}

Figure~\ref{fig:sim:phasespace1} shows the $x_B$ and $Q^{2}$ coverage in slices of $D^{0}$ $\eta$. Comparing to the middle plot of Fig.~\ref{fig:sim:phasespace} one can see that the largest $Q^2$ events appear at lower $|\eta|$ while the large-$x_B$ events at fixed $Q^2$  correspond to large values of $\eta$. The fact that high-$x_B$ $D^{0}$ mesons tend to be produced at forward rapidity will, in effect, truncate the measurable cross section at low $Q^{2}$ and high $x_B$. Therefore, careful planning of which beam energies are suited for particular physics goals are needed. For example, as demonstrated in Ref.~\cite{Aschenauer:2017oxs}, high--$x_{B}$ charm structure function measurements have the strongest constraint on the gluon nPDFs compared to inclusive measurements. Therefore, charm structure function measurements at lower collision energies would have a higher impact on the gluon nPDFs. Moreover, since at least two beam energies are needed to extract the charm structure functions at fixed $Q^{2}$ and $x_{B}$, good kinematic overlap in $Q^{2}$ and $x_{B}$ between the two energies would be needed to enable the measurements over a broad kinematic range.  More details are discussed in Sec.~\ref{sec:Charmf2}.

\begin{figure}[htb]
    \centering
    \includegraphics[width=0.32\textwidth]{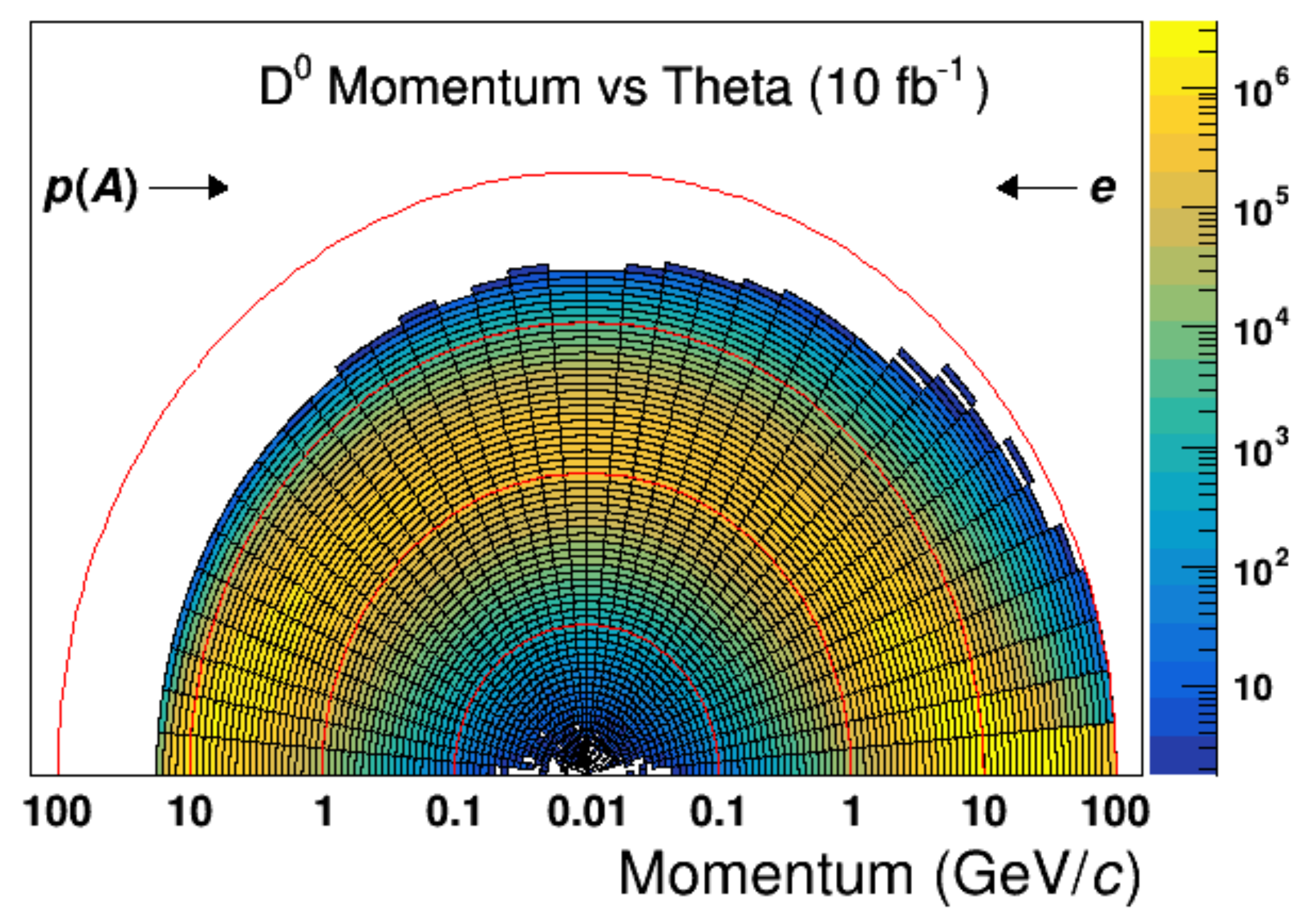}
    \includegraphics[width=0.32\textwidth]{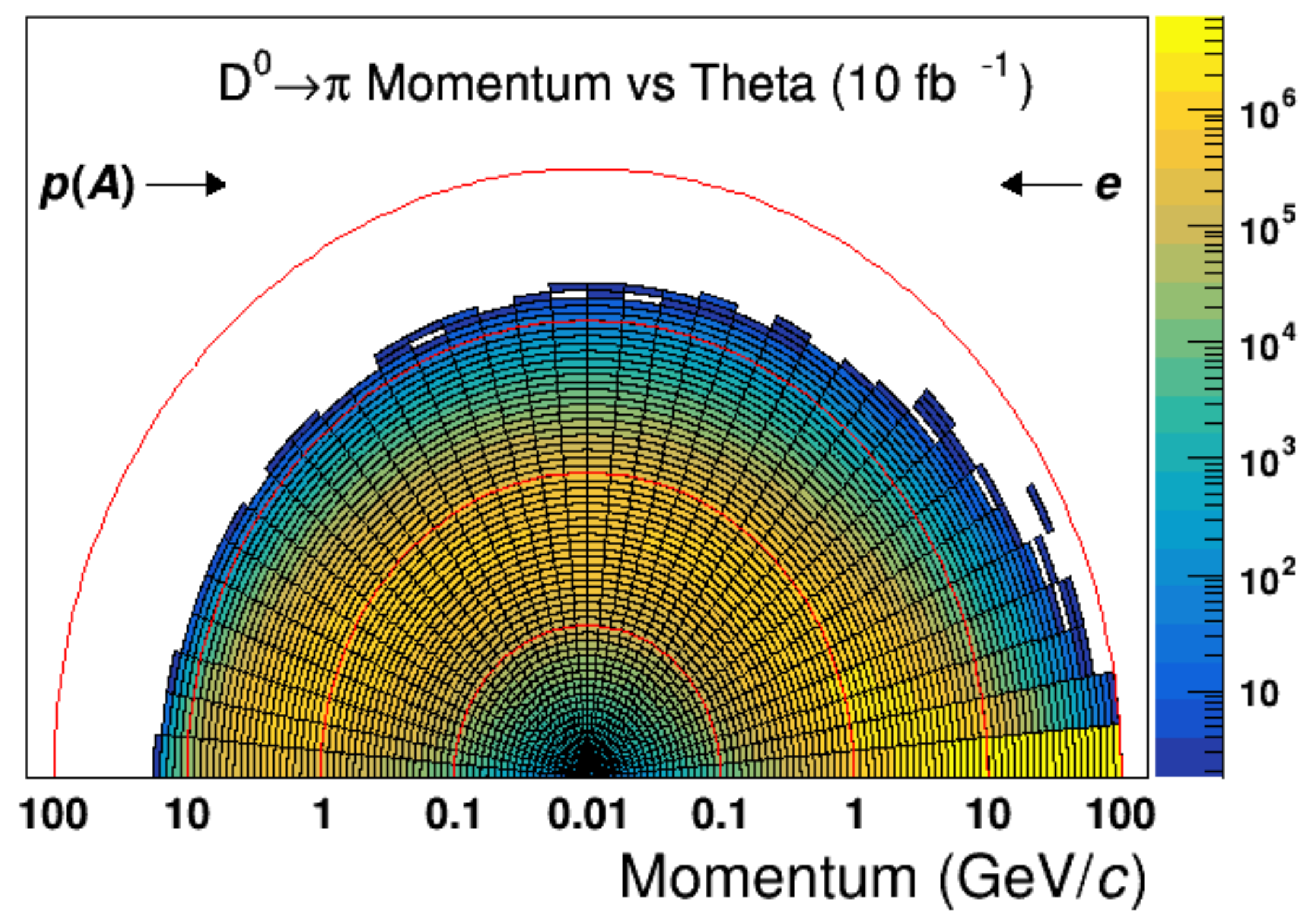}  
    \includegraphics[width=0.32\textwidth]{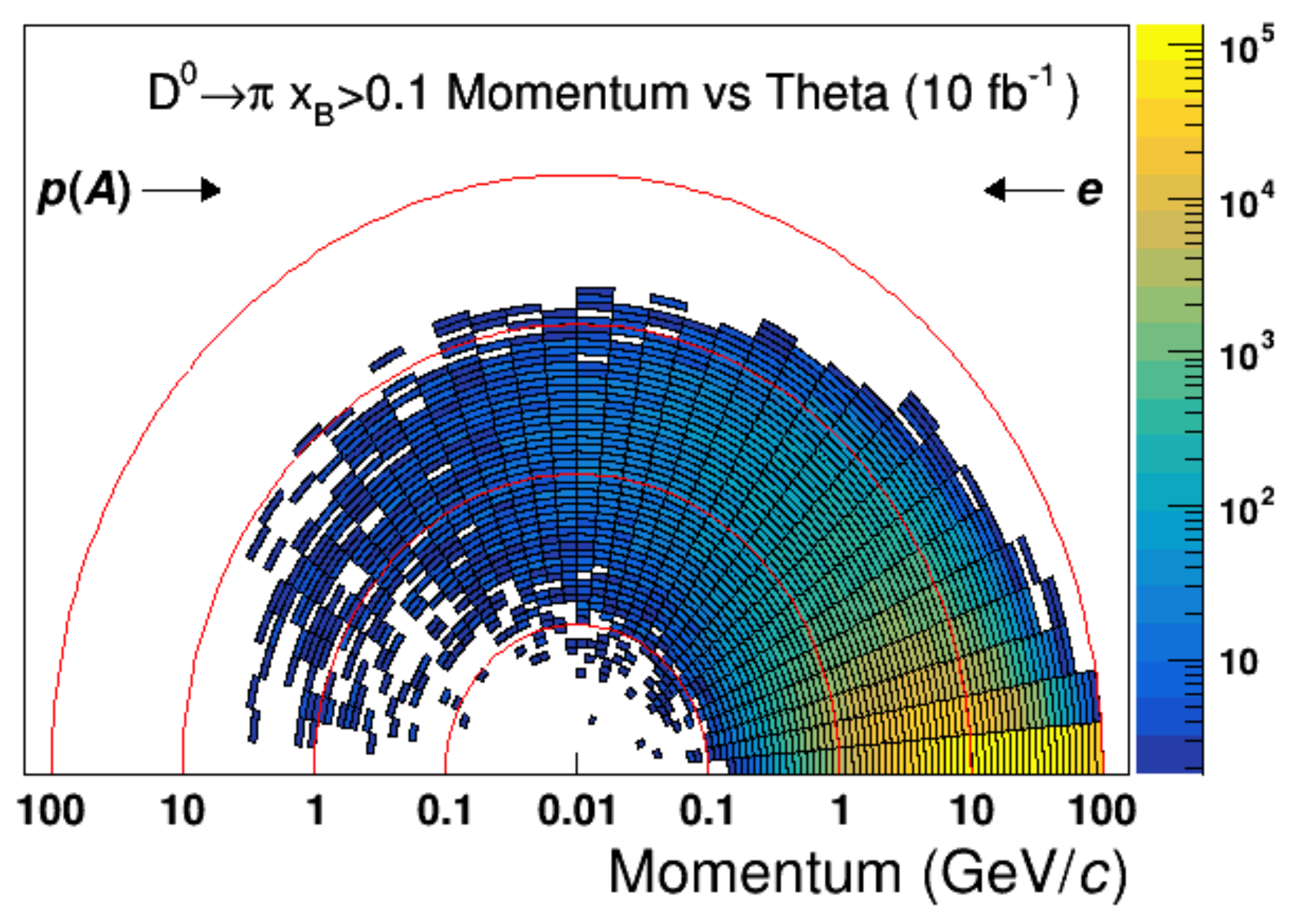}  
    \caption{Kinematic distributions, in polar coordinates, of $D^{0}$ mesons (left panel) and decay pions (middle and right panels) in 18$\times$275 GeV electron-proton collisions generated with PYTHIA 6. Each red semi-circle shows the absolute momentum scale at each order of magnitude as indicated by the $x$-axis intercept. The $z$-axis denotes the yield scaled to 10 fb$^{-1}$. The right panel shows the decay-pion distributions after applying an event-level cut requiring $x_B>0.1$.}
    \label{fig:sim:D0Kin}
\end{figure}

Charm hadrons and their decay particle distributions are studied in 18$\times$275 GeV electron-proton collisions. We show here those of $D^{0}$ mesons and $D^{0}$ decayed pions that are needed for reconstructing $D^0$ mesons experimentally; the general features of other charm hadrons (and bottom hadrons) and decay particles are similar. Figure~\ref{fig:sim:D0Kin} shows the momentum versus polar angle for both $D^{0}$ and decay pions. We additionally show the decay pion distributions after a event level cut of $x_B$$>$0.1. One can see that a large portion of charm hadrons (and decay pions) are produced with a pseudorapidity within -3 to 3, which is an approximate acceptance of the EIC detector design. It is also observed there is a slight rapidity asymmetry of the Charm cross section that is larger in the hadron-going direction. The pion distributions with and without an $x_B$ cut also show the correlation described in the text above, $i.e.$ high--$x_B$ events produce more charm hadrons in the forward region.

\paragraph{Fast Simulation for Charm-Hadron Reconstruction:}\label{sec:sim:fast}

The analyses and projections done in the next sections make use of a fast simulation to implement the detector responses and to allow generation of sufficient statistics to carry out  detailed studies. Our simulation of the heavy-flavor observables focuses primarily on the detector performance in the $|\eta|<3$ region that is enabled by the all silicon tracker described in Sec.~\ref{sec:sim:full}.

In this analysis, $D^0$ mesons are reconstructed through the decay channel $D^0\rightarrow K^-\pi^+$ with a branching ratio of $\sim$3.9\%. 
The fast simulation takes into account the track acceptance, momentum resolution, primary-vertex (PV) resolution and pointing (or DCA) resolution of tracks to properly reflect the detector conditions. The tracking acceptance and primary vertex resolution are taken from the full detector simulation described in Sec.~\ref{sec:performance}. The momentum resolution, pointing resolution and particle identification momentum limit parameters used in this analysis are taken from the latest detector-matrix table for the EIC detector requirements from the EIC Yellow Report Working Groups~\cite{DMtable:2020} and listed in Table~\ref{tab:sim:smearing}. The momentum resolution is implemented by smearing the momentum of the particles from the event generator with parametrized values. Two sets of smearing parameters are studied, corresponding to the resolution requirements under the 3.0~T and 1.5~T magnetic field configurations. 
The track DCA resolution is also included by smearing the DCA positions for charged particles from the event generator,
and the parameters are shown in Table~\ref{tab:sim:smearing}. We would like to point out that the momentum and DCA resolution performances from the proposed all-silicon tracker satisfy the physics requirement parameters used in these fast simulation studies, as shown in Sec.~\ref{sec:performance}.  For particle identification, we assume that charged kaon/pion/proton tracks have a clean separation up to a momentum limit listed in Table~\ref{tab:sim:smearing}.

\begin{table}[htb]
\centering
	\caption{Smearing parameters used in fast simulation in different $\eta$ bins: momentum resolution with two sets of magnetic-field configurations, DCA$_{r\phi}$ pointing resolution and particle identification (PID) momentum upper limits. All $p$ and $p_T$ values are in the unit of GeV/$c$.  \label{tab:sim:smearing}}
	\centering
	\begin{tabular}{c | c  c  c  c}
	$\eta$ & ~~~$\sigma_p/p$ - 3.0~T (\%)~~~ & ~~~$\sigma_p/p$ - 1.5~T (\%)~~~ & ~~~$\sigma(\rm DCA_{r\phi})$ ($\mu$m)~~~ & ~~~$p_{\rm max}^{\rm PID}$ (GeV/$c$)~~~ \\ \hline \hline
    ~~~(-3.0,-2.5)~~~ & 0.1$\cdot p$ $\oplus$ 2.0 & 0.2$\cdot p$ $\oplus$ 5.0 & 60/$p_T$ $\oplus$ 15 & 10 \\
	(-2.5,-2.0) & 0.02$\cdot p$ $\oplus$ 1.0 & 0.04$\cdot p$ $\oplus$ 2.0 & 60/$p_T$ $\oplus$ 15 & 10 \\
    (-2.0,-1.0) & 0.02$\cdot p$ $\oplus$ 1.0 & 0.04$\cdot p$ $\oplus$ 2.0 & 40/$p_T$ $\oplus$ 10 & 10 \\
	(-1.0,1.0) & 0.02$\cdot p$ $\oplus$ 0.5 & 0.04$\cdot p$ $\oplus$ 1.0 & 30/$p_T$ $\oplus$ 5 & 6\\
	(1.0,2.0) & 0.02$\cdot p$ $\oplus$ 1.0 & 0.04$\cdot p$ $\oplus$ 2.0 & 40/$p_T$ $\oplus$ 10 & 50 \\
	(2.0,2.5) & 0.02$\cdot p$ $\oplus$ 1.0 & 0.04$\cdot p$ $\oplus$ 2.0 & 60/$p_T$ $\oplus$ 15 & 50 \\
    (2.5,3.0) & 0.1$\cdot p$ $\oplus$ 2.0 & 0.2$\cdot p$ $\oplus$ 5.0 & 60/$p_T$ $\oplus$ 15 & 50 \\
	\end{tabular}
\end{table}

\begin{figure}[htbp]
    \centering
    \includegraphics[width=0.95\textwidth]{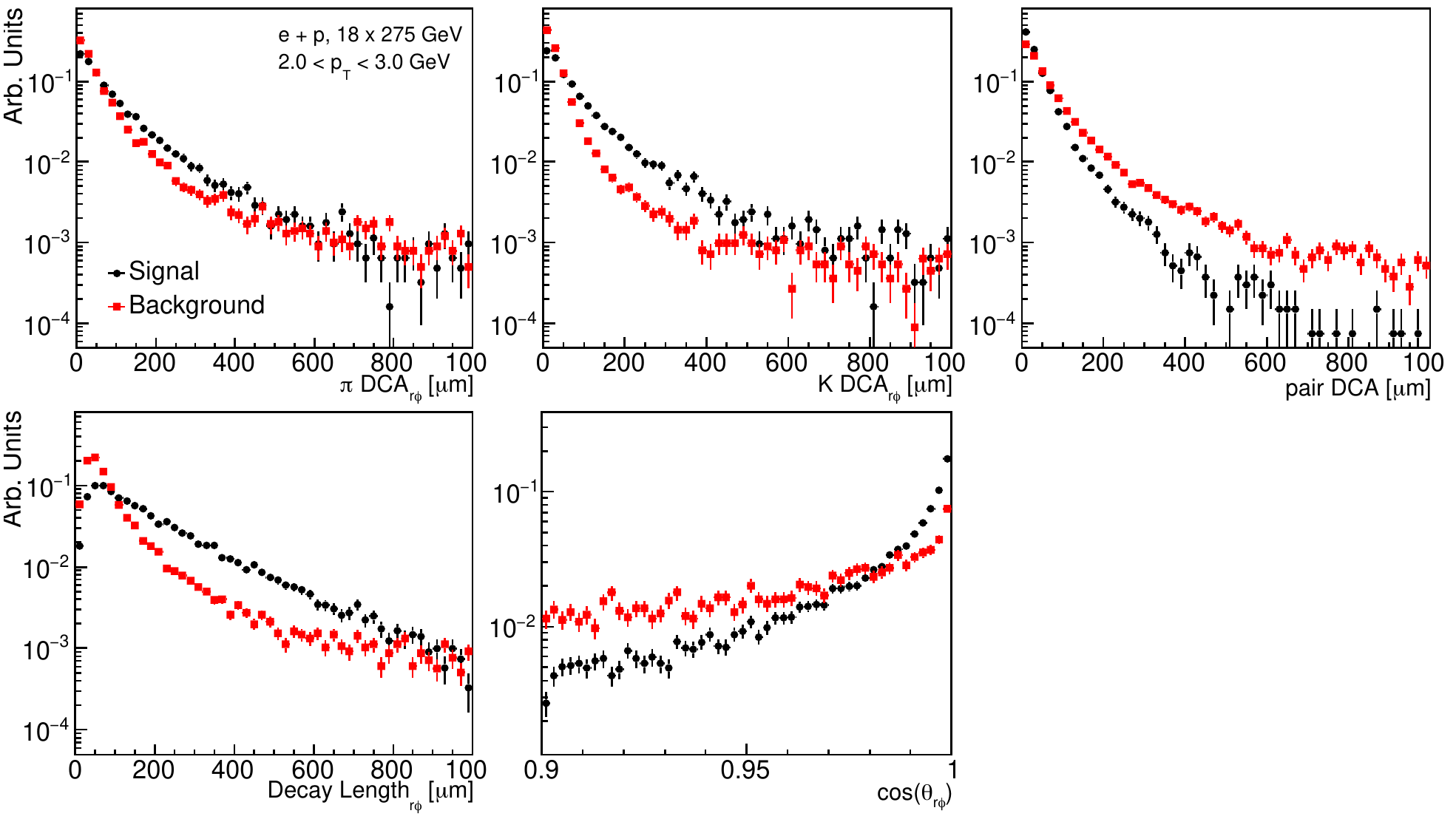}
    \caption{The distribution of variables characterizing the $D^0$-decay topology in the transverse direction, for signal (unlike sign) and background (like sign) candidates within 3$\sigma$ of the $D^0$ mass peak. The different panels from top left show distributions of DCA of pions to PV, DCA of kaons to PV, DCA between the $K\pi$ pairs, Decay Length and the cosine of the pointing angle to the PV, respectively. The candidate $D^0$ mesons have 2.0 $< p_T < 3.0$ GeV/$c$.}
    \label{fig:fs_topo}
\end{figure}

The distribution of different variables characterizing the $D^0$-decay topology is shown in Fig.~\ref{fig:fs_topo}, for tracks from signal (unlike sign $K^-\pi^+$ and $K^+\pi^-$ pairs) and background (like sign $K^+\pi^+$ and $K^-\pi^-$ pairs) candidates from within 3$\sigma$ of the $D^0$ mass peak. The distributions show a clear separation between the signal and background, allowing for an improvement in the signal-to-background ratio ($S/B$) and signal significance by placing cuts on the variables. The cuts used in the analyses in following subsections are shown in Table~\ref{tab:fs_cuts}.

\begin{table}[htb]
	\caption{Cuts on the decay-topology variables used for different $D^0$ $p_T$ bins. \label{tab:fs_cuts}}
	\centering
	\begin{tabular}{c | c  c  c }
	$p_T$ & ~~~Pair DCA$_{r\phi}$ ($\mu$m)~~~ & ~~~DecayLength ($\mu$m)~~~ & ~~~$\cos\theta_{r\phi}$~~~\\  \hline \hline
	0  $<p_T<$1.0 GeV/$c$ & $<$ 120 & - & - \\ 
	1.0$<p_T<$2.0 GeV/$c$ & $<$ 150 & $>$ 40 & -\\ 
	$p_T>$2.0 GeV/$c$ & $<$ 150 & $>$ 40 & $>$ 0.98\\
	\end{tabular}
\end{table}

\begin{figure}[htbpp]
    \centering
    \includegraphics[width=0.95\textwidth]{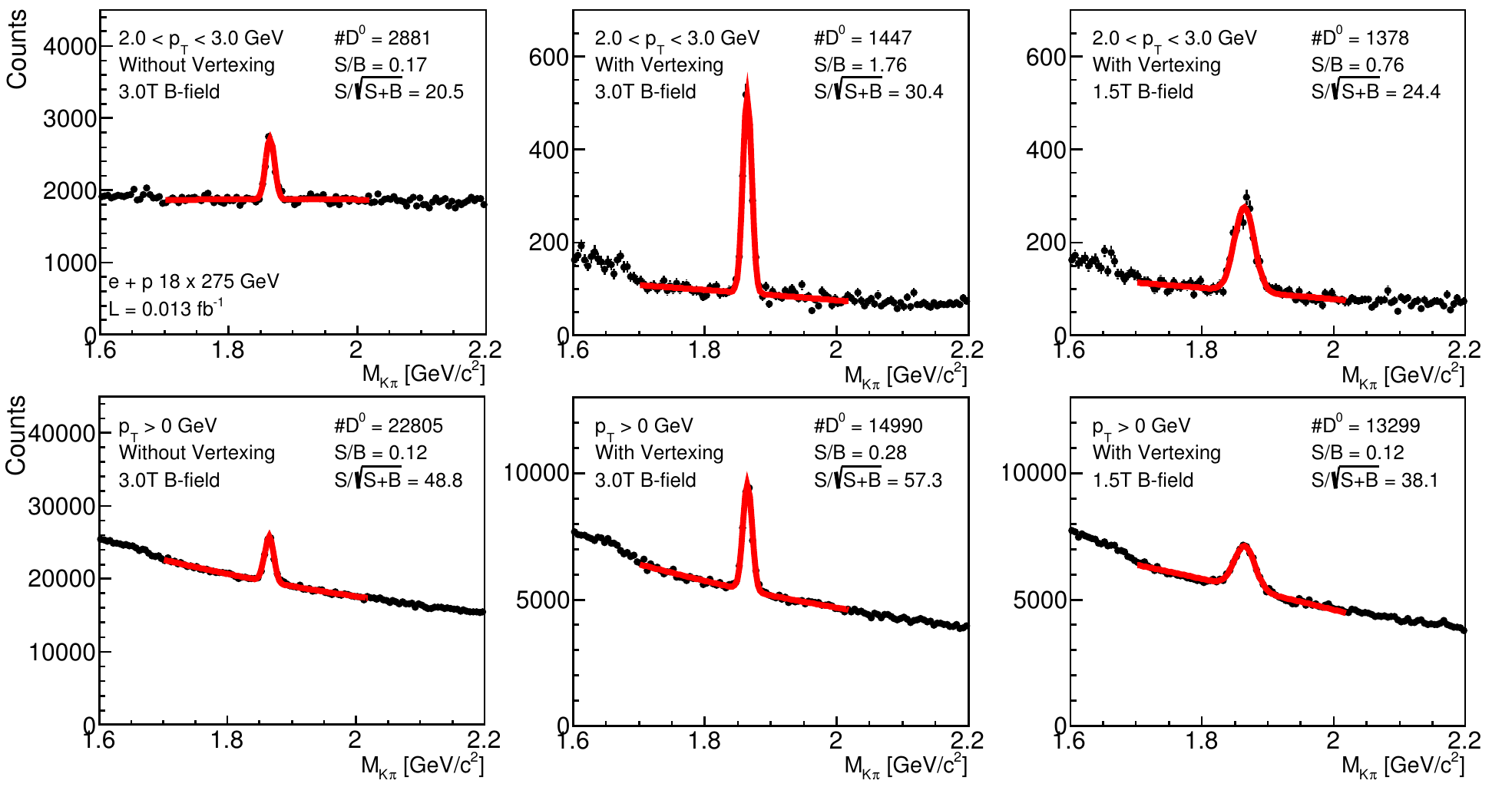}
    \caption{The $D^0$ meson invariant mass distributions, reconstructed through $K,\pi$ daughter pairs, for $2<p_T<3$\,GeV/$c$ (top row) and $p_T>0$ (bottom row). The plots are shown for cases: 3.0~T without topological cuts (left column), 3.0~T with topological cuts (middle column), and 1.5~T with topological cuts (right column)}
    \label{fig:fs_reco_vertexing}
\end{figure}

Figure~\ref{fig:fs_reco_vertexing} shows the reconstruction of $D^0$ mesons using tracks passed through the fast simulation without (left) and with selection cuts on the $D^0$-decay-topology variables with two momentum resolution requirements (middle and right). In all $p_T$ bins, applying cuts on $D^0$-decay-topology variables improves $S/B$ by a factor of $\sim10$ for $p_T > 2$\,GeV/$c$. The signal significance, defined as $S/\sqrt{S+B}$, is also improved for most bins, except for the lowest $p_T$ bin. For $p_T >$ 2 GeV/$c$, the improvement in significance is about 50\%, while the improvement is around 25\% for the $p_T$-integrated signal. 

The impact of different PID scenarios in $D^0$ reconstruction is also studied. 
The $S/B$ ratio increases by a factor of about 3.5 (2.5) and the signal significance by about 65\% (50\%) when going from a PID capability up to a momentum of 5\,GeV/$c$ within $-3<\eta<-1$ ($1<\eta<3$) to the PID scenario used in this simulation. The improvement for midrapidity $|\eta|<1$ region is minimal. The signal significance gains a further increase of about 35\% in backward rapidity region $-3<\eta<-1$ from the PID scenario used here to a perfect PID scenario while remains similar for mid- and forward rapidity regions.

The middle and right columns in Fig.~\ref{fig:fs_reco_vertexing} compare the reconstruction $D^0$ signal with two momentum resolution requirements (see Table~\ref{tab:sim:smearing}). As anticipated, the $D^0$ mass width with the 3.0~T resolution parameter is about half of that with 1.5~T resolution parameter. This results in the $p_T$-integrated $D^0$ significance with the 3.0~T configuration being about 50\% larger than the 1.5~T configuration.

Given that the all-silicon tracker has an outer radius of $r=43$\,cm, charged tracks must have at least $p_T \sim 0.2$\,GeV/$c$ to reach the last tracking layer in the central barrel. The impact of this threshold on the $D^0$ reconstruction is negligible, leading to a $<1\%$ reduction in $D^0$ significance. The $p_T$-threshold effect is implicitly included in the following physics simulation by applying the tracking acceptance obtained from the full simulation (right panel of Fig.~\ref{fig:fullsim:vtx}).

\begin{figure}[htbp]
    \centering
    \includegraphics[width=0.32\textwidth]{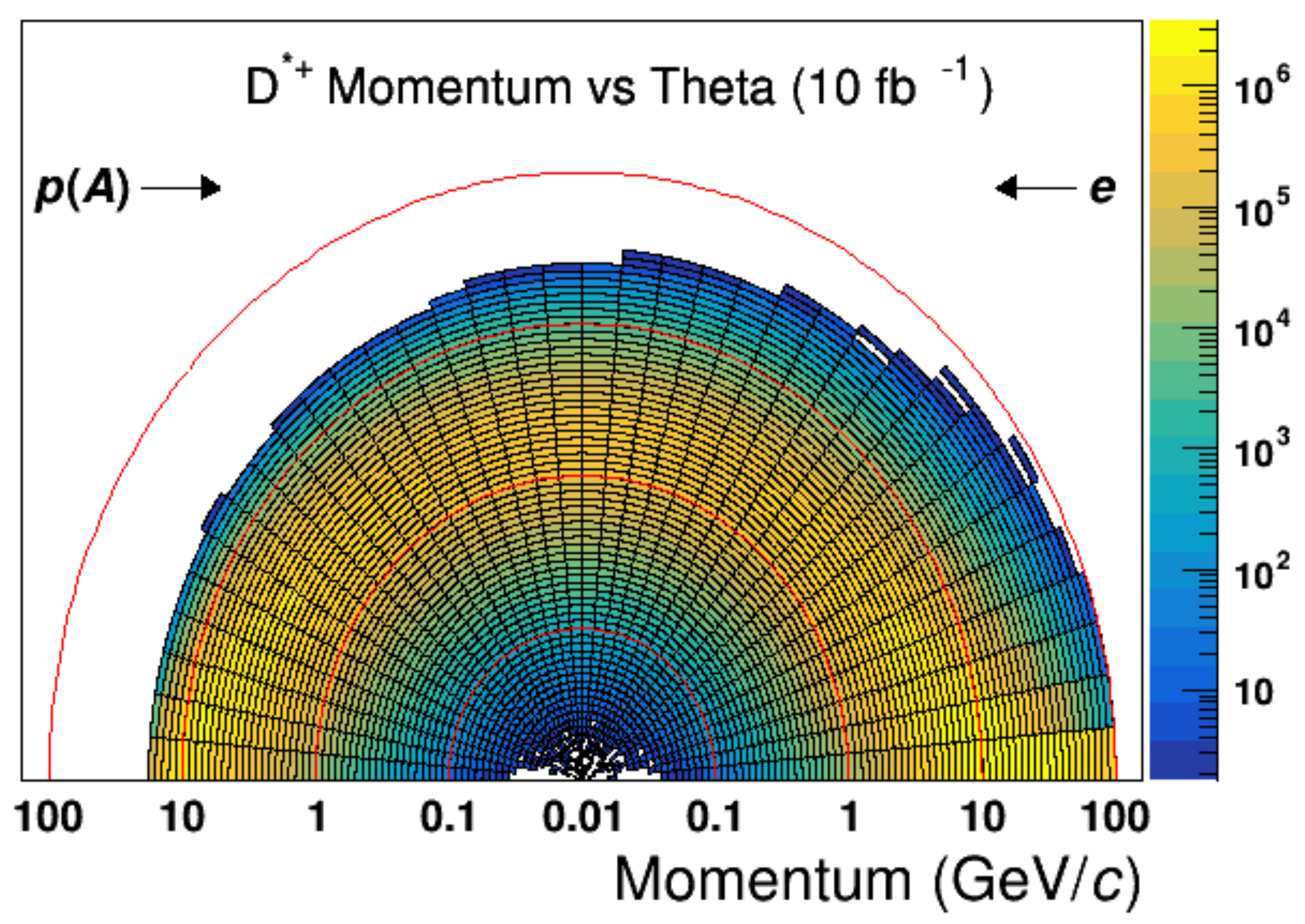}
    \includegraphics[width=0.32\textwidth]{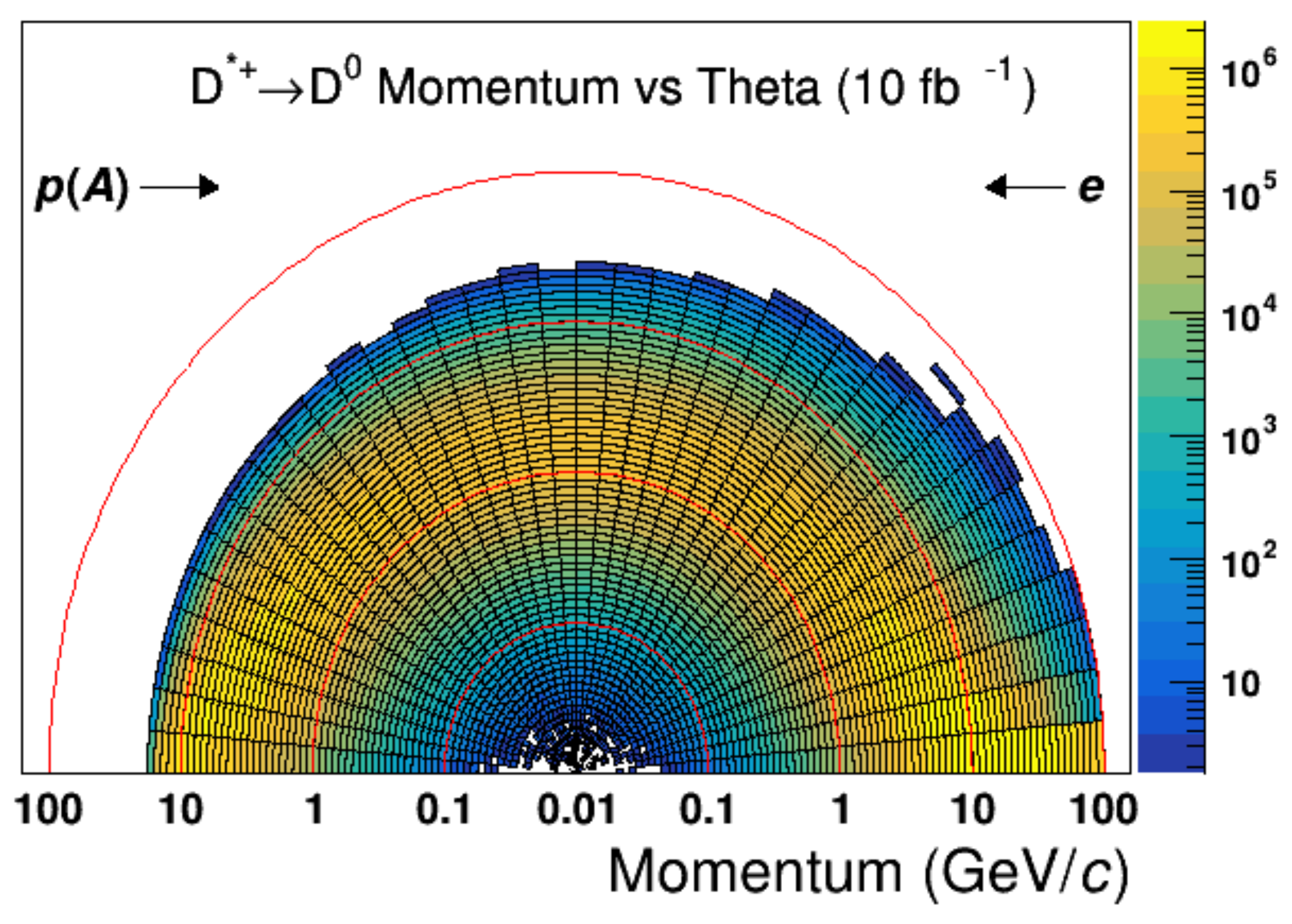}  
    \includegraphics[width=0.32\textwidth]{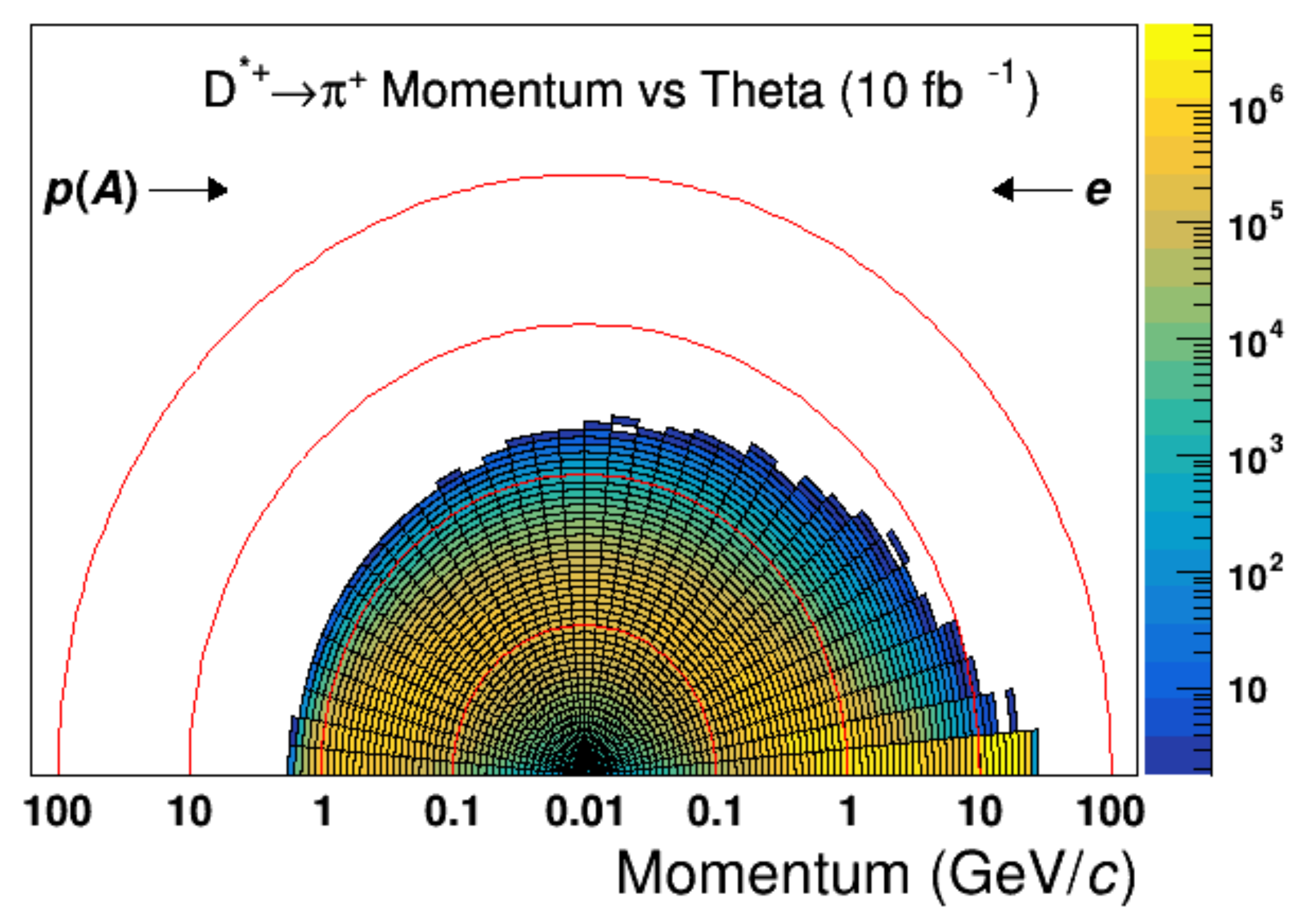}  
    \caption{Kinematic distributions, in polar coordinates, of $D^{*+}$ mesons (left), decay $D^{0}$ (middle) and pions (right) in 18$\times$275 GeV electron-proton collisions generated with PYTHIA 6. Each red semi-circle shows the absolute momentum scale at each order of magnitude as indicated by the $x$-axis intercept. The $z$-axis denotes the yield scaled to 10 fb$^{-1}$. 
    }
    \label{fig:sim:DstKin}
\end{figure}

\begin{figure}[htbp]
    \centering
    \includegraphics[width=0.45\textwidth]{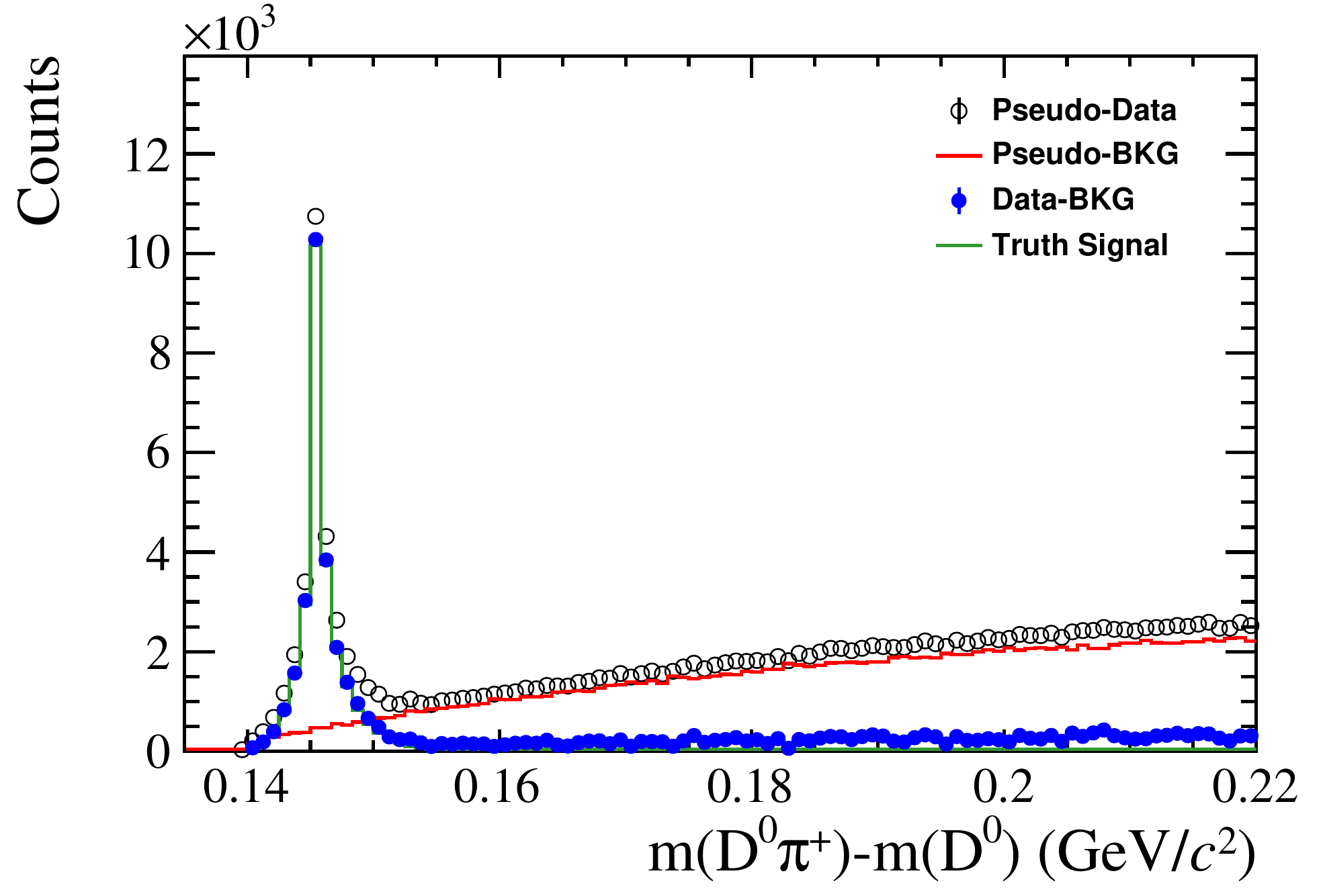}
    \includegraphics[width=0.95\textwidth]{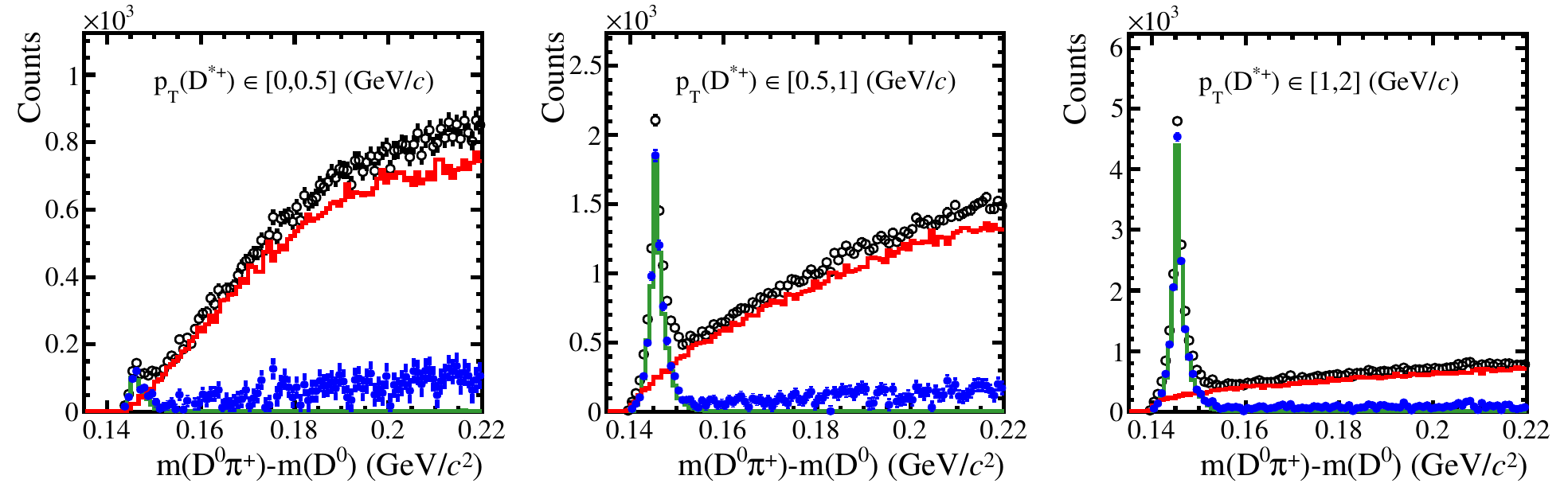}  
    \caption{$m(D^{0}\pi^{+})-m(D^{0})$ distributions for $D^{*+}$ candidates reconstructed using the method described in the text. The open black circles show all correct sign $D^{0}\pi^{+}$ combinations, and the red histogram shows a background estimation using like-sign $D^{0}$ candidates. The blue closed circles show the difference between the two, and the green histogram shows the truth-level $D^{*+}$ distribution. The top panel shows the $p_{T}$ integrated distribution, and the bottom three show three low $D^{*+}$ $p_{T}$ bins. The simulated integrated luminosity shown here is 0.056 fb$^{-1}$.}
    \label{fig:sim:DstDeltaM}
\end{figure}

Another charm channel studied with the fast simulation is $D^{*+}\rightarrow D^{0}\pi^{+}$. This decay is near threshold, and therefore the decay pion generally has a relatively low momentum (and is denoted as a $slow$ $pion$). Due to this feature, the quantity $\Delta m\equiv m(D^{0}\pi^{+})-m(D^{0})$ is typically used to extract the $D^{*+}$ yield as the signal peaks slightly above the pion mass, while combinatorial backgrounds peak at higher $\Delta m$. In this channel, the signal can be separated from background without the need for secondary vertex reconstruction. Conversely, the $p_{T}$ = 200 MeV/$c$ threshold for a track to hit all layers yields a 60\% efficiency for the slow pions from $D^{*+}$ decays. Figure~\ref{fig:sim:DstKin} shows the kinematic distributions of $D^{*+}$ mesons (left), decay $D^0$ (middle) and pions (rights) in $e$+$p$ 18$\times$275\,GeV collisions generated with PYTHIA 6.

We study the viability of reconstructing this channel in a scenario where the slow pion can be reconstructed below the 200 GeV/$c$ $p_{T}$ threshold using hits only within the first three barrel layers, which would lower the $p_{T}$ threshold down to about 100 MeV/$c$ and equate to a 90\% slow-pion acceptance efficiency. There are no current studies of the momentum resolution for such a scenario and for these studies we chose a conservative 10\%. $D^{*+}$ candidates are reconstructed in the simulation by first selecting $D^{0}\rightarrow K\pi$ combinations after the nominal fast-simulation smearing with decay tracks having $|\eta|<3$ and $p_{T}>$200 MeV/$c$, and with a pair invariant mass within 30 MeV of the $D^{0}$ PDG mass. $D^{0}$ candidates are paired with correct-sign slow pions to form $D^{*+}$ candidates. Slow pions with $p_{T}$ between 100 and 200 MeV/$c$ are smeared with the aforementioned 10\% momentum resolution. For slow pions with $p_{T}>$ 200 MeV/$c$, the nominal values are used. Figure~\ref{fig:sim:DstDeltaM} shows the $m(D^{0}\pi^{+})-m(D^{0})$ distributions after the fast-simulation smearing for a sample size corresponding to an integrated luminosity of 0.056 fb$^{-1}$. A background distribution is estimated using like-sign $K\pi$ pairs when reconstructing the $D^{0}$, and is shown as the red histogram. The difference between the signal and background distributions, shown as the blue data points, is compared to the true $D^{*+}$ decays shown as the green histogram. Besides a small residual background present within the peak of the distribution, the signal is well isolated. Also shown in Fig.~\ref{fig:sim:DstDeltaM} are the distributions in three low $D^{*+}$ $p_{T}$ bins, and it is observed that at very low $p_{T}$ there is still good separation between signal and backgrounds. The signal significance of the $D^{*+}$ channel is comparable to that of the inclusive $D^{0}\rightarrow K\pi$ channel with secondary vertex reconstruction when scaled to the same integrated luminosity, and will therefore be a viable channel for some charm-hadron measurements at the EIC.

\paragraph{Fast-Simulation Validation}\label{sec:sim:valid}

\begin{figure*}[htbp]
    \centering
    \includegraphics[width=0.95\textwidth]{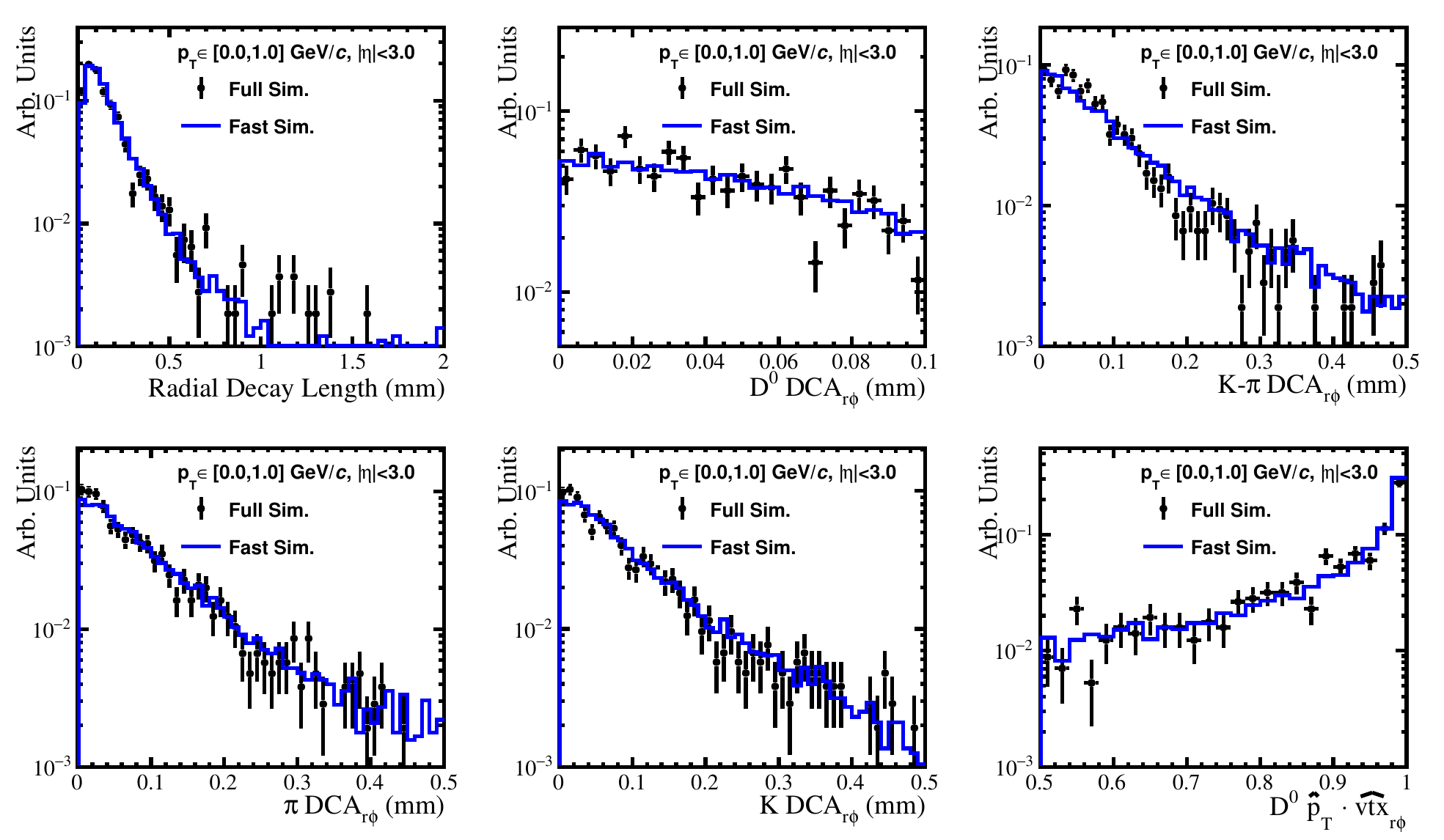}
    \caption{Comparison of the reconstructed $D^{0}$ topological variables in the GEANT4-based all-silicon simulation (data points) and in fast simulation (blue histograms). All distributions are normalized to have unit area. The $D^{0}$ candidates shown here are required to have a $|\eta|<3$ and $p_{T}<1$\,GeV/$c$.}
    \label{fig:valid:Topo}
\end{figure*}

The fast simulation procedure used for subsequent physics studies is validated by performing the same $D^{0}$ topological reconstruction in both the full simulation described in Sec.~\ref{sec:sim:full}, and the fast simulation using the single-track resolutions determined in the full simulation as input to the smearing routine. 

In the full-simulation setup, PYTHIA 8 $e$+$p$ events are embedded into the full simulator described in Sec.~\ref{sec:sim:full}. For these studies, we simulate the highest beam energies, $e$+$p$ 18$\times$275 GeV, and apply an event-level cut of $Q^{2}>1$ GeV$^{2}$ in both fast- and full-simulation setups.

\begin{figure}[htbp]
    \centering
    \includegraphics[width=0.6\textwidth]{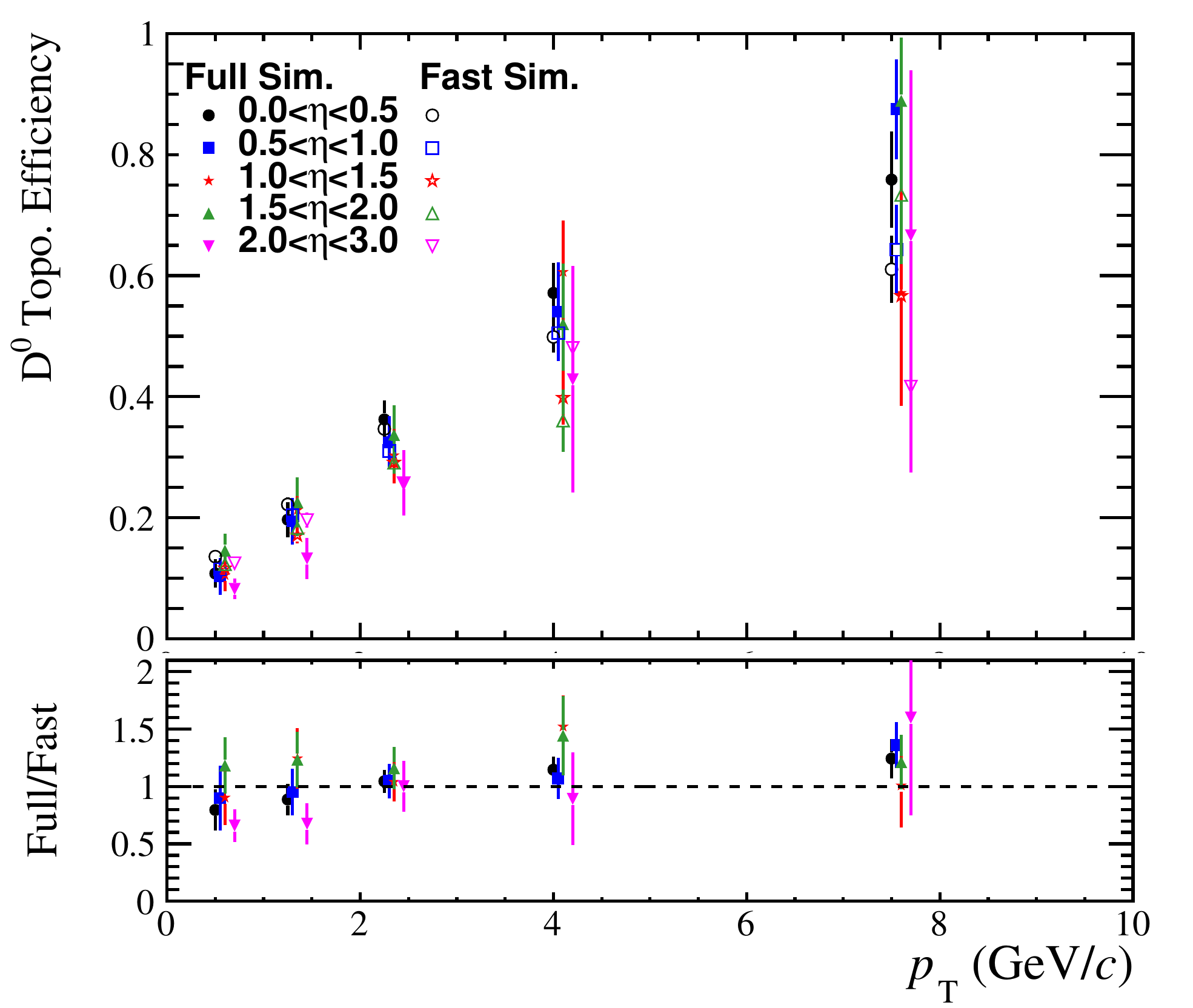}
    \caption{(Top) Comparison of the reconstructed $D^{0}$ topological reconstruction efficiency in the GEANT4-based all-silicon simulation (closed points) and fast simulation (open points). (Bottom) Ratio of efficiencies obtained from full and fast simulations.}
    \label{fig:valid:Eff}
\end{figure}

The smearing routine is applied in a similar way as described in Sec.~\ref{sec:sim:fast}, with some minor differences. To account for tails on the DCA distributions as well as charge-dependent differences, the smearing factors used are explicitly drawn from charge-dependent DCA distributions instead of assuming a pure Gaussian distribution. We additionally apply a pseudo-tracking efficiency as determined in the full simulation with truth particle seeding to keep the low-$p_{T}$ thresholds between full and fast simulation consistent (see Fig.~\ref{fig:fullsim:vtx} right plot). We determine our track multiplicities in the fast simulation (from which we determine the primary vertex smearing factor) by counting all tracks within $|\eta|<3$ passing the tracking acceptance. We have checked that track multiplicity in PYTHIA 6 and 8 are similar, and find that events with heavy-quark production have on average eight tracks within the acceptance of $|\eta|<3$.

Figure~\ref{fig:valid:Topo} shows the $D^{0}$ topological variables reconstructed in the full and fast simulation setups with $D^{0}$ candidates required to have $|\eta|<3$ and $p_{T}<2$\,GeV/$c$. It can be seen that all topological distributions agree quite well between the fast and full simulations. To further quantify the comparison across phase space we apply the topological cuts described in Sec.~\ref{sec:sim:fast} and plot the efficiency as a function of $D^{0}$ $\eta$ and $p_{T}$ in Fig.~\ref{fig:valid:Eff}. Within the statistical uncertainties of each respective sample there is general agreement between the efficiencies determined in the full and fast simulations, thus, validating that the fast-simulation smearing procedure gives an adequate description of a full GEANT4-based simulation.  

%% file: charm_physics.tex
\subsubsection{Charm Structure Function}\label{sec:Charmf2}

\begin{figure}[htb]
    \centering
    \includegraphics[width=0.5\textwidth]{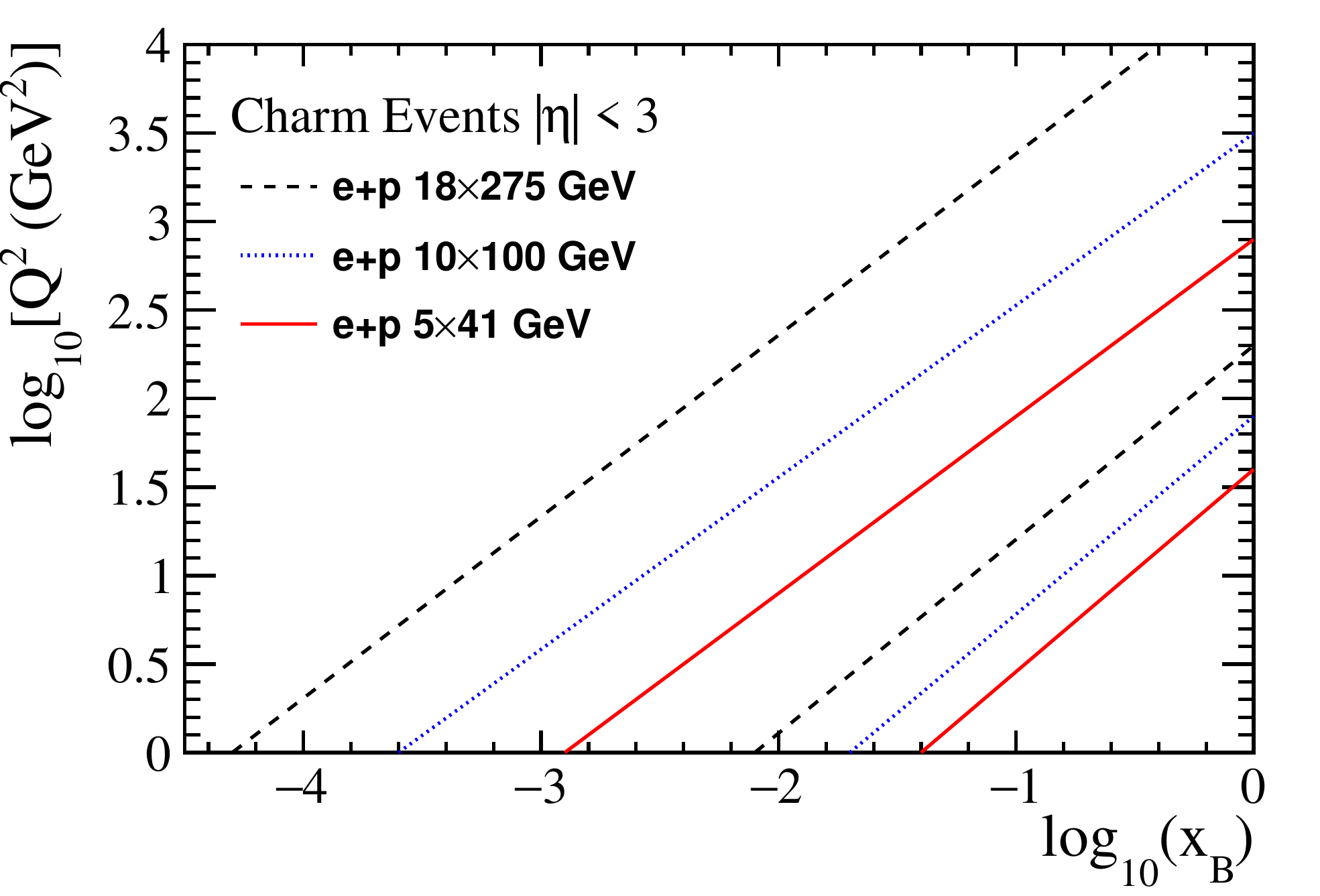}
    \caption{$Q^{2}$ vs. $x_B$ coverage of charm events with $|\eta|<3$ for three beam-energy configurations.}
    \label{fig:Charmf2:cq2x}
\end{figure}

To extract the statistical projections for the charm structure function, $F_{2}^{c\bar{c}}$, we first calculate the reduced charm cross sections in a two-dimensional grid of log$_{10}$($Q^{2}$) and log$_{10}$($x_B$) for 10$\times$100 GeV and 5$\times$41 GeV electron+proton collision configurations in PYTHIA 6 fast simulations. These two energies are chosen as they provide good coverage at high $x_B$ and low $Q^{2}$ (contrasted with higher energies such as $e$+$p$ 18$\times$275 GeV), and have relatively good overlap in $Q^{2}$ and $x_B$, as illustrated in Fig.~\ref{fig:Charmf2:cq2x}. Here, $Q^{2}$ and $x_B$ are taken directly from the generator level. As will be discussed below, to extract $F_2^{c\bar{c}}$ at fixed $Q^{2}$ and $x_B$, at least two energies are needed. Therefore, good overlap is desired and achievable with the selected beam configurations. We scale all uncertainties to an equivalent 10 fb$^{-1}$ worth of data for each beam-energy configuration as a baseline. 

The reduced cross section is explicitly defined as
%
\begin{equation}
\sigma_{r}^{c\bar{c}}(x_B,Q^{2}) = \frac{dN(D^{0}+\overline{D}^{0})/2}{\mathcal{L}\cdot \varepsilon\cdot \mathcal{B}(D^{0}\rightarrow K\pi)\cdot f(c\rightarrow D^{0}) \cdot \text{d}x_B\text{d}Q^{2}}
\times \frac{x_BQ^{4}}{2\pi\alpha^{2}[1+(1-y)^{2}]},
\end{equation}
%
where $y$ is the inelasticity, $\mathcal{L}$ is the integrated luminosity, $\varepsilon$ is the total efficiency (tracking, PID, reconstruction and acceptance),  $\mathcal{B}(D^{0}\rightarrow K\pi)$ is the $D^{0}$ branching ratio to $K\pi$, and $f(c\rightarrow D^{0})$ is the $D^{0}$ fragmentation fraction in PYTHIA (56.6\%). As can be observed from the latter quantity, for the purposes of these calculations we scale the measured $D^{0}$ yield to get the total charm cross section. The binning in log($Q^{2}$) and log($x_B$) is chosen to be five equal bins per decade along each dimension. 

The number of $D^{0}+\overline{D}^{0}$ candidates is determined by counting the number of true $D^{0}\rightarrow K\pi$ decays with invariant mass within $\pm$3$\sigma$ of the peak, daughter tracks within $|\eta|<3$, and that pass all topological reconstruction requirements outlined in Sec.~\ref{sec:sim:fast}. 
An additional tracking efficiency is applied using the efficiency curves shown in Fig.~\ref{fig:fullsim:vtx} (right) according to the decay daughter kinematics to simulate a realistic low-$p_{T}$ threshold. The background yields are counted similarly within the same mass window, and are composed of any unlike-charge-sign $K\pi$ pair when both hadrons have $p<$ 7 GeV/$c$ or any $K/\pi/p$ unlike-charge-sign combination when at least one hadron has as a $p>$ 7 GeV/$c$.  With this definition, background counts include combinatorial backgrounds, partially-reconstructed charm decays, and scenarios with a true $D^{0}$ decay hadron combined with a random hadron. We then take the statistical uncertainty of the counts as $\sqrt{N(S)+N(B)}$.   

\begin{figure}[htbp]
    \centering
    \includegraphics[width=0.6\textwidth]{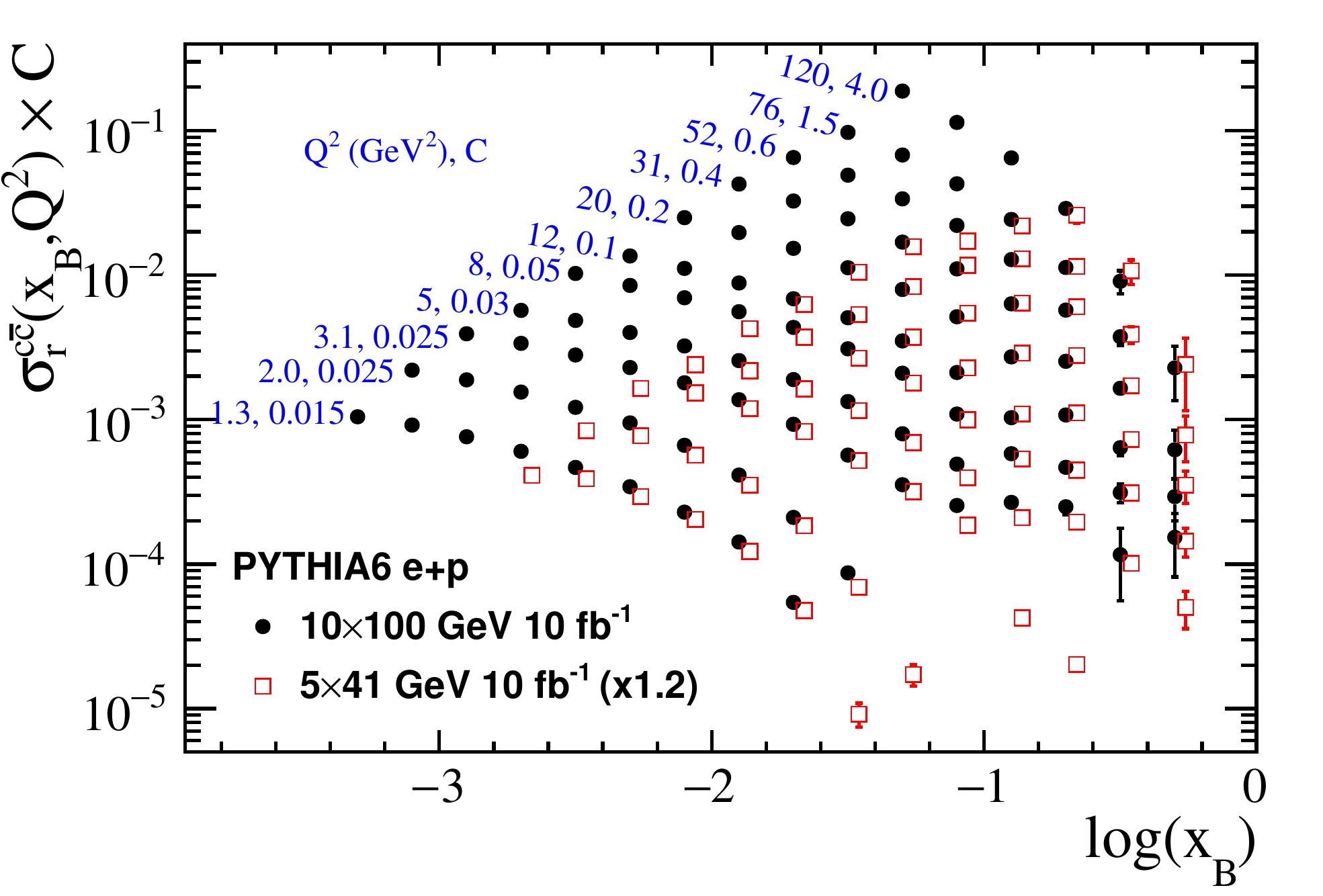}
    \caption{The reduced charm cross section in bins of log$_{10}$($x_B$) and log$_{10}$($Q^{2}$) for 10$\times$100 GeV (closed black circles) and 5$\times$41 GeV (open red squares) electron+proton collisions in PYTHIA 6 with the charm hadron reconstructed using the all-silicon detector. The vertical values in each $Q^{2}$ bin are scaled by the constant terms $C$ defined in the plot, and the 5$\times$41 GeV are further scaled up by twenty percent for clarity. The uncertainties shown are calculated using the signal significance as described in the text scaled to an integrated luminosity of 10 fb$^{-1}$ at each energy. The 10$\times$100 GeV data is placed at the $x$-axis bin centers while the 5$\times$41 GeV is displaced along the $x$-axis for clarity.}
    \label{fig:Charmf2:reducedCS}
\end{figure}

Figure~\ref{fig:Charmf2:reducedCS} shows the reduced charm cross sections in 10$\times$100 GeV and 5$\times$41 GeV electron+proton collisions with an integrated luminosity of 10\,fb$^{-1}$. At low $Q^{2}$, the cross section becomes truncated around $x_B$$\approx$0.01. This is due to the $D^{0}$ $\eta$ acceptance, as also illustrated in Fig.~\ref{fig:sim:phasespace1}. 

\begin{figure}
    \centering
    \includegraphics[width=0.48\textwidth]{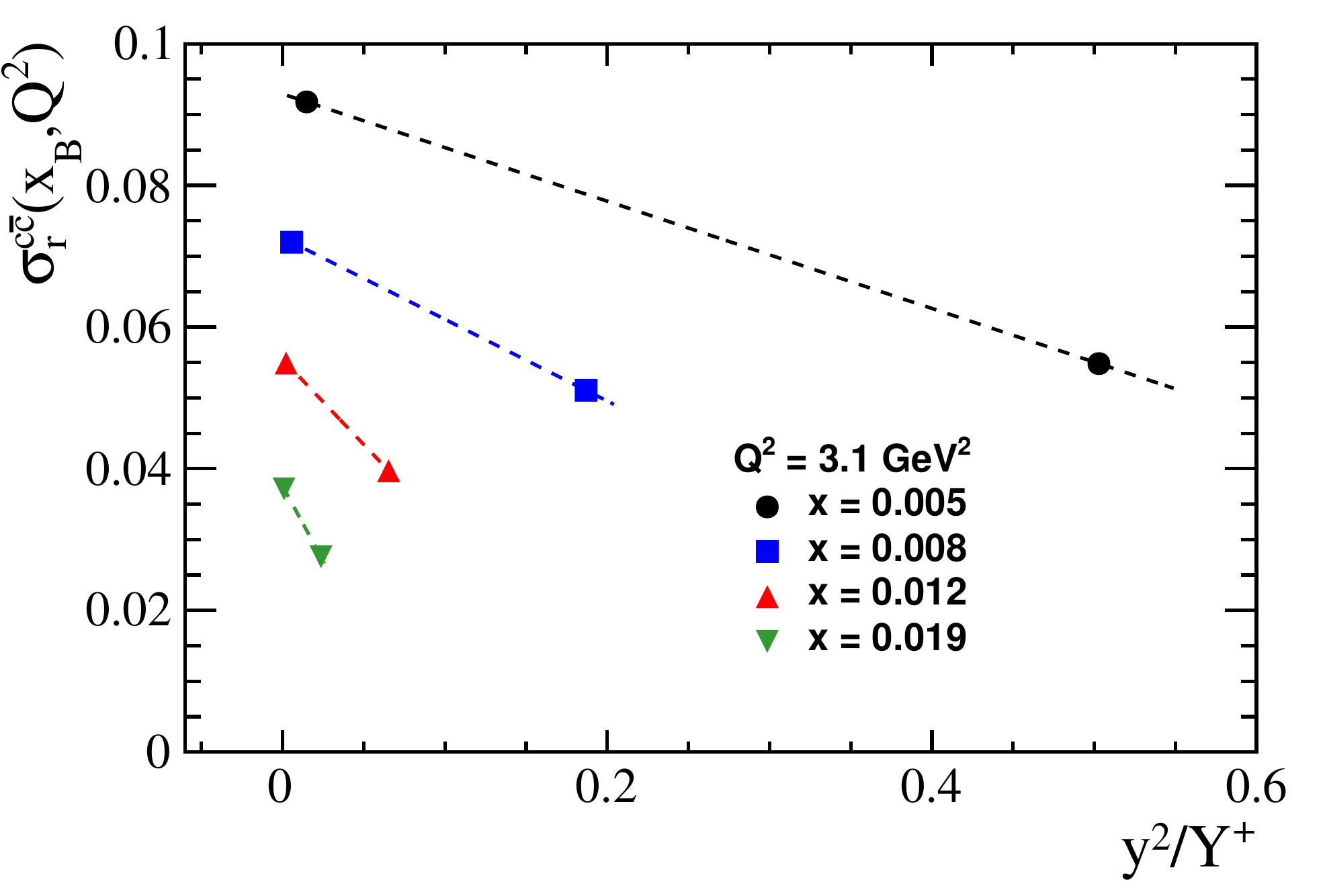}
    \includegraphics[width=0.48\textwidth]{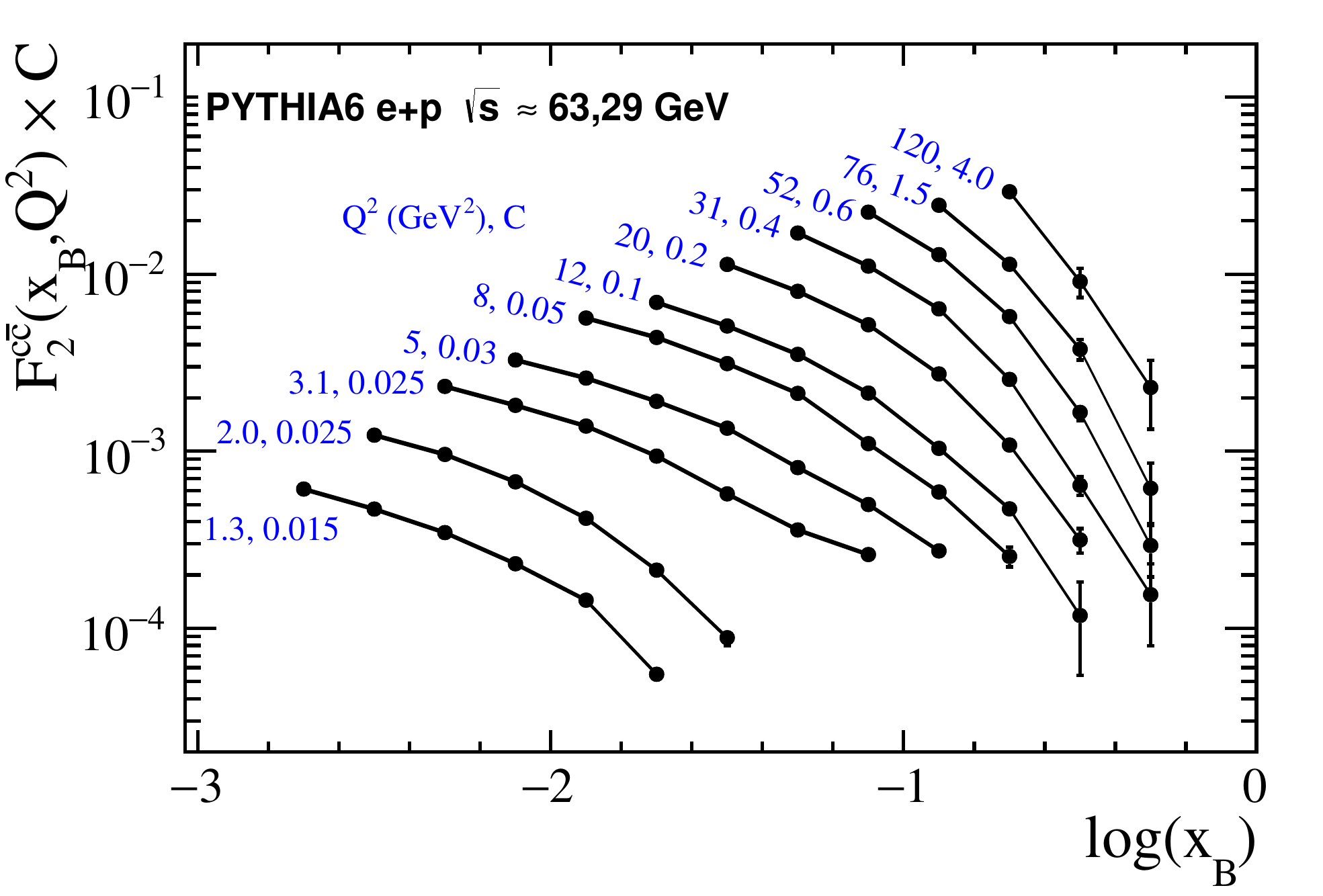}
    \caption{(Left) Example linear fits to the reduced charm cross sections versus $y^{2}/Y^{+}$ in four different slices of $Q^{2}$ and $x_B$. (Right) The measured $F_{2}^{c\bar{c}}$ in bins of log($x_B$) and $Q^{2}$. The data points in each $Q^2$ bin are scaled by a factor of $C$ for clarity. The statistical uncertainties are scaled to 10 fb$^{-1}$. The lines are added to the data to aid the reader.}
    \label{fig:Charmf2}
\end{figure}

\begin{figure*}
    \centering
    \includegraphics[width=0.55\textwidth]{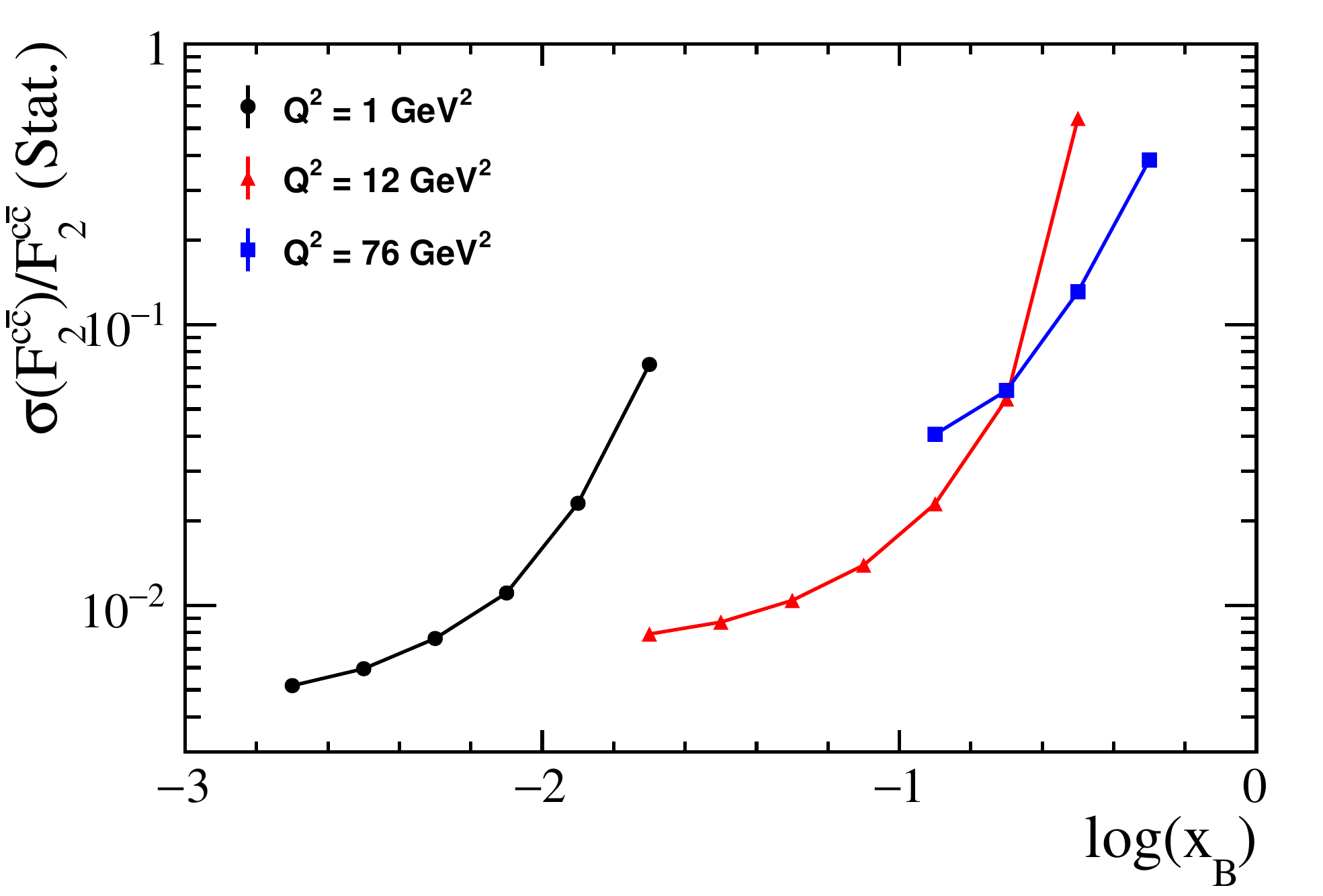}
    \caption{The projected $F_{2}^{c\bar{c}}$ relative statistical uncertainties scaled to 10\,fb$^{-1}$ for three representative $Q^{2}$ bins.}
    \label{fig:Charmf2:CharmF2Er}
\end{figure*}

To calculate the charm structure function $F_{2}^{c\bar{c}}$, we take the cross sections at 10$\times$100 GeV and 5$\times$41 GeV at fixed $x_B$ and $Q^{2}$ and fit the linear form:
\begin{equation}
\sigma_{r}^{c\bar{c}}(x_B,Q^{2}) = F_{2}^{c\bar{c}}(x_B,Q^{2}) - \frac{y^{2}}{Y^{+}}F_{L}^{c\bar{c}}(x_B,Q^{2}),
\end{equation}
where $Y^{+} = 1+(1-y)^{2}$. An example fit for four slices of $Q^{2}$ and $x_B$ are shown in Fig.~\ref{fig:Charmf2} (left). The extracted $F_{2}^{c\bar{c}}$ from the fits are shown in Fig.~\ref{fig:Charmf2} (right) and the relative statistical uncertainties in Fig.~\ref{fig:Charmf2:CharmF2Er}.

In the example linear fits in Fig.~\ref{fig:Charmf2} left panel, the data points closest to $y^{2}/Y^{+}$ = 0 correspond to the higher 10$\times$100 GeV $e$+$p$ collision cross section. We have estimated the uncertainty on the $F_{2}^{c\bar{c}}$ in scenarios where we vary the sample sizes of the 10$\times$100 GeV and 5$\times$41 GeV data sets. We find that reducing the lower energy integrated luminosity by up to a factor of ten has a small impact on the extracted $F_{2}^{c\bar{c}}$. Conversely, reducing the higher energy sample size by any factor increases the relative uncertainty on $F_{2}^{c\bar{c}}$ by the square-root of the scale factor, as this data point provides the best constraint on the $y$-axis intercept. Therefore, any EIC beam usage request could prioritize a larger integrated luminosity for the larger of the two beam energies.

Compared to the work in Refs.~\cite{Aschenauer:2017oxs,Chudakov:2016ytj}, our simulation studies represent a more accurate description of charm-reconstruction capabilities with the EIC detector as we have included PID, momentum and single track pointing resolutions guided by ongoing detector development/requirements and a full GEANT-based simulation. Furthermore, we have included for the first time the primary vertex resolution in the topological reconstruction of $D^{0}\rightarrow K\pi$ decays in an EIC simulation. 

In Ref.~\cite{Chudakov:2016ytj} the longitudinal charm structure functions $F_{L}^{c\overline{c}}$ are derived from simulation using charm events tagged by the identification of a displaced kaon vertex, and contain background levels that are less than 2\%. Comparing the kinematic coverage in $Q^{2}$ and $x_{B}$ of $F_{L}^{c\overline{c}}$, our derived $F_{2}^{c\overline{c}}$ has slightly better coverage, particularly in the high-$x_B$ region ($>$0.1). This difference is likely driven by the choice of beam energies used in the simulations and also from the need for three energies to reasonably constrain $F_{L}^{c\overline{c}}$, as opposed to two energies used for $F_{2}^{c\overline{c}}$. However, it should be noted in Ref.~\cite{Chudakov:2016ytj} single track and primary vertex resolutions are not folded into the kaon distributions. Incorporating these resolutions would significantly smear the charm and background kaon vertex distributions and in turn reduce the charm event purity and limit the kinematic coverage. Therefore, our studies show that now even with a realistic detector response measurements of $F_{2}^{c\overline{c}}$ are possible across a broad kinematic range.

\subsubsection{Gluon TMDs through Reconstruction of $D\overline{D}$ Pairs}\label{sec:DDbar}

Production of charm anti-charm hadron pairs offers an unique opportunity to study gluon Transverse Momentum Dependent (TMD) distributions. Since charm production in DIS proceeds primarily via the photon-gluon fusion (PGF) process (see Fig.~\ref{fig:diagram}), reconstruction of both the charm and anti-charm hadron allows to reconstruct the kinematics of the initial gluon, up to corrections from hadronization, initial- and final-state radiation effects. Gluon TMDs are hardly constrained with the present experimental data and are of fundamental interest to the physics of the EIC.

The gluon Sivers asymmetry and TMDs of linearly polarized gluons can be linked to azimuthal anisotropies of the produced charm anti-charm hadron pair. The Sivers asymmetry can be extracted from the measurements of transverse single-spin asymmetry, $A_{\rm UT}$, as a function of the azimuthal angle of the $c\bar{c}$ hadron pair relative to the direction of proton spin. $A_{\rm UT}(p_T)$ is defined as = $[\sigma_{L}(p_T) - \sigma_{R}(p_T)]/[\sigma_{L}(p_T) +\sigma_{R}(p_T)]$,
where $\sigma_{L(R)}$ are the cross sections for particle production of interest with spin polarized in the direction opposite to (same as) the spin of the proton, and $p_T$ is the transverse momentum of the heavy hadron pair. The $A_{\rm UT}$ is directly related to the Sivers asymmetry~\cite{Zheng:2018ssm},
\begin{equation}
    A_{\rm UT}(p_T) \propto \frac{\Delta f_{g/p^\uparrow}(x_g,k_T)}{f_g(x_g,k_T)},
\end{equation}
where $x_g$ is the momentum fraction, $k_T$ is the transverse momentum of the gluon and $\Delta f_{g/p^\uparrow}$ and $f_g$ are the gluon Sivers function and the unpolarized gluon TMD respectively. Note that the notation here is different from the Trento convention~\cite{Bacchetta:2004jz}.

The transverse momentum distribution of linearly-polarized gluons is related to the azimuthal distribution of the momentum of the $c\bar{c}$ hadron pair and can be accessed in unpolarized $e$+$p$ or $e$+$A$ collisions~\cite{Boer:2016fqd},
\begin{equation}
    \lvert \langle \cos(2\phi_T) \rangle \rvert \propto \frac{q_T^2}{2M_p^2}\frac{h_{1g}^{\perp}(x_g,k_T^2)}{f_g(x_g,k_T^2)},
\end{equation}
where 
$q_T$ is the sum of momenta of the heavy quarks in the pair, $\phi_T$ is the azimuthal angle of $q_T$ with respect to the leading charm meson, $M_p$ the proton mass, while $h_{1g}^{\perp}$ and $f_g$ are the linearly polarized gluon TMD and the unpolarized gluon TMD respectively. 

In this section we present studies on charm hadron pair reconstruction at an EIC experiment with the detector simulation settings described in Section~\ref{sec:sim}. The input distribution at the gluon level for the Sivers asymmetry is taken from previous simulation studies that use $D^0$ meson pair to study the gluon Sivers asymmetry at the EIC~\cite{Zheng:2018ssm}, while that for the linearly polarized gluon TMDs is taken from~\cite{Boer:2016fqd}. The simulations in this section were carried out for electron beam at 18 GeV and proton beam at 275 GeV.

\begin{figure}[htb]
    \centering
    \includegraphics[width=0.30\textwidth, trim={2mm 0mm 0mm 4mm}, clip]{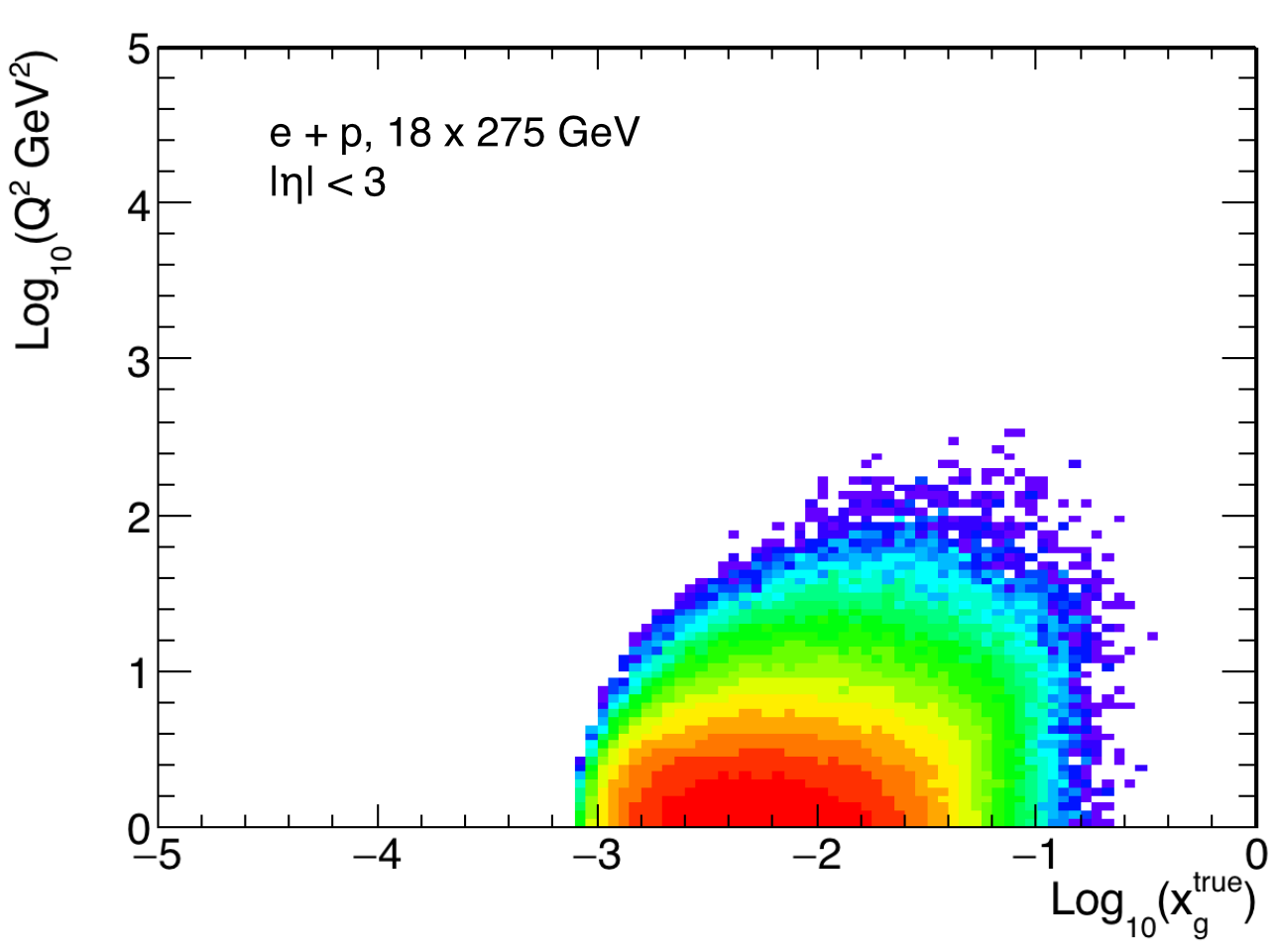}
    \includegraphics[width=0.34\textwidth, trim={6mm 2mm 0mm 0}, clip]{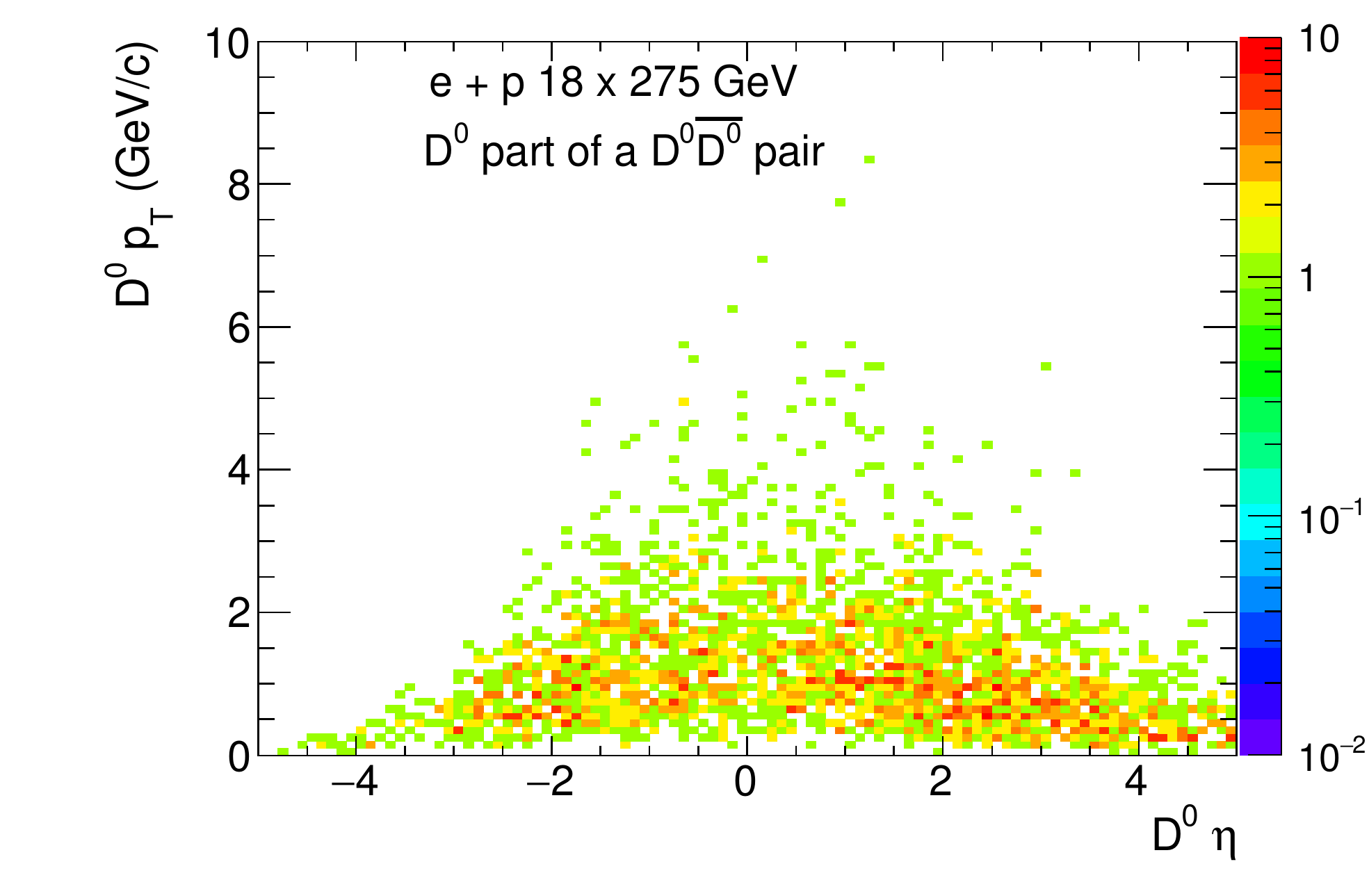}
    \includegraphics[width=0.34\textwidth, trim={6mm 2mm 0mm 0}, clip]{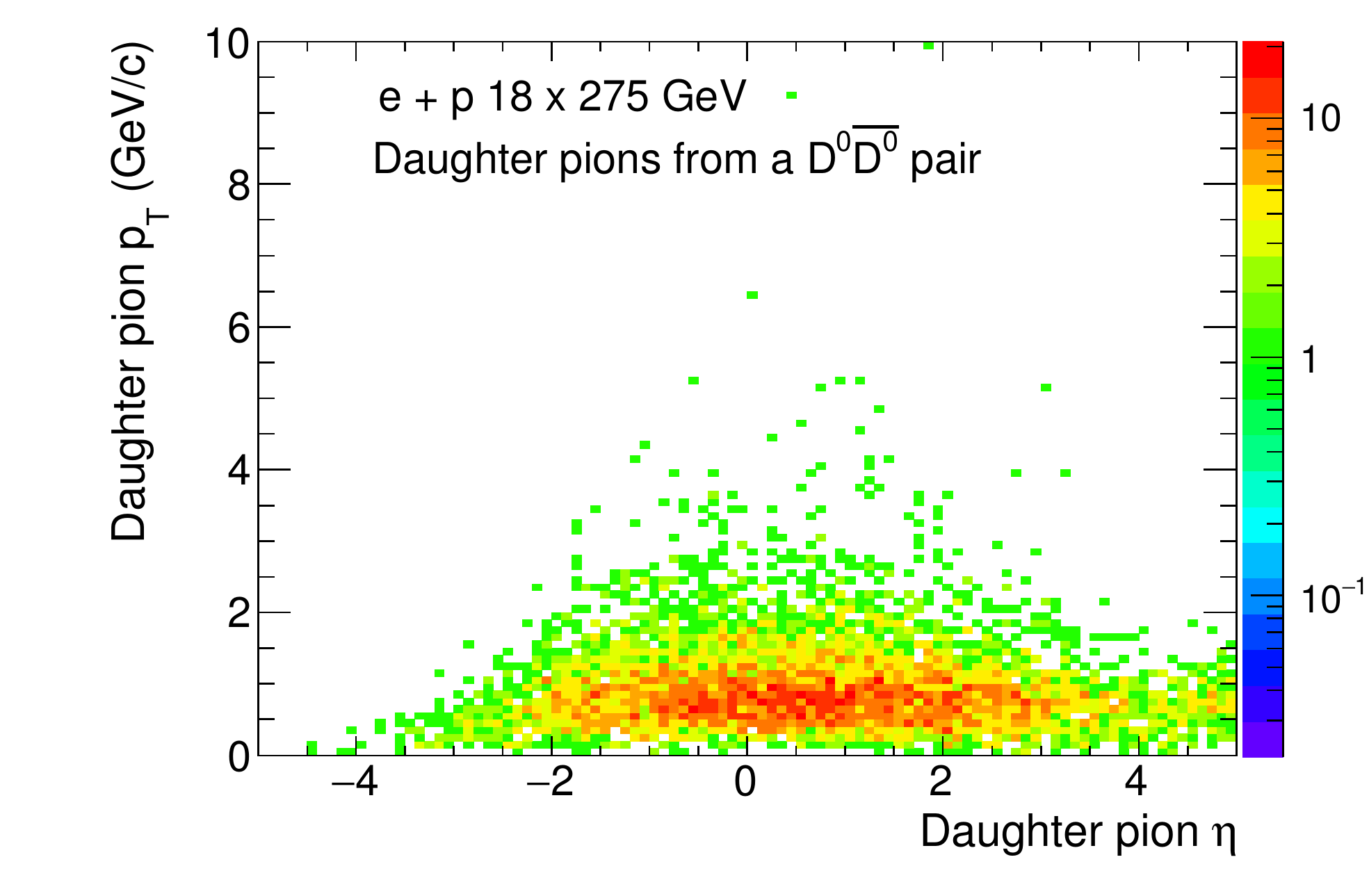}
    \caption{Event distributions based on a PYTHIA 6 simulation: (Left) $Q^2$-$x_g$ distribution of events with a $D^0\overline{D}^0$ hadron pair. (Middle) $p_T$-$\eta$ distribution of $D^0$ mesons that are part of a $D^0\overline{D}^0$ hadron pair and (Right) $p_T$-$\eta$ distribution of pions from decay of $D^0$ mesons that are part of a $D^0\overline{D}^0$ hadron pair.}
    \label{fig:TMD:kin}
\end{figure}

The $x_g$-$Q^2$ distribution of events with a $D^0\overline{D}^0$ hadron pair is shown in the left panel of Fig.~\ref{fig:TMD:kin}. The parton $x_g$ shown is the gluon momentum fraction, instead of the Bjorken-$x$. The middle and right panels of the figure show the $p_T$-$\eta$ distributions of $D^0$ mesons, and of decay pions from $D^0$ mesons, that are part of a $D^0\overline{D}^0$ pair.  From the middle and right panels of the figure it can be seen that the $p_T$ of both the $D^0$ meson and its decay daughters are rather small, up to $\sim$5\,GeV/$c$, and most of the daughter pions are have $\lvert \eta \rvert<3$, which is within our detector coverage. 

\begin{figure}[htb]
    \centering
    \includegraphics[width=0.34\textwidth, trim={15mm 5mm 0 0}, clip]{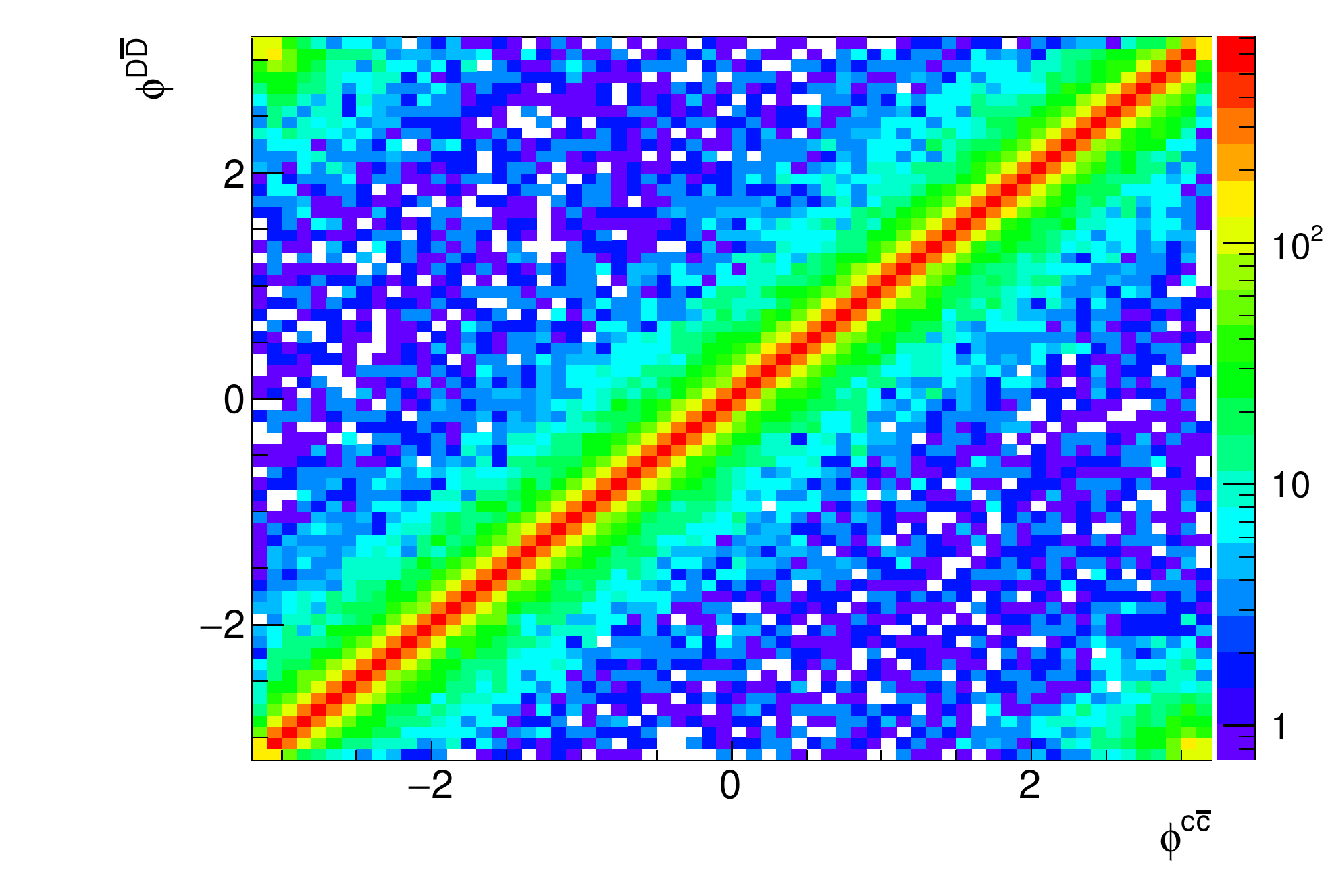}
    \includegraphics[width=0.64\textwidth, trim={0 0 0 0}, clip]{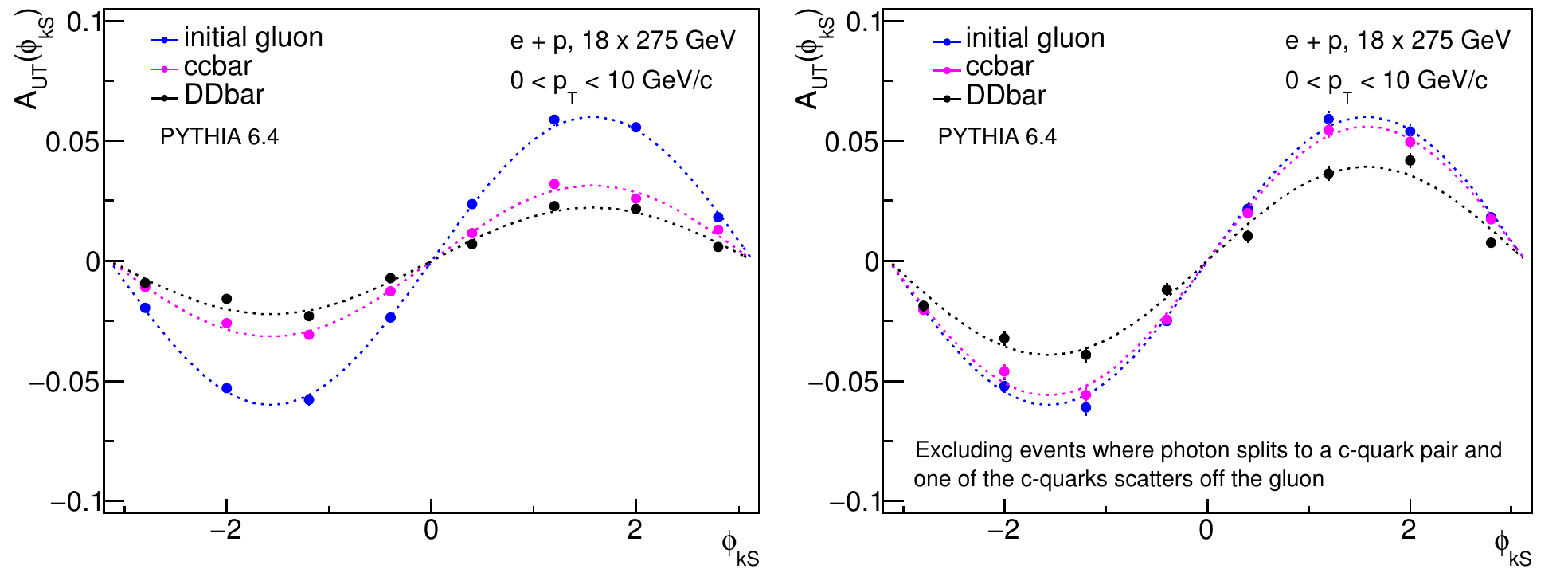}
    \caption{(Left) Correlation between the $\phi$ angle of the transverse momentum of the $c\bar{c}$ pair and that of the $D^0\overline{D}^0$ pair. (Middle) The evolution of signal strength for $A_{\rm UT}$ from initial gluon (input) to gluons reconstructed from $c\bar{c}$ pair and from $D^0\overline{D}^0$ pair. (Right) Same as middle panel, but excluding events in which the photon first split to a $c\bar{c}$ pair and then one of the charm quarks scatter off the gluon.}
    \label{fig:TMD:evo}
\end{figure}

As noted above, to probe the intrinsic transverse-momentum dependence in the gluon distribution we need to take into account initial- and final-state radiation and  hadronization effects in our analysis. The initial-state radiation can be included through the scale evolution of the TMD parton distribution~\cite{Collins:2011zzd}. Similarly, a relevant evolution can be carried out for the final-state gluon radiation by studying the soft factor associated with final state $c\bar c$ pair~\cite{Zhu:2013yxa, delCastillo:2020omr}. For the hadronization effects, on the other hand, we have to rely on Monte-Carlo simulations.  The left panel of Fig.~\ref{fig:TMD:evo} shows the correlation between the azimuthal-angle directions of the $c\bar{c}$ pair momentum and that of the corresponding $D^0\overline{D}^0$ hadron pair momentum, demonstrating that the angular correlations are well preserved during hadronization. The middle panel of Fig.~\ref{fig:TMD:evo} shows the evolution of the signal strength for the transverse single-spin asymmetry ($A_{\rm UT}$) from the initial gluon to the gluon reconstructed from the $c\bar{c}$ pair and the $D^0\overline{D}^0$ pair. The impact of hadronization is small, about a 30\% reduction in signal strength from the $c\bar{c}$ to the $D^0\overline{D}^0$ level. A larger dilution is seen when going from the initial gluon to the $c\bar{c}$ level and may be specific to the Monte Carlo used (PYTHIA v6.4). This impact is found not to arise from initial-state radiation. 
However, in this PYTHIA simulation, excluding events where the photon first splits to a $c\bar{c}$ pair, and then one of the charm quarks scatters off the gluon, results in a very small dilution from the initial gluon to $c\bar{c}$. This can be seen in the right panel of Fig~\ref{fig:TMD:evo}.

\begin{figure}[htbp]
    \centering
    \includegraphics[width=0.75\textwidth]{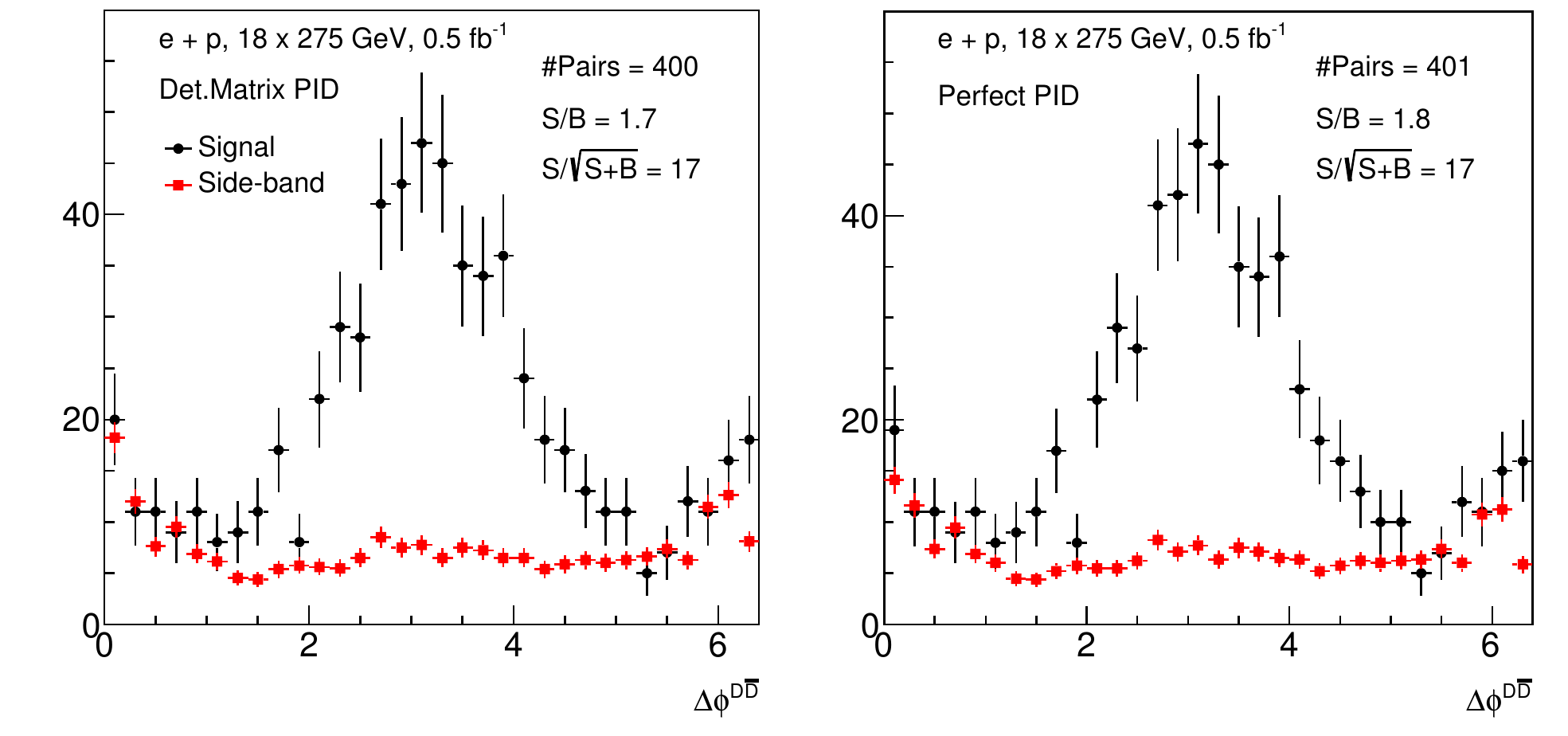}
    \caption{(Left) The $\Delta \phi^{D\overline{D}}$ distributions from $D^0$ and $\overline{D}^0$ meson candidates within 3$\sigma$ of the $D^0$ mass peak (solid black circles) and from those within a 12$\sigma$ mass window outside of the mass peak (solid red squares) for a PID scenario corresponding to that from the Detector Matrix (left) and a perfect-PID scenario (right).}
    \label{fig:TMD:reco}
\end{figure}

The $D^0\overline{D}^0$ meson pair is reconstructed using the azimuthal angle difference, $\Delta \phi^{D\overline{D}}$ between the $D^0$ and $\overline{D}^0$ meson in the pair. The distribution is expected to be predominantly back-to-back. Figure~\ref{fig:TMD:reco} shows the $\Delta \phi^{D\overline{D}}$ distribution from $D^0$ and $\overline{D}^0$ meson candidates within 3$\sigma$ of the $D^0$ mass peak in solid black circles and that from $D^0$ and $\overline{D}^0$ meson candidates in a 12$\sigma$ mass window outside of the mass peak in solid red squares. The two panels are for two different PID scenarios, a perfect PID case and a PID case corresponding to that from the Detector Matrix. The signal pairs show significant excess of candidates above the background. The number of signal $D^0\overline{D}^0$ pairs is obtained by subtracting the integrated counts over the full $\Delta \phi^{D\overline{D}}$ range from the side-band distribution to that from the signal distribution. The signal-to-background ratio ($S/B$) as well as the signal significance ($S/\sqrt{S+B}$) are also indicated in the figure.

The luminosity corresponding to the statistics used in Fig.~\ref{fig:TMD:reco} is 0.5 $\mathrm{fb}^{-1}$. Projections of statistical uncertainties can be estimated for $A_{\rm UT}$ and $\langle \cos(2\phi_T) \rangle$ from these significance numbers. As the $A_{\rm UT}$ is given by $A_{\rm UT} = (N_L - N_R)/(N_L + N_R)$, where $N_L$ and $N_R$ are the total number of signal $D^0\overline{D}^0$ pairs with momentum vector opposite to and aligned with the proton spin, respectively, 
we have for the statistical uncertainty in $A_{\rm UT}$ is equivalent to the inverse of
the projected significance ($\sigma$) for the combined $N_L + N_R$ candidates at the projected luminosity times the polarization fraction (P), 1/$\sigma$P. The estimation of statistical uncertainty for $\langle \cos(2\phi_T) \rangle$ also follows a similar procedure and goes as $1/\sigma$ with $\sigma$ being the significance for the $D^0\overline{D}^0$ pair signal at the projected luminosity.

\begin{figure}[htbp]
    \centering
    \includegraphics[width=0.45\textwidth,trim={10mm 0 0 0}, clip]{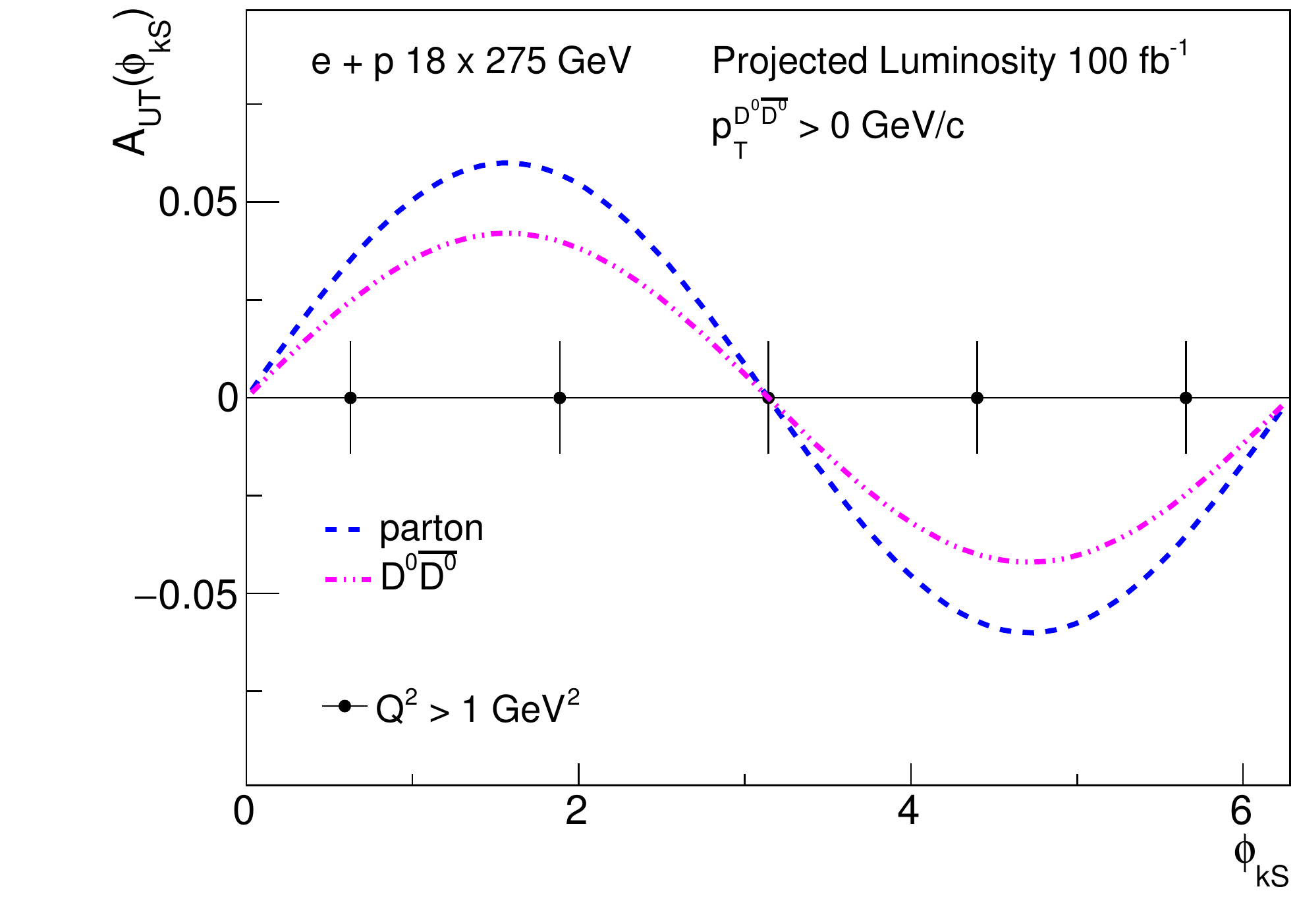}
    \includegraphics[width=0.47\textwidth,trim={10mm 0 0 0}, clip]{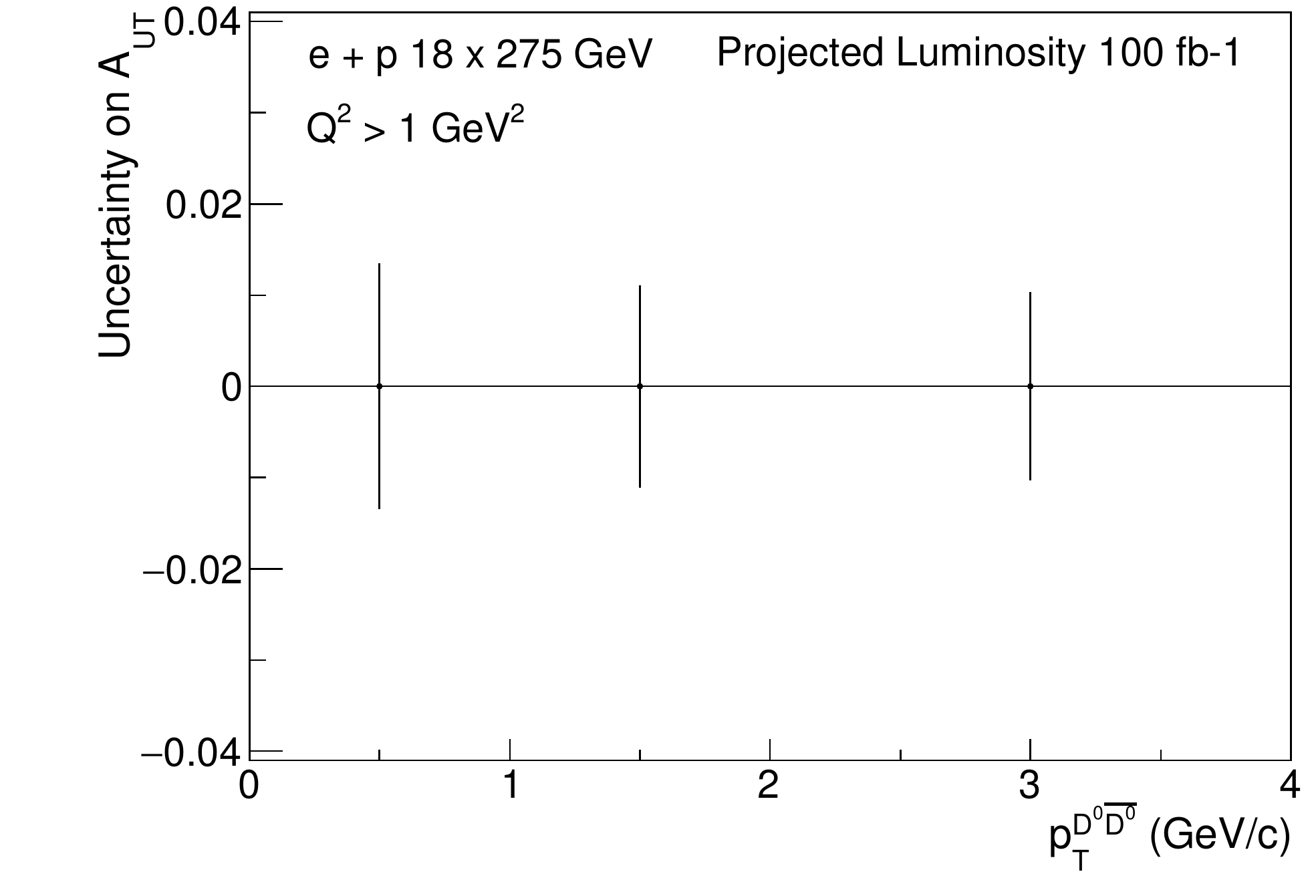}
    \caption{(Left) Statistical-uncertainty projections for $A_{\rm UT}$ in bins of azimuthal angle of the pair momentum of the $D^0\overline{D}^0$ pair relative to the spin of the proton ($\phi_{kS}$). The two curves indicate the signal strength at parton and $D^0\overline{D}^0$ levels. (Right) Statistical uncertainty projections for $\langle A_{\rm UT} \rangle$ in bins of $D^0\overline{D}^0$ meson pair $p_T$. A 70\% proton-beam polarization is included the uncertainty projections.}
    \label{fig:tmd_siversproj_100fb}
\end{figure}

The $D^0\overline{D}^0$ meson-pair reconstruction is studied for different pair $p_T$, event $Q^2$ and Bjorken $x_B$ bins, using a simulated sample of 0.5\,fb$^{-1}$ luminosity. The projected uncertainty on $A_{\rm UT}$ as a function of $\phi_{kS}$ is shown in Fig.~\ref{fig:tmd_siversproj_100fb}, along with the $A_{UT}$  signals at the parton level and for reconstructed $D^{0}\overline{D}^0$ pairs. The uncertainties are scaled by 1/$P$, where $P$ is the proton-beam polarization at the EIC experiment and is taken to be 70\%. The projected uncertainties on the $\phi_{kS}$ integrated $A_{\rm UT}$ as a function of the $D^0\overline{D}^0$ meson pair $p_T$ is shown in Fig.~\ref{fig:tmd_siversproj_100fb}, while those with projections for uncertainties in different $Q^{2}$ and $x_B$ bins for $A_{\rm UT}$ as a function of $\phi_{kS}$ are shown in Fig.~\ref{fig:tmd_siversproj_100fb} left plot. The projected uncertainty on the $\phi_{kS}$ integrated $\langle A_{\rm UT} \rangle$ is 0.57\%, which implies a 7$\sigma$ measurement for the projected signal corresponding to the 10\% positivity bound. The uncertainty on $\langle\cos(2\phi_T)\rangle$ does not get the scale contribution from polarization fraction, and is $\sim$0.4\% for the projected 100 f$\mathrm{b}^{-1}$ luminosity. For a projected signal of 2\%, this then implies a 5$\sigma$ measurement.

\begin{figure}[htbp]
    \centering
    \includegraphics[width=0.45\textwidth,trim={10mm 0 0 0}, clip]{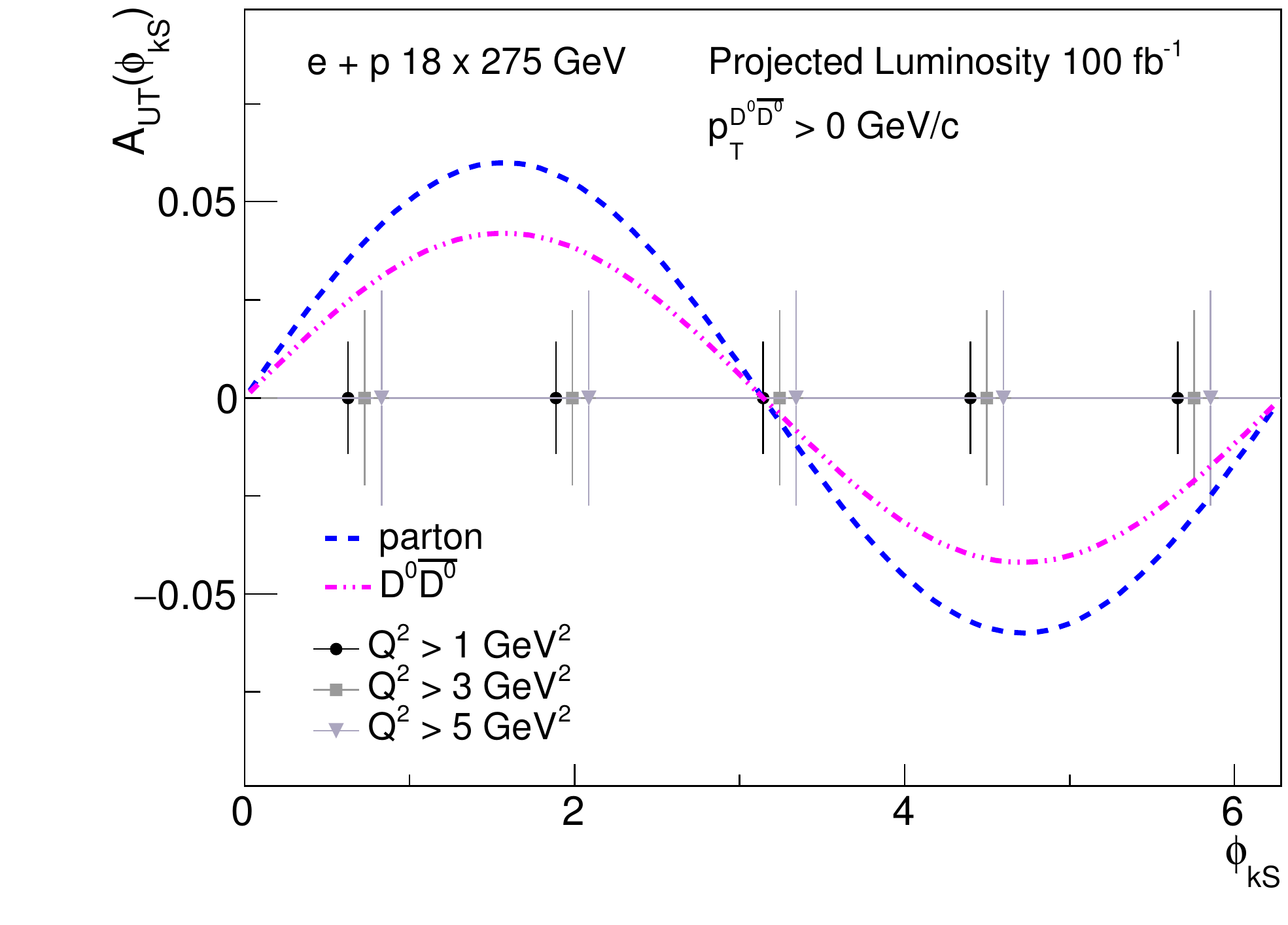}
    \includegraphics[width=0.45\textwidth,trim={10mm 0 0 0}, clip]{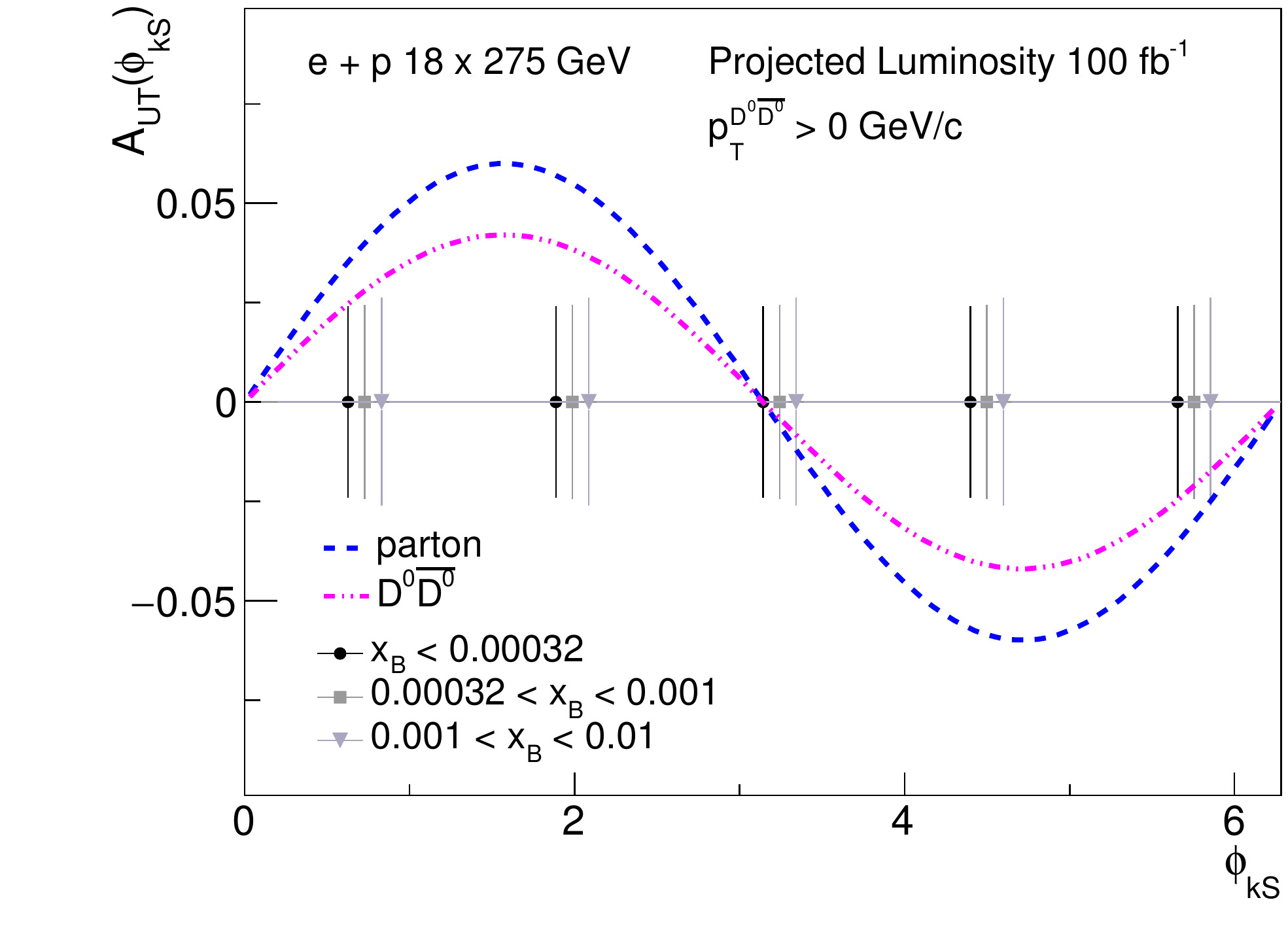}
    \caption{Statistical-uncertainty projections for $A_{\rm UT}$ in bins of azimuthal angle of the pair momentum of the $D^0\overline{D}^0$ pair relative to the spin of the proton ($\phi_{kS}$), for different $Q^{2}$ (left) and $x$ (right) selections. The two curves indicate the signal strength at parton and $D^0\overline{D}^0$ levels. A 70\% proton beam polarization is included the uncertainty projections.}
    \label{fig:tmd_siversproj_100fb_xQ2}
\end{figure}

The above analyses only use $D^0\overline{D}^0$ pairs, which is about $1/4$ of all charm hadron pairs. The event rates can be improved by looking at the reconstruction of all charm-hadron pairs as well as other decay channels.  A total luminosity of 100 $\mathrm{fb}^{-1}$ for the 18$\times$275 GeV collisions is feasible. For a 3\% signal this would imply a 9$\sigma$ and 7$\sigma$ measurements respectively for the two PID scenarios, making it possible to measure the signal associated with gluon TMDs. These measurements would be valuable to constrain the gluon TMDs along with measurements from di-jets~\cite{Zheng:2018ssm,Dumitru:2018kuw} while $D\overline{D}$ reconstruction offers a cleaner access to the initial gluon distributions.

To study the impact of these measurements on the gluon Sivers function and the linearly-polarized gluon distribution, we can combine the above simulations with the initial- and final-state radiation effects by applying the soft gluon resummation formalism~\cite{Zhu:2013yxa,delCastillo:2020omr}. 
The final-state gluon radiation can contribute to a nonzero $\cos(2\phi_T)$ asymmetry in heavy-quark pair production in DIS~\cite{Catani:2017tuc,Hatta:2020bgy}. All these effects should be included in the final results to extract these novel gluon TMD distributions.

\subsubsection{Gluon Helicity $\Delta g/g$ through Charm-Hadron Double-Spin Asymmetry}
\label{sec:DALL}

In electron-proton DIS processes, if both the electron and 
proton beams are longitudinally polarized, one can measure double-spin asymmetries $A_{\rm LL}$
in the inclusive or semi-inclusive channels, and thus extract the polarized structure function
$g_1$. With extensive measurements in a broad kinematics region, a comprehensive QCD fit can be performed to extract quark and gluon helicity distributions.

In addition to the aforementioned classic way to extract the gluon polarization within a longitudinally polarized nucleon, the heavy-flavor production can also contribute to the study.
For instance, $A_{\rm LL}$ measurements in the $e+p\to e' D^0 + X$ process can be linked
to $\Delta g/g$ assuming Photon-Gluon Fusion. At leading order, it can be
written as
$A_{\rm LL}=a_{\rm LL} \cdot \Delta g/g$,
where $a_{\rm LL}$ is double-spin asymmetry of partonic kinematics of the hard-scattering process. This has been studied in detail in references~\cite{Adolph:2012ca,Kurek:2011kka}. 

The COMPASS collaboration performed a pioneering study on the charm-hadron $A_{\rm LL}$ measurement in polarized $\mu p$ collisions~\cite{Adolph:2012ca}. Because of the lack of a vertex detector for the $D^0$-decay-topology study, the signal-to-background ratio for the $D^0$ sample is low. In addition, the luminosity and acceptance are limited at COMPASS relative to the EIC.
The study can be dramatically enhanced at the EIC with a good vertex detector to allow topological reconstruction of charm-hadron decays along with the high luminosity and large acceptance. 
This provides a direct way, in some sense, to measure $\Delta g/g$ over a broad kinematic range in $(x_B,Q^2)$. In the following, a simulation study at the EIC will be described in detail.

\begin{figure*}[htbp]
\centering
\includegraphics[width=0.32\textwidth]{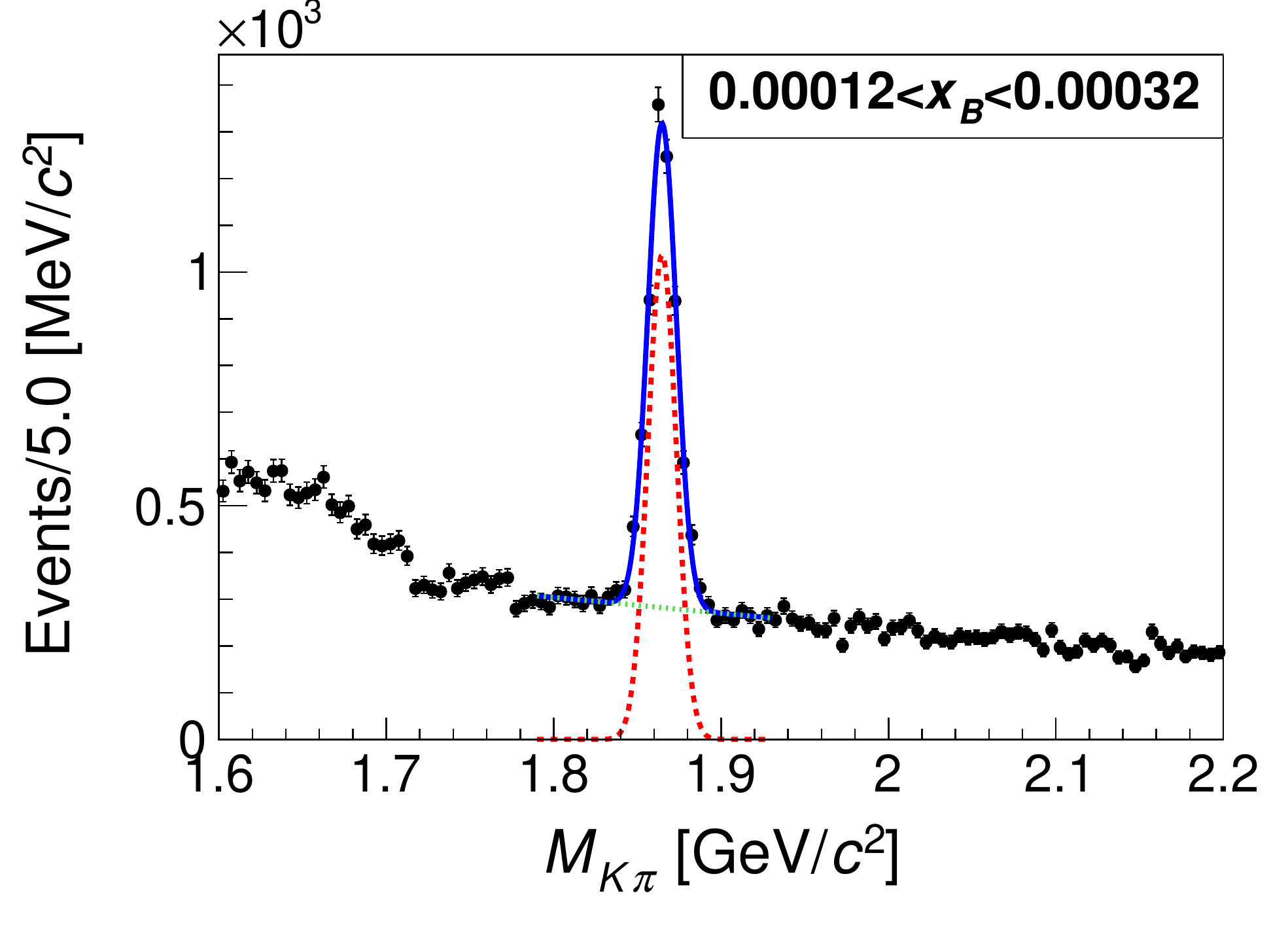}
\includegraphics[width=0.32\textwidth]{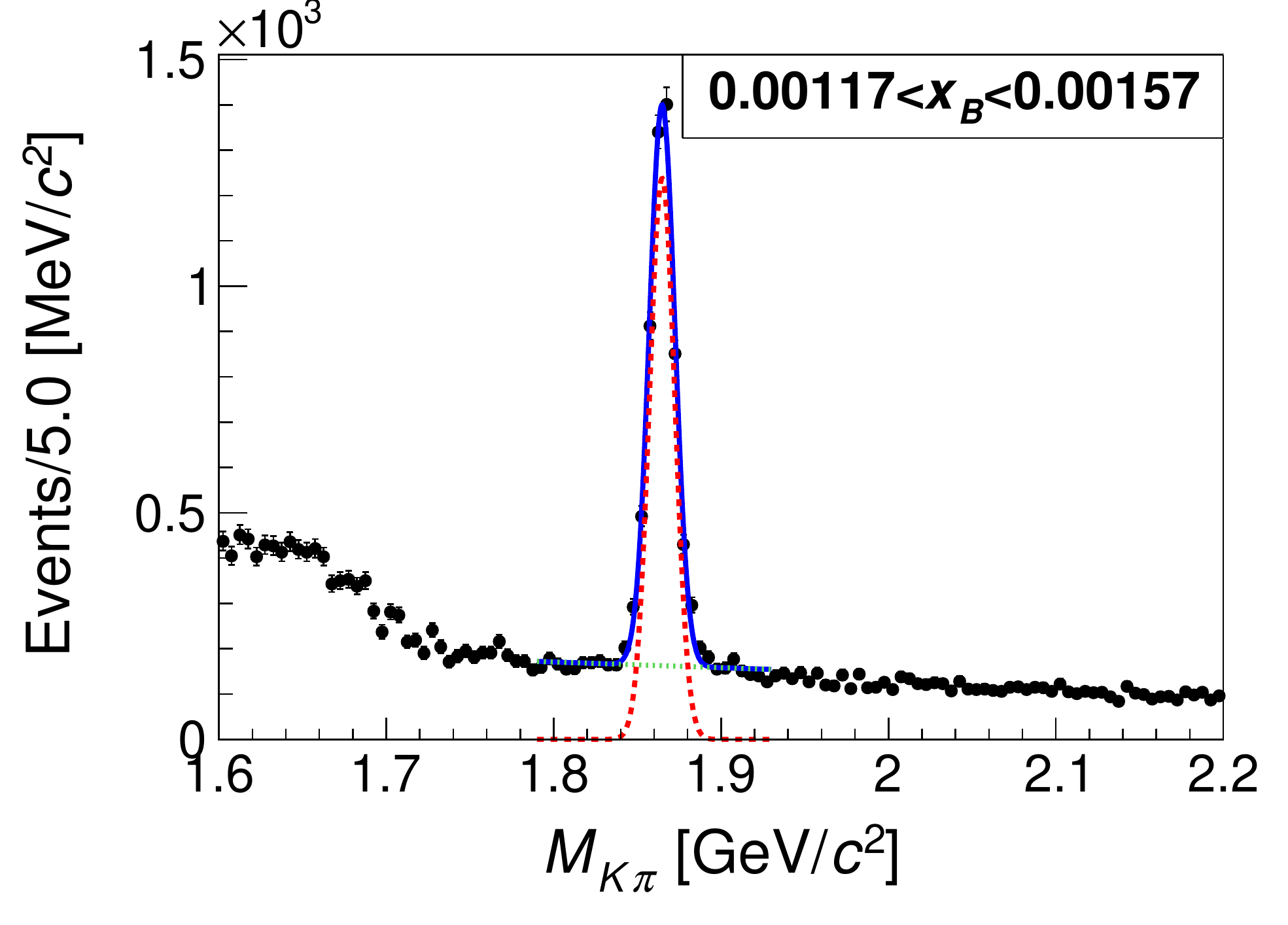}
\includegraphics[width=0.32\textwidth]{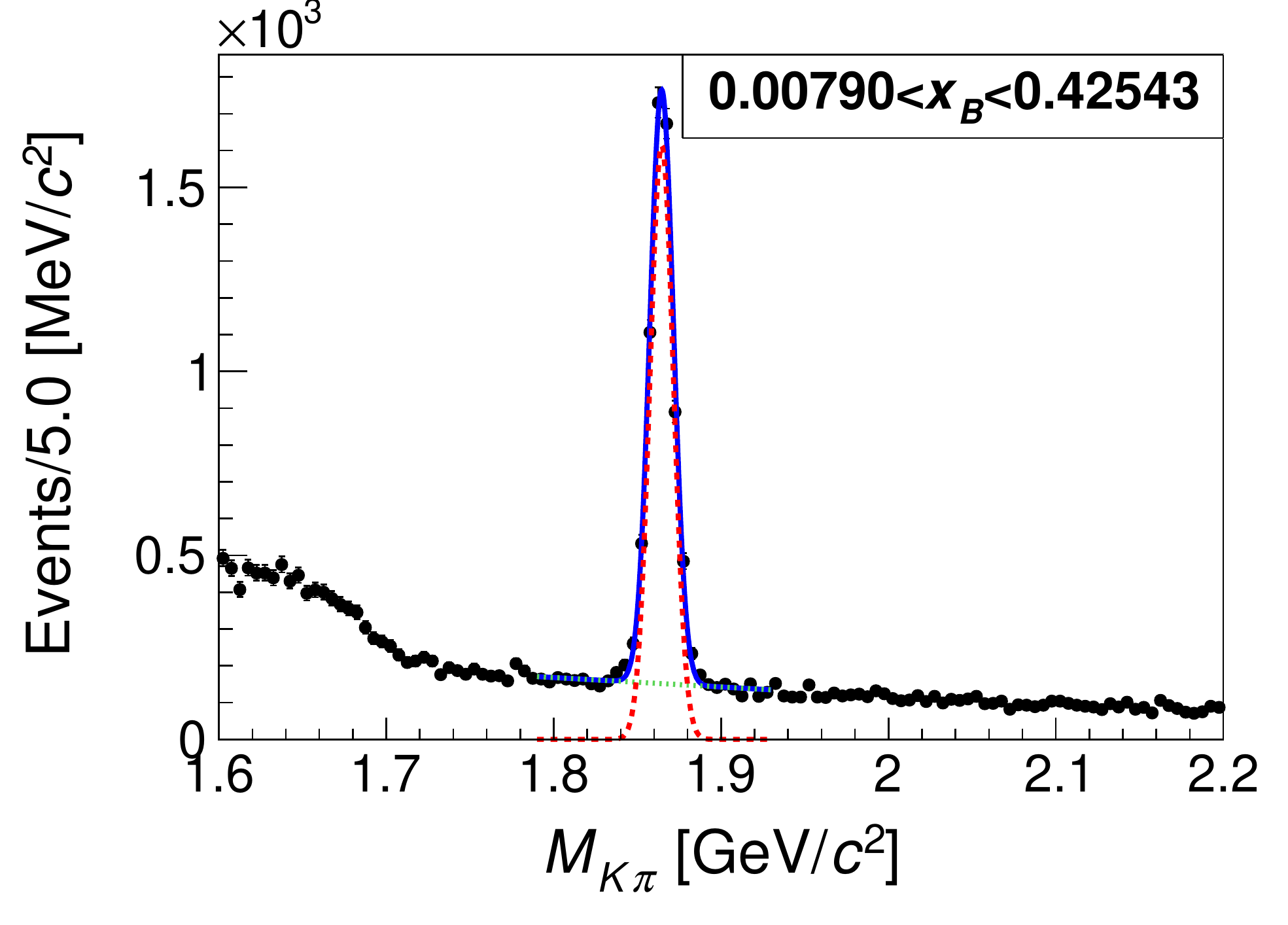}
\caption{Fits to the $K\pi$ invariant-mass distributions in a few different Bjorken-$x$ bins for 18$\times$275~GeV $e$+$p$ collisions. The red and green dashed curves are the signal (Gaussian) and background (linear) fits, and the blue curve is the sum. 
}  
\label{fig_fit_18_275}
\end{figure*}

The data was generated by pythiaeRHIC (PYTHIA v6.4) and then smeared according to the resolution listed in Table~\ref{tab:sim:smearing} via the fast-simulation setup described above.
To identify $D^0$ with a good signal-to-background ratio, an optimization on
$D^0$-decay topology cuts was performed. Three topological distributions (pair-DCA, Decay-Length$_{r\phi}$, and $\cos\theta_{r\phi}$) were investigated by using ``truth'' $D^0$ to $K\pi$ two-body decay and a background sample with $K\pi$ invariant mass outside of the $D^0$ mass window ($\pm$3$\sigma$), shown in Fig.~\ref{fig_fit_18_275}. 
The yield of the background sample chosen for the study depends on the selection range of the $M_{K \pi}$ distribution outside the $D^0$ peak, 
in order to be avoid of this artificial effect, the signal and background samples were separately self-normalized for the three
topological distributions and the crossing points between these two samples were chosen to be the analysis cuts.
Table~\ref{table_selection_criteria} lists the choices of these cuts for the different beam energy configurations 
used in the following projection estimation.
In addition to the $D^0$-decay topology cuts, the following kinematic cuts were used during the analysis: $Q^2>2$\,GeV$^2$, $0.05<y<0.8$ and $W^2>4$\,GeV$^2$. 

\begin{table*}[htbp]
\begin{center}
\caption{$D^0$ decay topology cuts for different beam-energy configurations for $A_{\rm LL}$ projection calculation.}
{\begin{tabular}{ c | c  c  c}
~~~Selection criteria~~~ & ~~~18$\times$275~GeV~~~ & ~~~5$\times$100~GeV~~~ & ~~~5$\times$41~GeV~~~\\
\hline \hline
$K\pi$ pair-DCA & $<80$~$\mu$m & $<80$~$\mu$m & $<80$~$\mu$m\\
Decay-Length$_{r\phi}$ & $>90$~$\mu$m & $>80$~$\mu$m & $>60$~$\mu$m\\
$\cos\theta_{r\phi}$ & $>0.983$ & $>0.982$ & $>0.982$\\
\end{tabular}}
\label{table_selection_criteria}
\end{center}
\end{table*}

\begin{table*}[htbp]
\begin{center}
\caption{Extracted number of $D^0$ candidate and background (estimated through a linear function fit) events within 3$\sigma$ of the peak in different Bjorken-$x$ bins. The numbers included here correspond to 0.24~fb$^{-1}$ 5$\times$41 GeV $e$+$p$ collisions. }
{\begin{tabular}{ c  c | c  c}
$x_B^{\rm min}$ & $x_B^{\rm max}$ & $N_{\rm Signal}$ & $N_{\rm Background}$\\
\hline \hline
~~~~~ $0.00306$~~~~~&~~~~~$0.00600$~~~~~&~~~~~$704_{-38}^{+39}$~~~~~&~~~~~$575_{-78}^{+79}$~~~~~\\
~~~~~ $0.00600$~~~~~&~~~~~$0.00808$~~~~~&~~~~~$651_{-36}^{+36}$~~~~~&~~~~~$514_{-72}^{+74}$~~~~~\\
~~~~~ $0.00808$~~~~~&~~~~~$0.01040$~~~~~&~~~~~$800_{-39}^{+39}$~~~~~&~~~~~$522_{-74}^{+76}$~~~~~\\
~~~~~ $0.01040$~~~~~&~~~~~$0.01324$~~~~~&~~~~~$731_{-36}^{+37}$~~~~~&~~~~~$422_{-66}^{+68}$~~~~~\\
~~~~~ $0.01324$~~~~~&~~~~~$0.01676$~~~~~&~~~~~$743_{-36}^{+37}$~~~~~&~~~~~$416_{-65}^{+68}$~~~~~\\
~~~~~ $0.01676$~~~~~&~~~~~$0.02150$~~~~~&~~~~~$860_{-38}^{+39}$~~~~~&~~~~~$444_{-68}^{+70}$~~~~~\\
~~~~~ $0.02150$~~~~~&~~~~~$0.02822$~~~~~&~~~~~$853_{-38}^{+39}$~~~~~&~~~~~$470_{-70}^{+72}$~~~~~\\
~~~~~ $0.02822$~~~~~&~~~~~$0.03928$~~~~~&~~~~~$1026_{-41}^{+42}$~~~~~&~~~~~$469_{-71}^{+73}$~~~~~\\
~~~~~ $0.03928$~~~~~&~~~~~$0.06180$~~~~~&~~~~~$1057_{-43}^{+43}$~~~~~&~~~~~$554_{-77}^{+79}$~~~~~\\
~~~~~ $0.06180$~~~~~&~~~~~$0.62764$~~~~~&~~~~~$1430_{-54}^{+55}$~~~~~&~~~~~$1016_{-106}^{+110}$~~~~~\\
\end{tabular}}
\label{table_5_41}
\end{center}
\end{table*}
 
\begin{figure*}[htbp]
\centering
\includegraphics[width=0.7\textwidth]{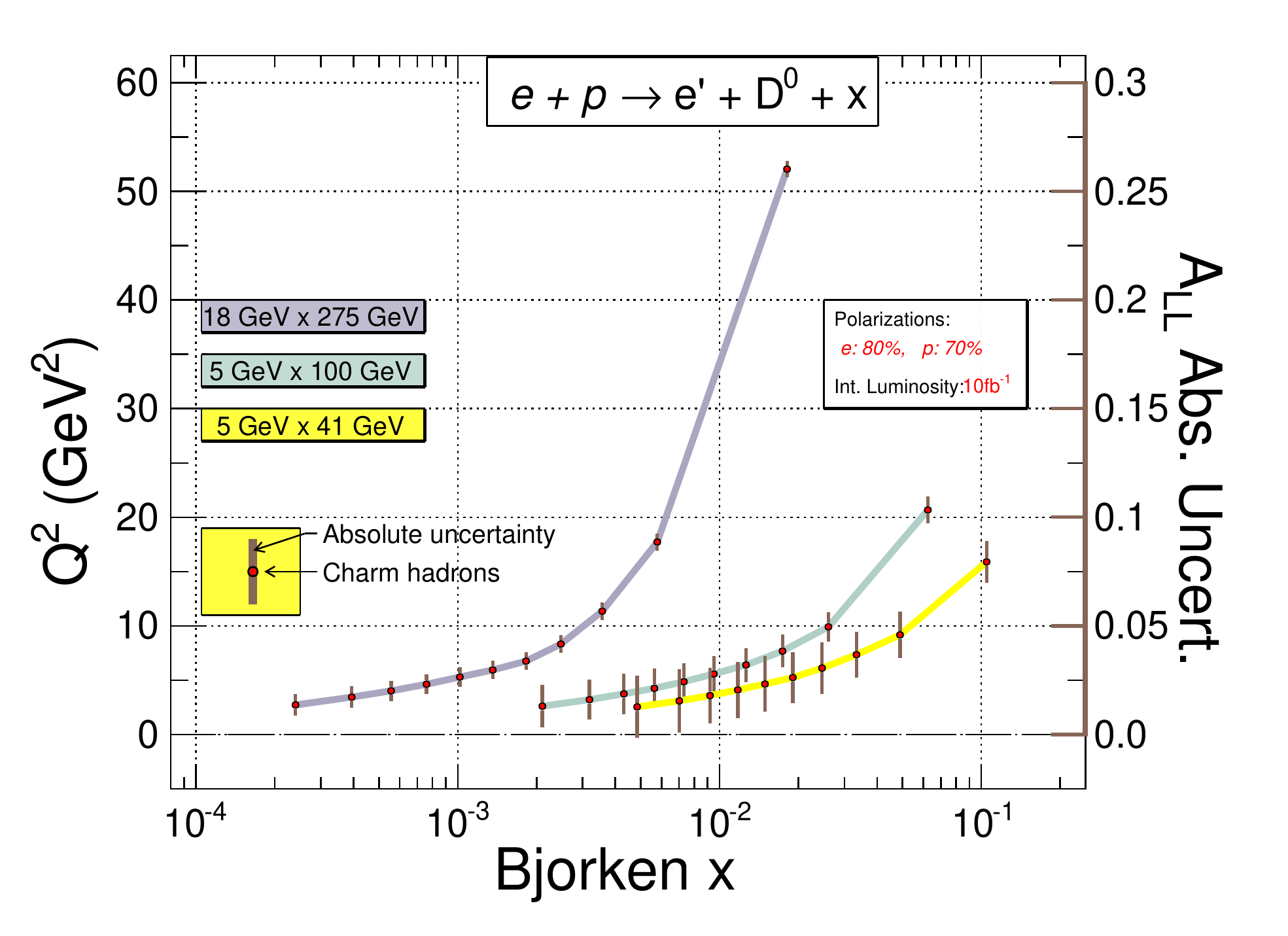}
\caption{Projections on double-spin asymmetry A$_{\rm LL}$ in the
$e+p \to eD^0 + X$ process for different beam-energy configurations.
The data was binned in Bjorken-$x$.
The position of each data point in the plot is defined by the weighted center
of Bjorken-$x$ and Q$^2$ for this particular bin. The uncertainty indicated for each data point should be interpreted using the scale shown on the right-side vertical axis of the plot.}
\label{fig_ALL_projection}
\end{figure*}

The data were binned in Bjorken-$x$ after all selection requirements. In each bin, the reconstructed $K\pi$
invariant-mass spectrum was fitted to a gaussian function for signal plus a linear background to extract the number of $D^0$ signal and background, as shown in Fig.~\ref{fig_fit_18_275}. In general, one can see that the signal is quite significant for all the bins. The fit results, as well as the binning information, for 5$\times$41 GeV $e$+$p$ collisions are summarized in Table~\ref{table_5_41}. These numbers were scaled to $10\,\mathrm{fb}^{-1}$ of integrated luminosity for each collision energy configuration in order to obtain the projected statistical uncertainties on the $A_{\rm LL}$ experimental observable, as shown in Fig.~\ref{fig_ALL_projection}.

After obtaining the projections on $A_{\rm LL}$, $a_{\rm LL}$ was calculated event by event using equations 5.8 and 5.9 at leading order from reference~\cite{Kurek:2011kka}. Unlike the COMPASS data analysis, where a neural network was employed to map the $a_{\rm LL}$ to measurable kinematic variables, we take event-generator-level information directly to calculate $a_{\rm LL}$. The mean value in each bin was used to extract the $\Delta g/g$ uncertainty from the $A_{\rm LL}$ uncertainty. 

\begin{figure}[htbp]
\centering
\includegraphics[width=0.46\textwidth]{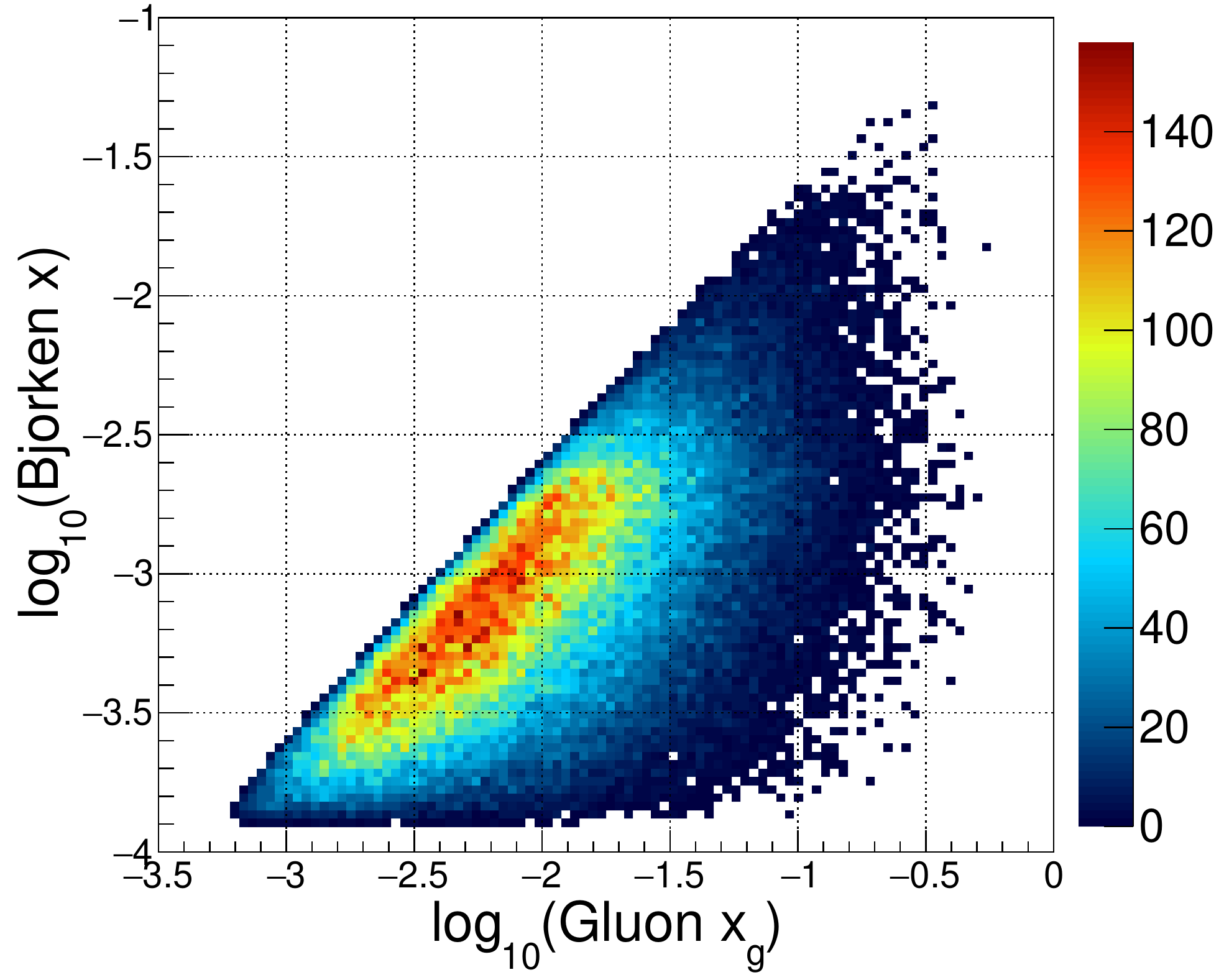}
\hspace{0.2in}
\includegraphics[width=0.45\textwidth]{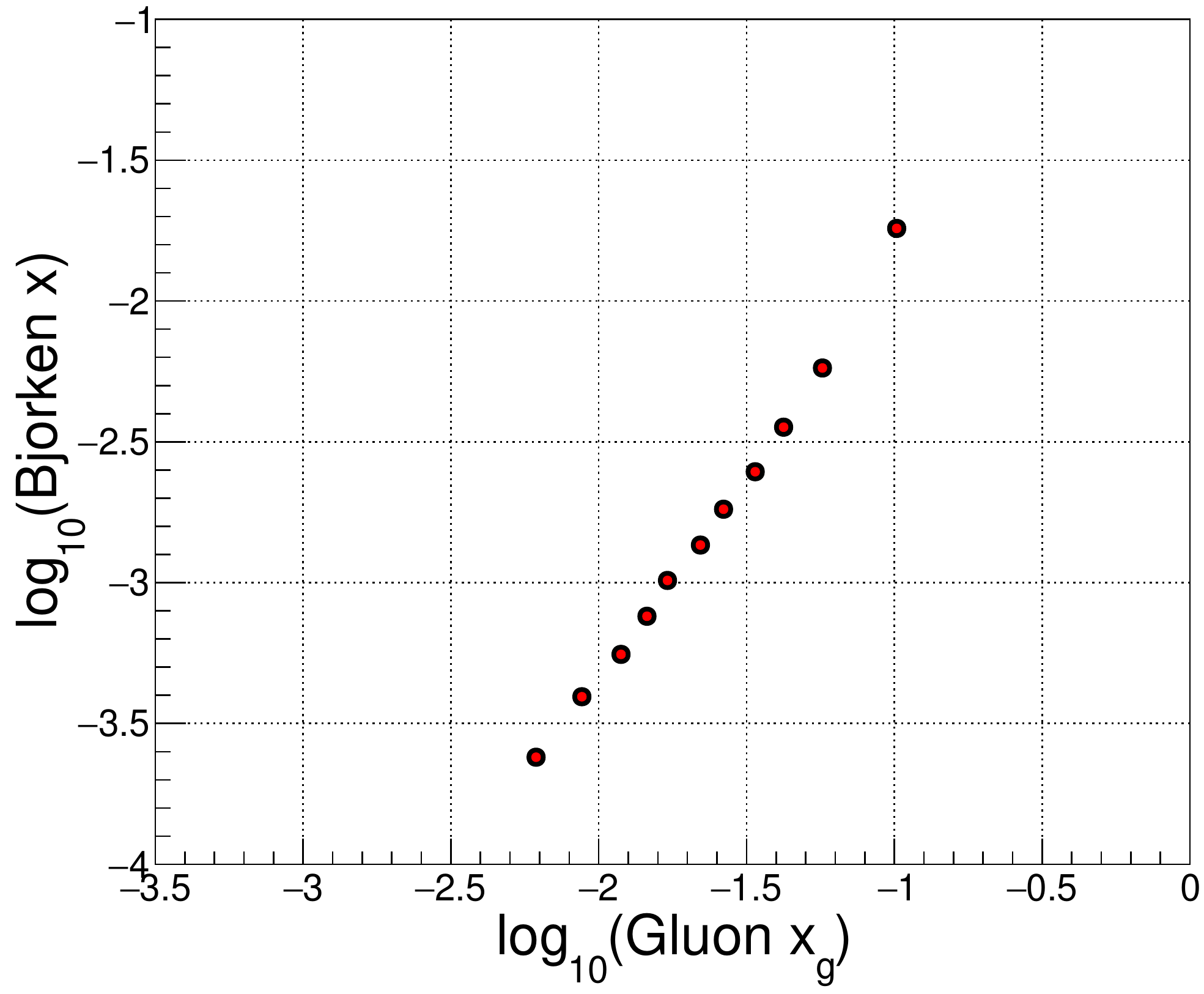}
\caption{The relation between Bjorken-$x$ ($x_B$) and gluon-$x$ ($x_g$) at event level (left) and average $x_g$ values in different $x_B$ bins (right) for 
energy configuration of 18$\times$275 GeV calculated from the PYTHIA v6.4 simulation.}
\label{fig:xg_x_18_275}
\end{figure}

Due to the nature of the photon-gluon fusion process, as shown in Fig.~\ref{fig:diagram}, the value of Bjorken-$x$ ($x_B$) obtained from the scattered electron differs from the gluon-$x$ ($x_g$), which describes the momentum fraction carried by the gluon inside the proton. A conversion from $x_B$ to $x_g$ is needed to interpret the measurement in terms of the partonic structure of gluon inside the proton, unlike the situation in virtual photon-quark scattering where $x_B$ is equal to $x$ of the quark. The event generator PYTHIA was used for the conversion bin-by-bin: in each $x_B$ bin, the photon-gluon fusion subprocess was identified, and then the target parton $x$ (here, it is $x_g$) distribution was drawn, from which the mean value can be obtained. As an example, Figure~\ref{fig:xg_x_18_275} shows the relation between Bjorken-$x$ and gluon-$x$ both at event level (left) and at bin-by-bin level (right) for the energy configuration 18$\times$275 GeV.

\begin{figure*}[htbp]
\centering
\includegraphics[width=0.7\textwidth]{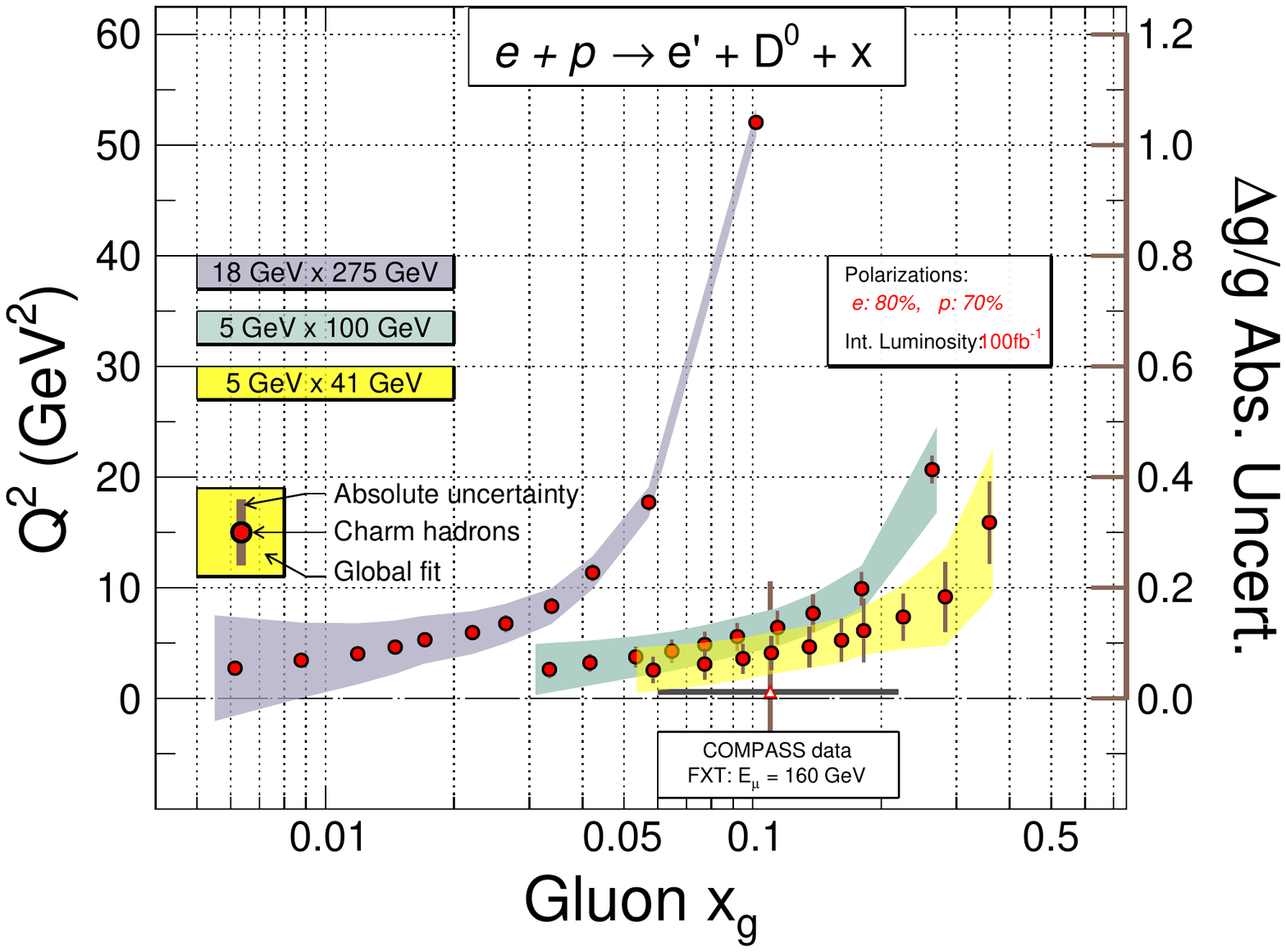}
\caption{Projections on $\Delta g/g$ as a function of gluon-$x$ ($x_g$)
and $Q^2$. The position of each data point is according to the mean
value of $x_g$ and $Q^2$ of the particular bin. The uncertainty for 
the data points corresponds to the scale shown in the vertical axis
on the right side of the plot. The colored band behind the data set of
each beam energy configuration is the uncertainty calculated
using NNPDF unpolarized and polarized PDFs~\cite{Ball:2012cx,Nocera:2014gqa}. The red triangle marker shows the existing measurement from COMPASS~\cite{Adolph:2012ca}.}  
\label{fig_delta_g}
\end{figure*}

Figure~\ref{fig_delta_g} shows the projections on $\Delta g/g$ as a function of $x_g$ and $Q^2$ for different energy configurations at the EIC. In addition, the calculations using NNPDF unpolarized and polarized PDFs at certain $x_g$ and $Q^2$ values are also shown as colored bands~\cite{Ball:2012cx,Nocera:2014gqa}. The only existing measurement in this channel from the COMPASS collaboration is also shown. As one can see from the plot, $\Delta g/g$ can be measured at high precision by taking advantage of open-charm production at the EIC. Although the precision is lower compared to the $g_1$ structure function measurements, this physics channel can provide an opportunity to access gluon distribution at a different angle, thus providing a crosscheck on the complicated QCD fits in order to extract the gluon information. Moreover, due to the shift between $x_B$ and $x_g$, the measurements at the EIC allow us to study gluon polarization in a relatively high-$x_g$ region, which is unique compared to the $g_1$ measurement at the EIC.
Especially, in the overlap region $0.03<x_g<0.3$ for different beam energy configurations, the uncertainty of $\Delta g/g$ can be significantly reduced by combining measurements in different beam-energy configurations.

We would like to emphasize that the inclusive measurements of the polarized structure functions over a wide range of kinematics at the EIC will play the dominant role in constraining the gluon helicity distribution and its contribution to the proton spin~\cite{Accardi:2012qut}. Nevertheless, the longitudinal double-spin asymmetries of heavy flavor production in DIS processes will provide complementary constraints on the gluon helicity distribution. As we show in the above, in some kinematics, e.g. the moderate-$x_g$ region, heavy flavor production may play a unique role. More importantly, this can be compared to similar measurements of inclusive jet and dijet production~\cite{Boughezal:2018azh, Page:2019gbf, Borsa:2020ulb,Borsa:2020yxh}. All of these studies should be carried out systematically in EIC experiments to answer the nucleon spin puzzle.

\subsubsection{Charm Baryon $\Lambda_c^{\pm}$ Production and Hadronization}\label{sec:Lc}

The hadronization process remains a challenging problem that is yet to be understood in QCD. Fragmentation functions (FFs) have been widely applied under the collinear factorization and are constrained via experimental data from $e^+$+$e^-$ or $e$+$p$ collisions and are expected to be universal and thus directly applicable to hadronic collisions. Many Monte Carlo event generators utilize the same or similar schemes for partons hadronizing into hadrons, e.g. Lund string fragmentation used in the PYTHIA generator.

Recently, data from $p$+$p$, $p$+$A$, and $A$+$A$ collisions at RHIC and LHC showed that 
the $\Lambda_c^+/D^0$ ratio is considerably larger than the fragmentation baseline~\cite{Acharya:2017kfy,Adam:2019hpq}. The new color reconnection (CR) scheme implemented in the PYTHIA8 generator ~\cite{Christiansen:2015yqa}, together with the baryon-junction scheme, increases the $\Lambda_c^+/D^0$ ratio at low $p_T$ and is comparable to the experimental data. A detailed investigation of the $\Lambda_c^+$ production at high-luminosity EIC collisions will offer an opportunity to enable detailed investigations to understand how the hadronization plays a role from $e^+e^-$ to hadronic collisions.

The $\Lambda_c^+$ baryon has an extremely short lifetime with a proper decay length $c\tau\sim$ 60\,$\mu$m (a factor of 2 smaller than $D^0$ and an order of magnitude smaller than $B$ hadrons). The decay vertex reconstruction of $\Lambda_c^+$ will place very stringent requirements on the detector pointing/vertexing/PID capabilities. In this section, we will describe how a silicon tracker can improve the $\Lambda_c^+$ signal significance and statistical uncertainties on physics observables at the EIC.

The $e$+$p$ collision events are generated with PYTHIA v6.4 using the EIC tune with 18$\times$275 GeV beam energies, and processed through the fast-simulation framework described in Sec.~\ref{sec:sim:fast}. Similar to the $D^0$ simulation, DCA and momentum resolutions based on the parametrizations listed in Table~\ref{tab:sim:smearing} are used in the fast simulation. 
We apply the primary vertex resolution as well as its multiplicity-dependence, and the tracking efficiency, evaluated from full GEANT4 simulation, as shown in Fig.~\ref{fig:fullsim:vtx}. 
We assume that $p/K/\pi$ tracks can be separated perfectly below the momentum limits listed in Table~\ref{tab:sim:smearing}. 

\begin{figure*}[htbp]
	\centering
	\includegraphics[width=0.48\textwidth, trim={15mm 0 0 0}, clip]{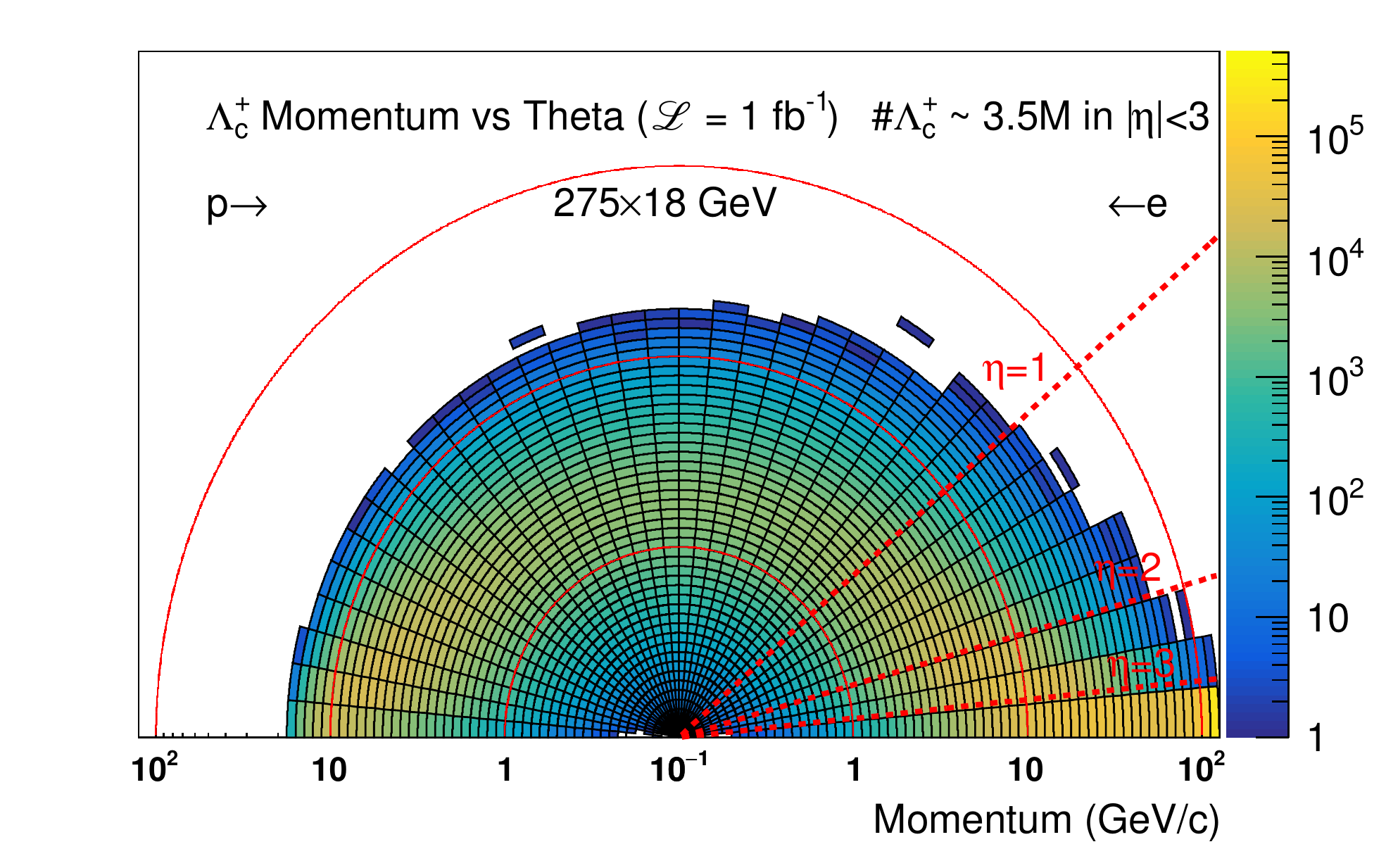}
	\includegraphics[width=0.48\textwidth, trim={15mm 0 0 0}, clip]{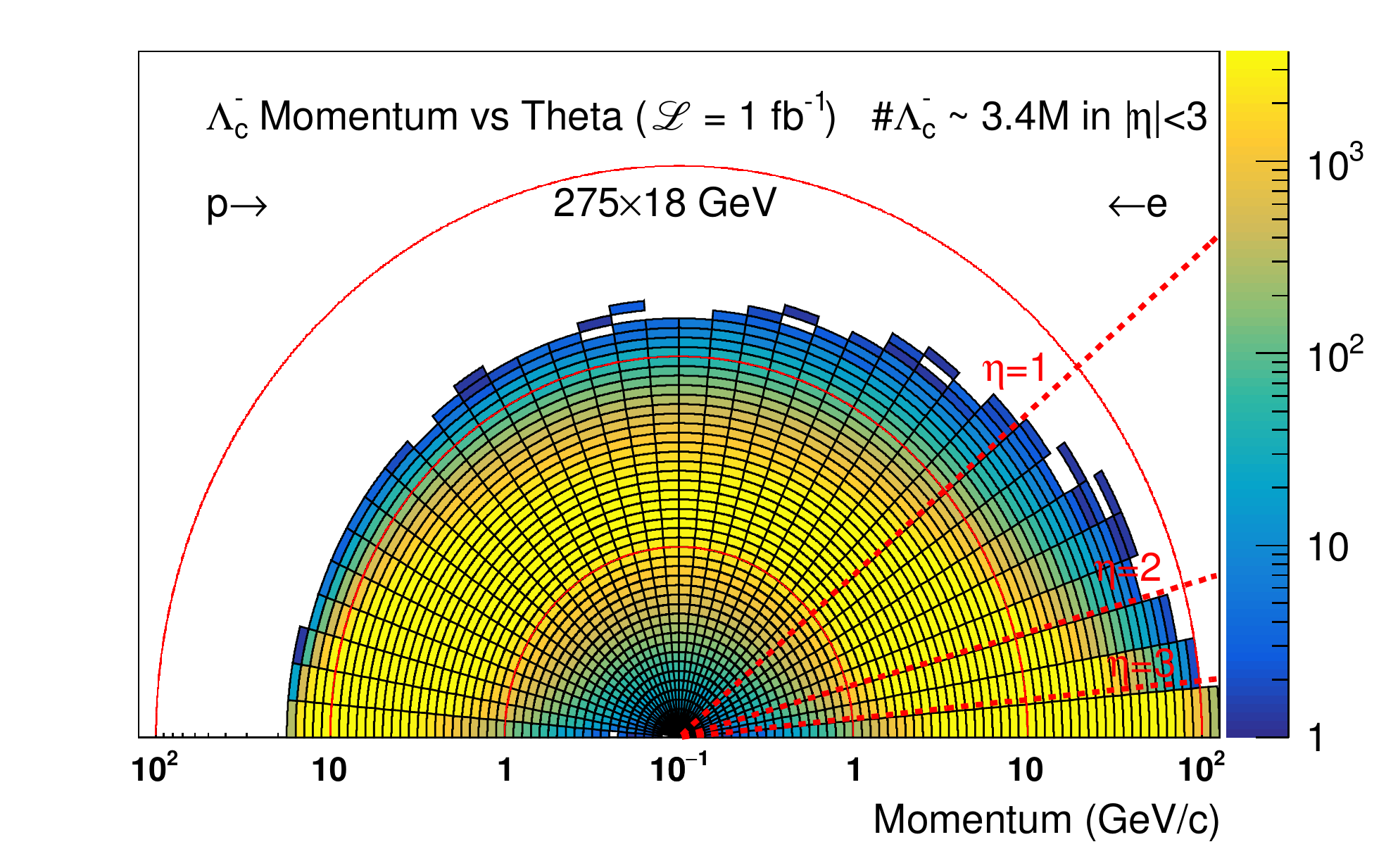}
	\caption{Kinematic distributions of $\Lambda_c^{+}$ (Left) and $\Lambda_c^{-}$ (right) from PYTHIA v6.4 with EIC tune in $e$+$p$ (18$\times$275 GeV) collisions as a function of momentum and polar angle. The total counts of $\Lambda_c$ within $|\eta|<$3 is about 3.5$\times 10^{6}$ at a integrated luminosity of $1$ fb$^{-1}$ in this calculation.}
	\label{fig:Lc:Accpt}
\end{figure*}

Fig.~\ref{fig:Lc:Accpt} shows kinematic distributions of produced $\Lambda_{c}^+$ and $\Lambda_{c}^-$ hadrons from PYTHIA v6.4 in $e$+$p$ 18$\times$275 GeV collisions as a function of momentum and $\theta$ (angle with respect to the beam line) in polar coordinates. There are more $\Lambda_{c}^{+}$ produced in the very forward region compared to $\Lambda_{c}^{-}$ in PYTHIA v6.4, which is due to processes in which charm quarks re-combine with the beam remnants. In the central acceptance region, e.g. $|\eta|<3$, the $\Lambda_{c}^{-}/\Lambda_{c}^{+}$ ratio is close to 1. 

\begin{figure*}[htbp]
	\centering
	\begin{minipage}[t]{0.48\linewidth}
	\centering
	\includegraphics[height=4.8cm,trim={0 0 0 5mm}, clip]{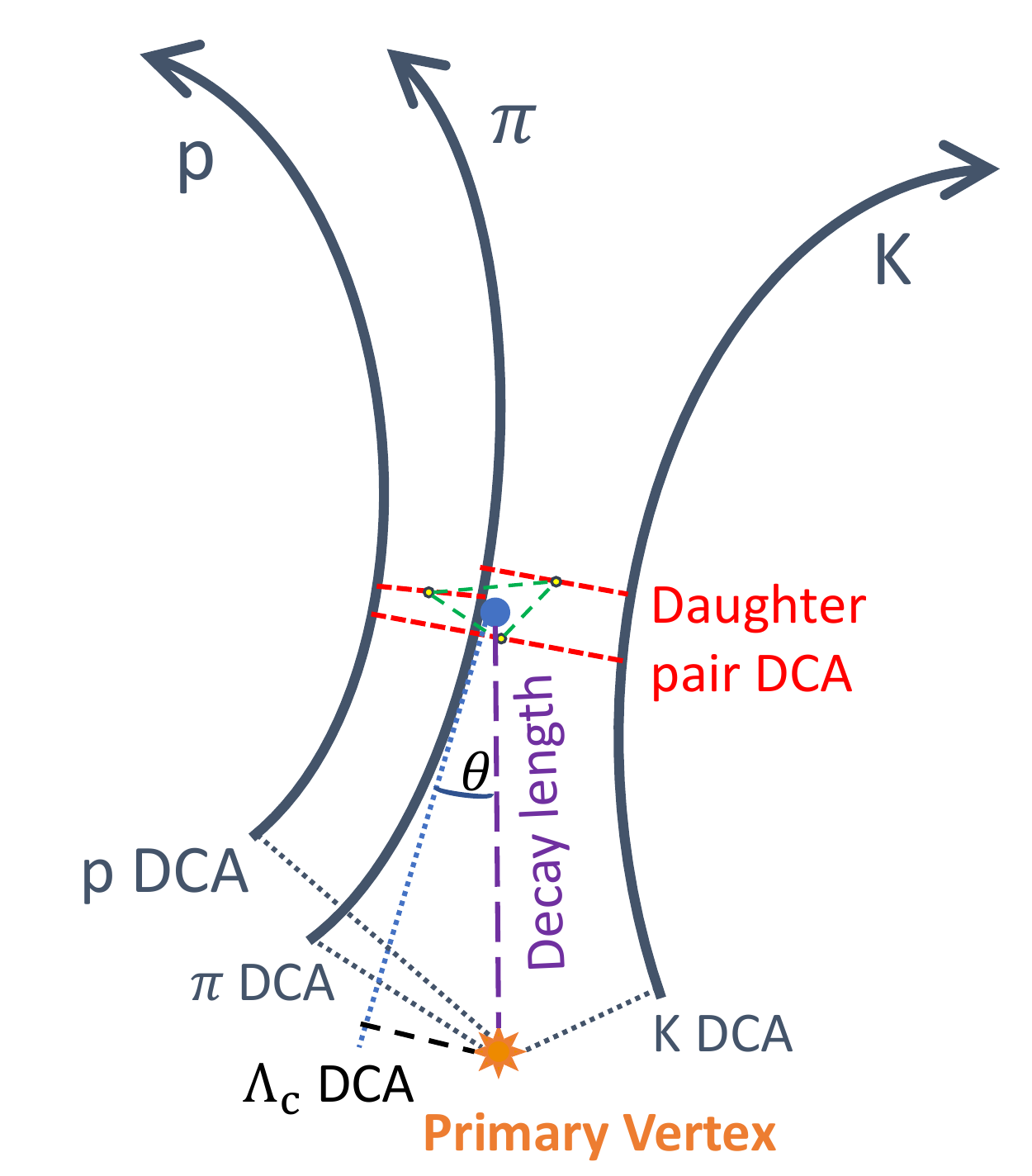}
    \end{minipage}
    \begin{minipage}[t]{0.48\linewidth}
    \centering
    \includegraphics[height=5.2cm,trim={0 0 0 0}, clip]{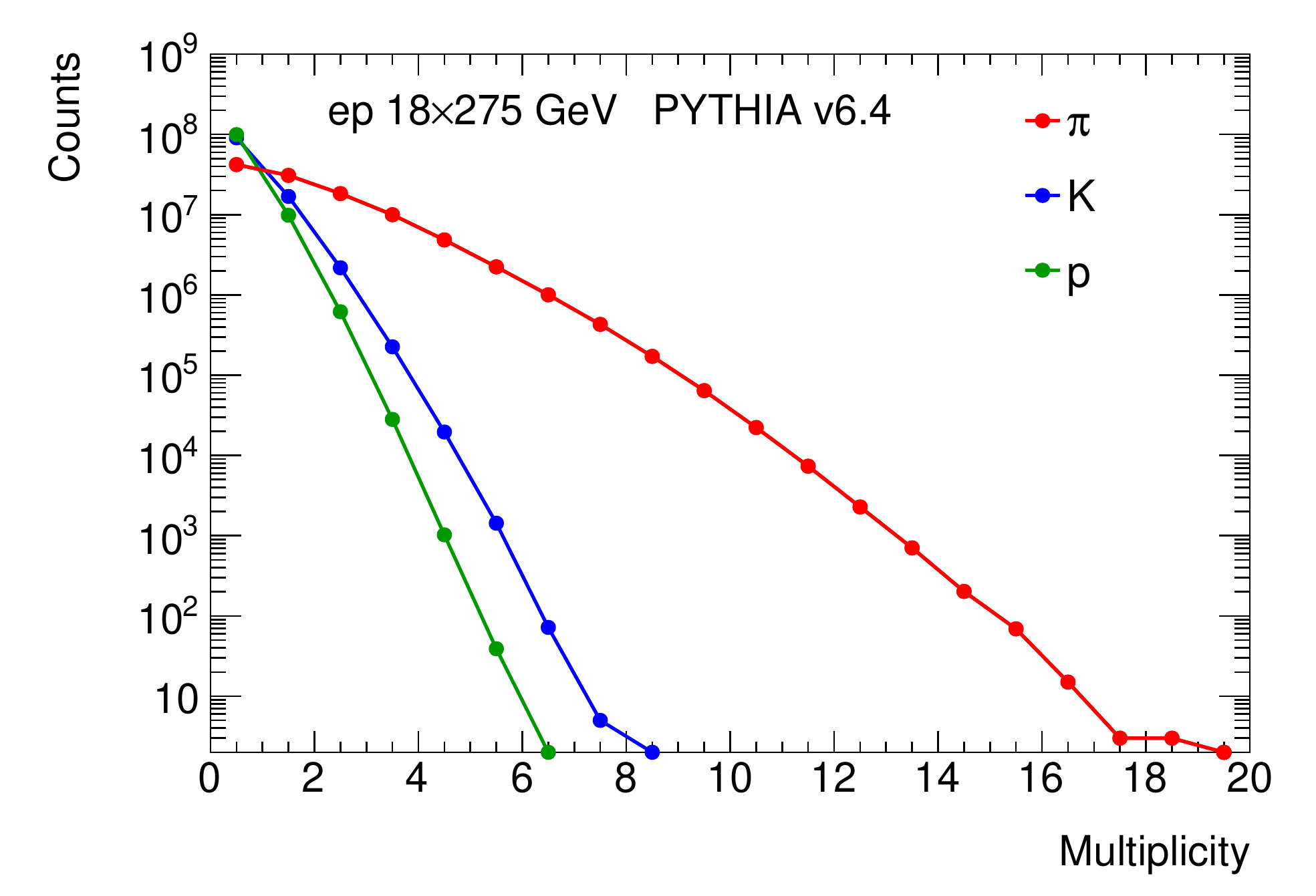}
    \end{minipage}
	\caption{(Left) A schematic cartoon of $\Lambda_{c}$-decay topology. (Right) Final-state identified hadron ($p/K/\pi$) multiplicity distributions in $e$+$p$ (18$\times$275 GeV) collisions from a PYTHIA v6.4 simulation.}
	\label{fig:Lc:reco}
\end{figure*}

In this simulation study, final state $\Lambda_c^{+}$($\Lambda_c^{-}$) hadrons are reconstructed via the decays of $\Lambda_{c}^{+}\rightarrow K^{-}p\pi^{+}$ (and its charge-conjugate channel), which include one non-resonant channel and three resonant channels~\cite{PDG2020}. In PYTHIA v6.4, branching ratios (B.R.) for these channels are not up-to-date and one resonant decay channel ($\pi^+\Lambda({\rm 1520})$) is missing. Table~\ref{tab:Lc:BR} compares the branching-ratio values of various channels used in PYTHIA v6.4 and in PDG 2020~\cite{PDG2020}. In this study, only the non-resonant channel is used for signal reconstruction while the final statistics for the $\Lambda_c^+$ signals are scaled to the total B.R. (6.28\%) for the $pK^-\pi^+$ channel.

\begin{table}
	\caption{ $\Lambda_c^+\rightarrow pK^-\pi^+$ decay channels including intermediate resonance channels and their branching ratios in PYTHIA v6.4 and PDG 2020. \label{tab:Lc:BR}}
	\centering
	\begin{tabular}{c | c  c}
	Decay Channel & ~~~B.R. (PYTHIA 6)~~~ & ~~~B.R. (PDG-2020)~~~ \\ \hline \hline
	$p\overline{K^*}\rightarrow p K^-\pi^+$ & 0.58\% & 1.96\%$\times$66.7\% \\	
	$K^-\Delta({\rm 1232})^{++}\rightarrow K^-p\pi^+$ & 0.66\% & 1.08\%$\times$99.4\% \\
    $\pi^+\Lambda({\rm 1520})\rightarrow \pi^+ p K^{-}$ & N/A & 2.20\%$\times$22.5\%\\
    non-resonant & 2.96\% & 3.50\% \\
    \hline
	Total $K^-p\pi^+$ & 4.20\% & 6.28\% \\
	\end{tabular}
\end{table}

We combine all three-track triplets $pK^-\pi^+$ and $\bar{p}K^+\pi^-$ with the right-sign combination for signal reconstruction. If they are not from the decay of the same $\Lambda_c^{\pm}$, the combinations are regarded as background. Figure~\ref{fig:Lc:reco} (left plot) shows a sketch of the $\Lambda_{c}$-decay topology, and the right plot shows the multiplicity distributions of identified particles ($p/K/\pi$) in $e$+$p$ events. Given that the multiplicity of produced particles from such kind of events is small, especially for protons, it is expected that the combinatorial background level is lower than $p$+$p$ collisions at similar energies.

\begin{figure*}
	\centering
	\includegraphics[width=0.95\textwidth]{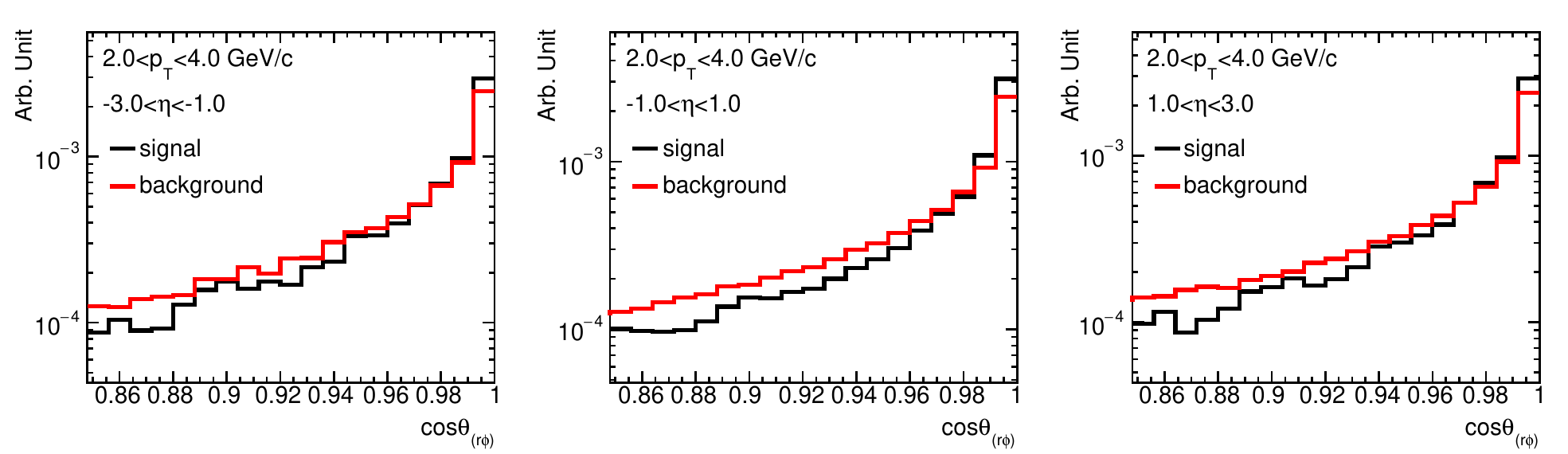}
	\includegraphics[width=0.95\textwidth]{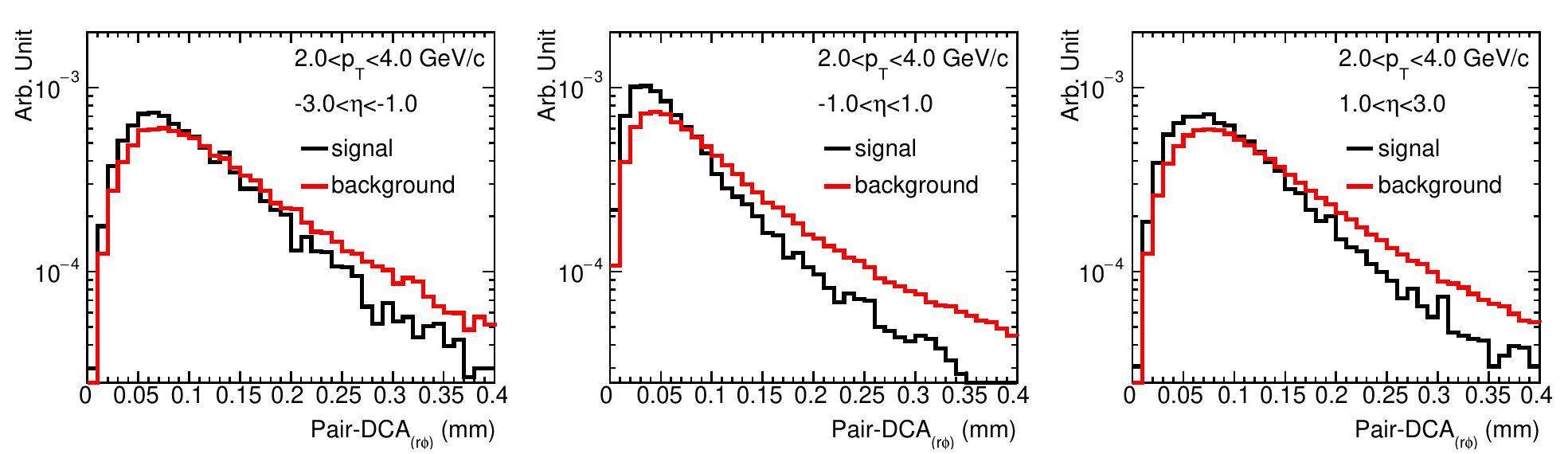}
	\caption{$\cos\theta_{r\phi}$ (upper row) and 
	Pair-DCA$_{r\phi}$ (lower row) distributions of signal and background in $\Lambda_c$ reconstruction for $-1<\eta<3$, $-1<\eta<1$ and $1<\eta<3$.}
	\label{fig:Lc:topo}
\end{figure*}

\begin{figure*}
	\centering
	\includegraphics[width=0.95\textwidth]{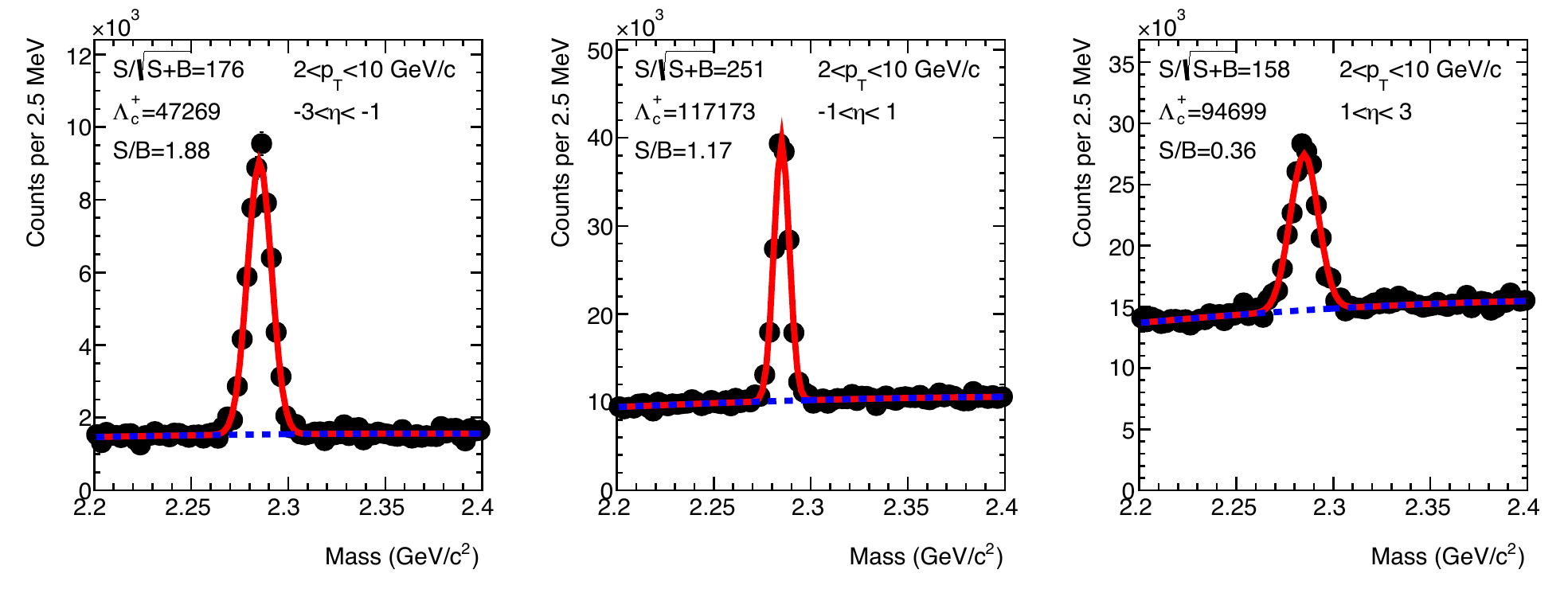}
	\caption{Invariant-mass distributions of $pK^-\pi^+$ triplets in $\Lambda_c^{+}$ reconstruction in $2<p_T<4$\,GeV/$c$ from DIS $e$+$p$ (18$\times$275 GeV) collisions based on a PYTHIA v6.4 + fast simulation in the $-3<\eta<-1$ (left), $-1<\eta<1$ (middle) and $1<\eta<3$ (right) regions, respectively.}
	\label{fig:Lc:mass}
\end{figure*}

Figure~\ref{fig:Lc:topo} shows the normalized distributions of selected topological variables in the $-1<\eta<3$, $-1<\eta<1$ and $1<\eta<3$ regions in the $p_T$ region of $2<p_T<4$\,GeV/$c$. The signal and background distributions are different in these variables, allowing the topological separation to enhance the $\Lambda_c$-signal reconstruction. A set of loose topological cuts is applied to keep a high reconstruction efficiency, which is listed in Table~\ref{tab:Lc:topocuts}. There is no minimum-$p_T$ cut applied on daughter tracks in the current calculation. The projected invariant-mass distributions of $\Lambda_{c}^{+}$ signals at in different $\eta$ ranges are shown Fig.~\ref{fig:Lc:mass}.

\begin{table*}[htbp]
	\begin{center}
		\caption{Decay topology cuts for $\Lambda_c^{\pm}$ signal reconstruction.}
		{\begin{tabular}{ c | c  c  c}
				~~~Selection criteria~~~ & ~~~0$<$$p_T$$<$4\,GeV/$c$~~~ & ~~~4$<$$p_T$$<$6\,GeV/$c$~~~ & ~~~6$<$$p_T$$<$10\,GeV/$c$~~~\\
				\hline \hline
				Pair-DCA$_{r\phi}$ & $<300$~$\mu$m & $<100$~$\mu$m & $<100$~$\mu$m\\
				$\Lambda_c$-DCA$_{r\phi}$ & $<150$~$\mu$m  &  $<100$~$\mu$m  & $<60$~$\mu$m \\
				Decay-Length$_{r\phi}$ & $>10$~$\mu$m & $>20$~$\mu$m & $>30$~$\mu$m\\
		\end{tabular}}
		\label{tab:Lc:topocuts}
	\end{center}
\end{table*}

\begin{figure*}
	\centering
	\includegraphics[width=0.48\textwidth]{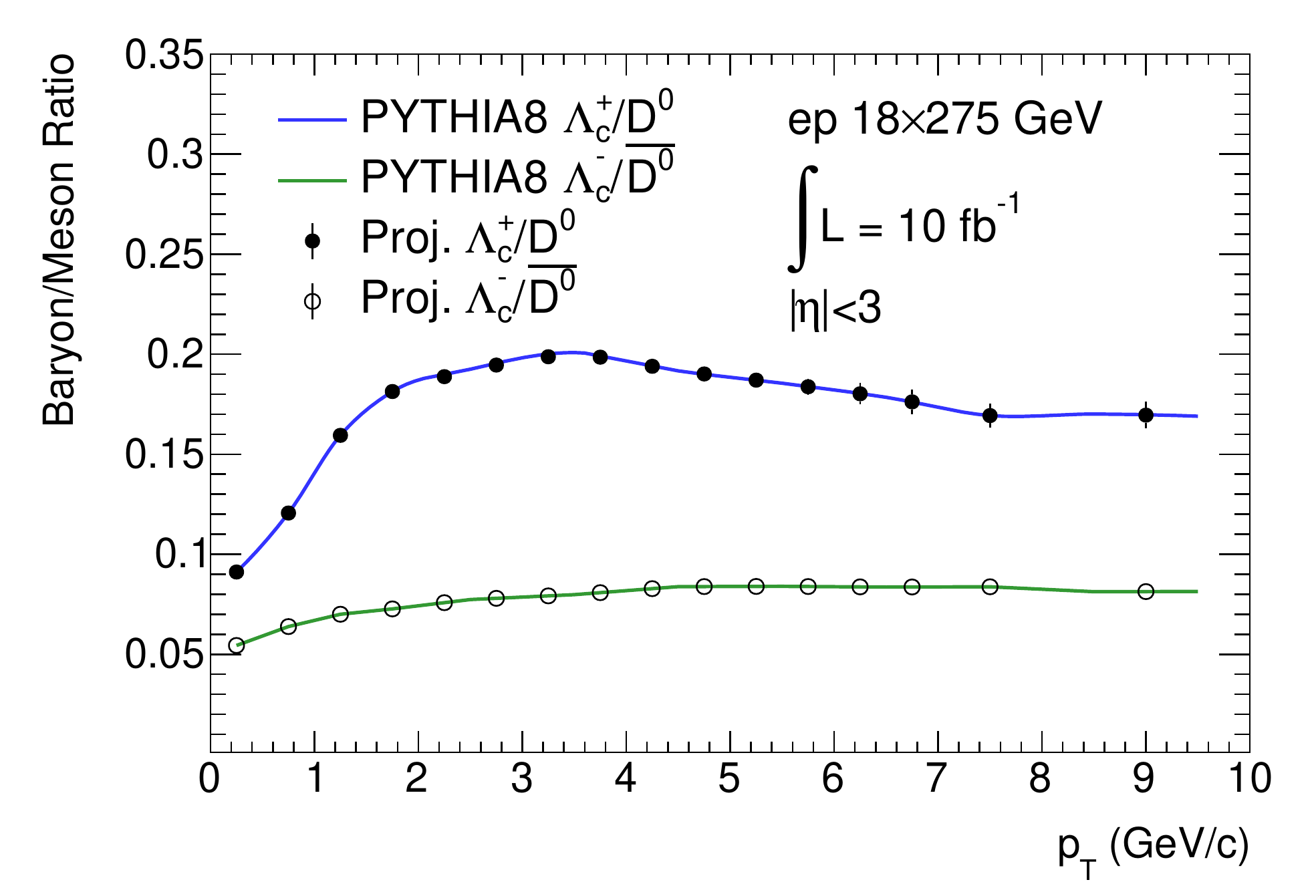}
	\includegraphics[width=0.48\textwidth]{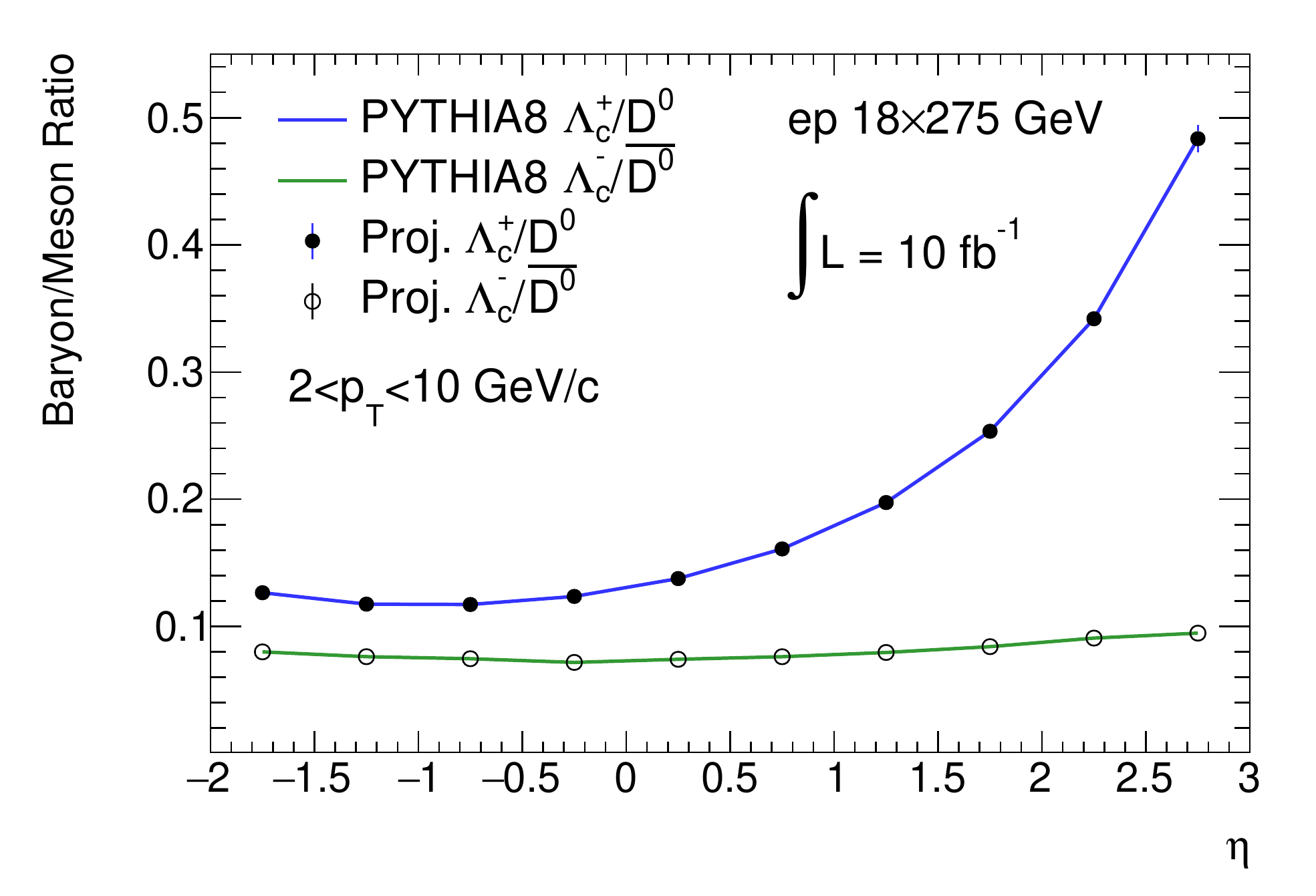}
	\caption{Projected statistical uncertainty of the $\Lambda_c^{+}/D^0$ and $\Lambda_c^{-}/\overline{D^0}$ ratios 
		as a function of $p_T$ for $|\eta|<3$ (left) and as a function of $\eta$ in $2<p_T<10$\,GeV/$c$ (right) in 10\,fb$^{-1}$ $e$+$p$ 18$\times$275 GeV collisions. The central values for the projected points are extracted from PYTHIA v8.3 simulations.
		}
	\label{fig:Lc:intproj}
\end{figure*}

\begin{figure*}
	\centering
	\includegraphics[width=0.5\textwidth]{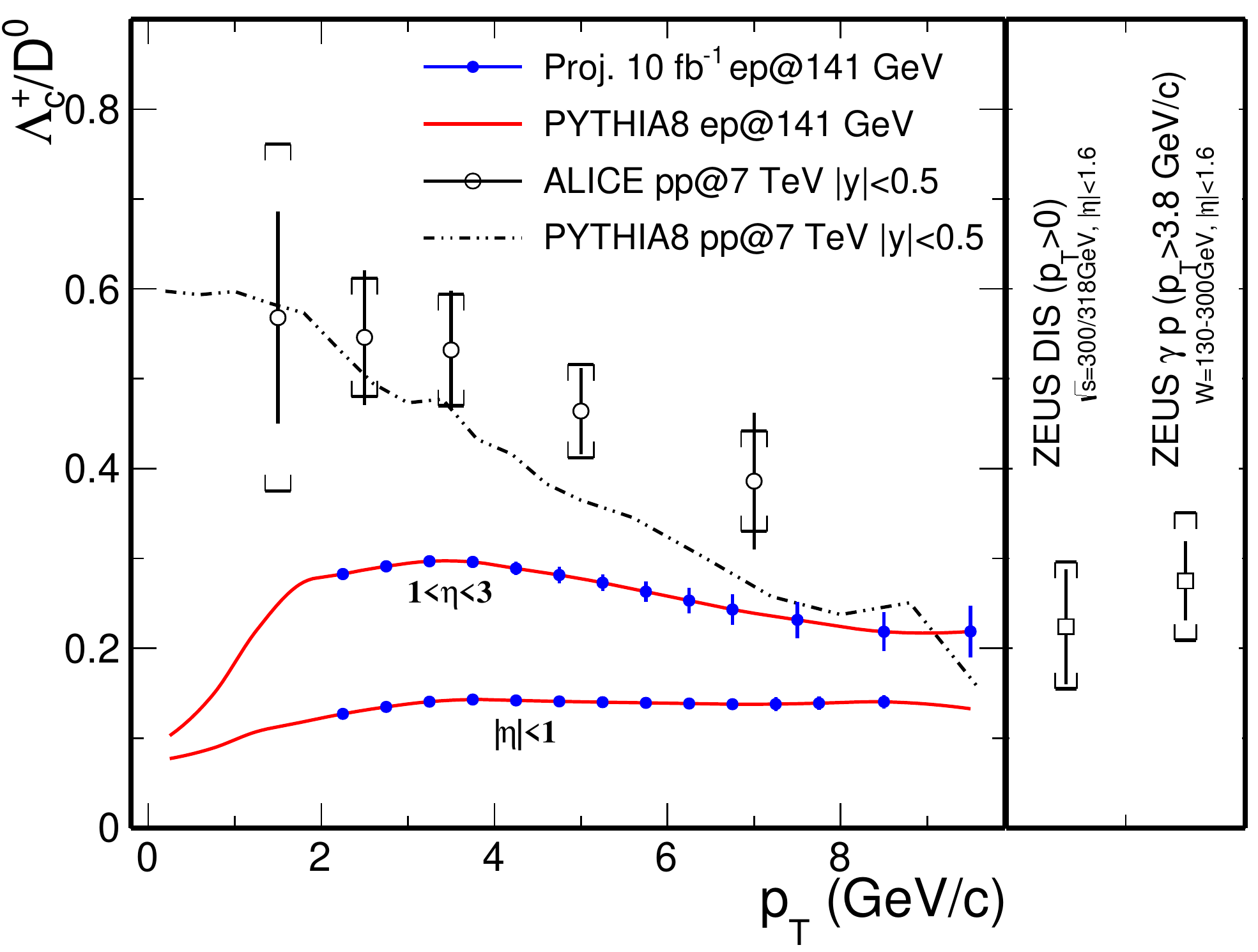}
	\caption{The projected statistical uncertainty of $\Lambda_c^{+}/D^0$ as a function of $p_T$ in $|\eta|<1$ and $1<\eta<3$ in $e$+$p$ 18$\times$275 GeV collisions. The mean values for the projected points are from PYTHIA8.3 calculations. Open circle and open square points are the measurements in $p+p$ $\sqrt{s}=7$\,TeV collisions from ALICE~\cite{Acharya:2017kfy} and $e$+$p$ collisions from ZEUS~\cite{Chekanov:2005mm,Abramowicz:2010aa}.}
	\label{fig:Lc:proj}
\end{figure*}

Figure~\ref{fig:Lc:intproj} shows projected statistical uncertainties of $\Lambda_{c}^+/D^{0}$ and $\Lambda_{c}^-/\overline{D^{0}}$ as a function of $p_T$ in $|\eta|<$3 (left) and as a function of $\eta$ in 2$<$$p_T$$<$10\,GeV/$c$ (right) with 10\,fb$^{-1}$ $e$+$p$ (18$\times$275 GeV) collisions. Figure~\ref{fig:Lc:proj} shows the projections of $\Lambda_{c}^+/D^{0}$ as a function of $p_T$ for two $\eta$ regions ($|\eta|<1$ and $1<\eta<3$). The $\Lambda_c^+$ cross section used in Fig.~\ref{fig:Lc:intproj} and ~\ref{fig:Lc:proj} is based on the recent PYTHIA v8.3 calculation which includes the latest development on the color reconnection scheme for baryon production at high energy $p$+$p$ collisions. Also shown in Fig.~\ref{fig:Lc:proj} are the existing measurements in $p$+$p$ collisions from ALICE~\cite{Acharya:2017kfy} and $e$+$p$ DIS and $\gamma p$ collisions from ZEUS~\cite{Chekanov:2005mm,Abramowicz:2010aa}. The projection shows that measurements at EIC $e$+$p$ DIS collisions would allow us to systematically investigate the $\Lambda_c$ production over a broad kinematic region, which will shed detail insights on charm hadrochemistry and charm-quark hadronization.

Recent measurements in open and closed charm hadron production in high multiplicity $p$+$p$ collisions at RHIC and LHC attracted lots of interests~\cite{Adam:2015ota,Acharya:2020pit,Adam:2018jmp}. Various models including multi-parton interaction, color reconnection implemented in PYTHIA, and coherent production~\cite{Kopeliovich:2019phc} have been exercised while the exact production mechanism is still under investigation. Measurements of various charm hadrons including $D^0$ and $\Lambda_c^+$ in $e$+$p$/A collisions at the EIC will give us an opportunity to study these non-perturbative features in detail.

High statistics will enable to perform the double-spin transfer $D_{\rm LL}$ measurement of $\Lambda_c$ baryon similar to $\Lambda$ in polarized $e$+$p$ collisions. Early prediction on $\Lambda_c$ $D_{\rm LL}$ for polarized $p$+$p$ collisions at RHIC suggested the sensitivity to the parton spin structure (gluon helicity) inside proton as well as polarized fragmentation function~\cite{Rykov:2004ss}. While the measurement turned out to be challenging at RHIC due to experimental limitations, a high-luminosity EIC will enable the $\Lambda_c$ $D_{\rm LL}$ measurement with the desired instrumentation capability for secondary vertex reconstruction. 


\subsubsection{Heavy Quark Conclusions}\label{sec:HFconclu}
We presented detailed simulation studies of heavy flavor measurements, focusing on charm hadron reconstruction using the all-silicon tracker
in electron-proton collisions. Precision measurements 
must separate charm hadron decay vertices from the collision vertex, which imposes stringent requirements on the tracking/vertexing capabilities for EIC experiments. The tracker concept presented here meets those requirements.

Our simulations of the inclusive charm structure function $F_2^{c\bar{c}}$ measurement in various ($x_B$,$Q^2$) bins are very encouraging.
High luminosity $e$+$p$/A collisions at EIC will enable unprecedented precision measurement of $F_2^{c\bar{c}}$, covering a $x_B$ region of $10^{-3}$ up to $>$0.1 and offering new insights into the gluon PDF and nuclear PDF's.

Reconstruction of $D^0\overline{D}^0$ pairs allows study of the gluon TMD functions including the gluon Sivers function in electron scattering on transversely polarized protons. 
The linearly polarized Boer-Mulders function in unpolarized $e$+$p$ collisions is also accessible. With 100\,fb$^{-1}$ integrated luminosity for 18$\times$275 GeV $e$+$p$ collisions, the projected uncertainties on $\langle A_{\rm UT}\rangle$ and $\langle\cos(2\phi_T))\rangle$ are 0.57\% and 0.4\% respectively using $D^0\overline{D}^0$ pair reconstructed via the $K\pi$ channel.

Charm hadron double-spin asymmetry ($A_{\rm LL}$) in longitudinally polarized $e$+$p$ collision provides a unique sensitivity to the gluon helicity $\Delta g/g$ contribution to the proton spin. We 
evaluated statistical uncertainties of $D^0$ hadron $A_{\rm LL}$ in various kinematic bins for
several collision energy configurations. We also estimated the impact on the uncertainty in  $\Delta g/g$ based on a leading order calculation. Compared to the existing measurement from COMPASS and current QCD analysis uncertainties, future measurements at the EIC will provide significantly improved precision in $A_{\rm LL}$, complementary to inclusive spin-dependent structure function measurements.

$\Lambda_c^+$ production has attracted significant
interest, inspired by recent findings in hadronic collisions at RHIC and the LHC. We investigated the charm hadron $\Lambda_c^+$ reconstruction capability 
and projected $\Lambda_c^+$ physics performance, in particular the measurement of $\Lambda_c^+/D^0$ over a wide kinematic region. This will open a great opportunity to characterize the $\Lambda_c^+$ production mechanism and gain insights into
hadronization, along with using $\Lambda_c^+$ as a tool to investigate the structure of nucleons and nuclei.

%% file: jets.tex
\subsection{Jets}

\subsubsection{Physics Introduction}
Partons from initial hard scatterings cannot be observed directly as final-state particles. Instead, they hadronize into a directional spray of final-state particles.
Consequently, jets are composite objects that relate such final-state particles measured in the detector to an initial parton, and so can serve as a powerful tool for probing QCD. Jets are measured experimentally by clustering the observed particles using a particular clustering algorithm (anti-$k_T$ \cite{Cacciari:2008gp} in this work) within a chosen jet resolution parameter.

\begin{figure}[htb]
\centering
	\includegraphics[width = 0.5 \textwidth]{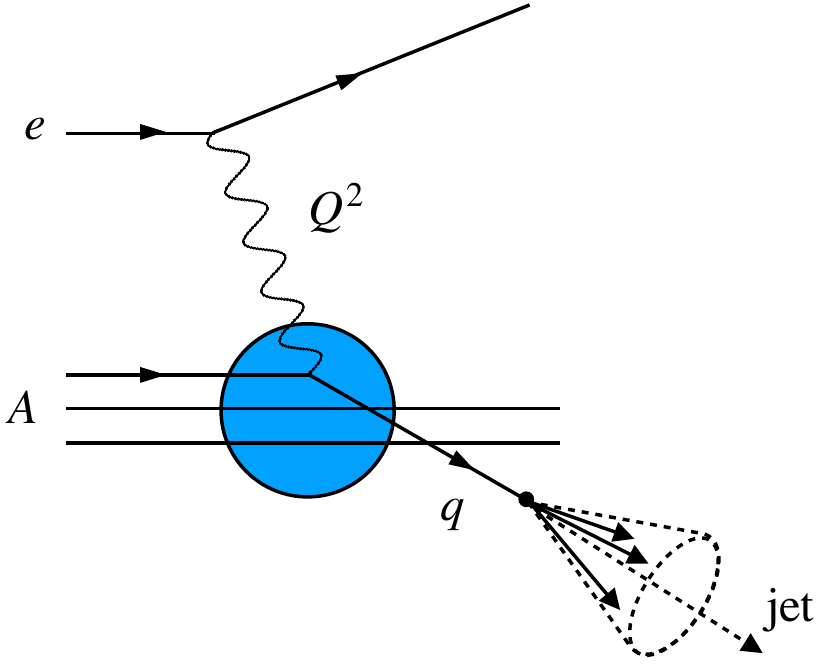}
	\caption{Leading order deep inelastic scattering diagram. The struck quark is observed as a jet of final state hadrons and serves as an excellent probe of the nucleus.}
\end{figure}

Earlier studies of $e$+$p$ collisions utilized jets, but in a limited fashion \cite{Wong:2020xtc}.
An extensive jet program has been proposed for the EIC, as follows: 
\begin{itemize}
\item Jets from electroproduction in DIS can be used to study parton energy loss and interactions in cold nuclear matter~\cite{Arratia:2019vju}.
\item Inclusive jet production in polarized electron-proton collisions constrains the helicity-dependent parton distribution function (PDF) of the proton at low $x$, complementary to existing measurements at high-$x$~\cite{Boughezal:2018azh}. 

\item Inclusive jet production in DIS off nuclei ~\cite{Klasen:2017kwb} and dijet quasi-real photoproduction~\cite{Klasen:2018gtb} can be used to advance our knowledge of nuclear PDFs. 
\item Dijet photoproduction gives access to the photon PDF~\cite{Chu:2017mnm}. \item Single-inclusive lepton scattering resulting in jets, where the scattered electron is not observed, has been proposed as an EIC measurement to study transverse spin effects in the nucleon~\cite{Hinderer:2015hra,Hinderer:2017ntk}.

\item Jets complement measurements of the three-dimensional structure of hadrons, encoded in transverse momentum-dependent (TMD) PDFs and fragmentation functions (TMD FFs).
Unlike in the semi-inclusive DIS case, jet measurements allow the extraction of these two quantities separately. Specifically, jet measurements at the EIC have been proposed to constrain the quark Sivers function, transversity distribution, and the Collins FF~\cite{Arratia:2020nxw}. 
\item Dijet production can be used to access gluon TMD functions at the EIC~\cite{Zheng:2018ssm,Dumitru:2018kuw}.

\item Charm-jet cross section measurements can be used to resolve the tension between different experimental results regarding the strangeness content of the nucleon~\cite{Arratia:2020azl}.

\item Substructure measurements can be used to tune parton-shower event generators and study cold nuclear matter effects as explored in~\cite{Aschenauer:2019uex}.
\end{itemize}

In this section we study the jet momentum and angular resolutions achievable with the all-silicon tracker described above. 
Then, we will focus on two specific observables, the azimuthal difference between jets produced in DIS and the scattered electron and the jet fragmentation function, to illustrate physics with jets reconstructed in the silicon tracker. 

\subsubsection{Charged Jet Reconstruction Performance}

Charged jets have been measured extensively in $p$+$p$ collisions with the ALICE detector at the large hadron collider~\cite{PhysRevD.91.112012}.
Track-only jets can often offer greater experimental precision, but are traditionally harder to compare to theoretical calculations. However, there has been progress in connecting experimental track-only jet observables with theoretical studies~\cite{PhysRevD.88.034030}. To quantify the jet reconstruction performance of the silicon tracker described in section~\ref{sec:sim:full},
electron-proton collisions are simulated with the PYTHIA8 Monte-Carlo generator and a full GEANT4 simulation with a 1.4\,T and 3.0\,T solenoidal magnetic field. Jets are reconstructed using the anti-$k_\mathrm{T}$ algorithm, with a large resolution parameter of $R= 1.0$. This is feasible due to the relatively low multiplicity of particles produced in $e$+$p$ collisions. 
Reconstructed jets are required to have 4 or more constituents and a minimum total energy of $4.0~\mathrm{GeV}$ in order to be considered an actual jet. Jets are reconstructed in the range $|\eta|<3.5$, according to the acceptance of the all-silicon tracker. Reconstructed jets within $\Delta R = 0.5$ of the highest energy electron in the event are omitted to ensure that the beam electron is not included as part of any jet.

Additional selections are made on the jet constituents.
They are required to have a $p_\mathrm{T} \geq$ 70 MeV/$c$, with a higher threshold depending on $\eta$ of the constituent.
The minimum $p_\mathrm{T}$ for different $\eta$ regions is shown in Table~\ref{tab:min_pt1}.
The values in Table \ref{tab:min_pt1} are extracted from Ref.~\cite{DMtable:2020} and exceed the values discussed in
section~\ref{sec:sim:full}. These are based on the need for three or more traversed barrel layers or disks in the all-silicon tracker 
in order to determine the track curvature for a charged particle and hence its transverse momentum.
Jets with constituents that hit the conical supports where the central barrel meets the forward and backward disks are omitted. 
Based on Fig.~\ref{fig:material}, we take this range to be $1.06 < |\eta| < 1.13$.

\begin{table}[htb]
\caption{Minimum $p_\mathrm{T}$-threshold (in MeV/$c$) for charged jet constituents.}
\begin{tabular}{  c | c | c | c | c | c  }
~~$B$ field [T]~~& ~~$ |\eta| < 1.0 $~~ & ~~$ 1.0 < |\eta| < 1.5  $~~ & ~~$1.5 < |\eta| < 2.0 $~~ & ~~$ 2.0 < |\eta| < 2.5 $~~ & ~~$2.5 < |\eta| < 3.5$~~\\
\hline \hline
1.4 & 200 & 150 & 70  & 130  & 100 \\
3.0 & 400 & 300 & 160  & 220  & 150 \\
\end{tabular}
\label{tab:min_pt1}
\end{table}

Reconstructed jets are matched to truth-level jets by requiring that one or more reconstructed tracks in the jet originate from a truth particle. If this truth particle is a constituent of a truth-level jet, the reconstructed and truth jet are matched. A unique reco-truth match is enforced for the following jet performance studies.
Once the jets are matched, the neutral components from the truth-level jets are subtracted to obtain charged truth jets. 
The 4-momenta of the neutral constituents are subtracted from the particle-level generated jets to obtain charged jet 4-vector:
\begin{equation}
\label{eq:neutral_subtraction}
p^{{jet, \ }\mu}_{\rm \ charged} = p^{{jet, \ }\mu}_{\rm \ total}  - p^{{jet, \ }\mu}_{\rm \ neutral}.
\end{equation}
Certain aspects of the original jet will be unaltered by the subtraction; the jet area, for example is not recalculated according to Eq.~\ref{eq:neutral_subtraction}. 
A negligible difference was found in the jet performance studies using particle-level jets that were originally charged-only, versus particle-level jets where the neutral components are subtracted.

\begin{figure}[htbp]
    \centering
    \includegraphics[width=0.49\textwidth]{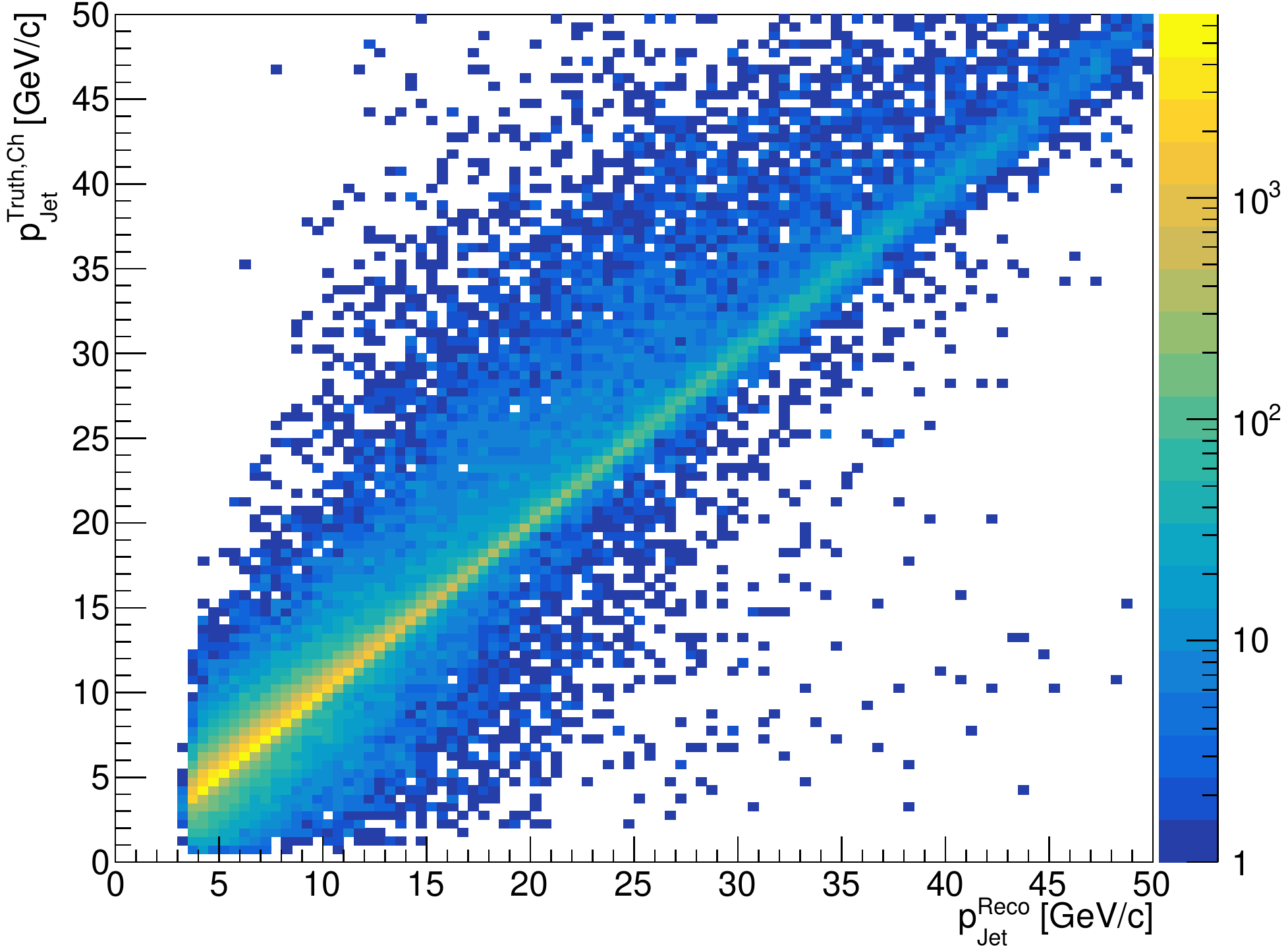}
    \includegraphics[width=0.49\textwidth]{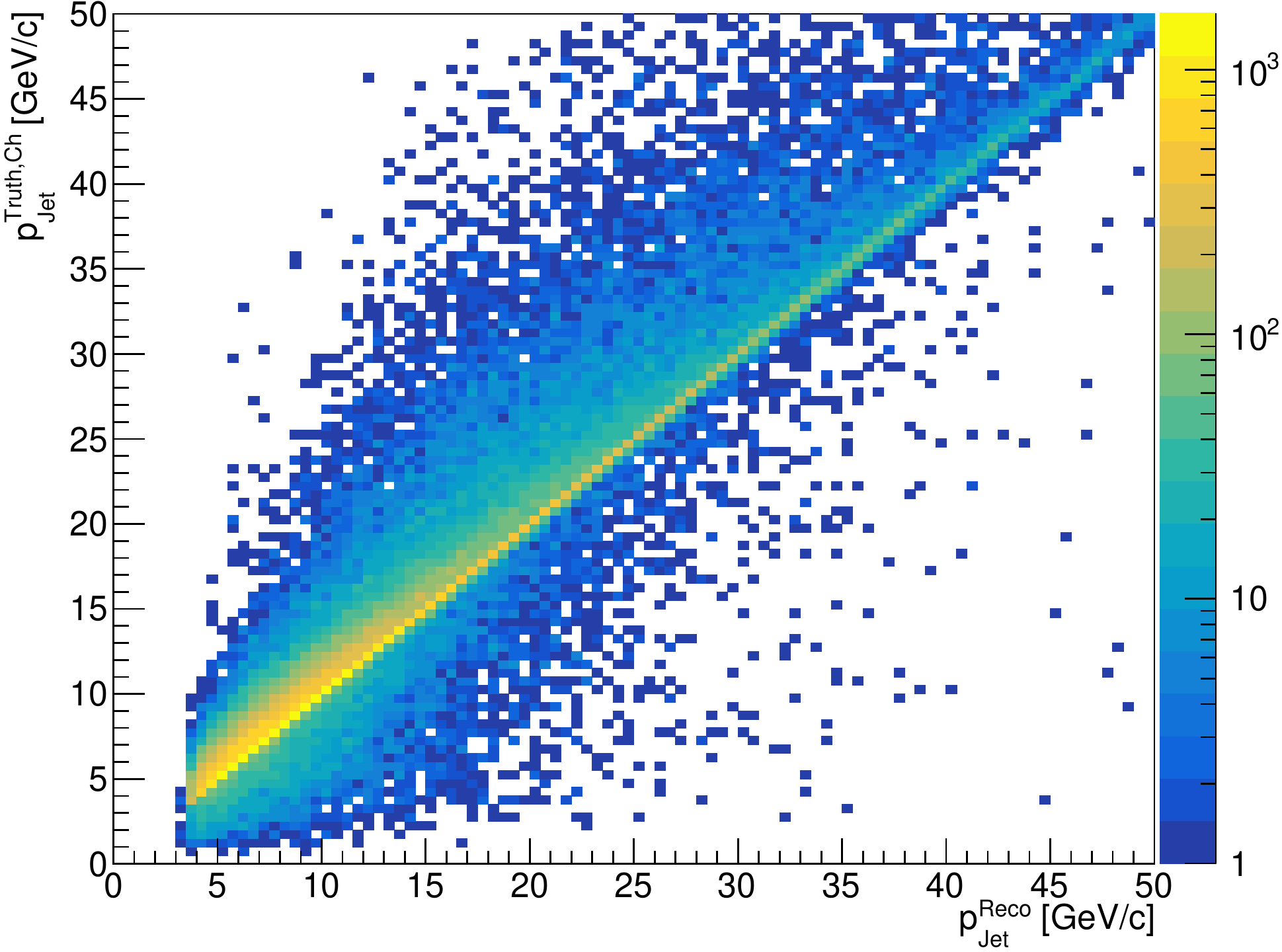}
    \caption{Charged-jet momentum response for jets with $N_\mathrm{constituent} \geq 4$ determined from PYTHIA $e$+$p$ events at 20$\times$100 GeV collisions in a 1.4\,T (left) and 3.0\,T (right) magnetic field.}
    \label{fig:jet_response}
\end{figure}

Figure~\ref{fig:jet_response} shows the momentum response matrix for charged jets passing all criteria, with $p_T \ge$ 4 GeV/$c$. There is a strong correlation between the reconstructed jet momentum, $p_\mathrm{Jet}^\mathrm{Reco}$, and the charged truth jet momentum, $p_\mathrm{Jet}^\mathrm{Truth,Ch}$, indicated by the prominent diagonal line in the histogram. There are, however, jets that fall outside of this strong linear correlation. The most important variable impacting the energy and position resolution of jets after all other selections are made is the number of 
particles not reconstructed as part of the jet (ie ``missing" particles).

\begin{figure}[htbp]
    \centering
    \includegraphics[width=0.42 \textwidth]{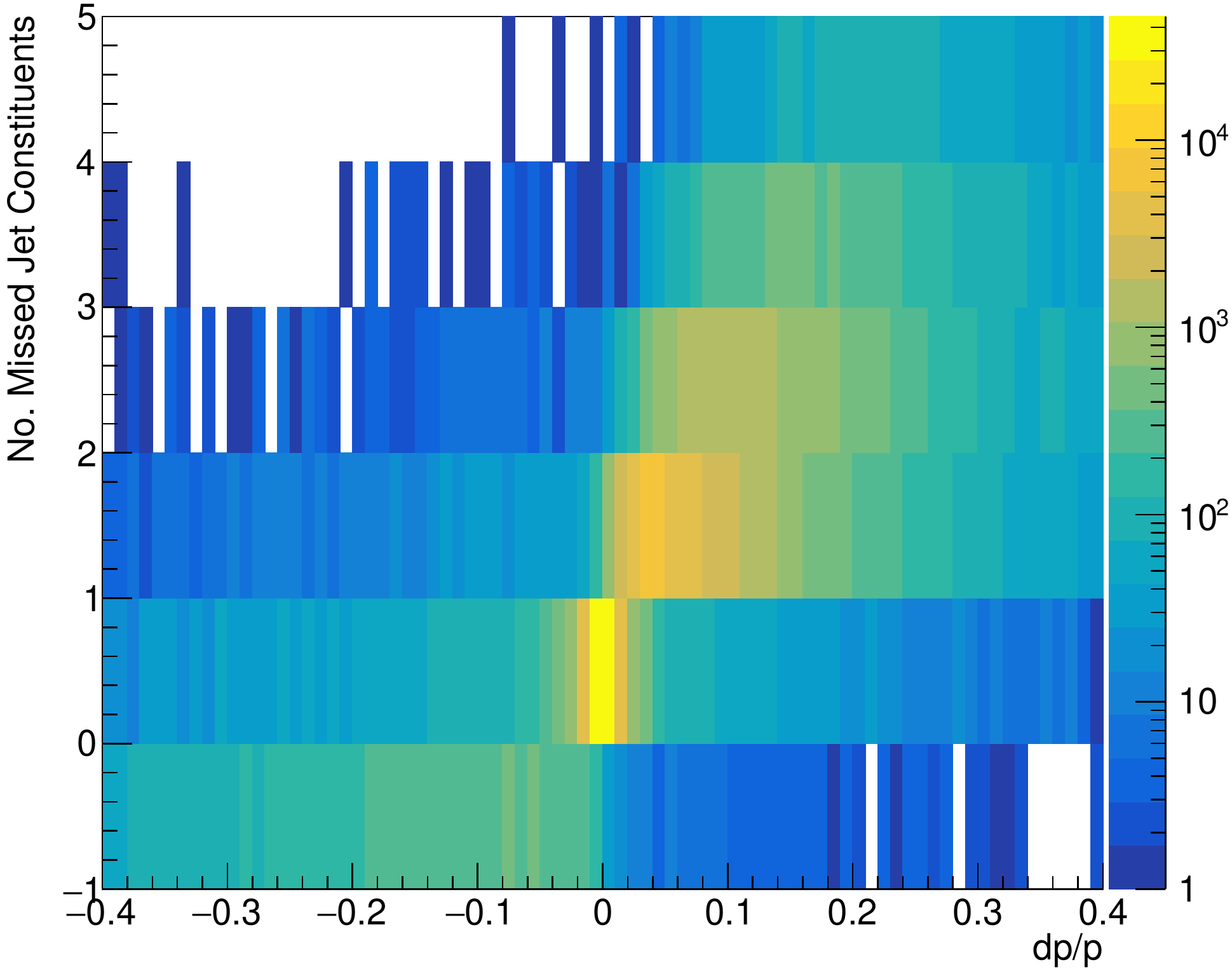}
    \includegraphics[width=0.42 \textwidth]{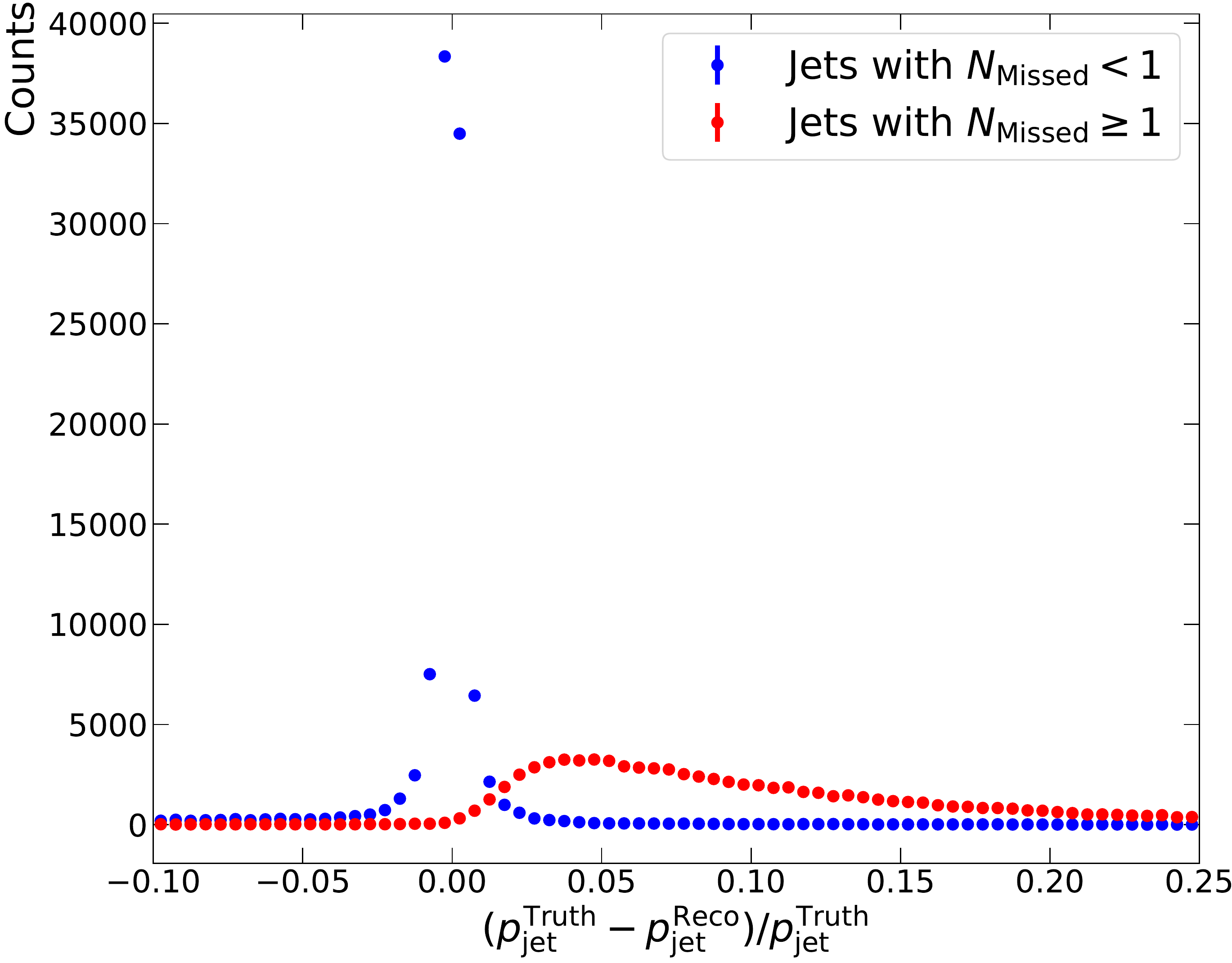}
    \caption{(Left) Number of missed jet constituents ($N_\mathrm{constituent}^\mathrm{truth} - N_\mathrm{constituent}^\mathrm{reco})$ vs. Jet d$p/p$. (Right) d$p/p$ distribution of jets with less than 1 missing constituent (blue), and jets with with one or more missing constituents (red).}
    \label{fig:n_missed_vs_dpp}
\end{figure}

Figure~\ref{fig:n_missed_vs_dpp} shows the energy resolution of jets vs. the difference between number of truth and reconstructed constituents, $N_\mathrm{Missed} = N_\mathrm{constituent}^\mathrm{truth,ch} - N_\mathrm{constituent}^\mathrm{reco}$. 
The left panel shows a distribution centered at 0 for jets with no missing constituents. As the number of missed jet constituents increases, this distribution broadens and shifts towards higher values of d$p/p$. The right panel shows the d$p/p$ distribution of two populations of jets: jets with less than 1 missing constituent and jets with one or more missing constituents. Jets in a magnetic field of 1.4\,T (3.0\,T) with less than 1 missing constituent make up approximately 58\% (49\%) of jets that pass all other selection criteria and have a narrow distribution centered at 0 that is well described by a gaussian fit. Jets with one or more missing constituents, however, have a much broader d$p/p$ distribution that is shifted toward higher values. These are best described by a Landau distribution.  Consequently, we characterize the performance of the silicon tracker for these two populations of jets separately; 
gaussian fits to the the combined distributions are dominated by the narrow peak at d$p/p$ $\approx$ 0, with the detrimental effect of poorly characterizing the non-gaussian ``shoulder" shown in red in Fig.~\ref{fig:n_missed_vs_dpp}.

\begin{figure}[htbp]
    \centering
    \includegraphics[width=0.96\textwidth]{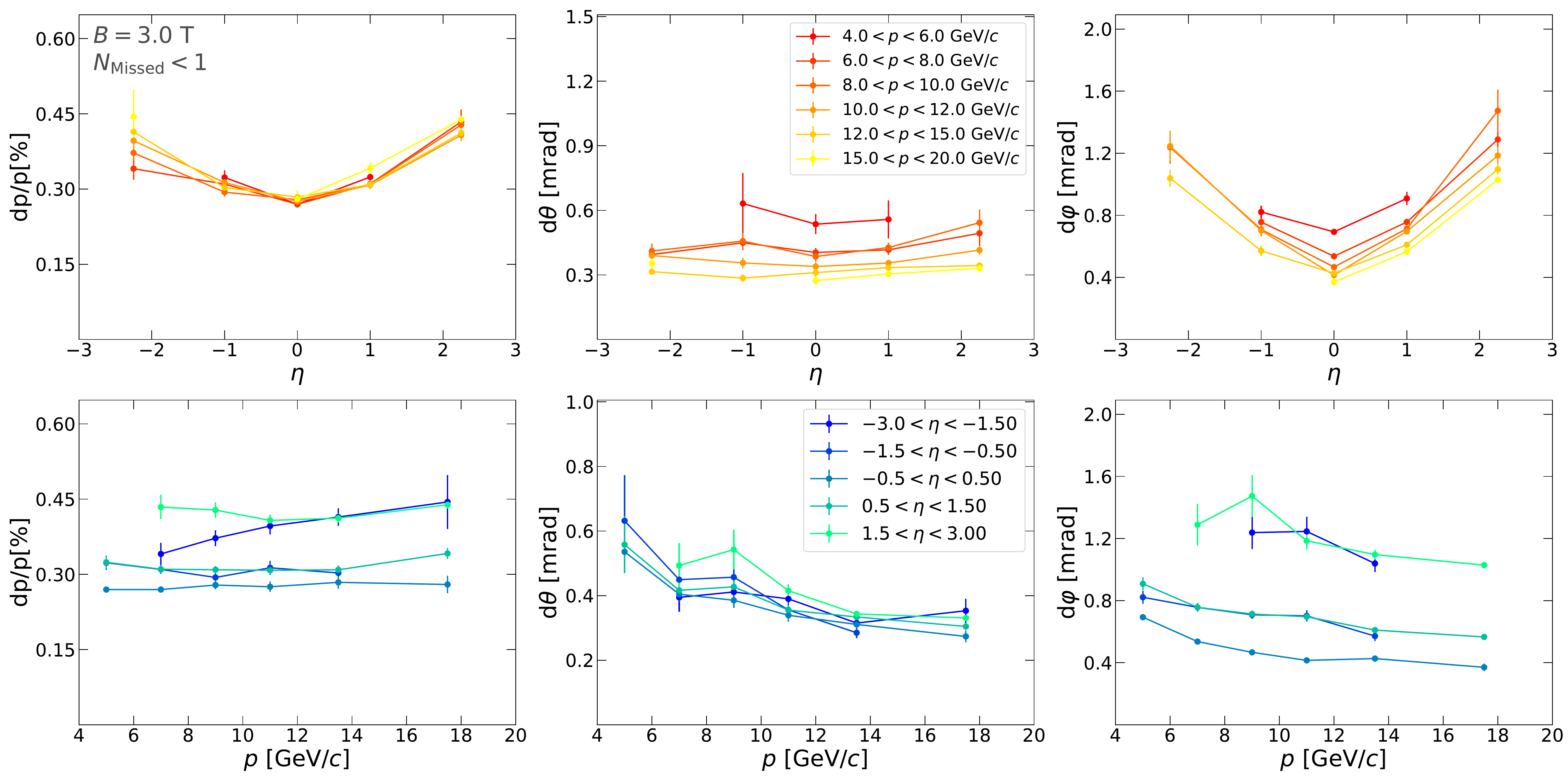}
    \caption{Momentum and angular resolutions of charged jets with no missing constituents, reconstructed with the all-Silicon tracker simulated in PYTHIA $e$+$p$ collisions at 20$\times$100 GeV with the 3.0\,T magnetic-field configuration.}
    \label{fig:no_miss_resolutions}
\end{figure}

\begin{figure}[htbp]
    \centering
    \includegraphics[width=0.96\textwidth]{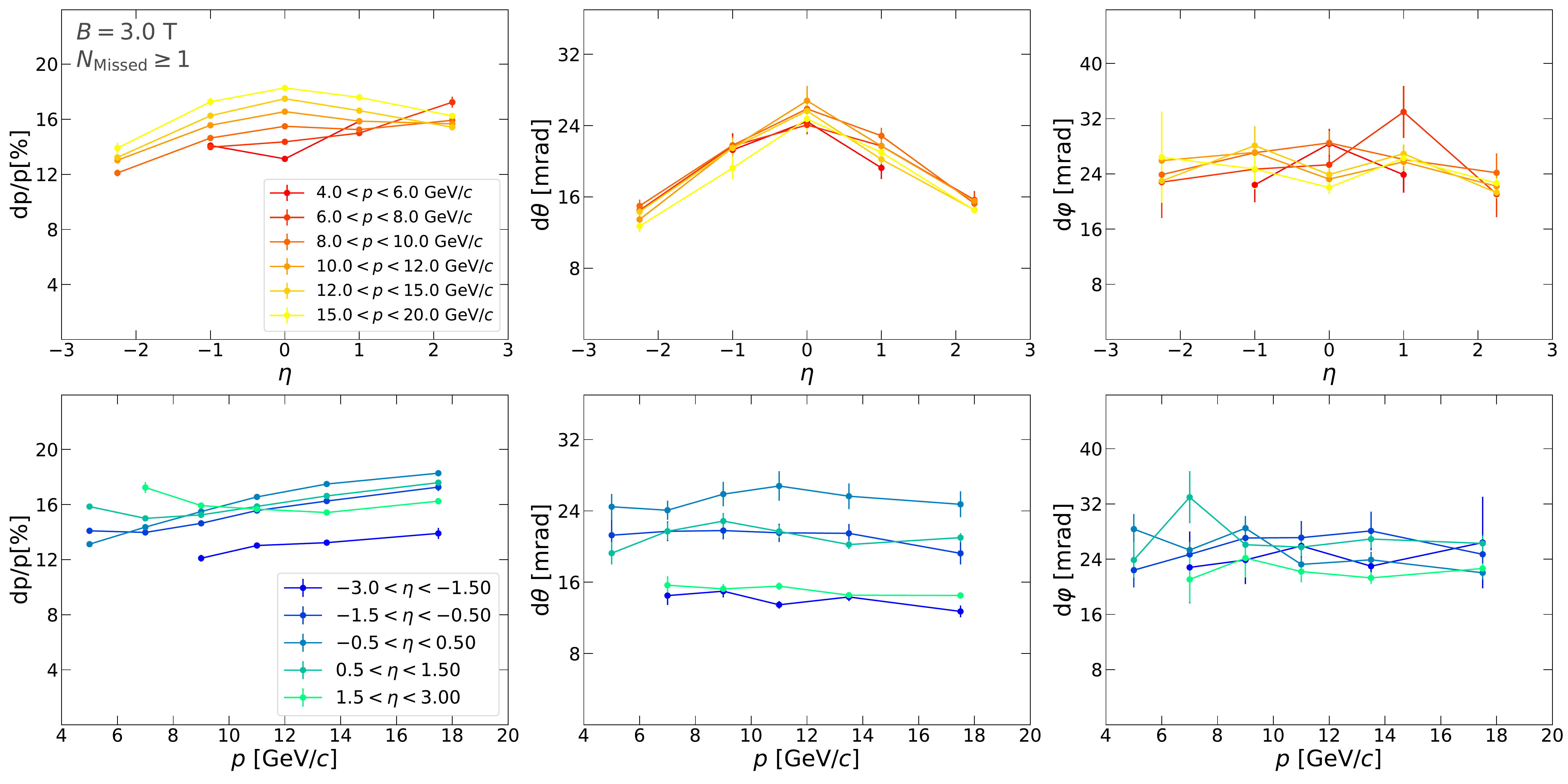}
    \caption{Momentum and angular resolutions of charged jets with one or more missing constituents, reconstructed with the all-Silicon tracker simulated in PYTHIA $e$+$p$ collisions at 20$\times$100 GeV with the 3.0\,T magnetic-field configuration.}
    \label{fig:miss_resolutions}
\end{figure}

\begin{figure}[htbp]
    \centering
    \includegraphics[width=0.96\textwidth]{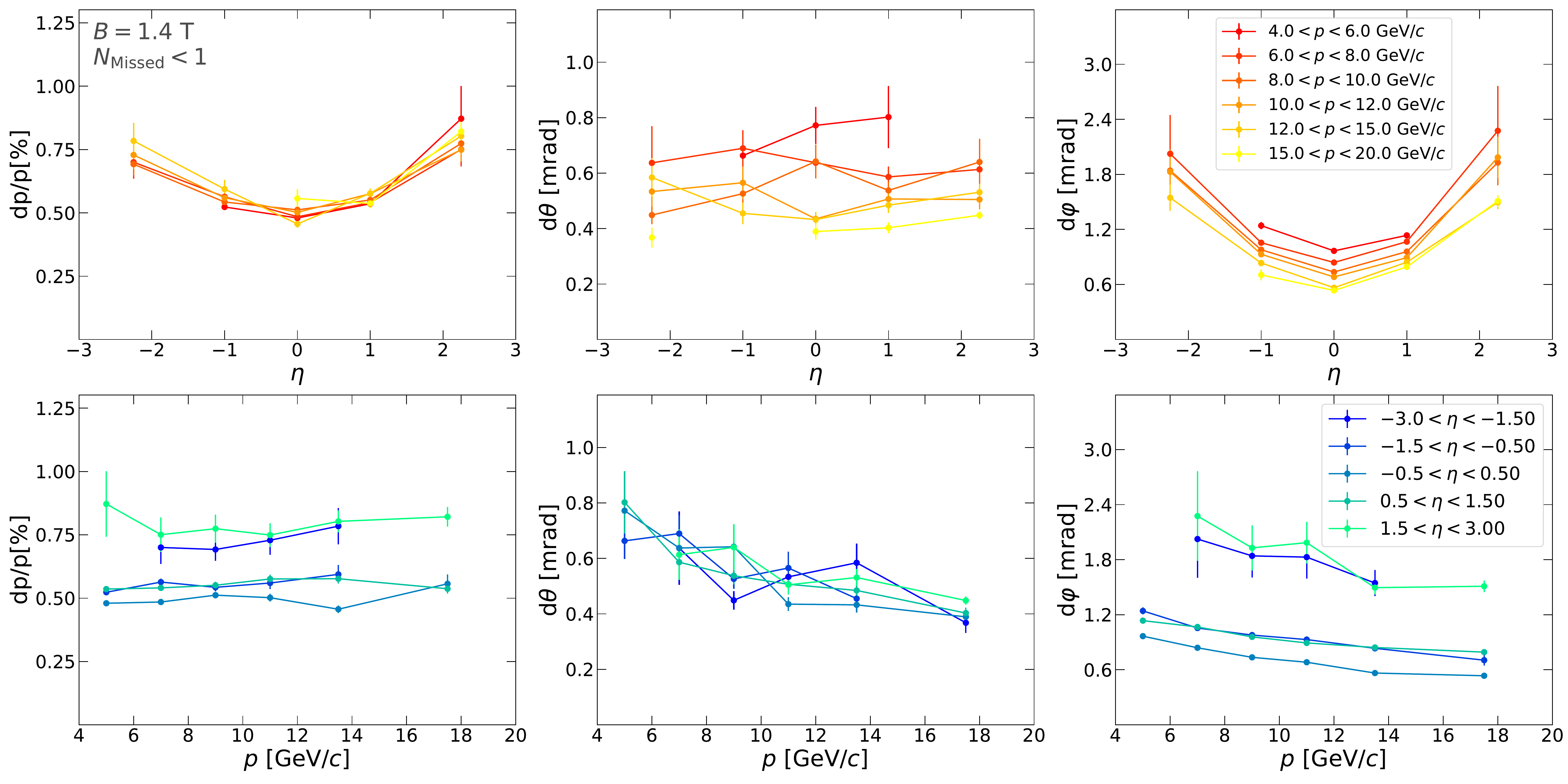}
    \caption{Momentum and angular resolutions of charged jets with no missing constituents, reconstructed with the all-Silicon tracker simulated in PYTHIA $e$+$p$ collisions at 20$\times$100 GeV with the 1.4~T magnetic-field configuration.}
    \label{fig:1.4_no_miss_resolutions}
\end{figure}

\begin{figure}[htbp]
    \centering
    \includegraphics[width=0.96\textwidth]{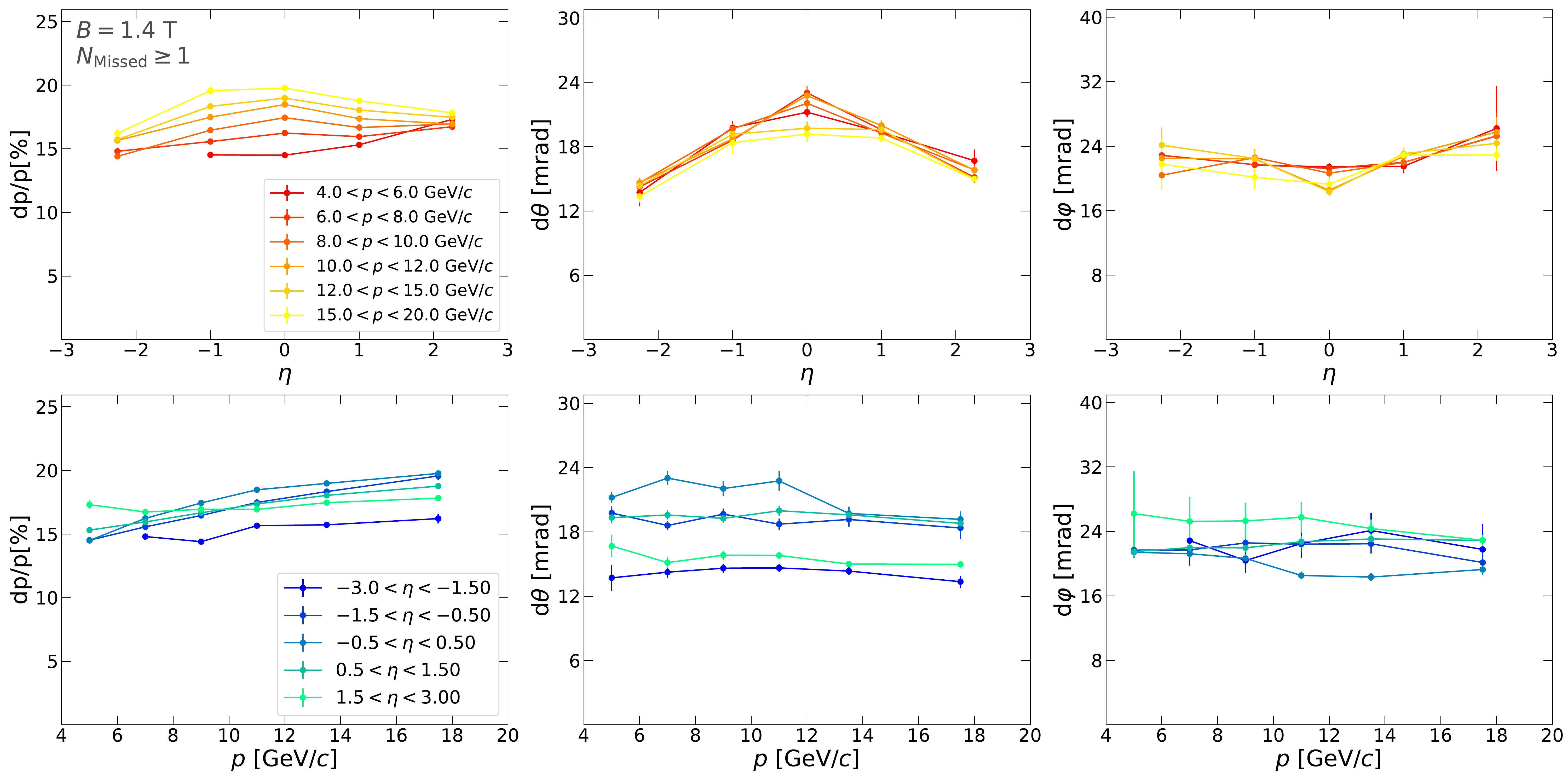}
    \caption{Momentum and angular resolutions of charged jets with no missing constituents, reconstructed with the all-Silicon tracker simulated in PYTHIA $e$+$p$ collisions at 20$\times$100 GeV with the 1.4\,T magnetic-field configuration.}
    \label{fig:1.4_miss_resolutions}
\end{figure}

Figure~\ref{fig:no_miss_resolutions} shows the momentum and angular resolutions of jets with no missing constituents. The resolutions are shown in bins of jet momentum and $\eta$. The resolutions are extracted by fitting the angular or momentum distributions in each bin to a Gaussian function, and extracting the standard deviation and its uncertainty from the fit. Figure~\ref{fig:miss_resolutions} displays the resolutions for jets with one or more missing constituents. The angular resolutions shown in Fig.~\ref{fig:miss_resolutions} are calculated using the same method as in Fig.~\ref{fig:no_miss_resolutions} as well as the method described in section~\ref{sec:performance} for single particles. The momentum resolution for jets with one or more missing particles is more difficult to describe. The d$p/p$ distribution shown in Fig.~\ref{fig:n_missed_vs_dpp} in red is best fit with a Landau distribution, where $\sigma$ is undefined. Therefore, instead of a fit, the simple numerical standard deviation is taken, and reported for the momentum resolution in Fig.~\ref{fig:miss_resolutions}. Figures \ref{fig:1.4_no_miss_resolutions} and \ref{fig:1.4_miss_resolutions} show the resulting resolutions of applying this procedure to jets simulated in a 1.4\,T magnetic field.

\subsubsection{Jet Observables}

\begin{figure}[htbp]
    \centering
\includegraphics[width=0.6\textwidth,height=0.5\textwidth]{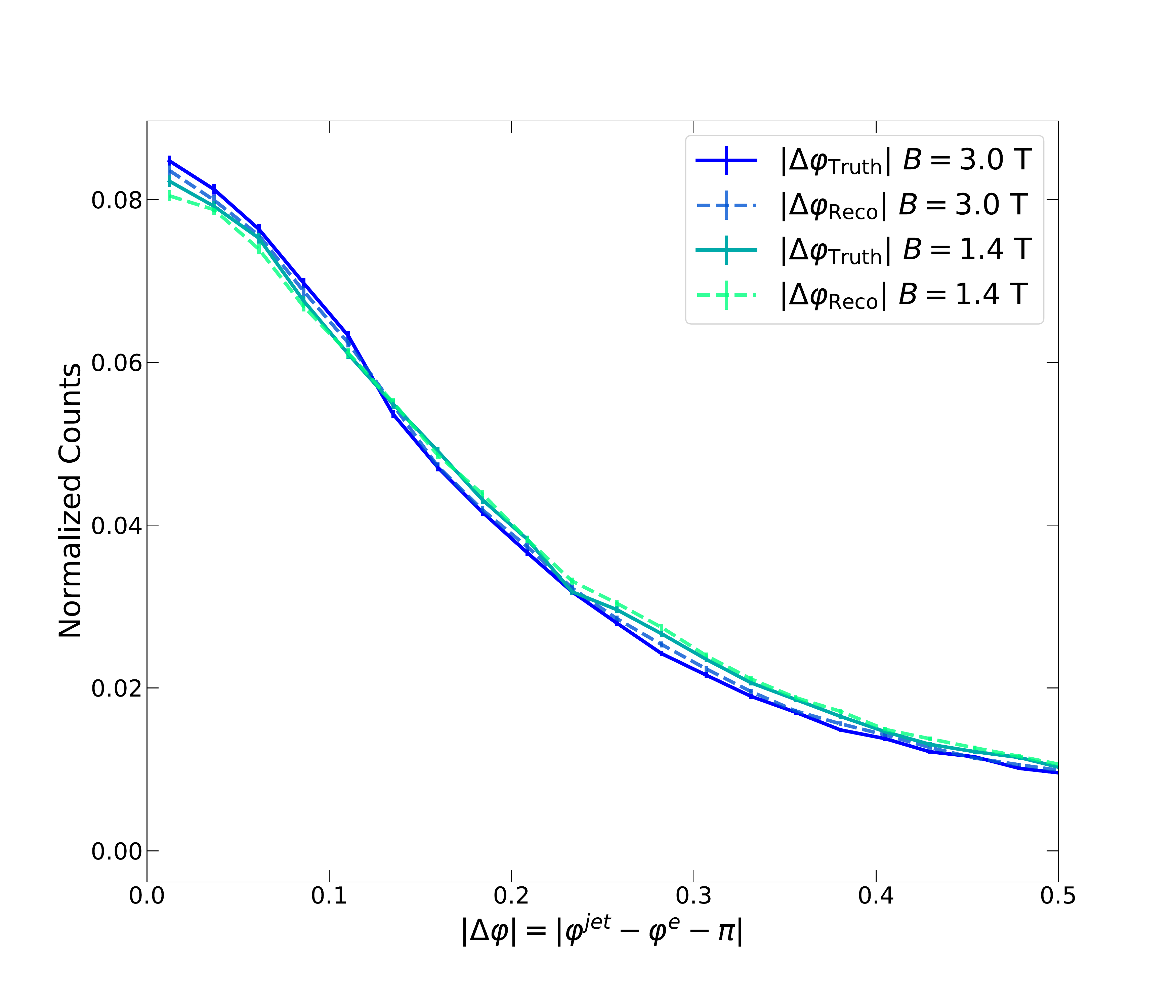}
    \caption{Azimuthal correlation between the scattered electron and jets in $e$+$p$ collisions simulated with a 1.4\,T and 3.0\,T magnetic field. Dashed lines display the correlation between particle-level scattered electron and jets. The solid lines display the correlation between the reconstructed electron and reconstructed jets.}
    \label{fig:az_corr}
\end{figure}

Figure~\ref{fig:az_corr} shows full simulation results for the azimuthal difference between jets and the scattered electron, $|\varphi^\mathrm{jet} - \varphi^{e} - \pi|$ in a 1.4 and 3.0\,T magnetic field. Jets with $N_\mathrm{Missing} < 1$ were used. Dashed lines display the correlation between the scattered electron and particle-level jets matched that are matched to reconstructed jets. The solid lines show the correlation between the scattered electron and jets reconstructed with the all silicon tracker. The figure shows a peak at zero as expected from LO DIS where the electron and jet are 
emitted back-to-back. In the limit that the transverse momentum imbalance, $p_\mathrm{T}^e / p_\mathrm{T}^{jet}$, is much smaller than the electron transverse momentum, this observable can provide clean access to the quark TMD PDF and the Sivers effect in transversely polarized scatterings in $e$+$p$ collisions~\cite{PhysRevLett.122.192003}.

\begin{figure}[htbp]
    \centering
    \includegraphics[width=0.6\textwidth,height=0.5\textwidth]{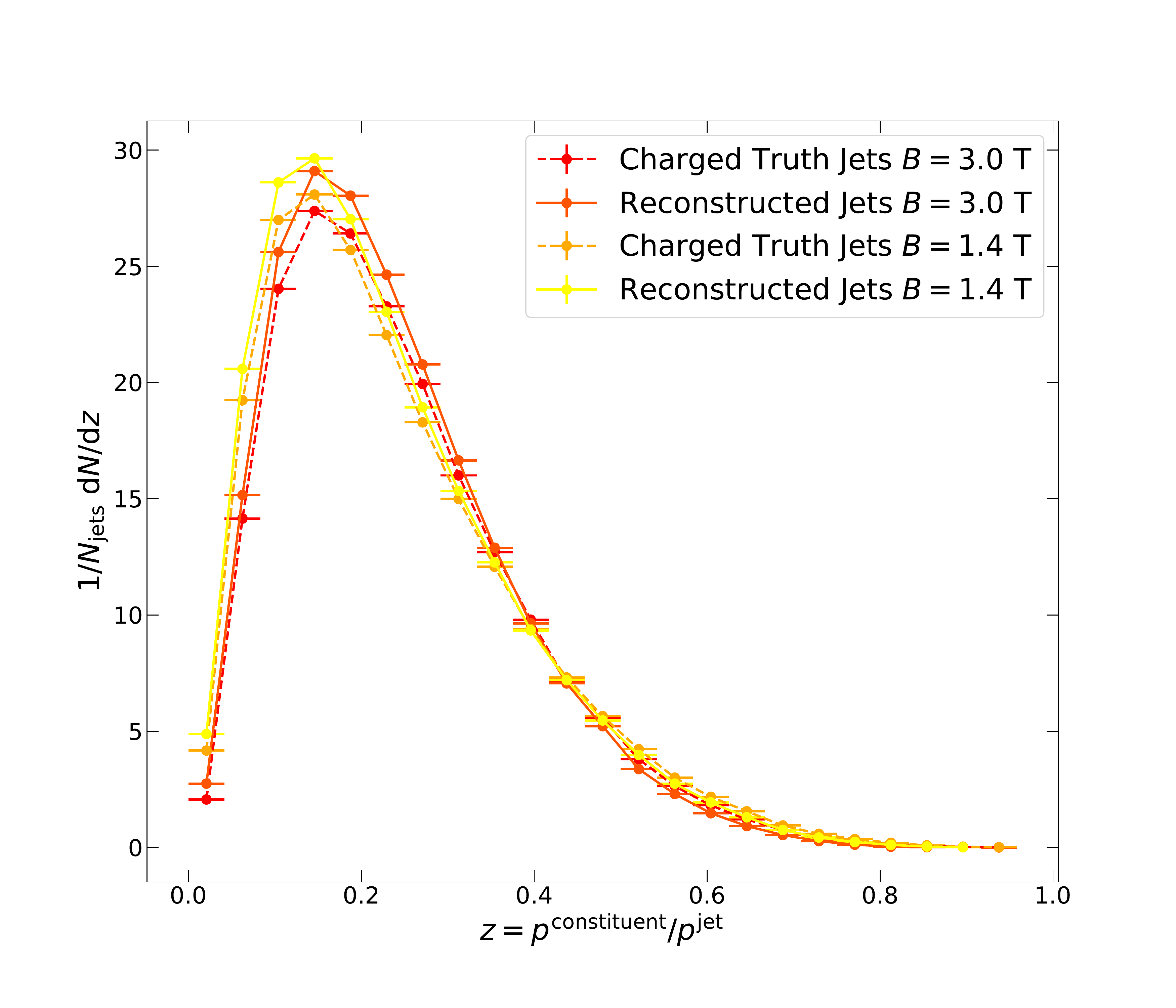}
    \caption{Truth-level (dashed lines) and reconstructed-level (solid lines) charged jet fragmentation as a function of $z = p^\mathrm{constituent}/p^\mathrm{jet}$ in 1.4\,T and 3.0\,T magnetic fields.}
    \label{fig:jet_frag}
\end{figure}

Figure \ref{fig:jet_frag} shows the particle-level and reconstructed-level charged jet fragmentation functions in the two magnetic fields. Measurements of the charged-jet fragmentation function in $e$+$p$ 
should provide sensitivity to the process of hadron formation. 
Comparisons of charged jet fragmentation functions measured in $e$+$p$ and $e$+A collisions can elucidate the effects of nuclear matter on the fragmentation process, and yield information on parton transport in a nuclear medium. For lower energy jets, which begin to fragment inside a nucleus in $e$+$A$ collisions, we can use the nucleus as a filter to probe hadronization. 

The choice of magnetic field is shown to have little effect on both observables shown. While the different  magnetic fields result in different angular and momentum resolutions, that will in turn affect the the azimuthal correlation and fragmentation function measurements, the magnitude of these changes is quite small. For example, the momentum resolution for reconstructed jets with less than 1 missing constituent is approximately between 0.3\% and 0.45\% in a 3.0\,T magnetic field. In a 1.4\,T magnetic field, the momentum resolution ranges between 0.5\% and 0.75\%. This change in the momentum resolution of a few fractions of a percent is not expected to significantly impact the fragmentation function measurements, with similar reasoning applying to the differences in angular resolution and their effect on the electron-jet correlation measurements.